**РОССИЙСКАЯ АКАДЕМИЯ НАУК**

**Федеральное государственное бюджетное учреждение науки**
**Институт биологии внутренних вод им. И.Д. Папанина**
**Российской академии наук**

**Д.П. Карабанов**

# ГЕНЕТИЧЕСКИЕ АДАПТАЦИИ ЧЕРНОМОРСКО-КАСПИЙСКОЙ ТЮЛЬКИ *CLUPEONELLA CULTRIVENTRIS* (NORDMANN, 1840) (ACTINOPTERYGII: CLUPEIDAE)

***Рецензенты:***

Впервые методами биохимической генетики изучено генетическое разнообразие и популяционная структура черноморско-каспийской тюльки по всему ареалу. Получены современные данные о состоянии вида в водохранилищах Верхней Волги. Изучены физиологические и экологические адаптации вида к условиям северных водохранилищ. Установлена селективная значимость ряда генетических локусов по отношению к важнейшим абиотическим факторам среды. Установлено, что на всем протяжении своего современного ареала черноморско-каспийская тюлька представляет собой генетически единую совокупность популяций *Clupeonella cultriventris* (Nordmann, 1840). Успешной экспансии вида по каскаду Волжских водохранилищ способствовал сложный комплекс генетико-биохимических адаптаций к обитанию в пресных водах. Предположено, что популяции тюльки волжских водохранилищ формировались на основе пресноводных популяций Саратовских затонов. Существенную роль в распределении генотипов тюльки в новообразованной популяции играют сезонные изменения абиотических и биотических факторов среды.

Для ихтиологов, генетиков, экологов, специалистов по охране и использованию природных ресурсов.

## Содержание

*Светлой памяти*
*Валентины Ивановны*
*КИЯШКО*

## Введение

За последние полвека крайне актуальной стала проблема проникновения и натурализации живых организмов за пределы их исторических ареалов. Постоянно усиливающееся антропогенное преобразование естественной среды в совокупности с глобальными геоклиматическими изменениями, резко активизировавшимися с последних десятилетий XX века, вызвали ускорение процессов трансформации ареалов многих видов растений и животных (Дгебуадзе, 2002; Gollash et al., 2003; Swaminathan, 2003; Павлов и др., 2005). Наиболее пристальное внимание, наряду с сокращением ареалов ряда видов, вызывают процессы взрывного расширения ареалов, приводящие к быстрым изменениям не только зоогеографического статуса регионов, но и всей организации взаимосвязей и отношений внутри экосистем. Пресноводные экосистемы континентальных водоемов Европы оказались в числе наиболее затронутых человеческой деятельностью (масштабное зарегулирование большинства крупных и средних речных систем). Также в водоёмах Европы ярко выражены и эффекты глобального потепления (повышение среднегодовых температур поверхностных вод суши, устойчивый рост минерализации и трофности) (Slynko et al., 2002; Leppakoski et al., 2002; Schindler, Parker, 2002; Huber, Knutti, 2012).

Вселение чужеродных видов потенциально опасно как из-за непосредственного экономического ущерба, наносимого вселенцем, так и через влияние адвентивных видов на биоразнообразие региона в целом. Вселенцы способны не только конкурировать и вытеснять коренных обитателей, но и успешно гибридизировать с ними, что ещё более затрудняет предсказание экономических и экологических последствий биологических инвазий (Heger, Trepl, 2003; Graziani et al., 2003; Largiader, 2007).

На современном этапе роль чужеродных видов в водоёмах мира всё более увеличивается. Так, в Великих озёрах количество адвентивных видов составляет более 4% (Fuller et al., 1999), в ихтиофауне различных водоёмов Европы - до 20% (Kottelat, Freyhof, 2007; Rivers…, 2009). Одним из самых распространённых и наиболее многочисленным представителем рыб-вселенцев в Европейской части

России является черноморско-каспийская тюлька *Clupeonella cultriventris* (Nordmann, 1840). С началом массового гидростроительства тюлька начала экспансию по водохранилищам Днепра (Амброз, 1956). С 1960-х годов данный вид начинает энергично расселяться в пелагиали водохранилищ Нижней и Средней Волги (Шаронов, 1971), а к началу XXI в. эта рыба освоила водохранилища Верхней Волги и Шексны вплоть до Белого озера (Slynko et al., 2002). Успешной экспансии тюльки при зарегулировании крупных рек способствовало значительное расширение пелагиали, увеличение трофности водоёмов и обильное развитие зоопланктона (Циплаков, 1974). Наблюдавшееся на современном этапе повышение среднегодовых температур воды и удлинение вегетационного периода также благоприятствовало натурализации тюльки в северных водохранилищах (Осипов, 2006).

Несмотря на ряд специальных исследований (Берг, 1948; Владимиров, 1950; Световидов, 1952) до сих пор обсуждаются видовой статус и таксономические отношения внутри рода *Clupeonella* (Kottelat, 1997; Богуцкая, Насека, 2004; Kottelat, Freyhof, 2007).

Относительно хорошо изучена биология тюльки естественноисторической части ареала - Азовского и Каспийского морей (Световидов, 1952; Луц, 1978; Приходько, 1979). В научной литературе имеется ряд работ, посвящённых биологическим особенностям тюльки в новой среде обитания. Такие исследования выполнены для популяций тюльки Днестра (Владимиров, 1951), Днепра (Владимиров и др., 1964; Сухойван, Вятчанина, 1989), рек Дон и Маныч (Троицкий, Цуникова, 1988). В Волжском каскаде изучена тюлька Куйбышевского (Кузнецов, 1973; Козловский, 1984, 1987; Kozlovsky, 1991), Камского и Воткинского водохранилищ (Пушкин, Антонова, 1977, Антонова, Пушкин, 1985; Горин, 1991). Подробно рассмотрены воспроизводство, рост, питание и роль тюльки в рыбном сообществе Верхневолжских водохранилищ (Кияшко, Слынько, 2003; Кияшко, 2004; Кияшко и др., 2006; Осипов, 2006). Имеются данные по особенностям липидного обмена (Халько, 2007), аллозимной изменчивости (Слынько, Лапушкина, 2003; Karabanov, Slynko, 2005) и RAPD-анализу (Столбунова, Слынько, 2005) тюльки в новых частях ареала.

Тем не менее, специальных исследований, посвящённых генетико-биохимическим адаптациям и популяционной структуре тюльки в пределах всего её современного ареала, не проводилось. Вместе с тем, лишь изучение особенностей внутриклеточного метаболизма и генетической детерминации этих процессов позволяет определить стратегию генетико-биохимических адаптаций при экспансии данного вида. Применение сравнительного анализа популяционной структуры по всему ареалу рыбы позволяет установить микрофилогенез отдельных популяций и определить иерархическую структуру крупных групп популяций этого вида.

Автор считает своим долгом выразить глубокую благодарность всем коллегам, с которыми на протяжении многих лет проводил совместные исследования. В первую очередь хочу поблагодарить к.б.н. Юрия Владиславовича Слынько (ИБВВ РАН) за столь интересную тематику, руководство, критические замечания и обсуждение результатов исследования. Определение места тюльки и её роли в пресноводных экосистемах было бы невозможно без помощи к.б.н. Валентины Ивановны Кияшко (ИБВВ РАН). За ценные замечания и консультации по всей работе хочу выразить огромную благодарность к.б.н. Александру Анатольевичу Махрову (ИПЭЭ РАН). Огромная благодарность всем участникам комплексных экспедиций ИБВВ РАН, особенно: д.б.н. Ю.В. Герасимову, к.б.н. А.Н. Касьянову, м.н.с. М.И. Малину, м.н.с. Д.Д. Павлову, д.б.н. В.И. Лазаревой, к.б.н. В.И. Мартемьянову, д.б.н. В.Г. Терещенко, к.б.н. А.В. Тютину и к.б.н. В.В. Халько. Сбор материала был бы невозможен без помощи: на Горьковском и Чебоксарском водохранилищах - к.б.н. А.А. Клевакина (Нижегородское отделение ГосНИОРХ); в Ростовской обл. - к.б.н. А.З. Витковского (АзНИИРХ); на Каспийском море - д.б.н. А.К. Устарбекова (ДагНЦ РАН); на Саратовском и Волгоградском водохранилищах - к.б.н. Ю.А. Малининой (Саратовское отделение ГосНИОРХ); в Пермском крае - Т.Л. Кулевской (Администрация г. Перми); на территории Украины - к.б.н. Р.А. Новицкого (ДНУ). Отдельную благодарность хочу выразить академику, д.б.н. Юрию Юлиановичу Дгебуадзе (ИПЭЭ РАН) неоднократно помогавшего мне в сложные периоды жизни.

Выполнение настоящей работы было бы невозможно без поддержки Совета по грантам Президента Российской Федерации для государственной поддержки молодых российских учёных - кандидатов наук, проект МК-1793.2011.4 «Эколого-генетические механизмы адаптаций чужеродных видов рыб в пресноводных экосистемах Европейской части России».

# Глава 1.
# Таксономический статус, современный ареал и биологическая характеристика черноморско-каспийской тюльки. Механизмы адаптаций к новым условиям обитания

## 1.1. Таксономический статус и исторический ареал рода *Clupeonella* Kessler, 1877

Большинство сельдёвых Clupeidae (150 из 190 видов) обитают в морях тропической и субтропической зоны. В умеренных широтах распространена сравнительно небольшая часть представителей семейства, в которую входят роды, распространенные в бореальной и большей части субарктической областях (около 20 видов) (Световидов, 1952).

Черноморско-каспийская тюлька относится к роду *Clupeonella* Kessler, 1877, в котором на настоящий момент выделяются 4 вида: черноморско-каспийская тюлька - *Clupeonella cultriventris* (= *C. delicatula*) (Nordmann, 1840), большеглазая тюлька *Clupeonella grimmi* Kessler, 1877, анчоусовидная тюлька *Clupeonella engrauliformis* (Borodin, 1904) и абрауская тюлька *Clupeonella abrau* (Maljatskij, 1930) (Световидов, 1952; Атлас…, 2003). Черноморско-каспийская, анчоусовидная и большеглазая тюльки живут в Каспийском море, Азовское и Черное моря населяет только черноморско-каспийская тюлька. Реликтовая абрауская тюлька живет лишь в оз. Абрау возле г. Новороссийск, к этому же виду, как правило, относят и тюльку из оз. Абулионд (Турция) (Атлас…, 2003).

Ранее черноморско-азовская и каспийская тюльки (кильки) обычно объединялись в литературе в один вид под названием *Clupeonella delicatula* (Nordmann, 1840) (Берг, 1948; Световидов, 1952). Старшим пригодным синонимом для черноморской тюльки является название *Clupea cultriventris* Nordmann, 1840 (Svetovidov, 1973). В настоящее время валидным считается название *Clupeonella cultriventris* (Nordmann, 1840) (Решетников и др., 1997; Аннотированный…, 1998; Атлас…, 2003; Богуцкая, Насека, 2004).

По Л.С. Бергу (1948) в северо-восточной части Черного моря, Азовском и Каспийском морях и в низовьях рек бассейна этой части Понто-Каспия обитает вид с номенклатурным названием обыкновенная тюлька, *Clupeonella delicatula* (Nordmann, 1840). На основании географической приуроченности и отдельных морфологических различий Л.С. Берг выделял также два подвида: чархальская тюлька (оз. Чархал и затоны Волги у Саратова) и черноморская тюлька (северо-западная часть Черного моря, дельта Дуная, Днестровский, Бугский, Кучурганский и Днепровский лиманы). А.Н. Световидов (1943; 1945), также преимущественно на основании географической приуроченности и некоторых морфологических различий подразделил вид на два подвида - черноморско-азовская (*C. d. delicatula*) и каспийская тюлька (*C. d. caspia*). В.И. Владимиров (1950) в свою очередь разделил черноморско-азовскую тюльку на два подвида - черноморскую (*C. d. delicatula*) и азовскую (*C. d. azovi*). В дальнейшем от разделения вида на подвиды было решено отказаться (Аннотированный…, 1998). Вместе с тем, азово-черноморские, волго-каспийские и чархальские популяции тюльки по-прежнему рассматриваются в статусе отдельных подвидов (Богуцкая, Насека, 2004). М. Коттела (Kottelat, 1997; Kottelat, Freyhof, 2007) на незначительных морфологических различиях обосновывает мнение, что существует четыре самостоятельных вида пресноводных тюлек: каспийская (*C. caspia*), черноморская (*C. cultriventris*), абраусская (*C. abrau*) и пресноводная (*C. tcharchalensis*).

К началу XXI в. ареал черноморско-каспийской тюльки охватывает Азовское море, опресненные участки Черного моря, низовья рек Дон, Кубань, водохранилища Днепра, реки Буг, Днестр, Дунай. Каспийская популяция тюльки встречается по всему Каспийскому морю от прибрежных районов Ирана, Туркмении, Азербайджана, Казахстана до низовий Волги. С начала 1950-х годов тюлька расселилась по акватории рек Волга и Кама (Световидов, 1952; Чижов, Абаев, 1968; Пушкин, Антонова, 1977; Слынько и др., 2000).

## 1.2. Формирование современного ареала черноморско-каспийской тюльки

Для всех популяций черноморско-каспийской тюльки исторически был характерен заход на нерест в низовья крупных рек понто-каспийского бассейна (Световидов, 1952). Зарегулирование и создание каскадов водохранилищ на реках Волга, Кама, Днепр, Дон стало решающим фактором расселения тюльки по этим рекам и возникновения самовоспроизводящихся пресноводных популяций. Важную роль в успешности освоения водохранилищ сыграли такие биологические свойства этого короткоциклового вида, как: высокая эффективность размножения, экологическая пластичность, обширная зона пелагиали новообразованных водоёмов с богатой кормовой базой и слабая конкуренция местных рыб-планктофагов (например, уклейки

*Alburnus alburnus*, синца *Ballerus ballerus*), а также относительно небольшой пресс хищников. Успеху инвазии также способствовали абиотические факторы: глобальное потепление климата, замедление течения, формирование обширных открытых участков, пригодных для обитания рыб пелагического комплекса (Кияшко, Слынько, 2003).

На черноморском побережье имеются крупные популяции тюльки в низовьях и лиманах р. Дунай (Смирнов, 1967). В Днестре тюлька образует три крупные популяции: Днестровского лимана, низовьев р. Днестр и Кучурганского лимана (Владимиров, 1951; Гидробиологический…, 1992).

После зарегулирования стока в результате сооружения каскада гидроэлектростанций на р. Днепр тюлька заселила все днепровские водохранилища. Впервые тюлька отмечена в 1956 г. в Каховском, а в 1958 г. - в Днепровском водохранилище. Оттуда этот вид распространился по среднему течению Днепра по Днепродержинскому, Кременчугскому и Каневскому водохранилищам (Владимиров и др., 1964; Булахов, 1966; Смирнов, 1967; Шевченко, 1991).

Современное распространение тюльки в бассейне Дона охватывает низовья реки, Цимлянское водохранилище и водохранилища Манычского каскада (Доманевский и др., 1964; Троицкий, Цуникова, 1988; Витковский, 2000).

В бассейне Волги до создания каскада водохранилищ тюлька отмечалась лишь в низовьях Волги и Ахтубы. Немногочисленные косяки тюльки поднимались вверх по рекам Волга и Урал. После падения уровня Каспия в 1940-х годах тюлька перестала подниматься выше г. Астрахань (Казанчеев, 1963). В оз. Чархал (бассейн р. Урал), в ильменях дельты Волги и в Волге в затонах у г. Саратов также имелась жилая пресноводная форма «чархальской селёдочки» - *C. delicatula caspia* morpha *tscharchalensis* (Берг, 1948; Казанчеев, 1963). После создания водохранилищ на Волге началось постепенное продвижение тюльки вверх по реке. Предполагается, что в формировании стада тюльки Волгоградского водохранилища приняли участие аборигенные популяции тюльки из саратовских затонов (Егерева, 1970; Шаронов, 1972; Циплаков, 1972). В Куйбышевском водохранилище данный вид впервые был обнаружен в 1963 г. К 1966 г. тюлька распространилась уже по всему Куйбышевскому водохранилищу и вверх по Волге до г. Чебоксары. Уже в 1967 г. в уловах пелагическим мальковым тралом тюлька в Куйбышевском водохранилище составляла более 70% от общего числа выловленных рыб (Шаронов, 1971; Кузнецов, 1973). Особенно крупные скопления формировались в заливах при устьях притоков (Козловский, 1984). По Волге в 1968 г. она поднялась до устья р. Суры (Шаронов, 1971). К 1974 г. на Каме тюлька заселила все подходящие участки Камского и Воткинского водохранилищ (Пушкин, 1975; Пушкин, Антонова, 1977). С середины 1980-х годов тюлька быстро оккупировала Горьковское водохранилище и начала экспансию по Верхней Волге. В 1994 г. тюлька отмечается в Рыбинском, 1999 - Иваньковском и Угличском, а с 2000 г. - в Шекснинском водохранилищах (Слынько и др., 2000; Экологические…, 2001).

Высокая экологическая пластичность и конкурентоспособность вида, обилие корма, слабый пресс хищников - всё это позволило тюльке в сравнительно короткий срок освоить обширную территорию и стать массовой рыбой во всех водохранилищах. Опасения (Шаронов, 1971; Циплаков, 1972), что она потеснит многих аборигенных рыб, в целом не оправдались. С появлением тюльки улучшились условия питания хищников, более полно используется кормовая база по зоопланктону (Козловский, 1987; Шевченко, 1991). При этом не произошло заметного угнетения аборигенных видов, хотя бы немного сходных по экологии (синца, уклейки, чехони *Pelecus cultratus*). Динамика популяций этих видов рыб лимитируется в первую очередь их собственными адаптационными способностями в достаточно специфичных условиях водохранилищ (Антонов, Козловский, 2003).

В.В. Осипов (2006) отмечает, что на современном этапе тюлька успешно встроилась в экосистемы Верхневолжских водохранилищ, изменила структуру сообществ пелагиали и оказывает существенное влияние на численность и биомассу популяций аборигенных пелагических видов рыб.

Большая работа по изучению питания, роста и биологической роли тюльки в водохранилищах Верхней Волги проведена В.И.Кияшко (Кияшко, 2004; Кияшко и др., 2006, 2007; Dgebuadze et al., 2008). В этих исследованиях показано, что вселившись в Рыбинское водохранилище, тюлька успешно натурализовалась, найдя здесь благоприятные условия для размножения и нагула. В пелагиали водоёма тюлька является доминирующим видом, где основным пищевым конкурентом ей служит синец и, в меньшей мере, молодь карповых и окунёвых рыб. Летом наиболее плотные скопления тюлька образует в речных плёсах водохранилища (Волжском, Моложском), значительно меньше её в Центральном плёсе. Осенью распределение этой рыбы более равномерно по всем плёсам.

В целом, полученные данные позволяют предположить, что процессы расселения южных видов на север и освоения ими северных водоемов сопровождаются процессом адаптации на популяционном уровне.

## 1.3. Особенности биологии черноморско-каспийской тюльки

Черноморско-каспийская тюлька - наиболее многочисленный стайный пелагический вид сем. Clupeidae в Каспийском море. Будучи обитателем прибрежной зоны, тюлька хорошо приспособлена к изменениям солености и температуры воды. Она встречается в водах различной мутности, обеспеченности кислородом и солёности (обнаруживается даже в заливе Мертвый Култук при солености 36‰) (Приходько, 1979). В Азовском море тюлька образует локальные популяции по всей акватории, а также обитает в северо-западной части Чёрного моря (Георгиев, Александрова-Колеманова, 1983).

В Каспийском море популяция тюльки состоит не менее чем из двух стад. Одно из них, проведя зиму в Южном Каспии в начале весны начинает двигаться на север преимущественно вдоль западного побережья Среднего Каспия, направляясь для нереста в Северный Каспий. Второе стадо, по-видимому, всю жизнь проводит в Южном Каспии совершая сезонные миграции в его пределах, а в конце зимы и весною эта южная тюлька подходит на нерест к западным и восточным берегам Южного Каспия (Казанчеев, 1963; Панин и др., 2005). В Азовском море тюлька зимует в открытой части акватории, а весной и в начале лета косяки половозрелой тюльки совершают миграции в опреснённые участки моря (Майский и др., 1950).

Размножается тюлька в Каспийском море весной в прибрежных мелководных участках в основном над глубинами менее 10 м. Нерестовое стадо состоит из 1-6-летних особей (в основном рыбы возрастом 2-3 года). Половое созревание наступает в возрасте 1-2 года при достижении длины 5-7 см. Плодовитость в Каспии 9,5-60 (в среднем 31,2) тыс. икринок (Приходько, 1979). В Азовском море абсолютная плодовитость тюльки колеблется от 3,9 до 28,2 (в среднем 10-11) тыс. икринок (Майский и др., 1950; Михман, 1970). Икрометание порционное, имеется несколько выметов с промежутком в несколько дней. Соотношение полов близко к 1:1, однако вследствие более раннего созревания самцов и меньшей продолжительности их жизни, они преобладают среди мелких особей, а с увеличением длины и возраста рыб наблюдается преобладание самок (Луц, 1981). Нерест тюльки в Каспии совершается с апреля по июнь при температуре воды +6…14°С. Основу нерестового стада составляют 2-3-4-летние особи (Устарбеков, Аджимуратов, 1982). В Азовском море нерест происходит главным образом в Таганрогском заливе, частично в дельте Дона и в многочисленных лиманах. Основной нерест начинается обычно в апреле–мае при температуре +6…8ºС и продолжается до августа–сентября (Луц, 1978). Икринки маленькие (0,5-0,6 мм), но имеется большая желточная капля (до 0,4 мм), за счёт которой в совокупности с яйцевыми оболочками обеспечивается высокая плавучесть икринки (Крыжановский, 1956).

Качественный анализ состава питания тюльки в Каспийском море показал наличие около 50 видов и групп пищевых организмов. Почти все группы организмов зоопланктона, встречающихся в море, представлены в пище тюльки. Основа питания рыбы состоит из Copepoda (56%) и Cladocera (20%), реже встречаются коловратки, личинки моллюсков и рыбы. Питается тюлька только в светлое время суток, ночью питание прекращается (Приходько, 1979; Атлас…, 2003). В Азовском море также основу питания тюльки составляют веслоногие ракообразные, в меньшей степени - коловратки, личинки рачков и червей (Майский и др., 1950).

В последнее десятилетие численность всех видов тюлек в Каспии значительно снизились: промысловый лов упал с 68,7 тыс. т. в 2001 г., до 0,73 тыс. т. в 2009 г. (Устарбекова, 2011). На снижение численности и биомассы тюльки, по всей видимости, влияет вселение гребневика *Mnemiopsis leidyi*, зарегистрированного в 1999 г. в Каспийском море (Daskalov, Mamedov, 2007). Уловы тюльки после вселения мнемиопсиса в 1998/99 гг. у побережья Ирана снизились почти на 50 % (Адели, 2005). На азовскую популяцию тюльки, так же как и в Каспийском море, тоже отрицательно сказалось вселение гребневика *M. leidy*. Негативные последствия от вселения гребневика связаны с сильным снижением продукции зоопланктона в морях, в результате чего рыбы вынуждены переходить на другие пищевые объекты. Возможно, вселение хищного гребневика *Beroe ovata* исправит эту проблему (Shiganova, 2005).

## 1.4. Механизмы адаптаций вселенцев к новым условиям обитания

### 1.4.1. Модели расселения и теория краевых популяций

Ещё в середине XX в. Ч.Элтон (1960) обратил внимание на проблему расширения ареалов некоторых видов. В результате успешного распространения особей, занесённых в экосистему извне, может произойти лавинообразное увеличение численности вселенца, что наносит ущерб всему сообществу. Но для того, чтобы понять взаимодействие вновь появившихся форм с аборигенами и роль инвайдеров в изменённой экосистеме, а также оценить экологические и экономические последствия вселения следует учитывать пути и механизмы реализации биологических инвазий.

Вселение чужеродных видов возможно либо за счёт постепенного расширения нативного ареала (экспансия по типу «шаг за шагом»), либо вселением в совершенно новые условия группы особей, достаточной для образования местной популяции. Примером успешной экспансии первого типа может служить постепенное расселение черноморско-каспийской тюльки в бассейне р. Волги, а второго типа - случайная интродукция амурского чебачка *Pseudorasbora parva* при акклиматизации дальневосточных растительноядных рыб. Однако в любом случае дальнейшее расселение из первичных центров интродукции происходит по типу «шаг за шагом». Таким образом, процессы адаптогенеза и закрепление генетико-биохимических адаптаций в новообразованных популяциях видов-вселенцев должны соотвествовать основным принципам генетических преобразований в популяциях, находящихся на периферии ареала (Johnson, Gerrish, 2002; Marco et al., 2002; Suchentrunk et al., 2003).

В природных условиях элементарной экологической единицей, способной поддерживать своё относительно самостоятельное существование в ряду поколений, обычно служит местная популяция. Она может включать в себя много родственных, но генотипически различных особей, между которыми происходят постоянные скрещивания, и, как правило, возникает «сетчатое родство». Вследствие панмиксии и постоянного обмена гаметами популяция приходит к такому состоянию, что в каждой особи сосредотачивается большинство «генетических возможностей» всей популяции. Постоянное поддержание гетерозиготности обеспечивает более широкую экологическую пластичность и является важным аппаратом, страхующим панмиктическую популяцию от обеднения наследственной основы во время периодических колебаний численности. Местная популяция оказывается объединенной процессами скрещивания в единое целое и приобретает значение элементарной размножающейся единицы вида (Завадский, 1967; Майр, 1974).

Локальные краевые популяции видов-вселенцев обычно основываются относительно небольшой по численности группой особей. Как в случае акклиматизации, а ещё более - в случае случайного заноса новых видов животных, сложно переместить значительную часть особей, репрезентативно отражающих генофонд вида. Таким образом, особи-основатели несут лишь часть генетической изменчивости родительской популяции. Соответственно, на первоначальном этапе существования новообразованная популяция должна быть менее разнообразной и генетически более однородной. К этому добавляется и эффект гомозиготизации, происходящий в результате неизбежного инбридинга в относительно небольшой группе особей. Однако одновременно с этим процессом происходит элиминация летальных либо неблагоприятных в данных условиях аллелей, за счёт чего уменьшается генетический груз популяции. Дальнейшая судьба новообразованной популяции может развиваться по двум сценариям. В первом случае продолжатся процессы гомозиготизации, генетическое разнообразие будет уменьшаться и популяция займёт специализированную экологическую нишу в экосистеме-рецепиенте. В ином случае возможны микроэволюционные преобразования, увеличивающие генетическое разнообразие в новой популяции (Майр, 1974; Левонтин, 1978; Грант, 1980; Яблоков, Юсуфов, 2006; Robert et al., 2003; Hufbauer, Torchin, 2007).

### 1.4.2. Направление и характер адаптаций. Адаптации на организменном и популяционном уровне

Все генетические процессы, описываемые любой из дистанционно-зависимых моделей расселения, справедливы лишь при условии отсутствия прямого действия отбора на генные частоты, то есть в случае отсутствия селективного значения аллелей. Однако, как правило, в процессе расселения видов происходят сложные адаптационные изменения. Одним из таких преобразований является акклиматизация как процесс возникновения адаптаций при освоении организмами новых условий среды. Именно в процессе акклиматизации происходят адаптационные процессы во всей новообразованной популяции по отношению к комплексу факторов среды. Адаптации, возникающие при акклиматизации животных, являются результатом интеграции всех уровней организации живого организма. Вследствие таких преобразований образуется определённый адаптивный фонд биологических систем, создаваемый в процессе отбора за счёт генетической изменчивости и зафиксированный в генотипе конкретных особей. Этот адаптивный фонд создаёт некий «запас прочности», многократно используемый биологическими системами при повторных изменениях условий среды (Шкорбатов, 1973; Хочачка, Сомеро, 1988).

Процессы акклиматизации включают сложные перестройки на всех уровнях организации живых систем. На популяционном уровне происходит образование новой, относительно устойчивой группы, способной существовать на протяжении многих поколений. На уровне живых организмов происходят изменения поведенческих и физиологических характеристик, обеспечивающих успешное выживание в новых условиях. Все эти изменения могут быть производными от адаптаций на уровне внутриклеточного метаболизма. В первую очередь это относится к функциональным изменениям ферментных систем, белковых комплексов и уровней экспрессии генов.

Следует отметить, что генетико-биохимические адаптации часто являются «крайним средством», к которому организм прибегает только тогда, когда у него нет поведенческих или физиологических способов избежать неблагоприятного воздействия среды. Механизмы адаптаций, как правило, разделяют на две группы: компенсаторные и эксплуативные (Хочачка, Сомеро, 1988). Первые направлены на компенсацию ущерба, вызванного воздействием неблагоприятных факторов среды. Сюда относится, например, изменение экспрессии генов, отвечающих за продуцирование белков «теплового шока», накопление антифризов в крови арктических рыб и т.д. Вторая группа адаптаций связана с возникновением новых признаков, способствующих новому использованию имеющихся ресурсов среды, либо освоению новых ресурсов. В отличие от компенсаторных адаптаций, восстанавливающих нарушенную приспособленность, эксплуатативная адаптация не является необходимой для существования организма. Однако с приобретением новых свойств организм способен значительно расширить свой ареал. Вероятно, для видов-вселенцев, последовательно расширяющих свой ареал, более характерны как раз адаптации эксплуативного типа. Для видов, интродуцированных в новые условия среды, в новые биотопы, на ранних этапах интродукции критическое значение имеет выработка компенсаторных механизмов. В случае успешного вселения начинают вырабатываться и эксплуативные адаптации.

Одним из примеров изучения акклиматизационных изменений у рыб служит сопоставление дегидрогеназной активности печени двух популяций прудового карпа: «южной» (Украина) и «северной» (Ленинградская обл.). После месячного содержания одновозрастных рыб в одинаковых условиях в температурном интервале +15…20°С дегидрогеназная активность печени северной популяции оказалась достоверно выше, чем у южных карпов. При температуре +30…40°С показатели ферментативной активности в обеих популяциях сближаются, а при +50ºС активней оказываются ферменты «южной» популяции (Шкорбатов, 1973). Подобные явления обнаруживались у многих животных (Кирпичников, 1987; Szumiec, Bialowas, 2003). Приспособительное значение всех этих различий, вероятно, связано с тем, что у северных популяций пойкилотермных животных в процессе отбора закрепляются ферментативные системы, активно работающие при низких температурах, а у южных - отличающиеся большей теплоустойчивостью.

Существенным фактором среды для гидробионтов служит показатель солёности и общей минерализации воды (Шкорбатов, 1973; Copp, 2003; Карабанов, 2011). Особенно он важен при переходе рыб от морского к пресноводному образу жизни, как это, например, происходило с исторически эстуарным видом - черноморско-каспийской тюлькой. Данных по адаптациям к солёности меньше, чем по температурным, что связано с большей сложностью проведения экспериментов и определения критериев адаптации. В настоящее время не вызывает сомнения факт, что адаптации к солёности происходят на основе широкой нормы реакции к воздействию данного фактора, сформировавшегося у исторически эстуарных форм. Так, достаточно хорошо изучены физиологические адаптации у сельди *Alosa pseudoharengus* к условиям морских и пресных вод (Stanley, Colby, 1971). Этот уроженец Атлантического океана из семейства сельдёвых, заселив многие пресноводные водоемы Северной Америки, включая Великие озера, сохранил поддержание концентрации ионов натрия, калия и кальция в плазме крови и мышечной ткани на уровнях, характерных для морских особей этого вида. Видимо, в нормальных условиях сельди превосходно регулируют свой ионный баланс, а физиолого-биохимические адаптации при переходе в пресные воды протекают на основе изменения проницаемости мембран и активного транспорта ионов.

Часто температурные, кислородные и солевые адаптации рассматриваются отдельно друг от друга. Однако, адаптогенез затрагивает целый ряд метаболических реакций. На организменном уровне комплекс адаптивных преобразований проявляется в «законах толерантности», являющимися следствием принципа Шелфорда (Пианка, 1981). Так, изменение характера осморегуляции непосредственно влияет на интенсивность энергетического обмена, когда приспособление к пониженной солёности среды вызывает повышение потребления кислорода. Изменение солёности среды приводит также к изменению терморезистентности организмов, в свою очередь сопряженной с содержанием кислорода в воде (Карпевич, 1998).

Процесс адаптогенеза при расширении ареала протекает в несколько фаз. Разные исследователи, основываясь на различных показателях (в основном на основании изменения численности популяций акклиматизантов) выделяли от трёх до семи этапов акклиматизации. В целом можно выделить два основных этапа акклиматизации. На первом этапе происходит реализация адаптивного фонда популяции на основе имеющейся широкой нормы реакции организмов. При этом имеющаяся норма реакции вида должна перекрывать пределы действия лимитирующих факторов. В таких случаях иногда говорят о «преадаптациях» живых систем. Однако ещё Дж. Симпсон (1948) рекомендовал использовать этот термин с осторожностью, отмечая, что адаптационные изменения способны происходить лишь на фоне высокой изменчивости живых организмов, что также неоднократно подчёркивал и Ф.Айала (1981).

На втором этапе акклиматизации формируются и, благодаря отбору, генетически закрепляются адаптивные признаки, возникает генетическая дифференциация между новообразованными популяциями, отражающая пути адаптаций групп особей к конкретным условиям среды.

Несмотря на столь значительный материал, работ посвящённых генетико-биохимическим адаптациям непосредственно для видов-вселенцев очень мало. Как правило, генетические процессы в популяциях инвазийных видов рассматриваются как пример пошаговой модели расселения (Muller, 2001), либо с акцентом на теоретические эволюционные последствия инвазий (Hufbauer, Torchin, 2007) в том числе при гибридизации с нативными видами (Largiader, 2007). В тоже время изучение генетико-биохимических адаптаций *in situ*, непосредственно в процессе вселения в новые водоёмы, имеет важное теоретическое и практическое значение. Очень удачным объектом для данной работы служит черноморско-каспийская тюлька, обладающая широким новоприобретённым ареалом, позволяющим проводить изучение адаптаций в водоёмах, различных как по условиям, так и относительно времени вселения.

# Глава 2.
# Материал и методы исследования

## 2.1. Характеристика материала исследования

Исследование проведено на представителях различных популяций черноморско-каспийской тюльки *Clupeonella cultriventris* (Nordmann, 1840) и двух популяциях анчоусовидной тюльки *Clupeonella engrauliformes* (Borodin, 1904). В основе работы лежат сборы, проведённые автором в экспедиционный сезон 2002-2011 годов. Сбор материала проводился с использованием малого пелагического трала с горизонтальным раскрытием 12 м, вертикальным - 2 м, ячеёй в крыльях 30-10 мм, в кутке - 5 мм. Время одного траления - 15 мин. На ряде водоёмов применялся мальковый невод с горизонтальным раскрытием 25 м, вертикальным - 2 м, ячёй в крыльях 20-10 мм, в кутке - 3 мм. В зимний период вылов тюльки осуществлялся ставной сетью с ячёй 10 мм высотой 3 м, длиной 12 м. Лов проводился в течение суток в пелагиали Волжского плёса (ст. Коприно, русло Волги) на глубине 2 м подо льдом.

Географическое положение исследованных водоёмов и станций отбора проб представлено на рис. 2.1 и 2.2, объём обработанного материала представлен в табл. 2.1.

С применением традиционных методик ихтиологических исследований (Правдин, 1966; Методическое…, 1974) изучалась видовая структура уловов, размер, возраст, пол, состав и питание рыб. Для генетико-биохимических исследований брались одноразмерные половозрелые особи, длина тела *l.*=60-80 мм, также брали и неполовозрелых особей, длина тела *l.*=20-30 мм.

## 2.2. Методы исследования

### 2.2.1. Методы популяционно-генетического анализа

В качестве основного метода исследований был выбран диск-электрофорез полипептидов в полиакриламидном геле (disc-PAGE). Разделение белков при PAGE основано на том, что смесь полипептидов в образце под действием электрического поля фракционирует в зависимости от соотношения заряда и молекулярной массы белков. Этот метод сильно чувствителен к условиям хранения образцов, а также довольно трудоёмкий и дорогой по стоимости реактивов. Методом электрофореза выявляется около 30% возможных аминокислотных замен в полипептиде, приводящих к изменению заряда макро-

**Таблица 2.1**. Характеристика собранного материала для генетико-биохимического анализа

| Год | Водоём | Экз. | Локусов |
|---|---|---|---|
| Черноморско-каспийская тюлька *Clupeonella cultriventris* | | | |
| 2002 | Рыбинское водохранилище (8 выборок) | 529 | 12 |
| 2003 | Рыбинское водохранилище (общая проба) | 156 | 12 |
| | Иваньковское водохранилище | 40 | 12 |
| | Угличское водохранилище | 39 | 12 |
| | Чебоксарское водохранилище | 39 | 12 |
| | Устье р. Дон | 34 | 12 |
| | Азовское море, Порт Кантон | 40 | 12 |
| 2004 | Рыбинское водохранилище (9 выборок) | 489 | 17 |
| | Куйбышевское водохранилище | 39 | 12 |
| | р. Волга у г. Саратов | 30 | 14 |
| | Волгоградское водохранилище | 20 | 14 |
| | Азовское море, Чумбур-Коса | 39 | 12 |
| | Днестровский лиман | 39 | 12 |
| | Манычский каскад водохранилищ (4 выборки) | 126 | 12-17 |
| 2005 | Рыбинское водохранилище (7 выборок) | 238 | 17 |
| | Шекснинское водохранилище | 16 | 17 |
| | Горьковское водохранилище (2 выборки) | 80 | 13 |
| | Сев. Каспий, устье р. Сулак | 160 | 17 |
| | р. Днепр (2 выборки) | 80 | 16 |
| | Азовское море, Чумбур-Коса | 76 | 12 |
| 2006 | Рыбинское водохранилище (общая проба) | 198 | 12 |
| 2007 | Рыбинское водохранилище (4 выборки) | 197 | 17 |
| 2008 | Рыбинское водохранилище (общая проба) | 160 | 12 |
| 2011 | Рыбинское водохранилище (4 выборки) | 160 | 8 |
| 2011 | Горьковское водохранилище (2 выборки) | 80 | 8 |
| 2011 | Камское водохранилище | 60 | 8 |
| Анчоусовидная тюлька *Clupeonella engrauliformes* | | | |
| 2004 | Сев. Каспий | 37 | 12 |
| 2005 | Сев. Каспий | 40 | 8 |
| Всего: | | 3241 | 6-17 |

макромолекул. Вместе с тем, изучение белков-ферментов позволяет выяснить не только микроэволюционные изменения в популяциях, но и проследить селективное значение и адаптивные возможности тех или иных форм ферментов. Ещё одно важнейшее свойство изоферментов, выделяющих их из прочих генетических и биохимических маркеров - это изменение относительного количества и активности изоферментов при различных физиологических и патологических состояниях (Райдер, Тэйлор, 1983; Глазко, 1988; Reinitz, 1977; Copeland, 2000).

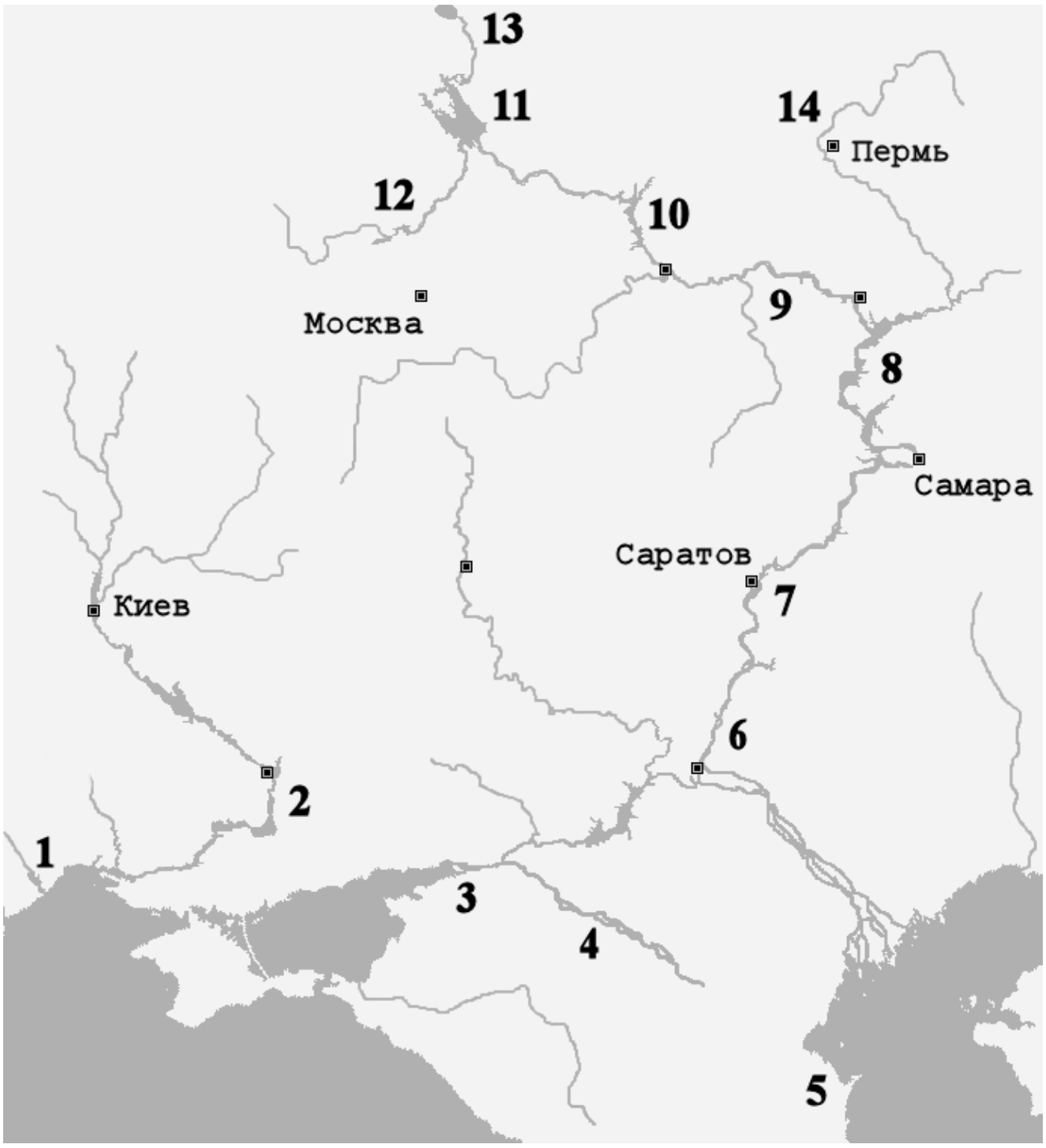

**Рисунок 2.1.** Места отбора проб. **1**– акватория Днестровского лимана, район г. Белгород-Днестровский; **2**– акватория Среднего Днепра: район устья р. Самара, Карачуновкое водохранилище, р. Днепр; **3**– акватория Азовского моря: район пос. Порт Кантон, Чумбур-коса, дельта р. Дон, тоня «Сельдёвая»; **4**– акватория Манычского каскада водохранилищ: канал ветви Азовской водораспределительной системы, Весёловское водохранилище, Пролетарское водохранилище, оз. Маныч-Гудило; **5**– северная часть Каспийского моря, район устья р. Сулак; **6**– средняя зона Волгоградского водохранилища; **7**– верхняя часть Волгоградского водохранилища у г. Саратов; **8**– Камский плёс Куйбышевского водохранилища; **9**– Чебоксарское водохранилище, г. Космодемьянск; **10**– Горьковское водохранилище возле г. Чкаловск, г. Юрьевец и г. Волгореченск; **11**– Рыбинское водохранилище (14 станций отбора проб); **12**– Угличское водохранилище: г. Углич и г. Калязин, Иваньковское водохранилище: г. Дубна и г. Конаково; **13**– Шекснинское водохранилище, Сизменское расширение; **14**– Камское водохранилище, г. Пермь.

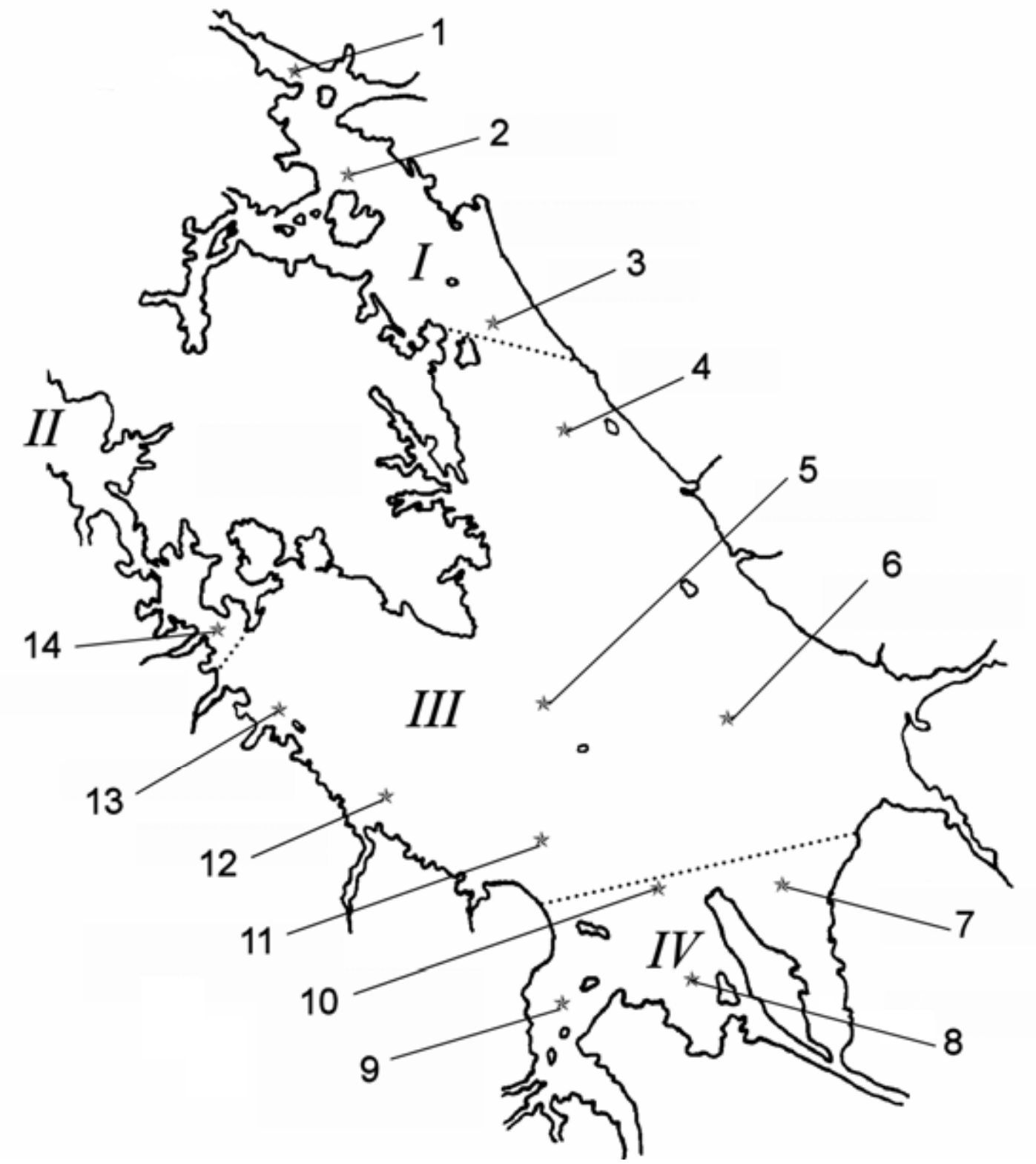

**Рисунок 2.2.** Подробная карта станций отбора проб на акватории Рыбинского водохранилища.

Плёсы водохранилища: ***I*** – Шекснинский, ***II*** – Моложский, ***III*** – Главный (Центральный), ***IV*** – Волжский.

Станции отбора проб: **1**– устье р. Суда, **2**– Любец, **3**– Мякса, **4**– Ягорба, **5**– Центральный мыс, **6**– Городок, **7**– Мелюшино, **8**– Переборы, **9**– Глебово, **10**– Бабьи Горы, **11**– Горькая Соль, **12**– Брейтово, **13**– Первомайка, **14**– Борок-Заповедник

Живую рыбу, непосредственно после отлова помещали в жидкий азот в сосудах Дьюара СК-50, либо замораживали целиком при температуре не выше -27ºС. Для предотвращения вымораживания тканей проэтикетированную общую пробу заворачивали в алюминиевую фольгу. В лабораторных условиях проводился биологический анализ рыб и отделялись образцы для биохимического исследования. Наиболее удобны для популяционно-генетических исследований методом PAGE водорастворимые ферменты белых скелетных мышц. Эта ткань имеет набор ферментов всех основных классов (Gilbert, 2000), легко отделяется от других тканей даже неспециалистами, а также способна довольно долго храниться в замороженном состоянии (до 90 суток) без особых потерь активности.

В качестве основных изучаемых ферментов использовались: α-глицерофосфатдегидрогеназа (aGPDH, E.C. 1.1.1.8), лактатдегидрогеназа (LDH, E.C. 1.1.1.27), малатдегидрогеназа NADP-зависимая (ME, E.C. 1.1.1.40), глюкозо-6-фосфатдегидрогеназа (G6PDH, E.C. 1.1.1.49), супероксиддисмутаза (SOD, E.C. 1.15.1.1), аспартатаминотрансфераза (AAT, E.C. 2.6.1.1); эстеразы эфиров карбоновых кислот: 2-нафтилацетат-зависимая эстераза и D-эстераза (bEST и D-EST, E.C. 3.1.1.x), спектр общего белка (GP), а также проводилось биохимическое изучение дополнительного набора ферментов: малатдегидрогеназа NAD-зависимая (MDH, E.C. 1.1.1.37), 6-фосфоглюконатдегидрогеназа (6PGDH, E.C. 1.1.1.44), щелочная фосфатаза (AP, E.C. 3.1.3.1). Особенности гистохимического выявления изученных изоферментов представлены в табл. 2.2, а их характеристика изложена в Главе 3.

**Таблица 2.2.** Некоторые особенности гистохимического выявления изоферментов тюлек рода *Clupeonella*

| Фермент | Инкубационная смесь | Время инкубации |
|---|---|---|
| LDH<br>E.C. 1.1.1.27 | Фосфатный буфер pH 7,2 - 74 мл<br>Лактат натрия 1M - 24 мл<br>$MgSO_4 \cdot 7H_2O$ - 80 мг<br>NaCl - 40 мг<br>NAD - 40 мг<br>п-Нитротетразолий синий (10 г/л) - 5 мл<br>Вода дистиллированная - до 200 мл<br>Натрия феназинметасульфат (3 г/л) - 4 мл | 40 мин. |
| MDH<br>E.C. 1.1.1.37 | Трис-HCl 0,05M буфер pH 8,6 - 200 мл<br>Малат натрия - 600 мг<br>NAD - 60 мг<br>п-Нитротетразолий синий (10 г/л) - 4 мл<br>Натрия феназинметасульфат (3 г/л) - 4 мл | 50 мин. |
| ME<br>E.C. 1.1.1.40 | Трис-HCl 0,05M буфер pH 7,8 - 200 мл<br>Малат натрия - 800 мг<br>$MgSO_4 \cdot 7H_2O$ - 400 мг<br>NADP - 40 мг<br>п-Нитротетразолий синий (10 г/л) - 4 мл<br>Натрия феназинметасульфат (3 г/л) - 4 мл | 1 ч. 10 мин. |
| aGPDH<br>E.C. 1.1.1.8 | Трис-HCl 0,05M буфер pH 7,8 - 200 мл<br>Натрия sn-глицерол-3-фосфат - 350 мг<br>NaCl - 10 мг<br>NAD - 100 мг<br>п-Нитротетразолий синий (10 г/л) - 4 мл<br>Натрия феназинметасульфат (3 г/л) - 4 мл | 3 часа |

| Фермент | Инкубационная смесь | Время инкубации |
|---|---|---|
| G6PDH<br>E.C. 1.1.1.49 | Фосфатный буфер pH 7,1 - 200 мл<br>NADP - 60 мг<br>Натрия глюкозо-6-фосфат - 400 мг<br>$MgSO_4 \cdot 7H_2O$ - 300 мг<br>п-Нитротетразолий синий (10 г/л) - 4 мл<br>Натрия феназинметасульфат (3 г/л) - 4 мл | 2 часа |
| 6PGDH<br>E.C. 1.1.1.49 | Трис-HCl 0,05М буфер pH 8,8 - 200 мл<br>6-фосфоглюконат ТСНА-соль - 200 мг<br>$Na_2SO_4 \cdot 10H_2O$ - 20 мг<br>$MgSO_4 \cdot 7H_2O$ - 300 мг<br>п-Нитротетразолий синий (10 г/л) - 6 мл<br>Натрия феназинметасульфат (3 г/л) - 6 мл | 2 часа |
| β-EST<br>E.C. 3.1.1.x | Фосфатный буфер pH 7,2 - 200 мл<br>Прочный синий RR-соль - 100 мг<br>Натрия 2-нафтилацетат - 40 мг<br>(раств. в ацетоне - 2 мл) | 30 мин. |
| D-EST<br>E.C. 3.1.1.x | Фосфатный буфер pH 7,2 - 200 мл<br>Натрия 4-метилумбеллиферил - 40 мг<br>(раств. в воде 40 мл) | 20 мин.<br>в УФ-свете<br>λ=400 нм. |
| AP<br>E.C. 3.1.3.1 | Трис-HCl 0,05М буфер pH 8,8 - 200 мл<br>Прочный синий RR-соль - 200 мг<br>$MgSO_4 \cdot 7H_2O$ - 50 мг<br>Натрия 1-нафтилфосфат - 200 мг<br>(раств. в ацетоне - 2мл) | 40 мин. |
| SOD<br>E.C. 1.15.1.1 | Трис-HCl 0,05М буфер pH 8,5 - 200 мл<br>$MgSO_4 \cdot 7H_2O$ - 100 мг<br>п-Нитротетразолий синий (10 г/л) - 12 мл<br>Натрия феназинметасульфат (3 г/л) - 6 мл | 8 часов |
| AAT<br>E.C. 2.6.1.1 | Трис-HCl 0,05М буфер pH 7,6 - 200 мл<br>L-аспартат - 460 мг<br>NaCl - 10 мг<br>Натрия 2-оксоглутарат - 200 мг<br>Прочный синий ВВ-соль - 400 мг<br>Пиридоксаль-5-фосфат - 10 мг | 30 мин. |
| Общий белок, GP | Инкубация геля в 12% водном растворе трихлоруксусной кислоты - 4 часа<br>Кумасси бриллиантовый синий - 20 мг<br>(раств. в уксусной кислоте - 30 мл) | 3 часа |

При разделении ферментов применялись вертикальные электрофоретические камеры с пластинами PAG, при этом одновременно исследовалось 40 образцов в блоке. Один образец составлял: для aGPDH, LDH и SOD по 25 мкл супернатанта, для ME и G6PDH - по 20 мкл, для EST, AAT и GP - по 15 мкл, для других ферментов - по 25 мкл супернатанта. Разделение изоферментов проводилось с ис-пользованием щелочной системы disc-PAGE (Davis, 1964; Ornstein, 1964; Гааль и др., 1982). Применялась буферная система Трис-HCl-$H_3BO_3$ pH 8,9 с 4% PAG концентрирующего и 6% PAG разделяющего геля для ME, LDH, G6PDH, AAT и 7% PAG разделяющего геля для других ферментов. В качестве индикатора использовалось внесение в первую лунку с пробой 5 мкл 1% раствора индикатора бромфенолового синего (БФС).

Для изучения скрытой изменчивости и определения оптимальной концентрации PAG использовались различные концентрации разделяющего геля: от 4% до 10% с шагом в 1% (Vesterberg, Hansen, 1978; Ayala, 1982).

Электрофорез проводился в два этапа - преэлектрофорез, до вхождения индикатора в разделяющий гель; стабилизация по силе тока 80 mA. Второй этап - основной электрофорез; стабилизация по силе тока 200 mA. Время основного электрофореза зависит от соотношения заряд:масса полипептида и для разных ферментов подбирается экспериментально. Для исследованных видов время основного электрофореза составило: EST, GP и AAT - до выхода БФС из разделяющего геля, для ME, aGPDH и G6PDH - 1 час 10 минут, для SOD и LDH - 1 час 40 минут, остальные ферменты - до выхода БФС из разделяющего геля. Проведение электрофореза проводилось при охлаждении камеры до температуры +4°С путём принудительной конвекции охлаждённого нижнего электродного буфера в электрофоретическом резервуаре (объём 4,5 л.).

Для выявления ферментов в сложной смеси белков после электрофореза используются реакции, специфичные для конкретной ферментной системы. При гистохимическом выявлении изоферментов пользовались общепринятыми методиками (Генетика…, 1977; Глазко, 1988; Walker, 2002; Smith, 2002), основанными на базовых руководствах по выявлению ферментативной активности (Бернстон, 1965; Manchenko, 2003) с некоторыми авторскими изменениями (Карабанов, Слынько, 2005б). Соответствующие буферные системы с заданным значением pH приготовлялись по стандартным прописям (Досон и др., 1991). Инкубация гелей в красящей смеси проводилась в темноте при температуре +37°С до развития окрашивания в зоне ферментативной реакции. Среду для окрашивания блока приготовляли перед началом окрашивания. Кофермент, катализатор или инициатор реакции добавлялся непосредственно перед укладкой геля в инкубационную среду.

Вслед за окрашиванием гели промывались дистиллированной водой и выдерживались в 7% водном растворе уксусной кислоты на протяжении 12 часов. После пластины гелей помещались в консервирующую спиртоглицериновую смесь (15% водный раствор этанола с 1% глицерина). Консервация шла не менее 7 суток. Затем пластины гелей высушивались в 40-мкм плёнках целлофана с последующим хранением под гнётом в сухом, прохладном, защищённом от света месте.

Денситометрический анализ активности изоферментов проводили по индивидуальным электрофоретическим трекам с использованием пакета RFLPscan Plus v.3.12 (CSP, Inc.). Популяционно-генетический анализ проводился с использованием программы BIOSYS r.2 (University of Illinois, USA). Графическое представление материалов и статистический анализ проводили с использованием пакетов программ MS Office 2003 (Microsoft Corp.) и STATISTICA v.6.1 (StatSoft, Inc.).

### 2.2.2. Методы генетико-биохимического анализа

Для изучения влияния основных абиотических факторов, которые могли бы оказывать воздействие на механизмы внутриклеточного метаболизма тюльки, был проведён ряд биохимических экспериментов *in vitro*. В качестве модельных ферментов были выбраны: лактатдегидрогеназа, аспартатаминотрансфераза, 2-нафтилацетатзависимая эстераза и α-глицерофосфатдегидрогеназа. Эти ферменты относятся к трём важнейшим классам ферментов (Оксидоредуктазы, Трансферазы, Гидролазы), их структура, функция и генетическая детерминация у разных животных относительно хорошо изучены, а также имеются данные по селективности, количеству и функциональным свойствам этих ферментов (Кирпичников, 1987; Мещерякова, 2004, и др.).

Всего было проведено 3 серии экспериментов по выяснению влияния различных концентраций неорганических солей и карбамида, а также температуры (рис. 2.3.). Каждая серия воспроизведена в трёх повторностях опытов. Всего проведён индивидуальный анализ не менее чем по 20 особям для каждой повторности.

В первой серии экспериментов было изучено влияние различных концентраций хлорида натрия и сульфата магния, а также их совместного воздействия. Общая схема эксперимента показана на рис. 2.3.*а*. Образцы тканей от каждой особи разделялись на 4 равные порции, одна из которых шла в качестве контроля, остальные поступали в эксперимент. Образцы размером около 5 мм$^3$ помещались в мешочки из диализной трубки ServaPor 4023, размер пор которой гарантированно не пропускает макромолекулы. Мешочки помещались в индивидуальные резервуары с заданной концентрацией солей (табл. 2.3.) и инкубировались 15 мин. при +4ºС при постоянном перемешивании. Далее мешочки извлекались и отмывались от солей в течение 30 мин. в физиологическом растворе Рингера в модификации Кребса-Хенселейта для холоднокровных животных (Досон и др., 1991).

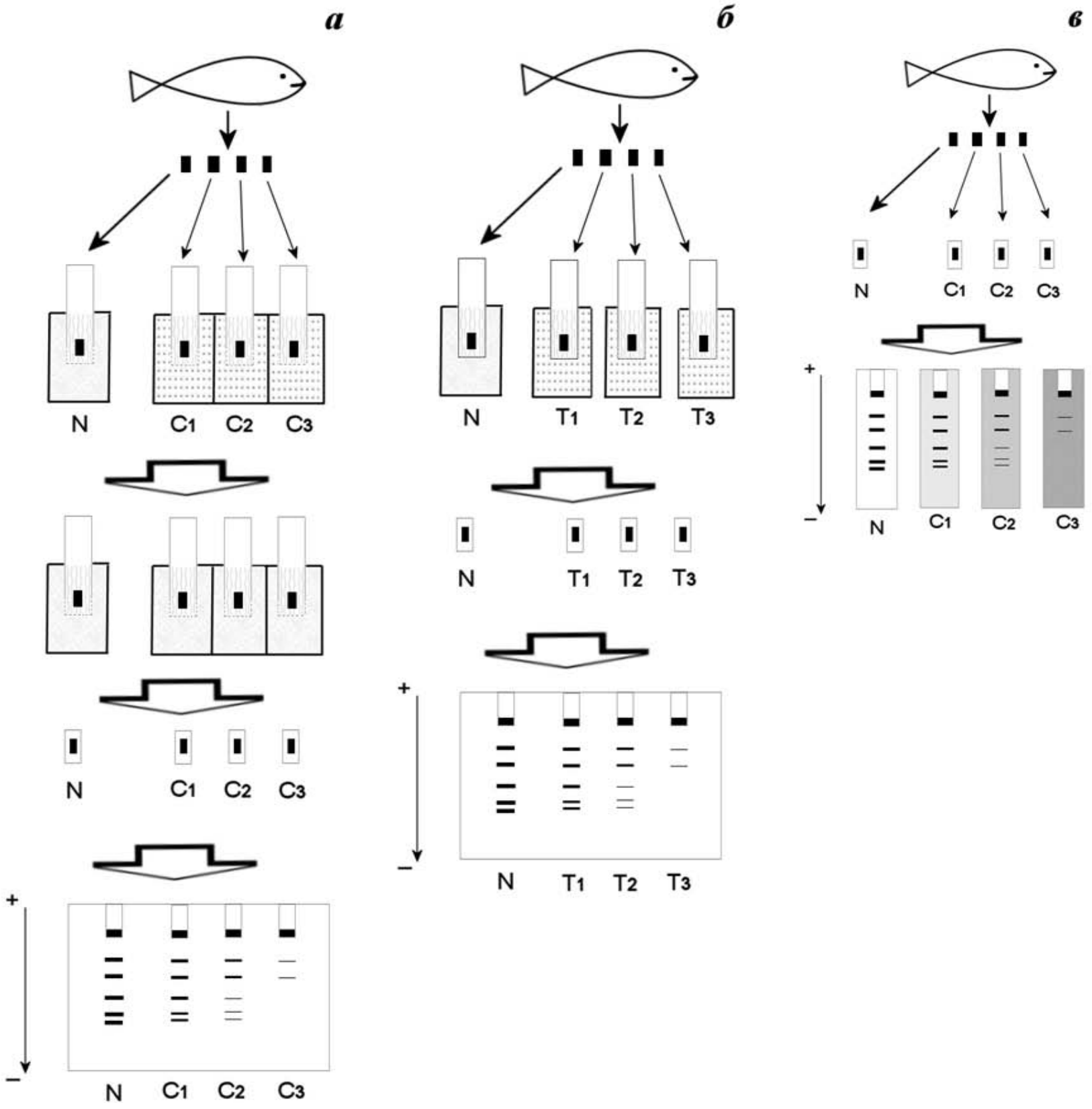

**Рисунок 2.3.** Схема биохимических экспериментов *in vitro* по воздействию различных концентраций неорганических солей (***а***), температурной устойчивости (***б***) и различных концентраций карбамида (***в***) на изоферменты черноморско-каспийской тюльки. Пояснения в тексте

Контрольный образец всё это время находился в диализном мешочке в физиологическом растворе. Применение именно физиологического раствора Рингера в модификации Кребса-Хенселейта для холоднокровных было обосновано после ряда экспериментов как обеспечивающее наибольшую сохранность ферментов: общие потери в контрольном образце по сравнению с нативными тканями составляли не более 7%, что позволяет рекомендовать этот раствор для биохимических и физиологических экспериментов для рыб *in vitro*. Далее образцы подвергались стандартной процедуре PAGE с последующей денситометрической обработкой электрофореграмм.

**Таблица 2.3.** Изучение влияния неорганических солей на изоферменты *C. cultriventris*

| Серия экспериментов | Навеска соли | Конц. ионов | Время инкубации | Кол.-во проб |
|---|---|---|---|---|
| Раздельное влияние солей | | | | |
| Хлорид натрия NaCl | 15 г/л | ($Na^+$ $Cl^-$) 15 г/л | 15 мин. | 5 проб, 3 пвт. |
| | 30 г/л | ($Na^+$ $Cl^-$) 30 г/л | 15 мин. | 5 проб, 3 пвт. |
| Сульфат магния $MgSO_4 \cdot 7H_2O$ | 30 г/л | ($Mg^{2+}$ $SO_4^{2-}$) 15 г/л | 15 мин. | 5 проб, 3 пвт. |
| | 60 г/л | ($Mg^{2+}$ $SO_4^{2-}$) 30 г/л | 15 мин. | 5 проб, 3 пвт. |
| Совместное влияние солей | | | | |
| NaCl + $MgSO_4 \cdot 7H_2O$ | 60 г/л<br>60 г/л | ($Na^+$ $Cl^-$) 60 г/л<br>($Mg^2$+ $SO_4^{2-}$) 30 г/л | 20 мин. | 5 проб, 1 пвт. |

Во второй серии экспериментов было изучено влияние различных температур на ферментативную активность водорастворимых белков тюльки. Общая схема эксперимента показана на рис. 2.3.*б*. За основу были взяты работы Р. Коена (Koehn et al., 1971) и В.С. Кирпичникова (1987). Всего было изучено 3 повторности по 5 индивидуальных проб в каждой. Образцы тканей от каждой особи разделялись на 6 равных порций, одна из которых шла в качестве контроля, остальные поступали в эксперимент. Образцы размером около 5 мм$^3$ помещались в пластиковые микропробирки с экстрагирующим раствором сахарозы. Контрольный образец выдерживался при температуре стандартной пробоподготовки (+4°С). Остальные образцы подвергались нагреванию в течение 15 мин. на водяной бане с конвекцией БВ-5 при необходимой температуре. Время инкубации вычислялось эмпирически - за этот промежуток ткани успевали равномерно прогреться до заданной температуры, но не наступала коагуляция белков. Было изучено влияние ряда температур: +20°С, +35°С, +50°С, +65°С и +80°С. После экспериментов образцы подвергались стандартной процедуре PAGE с последующей денситометрической обработкой электрофореграмм.

В третьей серии экспериментов исследовалось влияние различных концентраций карбамида на ферменты тюльки. За основу были взяты работы В.С. Кирпичникова (1987). Всего было изучено 3 повторности по 10 индивидуальных проб в каждой. Образцы тканей от каждой особи разделялись на 4 равных порций, одна из которых шла в качестве контроля, остальные поступали в эксперимент. Контрольные образцы подвергались обычной процедуре PAGE. Другие три повторности независимо исследовались методом PAGE, при котором в концентрирующий и разделяющий гели, а также в экстрагирующий раствор и электродный буфер добавлялся карбамид до концентрации 2М, 4М и 6М соответственно для каждой повторности (рис. 2.3.*в*).

В изучение популяционной структуры входят следующие задачи: оценка частот аллелей и генотипов, доли полиморфных локусов, гетерозиготности, приспособленности, определение равновесных частот аллелей. Расчёт этих параметров детально рассмотрен в ряде фундаментальных руководств (Ли, 1978; Nyquist, 1990; Хедрик, 2003) и здесь не представлен. В работе используются следующие сокращения: $p$ и $q$ - частоты аллелей; $\boldsymbol{P}_{ob}$ и $\boldsymbol{P}_{ex}$ - наблюдаемые и ожидаемые численности генотипов (при прямом подсчёте); $H_{ob}$ и $H_{ex}$ - наблюдаемая и ожидаемая гетерозиготность при k=0,95; $\chi^2$ - критерий Пирсона, $\chi^2_{(0.05;1)}$=3,84 (в случае недостоверности гипотезы соответствия распределения частот генотипов равновесию Харди-Вайнберга отмечен знаком «*»); $\lambda$ - показатель Колмогорова-Смирнова, $\lambda_{0,95}$=1,36 (в случае недостоверности гипотезы соответствия распределения частот генотипов равновесию Харди-Вайнберга отмечен знаком «*»); $w$ - усреднённая относительная приспособленность генотипов; W - общая приспособленность популяции; S - давление отбора на популяцию; ***F*** - индекс фиксации гетерозигот; $D_H$ - дефицит гетерозигот; коэффициенты инбридинга F-статистики: $F_{IT}$ - особь относительно целой популяции, $F_{IS}$ - особь относительно субпопуляции, $F_{ST}$ - субпопуляция относительной всей популяции. Для оценки генетических расстояний применялись дистанции Нея D (Nei, 1972, 1978). Также для сравнения двух выборок и установления сходства между популяциями применялся критерий ***r*** и генетические дистанции ***D***, предложенные Л.А. Животовским (1984).

## Глава 3.
## Краткая характеристика исследованных водоёмов

### Таганрогский залив Азовского моря

Азовское море - мелководный водоём, который можно рассматривать как большой причерноморский лиман реки Дон. Располагается между 45°16' и 47°17' с.ш. и 33°36' и 39°21' в.д. Площадь моря составляет 38 тыс. км$^2$, средняя глубина 8 м, максимальная - 14 м. Это самое мелководное море в мире, его объём составляет всего 320 км$^3$. В северо-восточной части моря лежит большой Таганрогский залив, вытянутый в направлении устья Дона. Формирование водной массы происходит в результате смешения черноморских вод, поступающих через Керченский пролив с речными водами и осадками (Рождественский, Цветков, 1983).

В зимнее время Азовское море покрывается льдом, толщина льда в Таганрогском заливе достигает 80 см. В летнее время воды моря прогреваются до +25-30°С, зимой температура воды у поверхности охлаждается до 0…–0,9°С. Обычно, вследствие мелководности, воды Азовского моря хорошо перемешиваются, солёность и температура одинаковы от поверхности до дна. Кислородный режим нарушается лишь в период зимней стратификации либо в продолжительные летние штили, что может приводить к дефициту растворённого кислорода и «заморам» рыб (Зенкевич, 1951).

Особенности гидрологического режима Азовского моря, и в особенности Таганрогского залива, обуславливаются большим притоком пресной воды и мелководностью бассейна. Солёность воды увеличивается от устья Дона в юго-западном направлении от 1 до 9‰, в центральной части моря - до 11‰, в северной части моря - до 18‰. В связи с зарегулированностью стока р. Кубань и влиянием стока р. Дон от уровня Цимлянского водохранилища солевой режим моря подвержен значительным сезонным колебаниям под воздействием хозяйственной деятельности человека.

Химические особенности азовских вод зависят в основном от состава поступающих речных и черноморских вод. Эта связь отражается не только в непосредственном опреснении водоёма при смешении морских и речных вод, но и в изменении общей увлажнённости бассейна, регулирующей балансовый уровень Азовского моря, его водо- и солеобмен с Чёрным (Воловик, 1985). Некоторое влияние, особенно заметное на мелководных участках, оказывают донные отложения. В Азовском море содержится в сред-нем 11,5‰ солей, а в Таганрогском заливе редко превышает 8‰ (от 3,7 до 8,1‰). По ионному составу основными являются ионы хлора ($Cl^-$), сульфаты ($SO_4^{2-}$), карбонаты ($CO_3^{2-}$) и гидрокарбонаты ($HCO_3^-$), а также ионы натрия ($Na^+$), магния ($Mg^{2+}$), кальция ($Ca^{2+}$) и калия ($K^+$). Перечисленные ионы в морской воде находятся в небольшом количестве, их соотношение под влиянием состава донских вод и испарения отклоняется от нормального для морских вод: содержание $Cl^-$ меньше, чем в океане (около 1%), $SO_4^{2-}$ немного больше (приблизительно 0,2%), а $HCO_3^-$ значительно больше, чем в океанической морской воде (1,2%). Среднее процентное содержание катионов также отличается от такового в океане (Рождественский, Цветков, 1983). Щёлочность азовской воды составляет около 3 мг-экв/л, что значительно меньше, чем щёлочность Чёрного моря. Однако отношение щёлочности к хлорности воды в азовских водах больше, чем в черноморских, вследствие малого количества ионов хлора. Это указывает на влияние речных вод, содержащих большие количества гидрокарбонатов и малое количество хлоридов. Влияние речных вод в Таганрогском заливе увеличивается направлении к устью Дона. Хлорность воды вблизи устья составляет в среднем 0,073‰, а количество гидрокарбонатов около 0,192‰ (щёлочность около 3,1 мг-экв/л).

Газовый режим азовских вод определяется небольшой глубиной и увеличенным перемешиванием водных масс при волнении, а также интенсивными биологическими процессами в толще воды и на дне моря. Летом в Таганрогском заливе на глубинах более 5 м иногда образуется сероводородный слой, сравнительно редко поднимающийся высоко от дна. Кислотность вод колеблется от pH 8,8 в период летнего цветения воды до pH 7,6 в конце осени.

Фауна Азовского моря обогащается в наши дни как за счёт черноморских видов, так и видов, проникающих в море из других водоёмов (часто - как результат хозяйственной деятельности человека). Этому способствуют несколько факторов - современный морской транспорт, деятельность человека по акклиматизации новых видов рыб, продолжающееся увеличение солёности моря. Можно отметить проникновение в Черноморско-Азовский бассейн рака *Rhitropanopeus harrisii* с Атлантического побережья Северной Америки; космополитичного трубчатого червя *Mercierella enigmatica*; голожаберного моллюска *Stiliger bellulos* из Средиземноморья. Наибольшее распространение в бассейне Азовского моря получила дальневосточная кефаль - пиленгас (*Mugil soiuy*), запасы которой в настоящее время оцениваются в 25 тыс. т. (Витковский, Богачев, 2005).

Биомасса зоопланктона достигает 1,5 г/м$^3$, большую часть её составляет солоноватоводная *Calanipeda aquae*, наиболее массово развивающаяся в Таганрогском заливе.

Рыбы Азовского моря представлены 79 видами, из них 20 видов проходные и полупроходные (осетровые, сельдёвые, окунёвые) и 13 пресноводные (карповые). Основными представителями рыбного Таганрогского залива являются (в процентах к общему запасу всех рыб): черноморско-каспийская тюлька *Clupeonella cultriventris* (29%), судак *Sander lucioperca* (8-16%), хамса *Engraulis encrasicholus* (12%), лещ *Abramis brama* (7-11%), бычки сем. Gobiidae (5-15%) и сельди р. *Alosa* (2-5%). Основной пищей (до 70%) тюльки и хамсы служат веслоногие рачки, а весной значительную часть пиши (до 30%) составляют коловратки. Из хищников важное значение имеет судак. Весной и летом он питается преимущественно хамсой и тюлькой, а осенью - почти исключительно бычками (Зенкевич, 1951; Воловик, 1985).

Нами производился лов тюльки на акватории Таганрогского залива Азовского моря севернее пос. Порт Кантон и в районе Чумбур-косы (рис. 2.1).

**Дельта реки Дон**

Дельта р. Дон - одна из самых крупных речных систем Юга России. Располагается на площади около 750 км$^2$ в координатах 47°05' – 47°16' с.ш. и 39°11' – 39°43' в.д. Средняя высота территории - от 0 до 6 м над уровнем моря. Дельта Дона занимает площадь около 750 км$^2$. Водоёмы (прилегающая часть Таганрогского залива, р. Дон и его гирла, Мертвый Донец, болота, пруды) и пойменные заливные луга занимают более 80% территории. Дельта представляет собой систему протоков, ериков и перекопов, отделяющих друг от друга множество островов. Общая протяжённость водотоков более 1700 км. В дельте находится сеть дренажных и других каналов, большое количество прудов воспроизводственных и товарных рыбных хозяйств (Природные…, 2002).

После зарегулирования Дона произошли не только сокращение и перераспределение стока воды по сезонам года, но и изменения в химическом составе донских вод (прежде всего растет их общая минерализация). В створе г. Ростова-на-Дону сумма солей в воде нередко достигает 0,9-1‰, в то время как в 1929 г. в среднем за год она была равна 0,45‰. Изменяется соотношение ионов и общая характеристика солевого состава во времени и в пространстве. При посто-янстве содержания ионов кальция и магния уменьшается концентрация гидрокарбонатов, но увеличивается доля сульфатов и хлоридов. Гидрокарбонатный класс речных вод, свойственный естественному гидрохимическому фону, изменяется на сульфатный и даже хлоридный. Среди катионов существенно возрастает роль натрия и калия (Земля…, 1975; Миноранский, 2006).

Разнообразие водоемов, ландшафтов, почв, растительных сообществ обусловили многообразие животного мира дельты. В зоопланктоне авандельты и дельты обнаружено 99 видов, из которых коловратки (*Rotatoria*) составляют 44, копеподы (*Copepoda*) - 23, кладоцеры (*Cladocera*) - 28, прочие - 4; в зообентосе - около 150 видов (Ресурсы..., 1980).

Дельта исторически славилась богатством рыбных ресурсов. Здесь встречаются практически все виды рыб, обитающие в донском бассейне (около 70 видов). Однако, начавшееся в середине XX в. сокращение рыбных запасов продолжается и до настоящего времени. За последние 70 лет занчительно сократился вылов ряда ценных промысловых рыб: осетра (*Acipenser giildenstddti*) в 7 раз, севрюги (*A. stellatus*) в 10 раз, белуги (*Huso huso*) в 25 раз, сельди (*Alosa caspia*) в 20 раз, чехони и леща в 10 раз. В целом почти за век общий вылов рыбы сократился примерно в 12 раз, при этом ценных промысловых рыб - в 17 раз (Миноранский, 2006). Многие в прошлом промысловые рыбы попали в Красные книги России и Ростовской области. Во второй половине XX в. не только резко сократились ресурсы и выловы рыбы в районе, но изменился и состав промысловых видов. Так, акклиматизированный в Азовском море пиленгас впервые на российском участке был выловлен в 1993 г. в количестве 63 т., а уже через 10 лет его вылов в дельте Дона и опреснённой части Таганрогского залива превысил 2,5 тыс. т. (Ресурсы..., 1980; Миноранский, 2002; Экологический…, 2003).

Нами производился лов тюльки в дельте р. Дон на левом берегу гирла, тоня «Сельдёвая» в 2003/05 гг. (рис. 2.1).

**Водохранилища Манычского каскада**

Водохранилища Манычского каскада представляют собой отдельные водоёмы с оригинальным водным режимом. Координаты каскада водохранилищ 46°05’ – 47°16’ с.ш. и 40°05’ – 43°35’ в.д.

Усть-Манычское водохранилище представляет собой довольно узкий извилистый водоём с двумя лиманами - Западенским и Шахаевским. Протяжённость водохранилища 60,5 км, площадь около 73

км$^2$. Перед впадением в р. Дон находится Усть-Манычский гидроузел. Максимальная глубина - 3,5 м, средняя - 1,2 м. Вода Усть-Манычского водохранилища принадлежит к хлоридному классу группы кальция. Минерализация в последние годы стабилизировалась на уровне 0,9-1,1 г/л. Кислородный режим водоема, как правило, благоприятен для развития гидробионтов, в зависимости от сезона колеблется от 7,4 до 6,9 мг/л. Заморных явлений обычно не наблюдается из-за достаточно высокой проточности водоёма (Витковский, 2000).

Биомасса зоопланктона Усть-Манычского водохранилища в разные годы колеблется от 0,025 г/м$^3$ до 0,123 г/м$^3$, а количество видов его составляющих - от 8 до 15. В настоящее время в Усть-Манычском водохранилище обитают 35 видов рыб; наиболее массовые виды - окунь (*Perca fluviatilis*), густера (*Blicca bjoerkna*), лещ, тарань (Витковский, 2000).

Весёловское водохранилище - водоём 93,2 км длиной, шириной 1,5-3,0 км, проектная площадь равна 300 км$^2$ при мелководной акватории в 80 км$^2$. Максимальные глубины не превышают 7,5 м. Минерализация воды из-за сильного испарения в летнее время повышалась в нижней части водоёма с 4,3 до 12,0 г/л, а в верхней - с 3,6 до 24,3 г/л. Вследствие этого водоём превратился в солоноватоводный с достаточно устойчивым гидрологическим режимом. Содержание кислорода в поверхностных слоях в вегетационный период обычно бывает близким к насыщению (Витковский, 2000). В придонных горизонтах этот показатель может снижаться до 2,8-3,0 мг/л. В придонных слоях показатель pH 7,5. В последние годы минерализация воды Весёловского водохранинилища возросла до 1,3-2,4 г/л, что обусловливалось увеличением поступления высокоминерализованных дренажно-сбросных вод с орошаемых земель Ставропольского края и Ростовской области в бассейне реки Б. Егорлык (Витковский, 2000; Казаков, Ломадзе, 2006).

Средняя биомасса зоопланктона в Весёловском водохранилище составляет 2,3 г/м$^3$. Больше всего в зоопланктоне коловраток (56,7% от общей массы зоопланктона) и ветвистоусых рачков (37,8%). В водохранилище на начальном этапе развития (1933-1938 гг.) обитало 24 вида рыб. В 1979-1985 гг. число видов увеличилось до 46. Появились как виды, самостоятельно проникшие в водоем (вырезуб *Rutilus frisii*, голавль *Leuciscus cephalus*, горчак *Rhodeus sericeus*, шемая *Chalcalburnus chalcoides*, подуст *Chondrostoma variabile*), так и искусственно интродуцированные (рыбец *Vimba vimba*, бестер, большеро-тый буффало *Ictiobus cyprinellus*, белый амур *Ctenopharyngodon idella*, белый *Hypophthalmichthys molitrix* и пестрый *Aristichthys nobilis* толстолобики). Однако к середине 1990-х гг. видовой состав ихтиофауны значительно сократился. В 1993-2000 гг. обнаружено только 33 вида рыб, в том числе стерлядь (*Acipenser ruthenus*), а также появился ещё один вселенец - пиленгас. Основными промысловыми видами рыб в настоящее время являются: лещ, толстолобик, густера, тарань, судак (Витковский, 2000; Казаков, Ломадзе, 2006).

Озеро Маныч-Гудило располагается на границе Ростовской области и Калмыкии. В пределах Ростовской области площадь водоёма составляет 250 км$^2$, основные глубины 1,5-2,5 м, предел их колебания 0,5-4,5 м. Ширина водохранилища на различных участках изменяется в пределах 1,5-3 км. Минерализация воды возросла от 70-х к 80-м годам с 13,4-22,4 г/л до 18,0-30,0 г/л, средний уровень содержания солей составил 35 г/л (Витковский, 2000; Миноранский и др., 2006).

Подавляющее большинство в биомассе зоопланктона озера составляют копеподы (99%). Как следствие высокого осолонения из зообентоса присутствуют лишь остракоды и два вида хирономид. Характерной особенностью данного водоёма является отсутствие моллюсков (Витковский, 2000). Ихтиофауна крайне бедная, представлена в основном колюшками (*Pungitius*) и черноморско-каспийской тюлькой.

Во всех водохранилищах Манычского каскада нами совместно с к.б.н. А.З. Витковским производился лов тюльки в 2004 г. (рис. 2.1).

**Днестровский лиман**

Водоёмы вдоль побережья Чёрного моря в большинстве случаев относятся к категории лиманов - затопленных устьев и долин рек, расположенных обычно перпендикулярно к береговой линии. Днестровский (Днестрово-Бугский лиман) - самый большой из черноморских лиманов, образованный из слияния лиманов Днепра и Южного Буга. Лиман находится между 46º07’ и 46º23’ с.ш. и 30º10’ и 30º31’ в.д. Вход в лиман располагается между пос. Очаков и Кирнбурнской косой, шириной до 4,2 км. Длина лимана до устья реки Днепр составляет 61 км, а ширина колеблется от 10 до 15 км. В середине северного берега отделяется Бугский лиман длиной 37 км. Глубина в Днепровском лимане в основном 5-6 м, по фарватеру - до 12 м.

Солёность в устье лимана изменяется от 0,5 до 15‰ (осенью при сильных южных ветрах). Морская вода обычно задерживается на большой глубине, в ней расходуется кислород и возникает серо-водородная зона. После создания Днепровских водохранилищ и Северо-Крымского канала сток Днепра в лиман уменьшился, а поступление морской воды, особенно в придонные слои, увеличилось (Вылканов и др., 1983). В настоящее время даже весной общая минерализация лиманных вод колеблется около 5‰. В ионном балансе лимана преобладают хлориды, сульфаты, карбонаты и гидрокарбонаты, их катионов - ионы кальция, магния, натрия и калия. Относительно большие концентрации гидрокарбонатов кальция связано с реабсорбцией карбоната кальция из мелководных илов, сильно насыщенных раковинами моллюсков (Зенкевич, 1951). Установлено наличие связи между развитием апвеллинга в прибрежной зоне Одесского региона и обнаружением гипоксии в придонном слое на свале глубин. Очагам гипоксии в придонном слое прибрежных областей акватории в большинстве случаев соответствовали очаги минимальной температуры воды в поверхностном слое (Тучковенко и др., 2004).

Планктон Днестровского лимана образован представителями как морской, так и пресноводной фауны. Из зоопланктона значительное количество составляют веслоногие ракообразные (отр. Copepoda), также много коловраток и ветвистоусых ракообразных (26 видов). Встречаются медузы аурелия (*Aurelia*) и корнерот (*Pilema*), потребляющие большое количество планктона. Из вселенцев в бассейн Черного моря следует отметить представителей средиземноморской фауны. Они нашли здесь благоприятные условия для своего развития и, несмотря на более мелкие размеры, образуют здесь довольно плотные популяции. Из таких вселенцев можно выделить корабельного червя (*Toredo*), сердцевидку (*Cardium edule*) и других моллюсков (*Corbulomya*, *Phaseolina*).

Среди рыб Днестровского лимана выделяются три экологические группы - представители морского комплекса, пресноводные и эстуарные рыбы. Значительная часть первично пресноводных рыб выносит приемлемые уровни осолонения Днестровского лимана (некоторые сельди, бычки, лещ). Большая часть солоноводных рыб (тюлька, бычки) выносят значительные колебания солёности воды в лимане. Периодически в лиман заходят типично морские рыбы, часть которых переселилась из Средиземного моря в Чёрное. Черноморско-каспийская тюлька обитает по всей опреснённой части Чёрного моря, и в Днестровском лимане образует большую популяцию (Георгиев, Александрова-Колеманова, 1983).

Лов тюльки в Днепровском лимане производился местной малой рыболовецкой бригадой осенью 2004 г. на правом берегу лимана в районе г. Белгород-Днестровский (рис.2.1).

**Река Днепр и среднеднепровские водохранилища**

Днепр - крупнейшая река на территории Украины. Площадь водосбора Днепра около 500 тыс. км$^2$, годовой сток около 52 км$^3$ (Денисова, 1979). Среднеднепровские водохранилища расположены между 48°25' и 48°45' с.ш. и 34°20' и 35°31' в.д.

Днепродзержинское водохранилище находится в долине Днепра на участке между створом Кременчугской ГЭС и с. Романково. Наиболее широкое место (до 19 км) находится в центре водохранилища. Около трети площади водохранилища занимают мелководья с глубинами до 2 м. (Денисова, 1979).

Запорожское водохранилище расположено на Среднем Днепре и простирается от плотины Днепродзержинской ГЭС до Запорожья. Мелководья составляют 34% площади водохранилища. Запорожское водохранилище характеризуется ассиметричным строением котловины и значительной вытянутостью в продольном направлении. В водохранилище близ плотины наблюдается летняя температурная стратификация, разница в температуре поверхностных и придонных слоёв может достигать 10°С (Мельников, 1955; Денисова, 1979).

Гидрохимический режим днепровских водохранилищ определяется в первую очередь гидрохимическим режимом рек, за счёт которых происходит их наполнение и питание (Воронков, 1958). Для речных водохранилищ характерны очень большие сезонные колебания минерализации, что обусловлено соответствующим притоком речных вод и каскадным стоком (Денисова, 1971). Химический состав основных водных масс каскада днепровских водохранилищ формируется в верхнем Киевском водохранилище под влиянием стока Верхнего Днепра и Припяти. К среднему течению Днепра их воды полностью перемешиваются. По величине минерализации и ионному составу днепровские и припятские воды по классификации О.А. Алекина (1952) относятся к гидрокарбонатному классу группы кальция ($C_{II}^{Ca}$). Вследствие этого вода всех днепровских водохранилищ относится к гидрокарбонатному классу группы кальция. В результате аккумуляции в водохранилищах весенней маломинерализованной воды и смешения её с поступающей более минерализованной речной имеются колебания минерализации и концентрации отдельных ионов, амплитуда которых невелика и носит ровный характер (Денисова, 1979). Величины минерализации и концентрации главных ионов в Среднем и Нижнем Днепре антибатны величине водного стока и обусловлены режимом верхних водохранилищ. Минимальная минерализация (около 0,2‰) наблюдается в период весеннего половодья (апрель, май). В летний период она повышается до 0,25-0,3‰, и максимальных значений (до 0,4‰ и более) достигает в зимний период и перед половодьем. В балансе основных ионов Среднего Днепра основными являются гидрокарбонаты ($HCO_3^-$) и карбонаты ($CO_3^{2-}$) - около 64%, сульфаты ($SO_4^{2-}$) - около 8%, ионы хлора ($Cl^-$) - до 5%, а также ионы кальция ($Ca^{2+}$) - до 18%, натрия ($Na^+$) и калия ($K^+$) - до 4% ионы магния ($Mg^{2+}$) - 3,5%.

Животный мир Днепра, днепровских водохранилищ и прилегающих к ним территорий богат и разнообразен. Особенно большим многообразием характеризуется фауна низших беспозвоночных организмов. В составе зоопланктона днепровских водохранилища выделено в настоящее время 169 видов животных, в том числе 100 видов планктонных и около 500 видов донных инфузорий, 44 вида ракообразных, 37 видов коловраток (Дехтяр, 1985; Зимбалевская, 1990).

Ихтиофауна днепровских водохранилищ представлена 61 видами рыб, относящихся к 12 семействам. Наибольшим разнообразием представителей отличаются карповые (31 вид), бычковые (9 видов) и окуневые (6 видов). Сельдёвые представлены 3 видами (черноморско-азовская сельдь *Alosa pontica*, черноморский пузанок *Alosa caspia*, черноморско-каспийская тюлька). В результате искусственного вселения в водохранилища ценных представителей ихтиофауны с целью повышения их рыбопродуктивности в Днепре появились три новых вида - представители дальневосточной ихтиофауны: белый и пестрый толстолобики и белый амур. Широкое распространение получили черноморско-каспийская тюлька, некоторые виды бычков и колюшки (Амброз, 1956).

Черноморско-каспийская тюлька, которая раньше поднималась из Днепровского лимана вверх по Днепру не выше г. Никополя стихийно проникла через судоходные шлюзы плотин гидроэлектростанций и в массовом количестве развилась во всех днепровских водохранилищах, образовав местные популяции и став объектом промысла. Черноморские бычки также широко распространились по Днепру и теперь встречаются во всех водохранилищах. В Днепровско-Бугском лимане и дельте Днепра образовалась самостоятельная популяция пиленгаса.

Лов черноморско-каспийской тюльки на акватории Среднего Днепра проводился в 2005/06 гг. (рис. 2.1) в Карачуновском водохранилище и на речном участке в районе г. Днепропетровск совместно с к.б.н. Р.А. Новицким (ДНУ).

**Северная часть Каспийского моря**

Уникальный и весьма богатый природными ресурсами внутренний водоём нашей планеты - Каспийское море - не имеет естественной связи с Мировым океаном и, по географическому определению, это самое крупное озеро Земли. Однако по всем своим характеристикам Каспий соответствует морской водной системе. Каспийское море вытянуто в меридиональном направлении и расположено между 47°07’ и 36°33’ с.ш. и 45°43’ и 54°03’ в.д. Площадь водосборного бассейна 3,5 млн. км$^2$. Физические параметры Каспия меняются в зависимости от уровня моря. Так, в настоящее время, когда уровень моря лежит на отметке -27 м его площадь равна 392,6 тыс. км$^2$. Значительно изменяется при изменении уровня моря площадь именно Северного Каспия. Каспийское море - глубоководный водоём с хорошо развитой шельфовой зоной. Наибольшая глубина зафиксирована в Южном Каспии (1025 м), средняя глубина моря, рассчитанная по батиграфической кривой, равна 208 м. На основании особенностей морфологического строения и физико-географических условий Каспийское море принято делить на три части: Северный, Средний и Южный Каспий. В качестве природной границы Северный Каспий отделяет порог Мангышлаг, протягивающийся в виде мелководья (глубины до 10 м) от п-ова Тюб-Караган к банке Кулалинской и далее к п-ову Чечень (Зонн, 1999; Панин и др., 2005).

Среднемноголетняя солёность поверхностных вод Северного Каспия изменялась от 1 до 13‰, что связано с мощным воздействием речного стока. Начиная с последней четверти XX в. наблюдается достоверное повышение уровня моря на 2,5 м, что привело к увеличению объёма воды примерно на 700 км$^3$ (Голицын, Панин, 1989). Сообразно этому, если главной причиной увеличение уровня моря служит положительное приращение водного баланса, то дополнительный объём воды должен приводить к снижению общей минерализации моря. Это подтверждается многолетними наблюдениями: за последние 25 лет средняя солёность Северного Каспия снизилась примерно на 1‰ (Панин и др., 2005). Основную площадь Северного Каспия (в среднем 60%) занимает зона с солёностью 2-8‰; зона метаморфизации речных вод (минерализация менее 2‰) составляет около 25%; наименьшую площадь (около 15%) занимают воды с солёностью более 10‰. По ионному составу основными являются ионы хлора ($Cl^-$) и сульфаты ($SO_4^{2-}$), в меньшем количестве гидрокарбонаты ($HCO_3^-$) и карбонаты ($CO_3^{2-}$). Из катионов присутствуют ионы натрия ($Na^+$), магния ($Mg^{2+}$), кальция ($Ca^{2+}$) и калия ($K^+$). Различия в солёности поверхностных и глубинных вод незначительны и не препятствуют вертикальной циркуляции водных масс, перемешиванию которых также способствуют частые волнения. Лишь на глубинах более 400 м наблюдается незначительный дефицит кислорода, а в придонном слое возможно появление незначительной сероводородной зоны (Каспийское…, 1986; Семеняк, 1996).

Зоопланктон Каспийского моря представлен 592 видами, из них в Северном Каспии обитает 216 видов. В зоопланктоне большое значение имеют коловратки, ветвистоусые и веслоногие рачки. В прибрежных мелководных зонах обитают в основном *Calanipeda aquae* и *Acartia clausi*, которые хорошо переносят большие изменения температуры и солёности воды (Касымов, 1987; Полянинова, 1998). В последнее время из Чёрного в Каспийское море проник гребневик *Memiopsis leydyi*, значительно повлиявший на местное планктонное сообщество, что отрицательно сказалось на кормовой базе рыб (Панин и др., 2005; Daskalov, Mamedov, 2007; Устарбекова, 2011).

Рыбы Каспийского моря довольно разнообразны (101 вид) и входят в состав 14 семейств, из которых 2 привнесены в Каспий человеком (камбаловые и кефалевые), а 2 проникли в относительно недавнее время (морские иглы и атериновые). Наибольшее количество видов входит в семейства бычковых (30 видов), карповых (14 видов), сельдёвых (9 видов) и осетровых (5 видов). Каспийские рыбы имеют разный генезис - главным образом морской и пресноводный, причём и те и другие рыбы входят в состав однородных биологических групп, сформировавшихся в результате длительного совместного существования (Казанчеев, 1981).

Каспийские сельди совершают в пределах моря регулярные и довольно сложные миграции. Зимой большинство сельдей концентрируются в южной и средней частях моря. К лету они перемещаются к северу, совершая довольно длительные миграции (Приходько, 1979). Откорм каспийских сельдёвых за счёт планктона происходит в основном в Северном Каспии (сельди *Caspialosa*, тюльки *Clupeonella*).

Процесс проникновения новых видов в каспийский бассейн продолжается на протяжении многих тысяч лет. За последние полвека в бассейн каспийского моря были произведены попытки интродукции 24 видов рыб. Относительно удачными можно считать только вселение белого и пёстрого толстолобиков, а также белого амура. Из непреднамеренных интродуцентов в бассейне Каспийского моря можно отметить востробрюшку (*Hemiculter leucisculus*), гамбузию (*Gambusia affinis*) и амурского чебачка (*Pseudorasbora parva*). Среди промысловых рыб особо следует отметить всех осетровых, безжалостный промысел которых ставит под угрозу исчезновения само существование местных популяций (Панин и др., 2005).

Отбор проб черноморско-каспийской и анчоусовидной тюльки на акватории Каспийского моря проводился с привлечением местных малых рыболовных бригад в районе устья р. Сулак в 2004-2005 гг., а также с привлечением промыслового лова на акватории Северного Каспия (рис. 2.1).

**Каскад водохранилищ р. Волги**

Бассейн Волги и её притоков вытянут преимущественно в меридиональном направлении и расположен в средней части Русской равнины. Длина Волги в настоящий момент составляет 3530 км, это крупнейшая река Европы. Площадь водосбора (1360 км$^2$) составляет около четверти Европейской части России

На самой Волге в настоящее время уже не осталось участков с естественным гидрорежимом, незатронутым регулированием. Регулирование стока Волги началось в 1937 году после строительства Иваньковского водохранилища. С 1940 по 1986 годы было построено дополнительно еще 10 плотин вдоль русла реки, сопровождаемых крупномасштабным расширением деятельности человека в бассейне Волги. Самые крупные водохранилища простираются на сотни километров, но их ширина в целом не превышает несколько десятков километров. Геоморфологически Волга разделяется на Верхнюю, Среднюю и Нижнюю. Верхняя Волга включает в себя верховья, Иваньковское, Угличское, Шекснинское и Рыбинское водохранилища, Средняя Волга - Горьковское, Чебоксарское и Куйбышевское водохранилища, Нижняя Волга - Саратовское и Волгоградское водохранилища, а также низовья реки (Буторин, 1969; Волга…, 1979).

Традиционно обзор речных водохранилищ проводят вниз по течению, от истоков к устью. Однако, в связи с распространением черноморско-каспийской тюльки вверх по Волге, логичнее рассматривать каскад Волжских водохранилищ от низовьев к истоку.

**Нижняя Волга**

В пределы Нижней Волги входят дельта Волги, Волго-Ахтубинская пойма, Волгоградское и Саратовское водохранилища. Северной границей Нижней Волги географически было устье Камы, сейчас считается плотина Волжской ГЭС (Фортунатов, 1971). Общий сток Волги в Каспийское море в разные годы составляет от 160 до 390 км$^3$. Паводок начинается в конце апреля - начале мая и продолжается до середины июня. Усреднённые многолетние основные гидрохимические показатели в водоёмах Нижней Волги приведены в табл. 3.1. (по: Волга…, 1979; Бикбулатова, Бикбулатов, 1982; Комплексный …, 2004).

Очень динамичным компонентом водных экосистем Нижней Волги является зоопланктон, в составе которого насчитывается 828 систематических групп. Во временных водотоках доминируют виды собственно планктона, представленные родами *Brachionus*, *Bosmina*, *Euchlanis*, *Chydorus* и др. Важное значение имеют также виды, временно живущие как планктон (организмы из других экологических групп - перифитон, бентос, и др.). Вспышки численности зоопланктона наблюдаются в мае-июне и в августе (Мордухай-Болтовской, 1963; Волга…, 1979; Комплексный…, 2004).

**Таблица 3.1.** Гидрохимические характеристики вод Нижней Волги

| Участок | Общая минерализация, мг/л | pH | Раств. $O_2$ (мг/л) | Раств. $O_2$ (%) |
|---|---|---|---|---|
| Низовья | 291 | 8,3 | 11,2 | 97 |
| Волгоградское | 262 | 8,2 | 10,6 | 89 |
| Саратовское | 315 | 8,1 | 10,8 | 90 |

| Участок | Баланс ионов (мг/л) | | | | | |
|---|---|---|---|---|---|---|
| | $Ca^{2+}$ | $Mg^{2+}$ | $Na^{+}+K^{+}$ | $HCO_3^{-}+CO_3^{2-}$ | $SO_4^{2-}$ | $Cl^{-}$ |
| Низовья | 52 | 10 | 20 | 116 | 55 | 37 |
| Волгоградское | 39 | 15 | 15 | 94 | 58 | 37 |
| Саратовское | 47 | 13 | 24 | 120 | 59 | 40 |

До строительства каскада водохранилищ предусматривалось, что вновь построенные водохранилища будет успешно использоваться для производства большого количества важных промысловых рыб. Однако водохранилища были не только не способны компенсировать потерю продуктивности Нижней Волги и Каспийского моря, более того, они стали одной из главных причин этого процесса (Бердичевский и др., 1972). Исследования показали, что большие внутригодовые сезонные вариации уровня воды в водохранилищах создали аномальные условия среды обитания для многих видов рыб, что в результате привело к массовой гибели икры и нарушению вос-производства. В данном случае следует особо отметить влияние режима работы гидроузлов. Ускоренное повышение уровня воды в начале регулирования паводка приводит к более раннему началу развития условий поверхностного склонового стока на затопляемой наземной территории, что негативно сказывается на смертности рыбной молоди, вынужденной на более ранней стадии жизни противостоять растущей скорости течения и вымываемой раньше времени с затопленных лугов в ручьи. Раннее начало поверхностного склонового стока также снижает первичную продукцию планктона, определяющего трофическое состояние экосистемы. Таким образом, и успех нереста, и развитие кормовой базы подвергаются негативному воздействию ступенчатых изменений объемов стока и связанных с ними изменений уровня воды, изменений гидрохимических параметров и температуры. Более быстрое понижение уровня воды приводит к тому, что большая часть молоди остается в отшнурованных водоемах и водотоках (Волга…, 1979; Комплексный …, 2004).

За последние 50 лет в водохранилища Нижней Волги проникли 17 новых видов рыб, большая часть из которых представлена мелкими непромысловыми особями. Самовоспроизводящиеся популяции отмечены для рыбца, ротана (*Perccottus glenii*), бычка-головача (*Neogobius kessleri*), бычка-цуцика (*Proterorhinus marmoratus*), звёздчатой пуголовки (*Benthophilus macrocephalus*), черноморской иглы-рыбы (*Syngnathus nigrolineatus*), и малой южной колюшки (*Pungitius platygaster*) (Шашуловский, Ермолин, 2005).

Лов тюльки на Нижней Волге производился малым пелагическим тралом с использованием судна-лаборатории Саратовского отделения ГосНИОРХ в средней части Волгоградского водохранилища (2004 г.) и в Саратовском водохранилища в районе г. Саратов (2004/05 гг.) (рис.2.1).

**Средняя Волга**

В пределы Средней Волги входят Куйбышевское, Чебоксарское и Горьковское водохранилища. Южной границей Средней волги служит плотина Волгоградской ГЭС, а северной - плотина Рыбинской ГЭС. Средняя Волга является той частью реки, в пределах которой наиболее сильно увеличивается площадь водосбора и расход воды. Основные гидрохимические показатели в водоёмах Средней Волги приведены в табл. 3.2. (по: Волга…, 1979; Экологические…, 2001).

Зоопланктон различных водохранилищ Средней Волги значительно отличается как в зависимости от водоёма, так и конкретного биотопа. Зоопланктон речных участков, как правило, представляет собой трансформированный зоопланктон вышележащего водохранилища. В озёрной же части восстанавливается пелагический комплекс, своеобразный в каждом конкретном водоёме. Среди видов-вселенцев в зоопланктоне Средней Волги встречаются 1 вид коловраток (*Notholca acuminata*), 8 видов ветвистоусых рачков и 7 видов веслоногих рачков (Попов, 2005).

**Таблица 3.2.** Гидрохимические характеристики вод Средней Волги

| **Участок** | **Общая минерализация, мг/л** | | | **pH** | **Раст. $O_2$** | |
|---|---|---|---|---|---|---|
| | | | | | **(мг/л)** | **(%)** |
| Куйбышевское | 314 | | | 7,5 | 9,98 | 87 |
| Чебоксарское | 214 | | | 7,4 | 9,88 | 88 |
| Горьковское | 181 | | | 7,5 | 9,56 | 81 |
| **Участок** | **Баланс ионов (мг/л)** | | | | | |
| | $Ca^{2+}$ | $Mg^{2+}$ | $Na^{+}+K^{+}$ | $HCO_3^{-}+CO_3^{2-}$ | $SO_4^{2-}$ | $Cl^{-}$ |
| Куйбышевское | 46 | 11 | 26 | 126 | 58 | 36 |
| Чебоксарское | 36 | 9 | 9 | 107 | 39 | 10 |
| Горьковское | 30 | 8 | 7 | 101 | 26 | 7 |

В ихтиофауне Средней Волги преобладают бентофаги (около 50% видов) и хищники (около 20%). Основу промыслового лова составляют лещ, плотва, густера, судак и щука (*Esox lucius*). Популяция волжской стерляди своим существованием обязана искусственному воспроизводству на рыбоводных заводах.

Из инвазийных видов в водоёмах Средней Волги встречается 21 вид, из них самовоспроизводящиеся популяции образуют 11 видов: ряпушка (*Coregonus albula*), корюшка (*Osmerus eperlanus*), гуппи (*Poecilia reticulata*), тюлька, девятииглая колюшка, ротан, звёздчатая пуголовка и 4 вида бычков (песочник, головач, кругляк, цуцик) (Клевакин, 2005). Наибольшую численность имеет черноморско-каспийская тюлька, прочие виды образуют немногочисленные популяции, не имеющие большого влияния на экосистемы водохранилищ.

Лов тюльки на Средней Волге производился малым пелагическим тралом с использованием судна-лаборатории ИБВВ РАН и Нижегородского отделения ГосНИОРХ в результате совместных работ в 2003 и 2005 гг. (рис.2.1).

**Верхняя Волга**

К водохранилищам Верхней Волги относят Верхневолжское, Иваньковское, Угличское и Рыбинское водохранилища. К Верхневолжскому бассейну относят так же и Шекснинское водохранилище, созданное в 1964 г. на р. Шексна. Бассейн Верхней Волги имеет дос-таточно густую речную сеть. Основная роль в питании рек принадлежит снежному покрову. Малая проницаемость подстилающих грунтов и избыточная увлажненность территории способствует заболоченности бассейна (Буторин, 1969; Экологические…, 2001).

**Таблица 3.3.** Гидрохимические характеристики вод Верхней Волги

| **Участок** | **Общая минерализация, мг/л** | **pH** | **Раст. $O_2$** | |
|---|---|---|---|---|
| | | | **(мг/л)** | **(%)** |
| Рыбинское водохранилище (в целом) | 191,3 | 7,4 | 9,9 | 85 |
| Волжский плёс | 217 | 7,4 | 9,2 | 85 |
| Моложский плёс | 210 | 7,5 | 8,9 | 77 |
| Шекснинский плёс | 166 | 7,2 | 10,8 | 90 |
| Центральный плёс | 172 | 7,5 | 10,6 | 88 |
| Угличское водохранилище | 234 | 7,3 | 9,7 | 83 |
| Иваньковское водохранилище | 244 | 7,3 | 8,9 | 81 |
| Шекснинское водохранилище | 125 | 7,2 | 10,8 | 91 |

| **Участок** | **Баланс ионов (мг/л)** | | | | | |
|---|---|---|---|---|---|---|
| | **$Ca^{2+}$** | **$Mg^{2+}$** | **$Na^{+}$+$K^{+}$** | **$HCO_3^{-}$+$CO_3^{2-}$** | **$SO_4^{2-}$** | **$Cl^{-}$** |
| Рыбинское водохранилище (в целом) | 32,3 | 9 | 4,5 | 110,8 | 31,3 | 4,3 |
| Волжский плёс | 36 | 9 | 7 | 127 | 27 | 7 |
| Моложский плёс | 34 | 11 | 4 | 125 | 31 | 3 |
| Шекснинский плёс | 30 | 8 | 3 | 91 | 31 | 3 |
| Центральный плёс | 29 | 8 | 4 | 100 | 36 | 4 |
| Угличское водохранилище | 40 | 9 | 9 | 137 | 30 | 8 |
| Иваньковское водохранилище | 42 | 10 | 9 | 152 | 26 | 8 |
| Шекснинское водохранилище | 23 | 6 | 3 | 88 | 24 | 3 |

Так как большая часть работы основывается на использовании Рыбинского водохранилища в качестве модельного водоёма, то его описание будет дано более подробно. Рыбинское - наиболее крупное из водохранилищ Верхней Волги. По объёму воды оно занимает второе место среди водохранилищ Волжского каскада после Куйбышевского водохранилища. По распределению глубин и морфометрическим особенностям ложа в водохранилище выделяют четыре основных района (плёса): Волжский, Моложский, Шекснинский, Главный (или Центральный) (Волга…, 1979). Основное назначение Рыбинского водохранилища - обеспечение нужд энергетики и водного транспорта. Оно является так же источником водоснабжения, в нем развито промысловое и любительское рыболовство. Шекснинское водохранилище в целом по своим морфометрическим характе-ристикам схоже с Горьковскоим водохранилищем. Угличское и Иваньковское - небольшие водохранилища речного типа.

Бассейн Верхней Волги расположен в пределах зоны умеренного климата и характеризуется продолжительным (около 7 месяцев) периодом положительных температур воздуха и с зимним периодом устойчивой отрицательной температуры воздуха, которая удерживается в течение 3-4 месяцев (с конца ноября до середины марта). Среднегодовая температура воздуха уменьшается по территории бассейна с запада на восток.

Определяющим фактором в распределении температуры воздуха в зимнее время является циркуляция атмосферы. Средняя температура самого холодного месяца (января) изменяется от -10°С до -14°С. Средняя температура июля составляет +16,5…18,5°С. В целом, с середины 90-х годов XX века наблюдается некоторое повышение среднегодовых температур воздуха (Экологические…, 2001), как и в целом по планете вследствие глобального потепления (Huber, Knutti, 2012).

Усреднённые многолетние основные гидрохимические показатели по водохранилищам Верхней Волги приведены в табл. 3.3. (по: Волга…, 1979; Былинкина, Трифонова, 1982; Экологические…, 2001). Для бассейна Верхней Волги в целом характерно относительно малое содержание растворённых минеральных соединений (как правило, даже в зимний период менее 400 мг/л), преобладают карбонаты кальция и магния. Иваньковское и Угличское водохранилища выделяются большими концентрациями ионов $Cl^-$ и $HCO_3^-$. В Рыбинском водохранилище химические показатели воды изменяются в течение года незначительно из-за слабого водообмена и большого объёма водоёма. Необходимо также отметить высокое содержание сульфатов в Шекснинском водохранилищи и Шекснинском плёсе Рыбинского водохранилища. Этот феномен обусловлен природными свойствами р. Шексны, водосбор которой отличается сильной заболоченностью и повышенным содержанием гипса, а также имеется влияние сточных вод металлургических и химических предприятий г. Череповца (Рыбинское…, 1972; Экологические…, 2001).

В Рыбинском водохранилище с 1980-х по 2000-е года обнаружено 79 видов ракообразных и 59 видов коловраток. К доминантным видам, распространённым во всех частях Рыбинского водохранилища, относились рачки *Mesocyclops leuckarti* (11–66% общей численности ракообразных), *Termocyclops oithonoides* (5–38%), *Eudiaptomus* *gracilis* (5–36%), *Daphnia galiata* (5–28%). Среди коловраток это были: *Synchaeta pectinata* (5–80% к численности коловраток*), Conochilus unicornis* (6–68%), *Keratella quadrata* (8–52%), *Polyarthra major* (5–50%) (Лазарева, 2005, 2010).

К началу XXI в. для бассейна Верхней Волги отмечено 69 видов рыб (включая круглоротых), относящихся к 52 родам, 23 семействам, 12 отрядам, 2 классам. Наибольшее число таксонов приходится на долю карпообразных - 36 видов, окунеообразных - 9 видов и лососеобразных - 8 видов. До зарегулирования стока ихтиофауна региона была представлена 38 видами жилых рыб и 6 видами мигрантов. После зарегулирования каспийские мигранты практически полностью выбыли из состава ихтиофауны бассейна. Однако, в связи с созданием системы водохранилищ существенно возросла численность популяций таких озерно-речных видов рыб как лещ, плотва, синец, густера, чехонь, окунь и судак. В связи с колонизацией дрейссеной бассейна Верхней Волги, в водохранилищах (преимущественно в Рыбинском и Горьковском) сформировалась моллюскоядная форма плотвы. Многие реофильные виды, исчезнувшие из зоны водохранилищ на первом этапе формирования, сохранились в притоках. Начиная с 30-х годов в бассейне Верхней Волги неоднократно предпринимались попытки акклиматизации, зарыбления и разведения целого ряда видов, большинство из которых исторически чужды бассейну (Экологические…, 2001). Большинство этих видов изредка встречаются в уловах в естественных водоемах, однако только сазан (*Cyprinus carpio*) представлен немногочисленными самовоспроизводящимися популяциями. Более результативной оказались случайные акклиматизации ротана-головешки и гуппи. Ротан освоил ряд естественных водоемов на водосборе и даже обнаруживается в настоящее время в литоральной зоне Иваньковского и Горьковского водохранилищ. Многочисленные популяции гуппи появились в прудах-отстойниках очистных сооружений бытовых стоков ряда крупных городов региона (Тверь, Рыбинск, Ярославль) и в местах сброса техногенных теплых вод. С 1980-х годов в бассейне Верхней Волги появляются саморасселяющиеся беломоро-балтийские (девятииглая колюшка) и эвригалинные понто-каспийские виды (черноморско-каспийская тюлька, бычок-цуцик, звездчатая пуголовка) (Экологические…, 2001).

Места отбора проб на акватории Верхней Волги и подробная сеть станций на акватории Рыбинского водохранилища показаны на рис. 2.1 и 2.2.

**Камское водохранилище**

Камское водохранилище - водохранилище Камской ГЭС на реке Каме. Водохранилище расположено с 57°30’ до 59°50’ с.ш., и с 50°20’ до 57°15’ в.д. Создано в 1954 году. Максимальная глубина - до 30 м, подпор уровня воды у плотины - 22 м. Территория водосбора водохранилища охватывает бассейн Верхней и Средней Камы и расположена на северо-востоке Европейской части России. Камское водохранилище относится к водоемам с сезонным регулированием, наполняется до проектной отметки за один сезон. Каждую весну в чаше водохранилища задерживается 1/3 весеннего стока. К концу зимы в результате работы гидроэлектростанции уровень воды в водохранилище понижается на 7-8 м, площадь водохранилища сокращается почти в 3 раза, а объем становится меньше в 4 раза (Дубровин и др., 1959; Двинских, Китаев, 2008).

Химический состав вод водохранилища определяется как речным стоком (прежде всего рек Кама, Чусовая, Сылва) и результатом растворения и выщелачивания береговых обнажений, так и химическим составом вод, поступающих с крупных промышленных предприятий Пермского края. Общая минерализация вод составляет на разных участках от 50 до 500 мг/л. В приплотинном участке водоема минерализация всегда выше (70-450 мг/л), что вызвано смешением камских вод с сылвенско-чусовскими. Во внутригодовом ходе минерализации вод отмечаются четко выраженные периоды, соответствующие фазам гидрологического режима. Максимальные в году значения суммы ионов наблюдаются в период зимней сработки водохранилища; во время прохождения весеннего половодья минерализация вод водоема резко снижается; в летне-осенний период она заметно возрастает по сравнению с весенним периодом, но остается ниже, чем зимой. Во время прохождения осенних дождевых паводков возможно некоторое уменьшение суммы ионов (Китаев, Рочев, 2008).

В балансе основных ионов водоёма основными являются гидрокарбонаты ($HCO^{3-}$) - до 350 мг/л, сульфаты ($SO_4^{2-}$) - до 100 мг/л, ионы хлора ($Cl^-$) - до 120 мг/л, а также ионы кальция ($Ca^{2+}$) - до 60 мг/л и ионы магния ($Mg^{2+}$) - до 12 мг/л. Внутригодовой режим гидрохимических показателей обусловлен в основном гидрологическими особенностями водоема, а именно со смешением камских вод с более минерализованными водами Сылвенско-Чусовского плёса (Китаев, Рочев, 2008).

В зоопланктоне Камского водохранилища имеется более 70 видов и подвидов беспозвоночных, из них около трети составляляют коловратки, по четверти - копеподы (*Mesocyclops*, *Eudiaptomus*, *Eurytemora*, *Heterocope*) и ветвистоусые ракообразные (*Daphnia*, *Bosmina*, *Diaphanosoma*, *Sida*, *Chydorus*). Численность и биомасса организмов зоопланктона, также как и в других водоёмах, значительно изменяется по годам и зависит от конкретных биотопов (Хозяйкин, 2011).

Рыбное население Камского водохранилища представлено 42 видами рыб из 10 отрядов, 15 семейств и 7 разных фаунистических комплексов (Зиновьев, 2007). Основными объектами промысла служат лещ (более 40%), значительно меньше уловы чехони, плотвы и судака (порядка 10-15% каждого вида). Доля в уловах других видов значительно меньше, а мелкие частиковые рыбы и вовсе практически не добываются (Ельченкова, Светлакова, 2001). Также следует отметить, что в Камском водохранилище, особенно в первые 20 лет после его образования в больших количествах присутствовали гибриды карповых рыб (Пушкин, Светлакова, 1992; Зиновьев, 2007). Из рыб, расширяющий свой ареал по Каме можно выделить черноморско-каспийскую тюльку, иглу-рыбу, ротана, сома (*Silurus glanis*) и подкаменщика (*Cottus gobio*). Черноморско-каспийская тюлька, проникшая в водохранилища Камы во второй половине прошлого века, сформировала многочисленные популяции, но промышленный вылов происходит только в нижнем бьефе Камской ГЭС (Зиновьев, 2007). Также в ихтиофауне Камского водохранилища в последнее время происходят структурные перестройки в рыбном сообществе: снизилась численность леща, увеличилась доля прибрежных и прирусловых хищных рыб, возросло значение планктоноядных рыб пелагического комплекса. На фоне этого в питании хищных рыб основное значение приобрели малоценные короткоцикловые виды, в том числе и тюлька, составляющие в сумме более 90 % (Коняев, Костицын, 2001).

Лов тюльки в нижнем бьефе Камского водохранилища производился местной малой рыболовецкой бригадой осенью 2011 г. (рис.2.1).

# Глава 4.
# Биологическая характеристика тюльки при её натурализации в водохранилищах Верхней Волги

Среди водохранилищ Верхней Волги наибольшую по численности популяцию тюлька образовала в Рыбинском. Менее чем за 10 лет она распространилась по всему водоёму и освоила биотопы пелагиали, на которых до середины 1990-х годов господствовали синец и корюшка. Такое активное вселение позволяет выдвинуть предположение об окончательной успешной натурализации тюльки. Признаками этого процесса служит высокая численность, особое пространственное распределение стад, размерно-возрастной состав популяции, успешное размножение, а также ряд адаптивных изменений внутриклеточного метаболизма.

## 4.1. Распределение в водоёмах и размерно-возрастной состав уловов

Впервые тюлька была зарегистрирована в Рыбинском водохранилище в 1994 г. (Слынько и др., 2000). Одновременно с её вселением в водоёме стала сокращаться популяция корюшки *Osmerus eperlanus,* которая была одним из доминирующих видов пелагиали (рис. 4.1). За последующие пять лет численность тюльки была минимальна: летом 2000 г. уловы тюльки были невелики (в среднем по водохранилищу 11 экз. на 15-минутное траление), тогда как корюшка в уловах уже отсутствовала (Кияшко и др., 2006, 2012). Доминировала тюлька только на небольшом количестве станций, в большинстве случаев основу скоплений составляла молодь карповых, а также молодь и взрослые особи уклейки *Alburnus albunnis*, ряпушки *Coregonus albula* и чехони *Pelecus cultratus*. Тюлька предпочитала пелагические биотопы речных плёсов, где она нагуливалась и размножалась. В Центральном плёсе её уловы были нерегулярны и малочисленны. В последующие годы летом распределение тюльки оставалось прежним, однако уловы значительно возросли, их средние величины по плесам составили 300-500 экз. за 15 мин. лова. Сократилось абсолютное и относительное количество молоди карповых и окуневых, тюлька стала доминантом на большинстве обследованных участков (рис. 4.1).

Так же, как и в материнском водоеме (Каспийское море), в новых для неё условиях тюлька осталась короткоцикловым рано созревающим видом, что определяет возрастную структуру и динамику численности популяции. Поэтому в течение ряда лет на-блюдались значительные межгодовые флуктуации её численности как в летний, так и в осенний периоды, что характерно для многих видов с коротким жизненным циклом (Иванова, 1982; Криксунов, 1995). Особенно ярко эта закономерность прослеживалась на раннем этапе заселения водохранилища. Так, изначально небольшая по количеству популяция тюльки летом 2000 г. дала мощное потомство, и осенью за счет сеголетков уловы увеличились в десятки раз. Особи этого поколения созрели на следующий год (в возрасте 1+) и составили основу нерестового стада 2001 года. Однако затем численность популяции пошла на убыль и к 2004 г. достигла многолетнего минимума. Вероятными причинами такого снижения могут быть как выедание основных кормовых объектов, так и очень холодная зима 2003/04 годов, сопровождавшаяся многочисленными заморами рыб. После этой депрессии тюлька снова начала увеличивать свою численность и вновь заняла доминирующее положение в комплексе пелагических рыб (рис. 4.1). В 2010 г. вновь наблюдалась депрессия популяции тюльки, причиной чему, вероятно, стало аномально жаркое лето, что будет подробнее обсуждено далее.

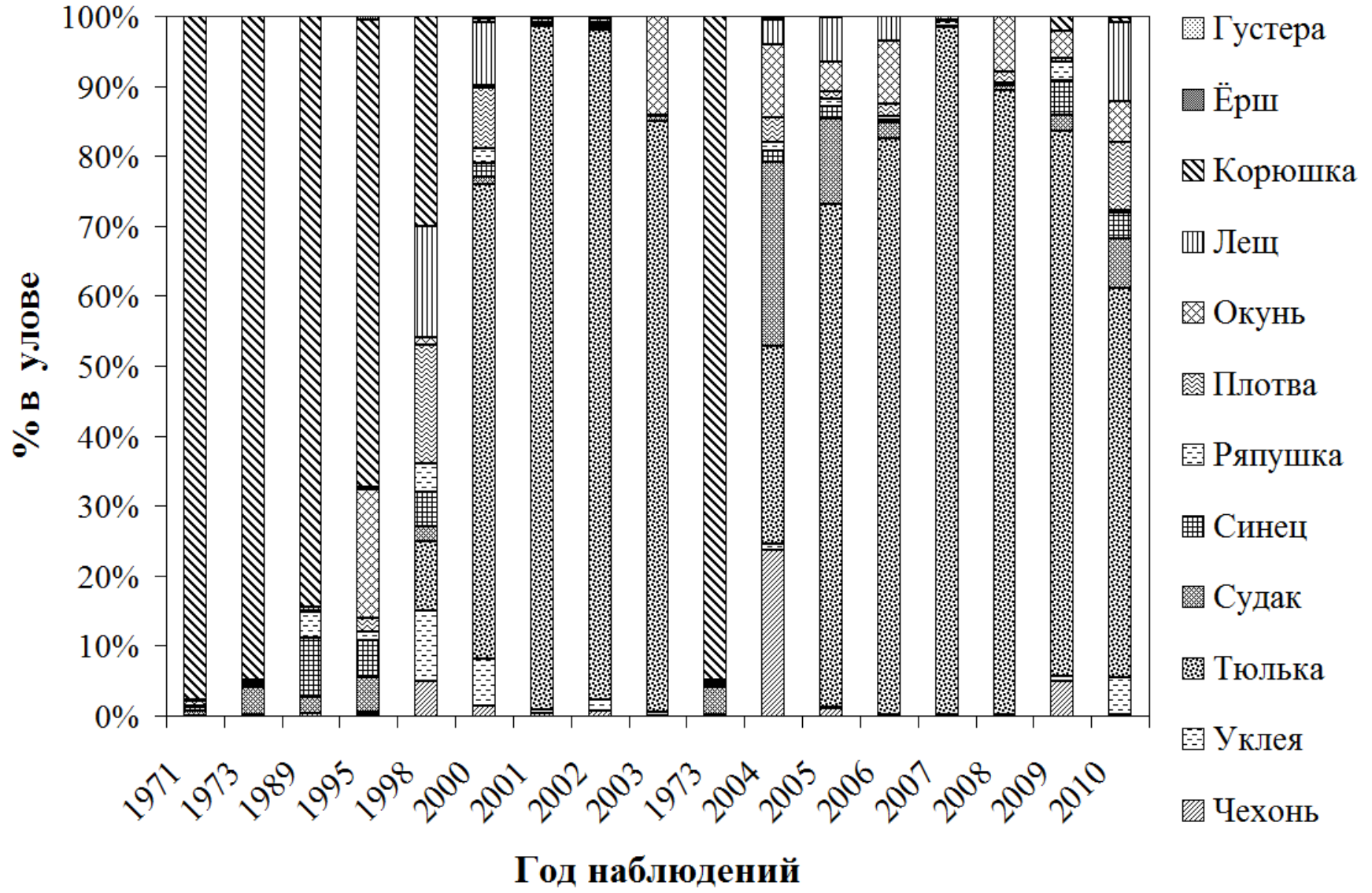

**Рисунок 4.1.** Относительная численность различных видов рыб в осенних уловах пелагического трала в Рыбинском водохранилище

Как отмечается в работе В.И. Кияшко с соавт. (2006), тюлька по своим трофоэкологическим характеристикам, также как и корюшка,

является типичным планктофагом. Биотопы, которые предпочитала корюшка, и биотопы, которые реально в настоящее время освоила тюлька, совпадают лишь частично. В летний период б*о*льшая часть популяции тюльки сосредоточена в верховьях речных плёсов, а также на участках Центрального плёса, которые мало подвержены ветровым волнениям, тогда как основные нагульные скопления корюшки были сосредоточены в Центральном плёсе (Пермитин, Половков, 1978). Наибольшие нерестовые скопления тюльки отмечены в речных плёсах и на некоторых защищённых от волнения биотопах Главного плёса. Следует отметить, что осенью численность тюльки в уловах значительно увеличивается не только в речных плёсах, но и в Центральном. В уловах доминируют подросшие к этому времени сеголетки, составляющие до 80% улова (Карабанов и др., 2010).

В других северных водохранилищах динамика численности популяций черноморско-каспийской тюльки имеет свои особенности. Так, в Горьковском водохранилище популяция тюльки подвержена колебаниям численности, аналогичным таковым для Рыбинского водохранилища. В русловой части Шекснинского водохранилища после вселения в 2001 г. популяция тюльки сильно уменьшилась и к настоящему времени её доля не превышает 1% от всего рыбного населения пелагиали. Вероятно, в этом самом северном водоёме распространения тюльки её самовоспроизводство затруднено, а высокая численность пищевых конкурентов (мелкий окунь) и хищников (судак) ещё более усугубляет ситуацию. В Угличском и Иваньковском водохранилищах тюлька заняла свою экологическую нишу и не испытывает значительных колебаний численности. Вероятно, это объясняется небольшим размером пелагиали водоёмов, что географически ограничивает распространение данной рыбы.

За многолетний период наблюдений в уловах пелагического трала длина тела тюльки составляла от 18 до 127 мм. Кривая распределения особей по размерам имеет один, четко выраженный, и второй, менее выраженный, пики. На начальном этапе заселения тюлькой Рыбинского водохранилища в мае-июне 2000/02 гг. в целом по водоёму модальную группу первого пика, с некоторыми вариациями модальных классов в разные годы, составляли особи длиной 40–60 мм. Второй пик был представлен особями длиной более 80 мм, относительное количество которых в исследованные годы варьировало от 5 до 15% (рис. 4.2.*а*). Среди рыб первой размерной группы обычно наблюдается равное соотношение полов, а во второй группе, как правило, преобладают самки.

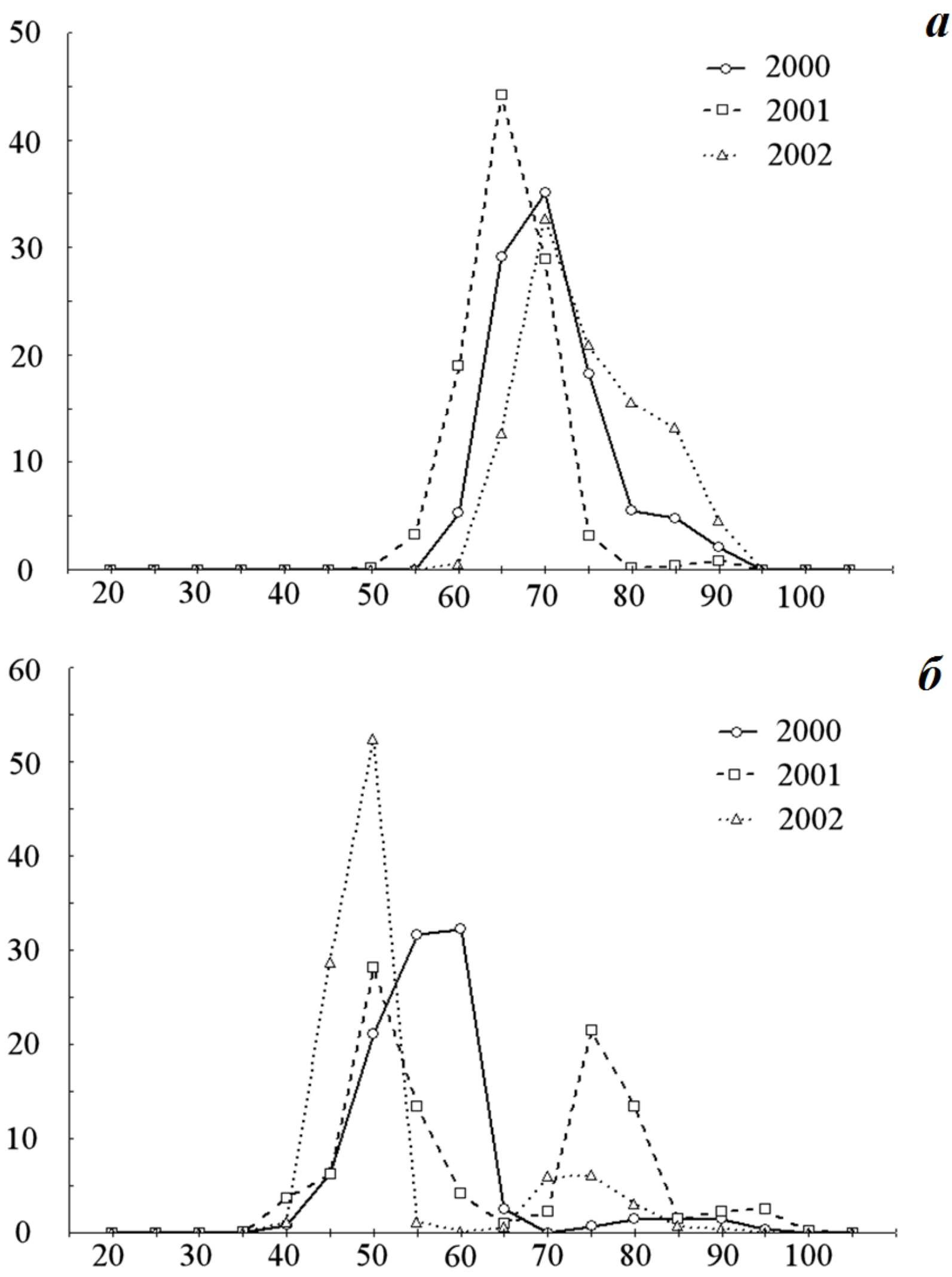

**Рисунок 4.2.** Размерный состав уловов тюльки *C. cultriventris* в Рыбинском водохранилище. Период 2000-2002 гг.: ***а***– июнь, ***б***– сентябрь. По оси абцисс - длина тела рыбы до конца чешуйного покрова, мм; по оси ординат - процент рыб в улове (по данным В.И. Кияшко)

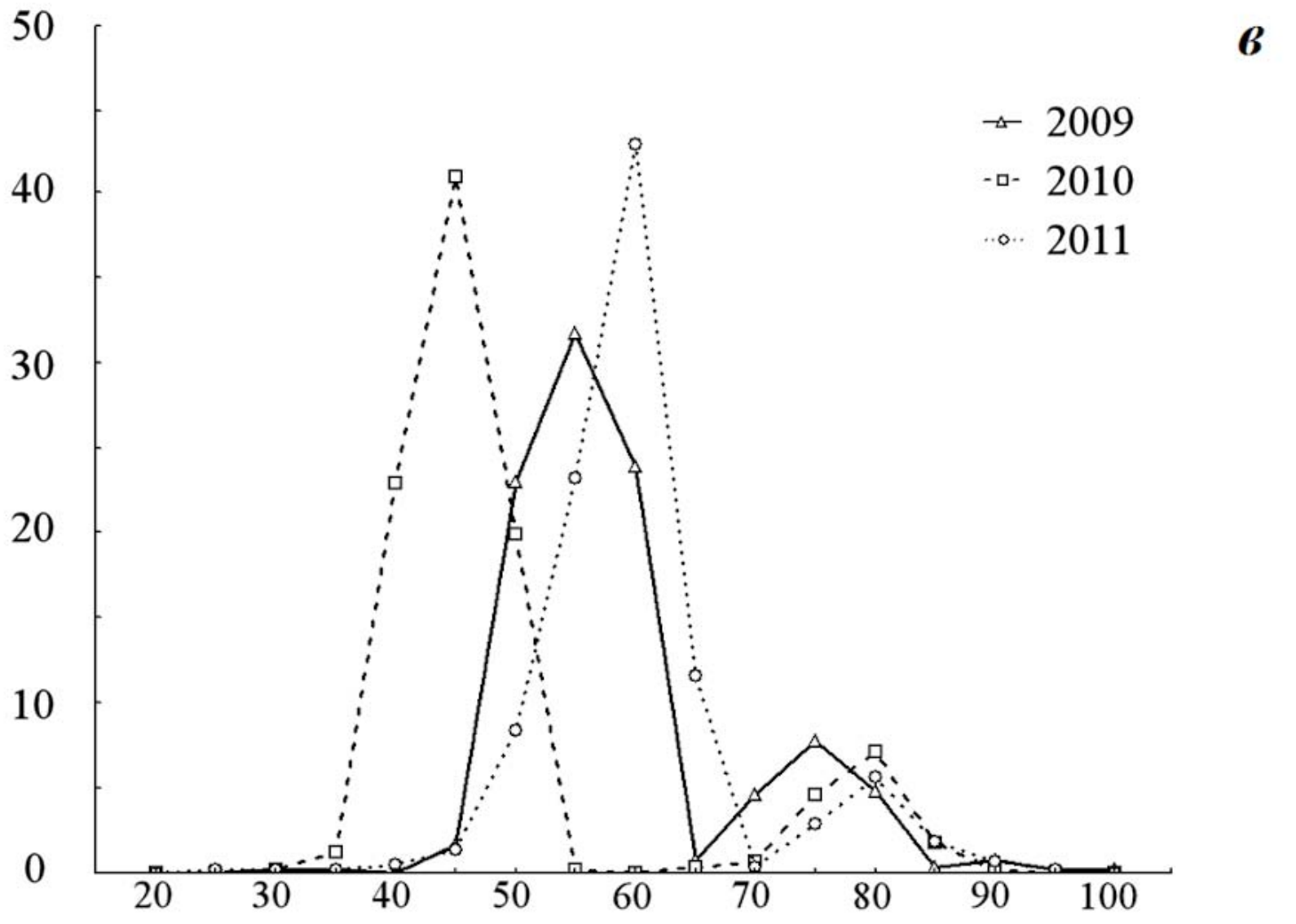

**Рисунок 4.2.** Продолжение. ***в*** – размерный состав уловов тюльки *C. cultriventris* по результатам осенних тралений в 2009-2011 гг. По оси абцисс - длина тела рыбы до конца чешуйного покрова, мм; по оси ординат - процент рыб в улове

Во второй половине августа и в сентябре в уловах тюльки появляются сеголетки, длина тела которых к этому времени варьирует от 20 до 50 мм. Кривая размерного состава уловов имеет четко выраженный двухвершинный характер. Первый пик в 2000/03 гг. был представлен доминирующей группой 35–55 мм, второй - 70–85 мм (рис. 4.2.*б*). Рыбы второй размерной группы имеют гонады VI-II стадии, что свидетельствует о завершении нереста тюльки к этому времени.

В сентябре-октябре размерный состав уловов также описывается двухвершинной кривой. В уловах возрастает доля сеголеток, размеры которых увеличиваются на 10–15 мм.

На современном этапе размерный состав уловов в августе-сентябре также описывается двухвершинной кривой (рис. 4.2.*в*). Однако размерные характеристики каждой группы могут существенно варьировать в разные годы. Так, в 2010 г. первая группа была представлена в массе особями длиной 40-50 мм, вторая - 75-85 мм. В следующем, 2011 г. средний размер особей сместился на 2 размерных класса: первая группа была представлена рыбами с длиной 55-65 мм, при этом размеры второй (старшей) группы остались такими же - 75-85 мм. Вероятно, такое отставание в размерах сеголеток в 2010 г. связано с очень жарким летом, когда происходил значительный прогрев воды в водохранилище (более +25°С), а в результате развития сине-зелёных водорослей и дефицита растворённого кислорода происходили даже «летние заморы» рыбы.

Таким образом, за десять лет с момента вселения в Рыбинском водохранилище произошло незначительное увеличение средних размеров как сеголетков, так и половозрелых особей тюльки. Вместе с тем, размерные характеристики сеголеток больше всего зависят от условий нереста и нагула (для данного периода, прежде всего - от температурного режима в водоёме) (Kiyasko et al., 2010).

В соотношении между запасом и пополнением в основном преобладают особи генерации текущего года (пополнение). Иногда доли запаса и пополнения примерно равны. Известно, что соотношение между численностью родителей и потомков не выражается прямой пропорциональной связью и зависит от многих биотических и абиотических факторов (Мина, Клевезаль, 1976; Криксунов, 1995). В 2000 г. при малочисленном нерестовом стаде (уловы в период размножения составляли в среднем 11 экз./траление) количество сеголеток в осенних уловах было сопоставимо с таковым в 2001-2002 гг., когда численность родительского стада по сравнению с 2000 г. возросла в 50 раз (Кияшко и др., 2006). Осенью 2004 г. сеголетки тюльки преобладали над половозрелыми особями более чем в 3 раза, в 2007 г. - в 2 раза, в 2009 г. - в 2,5 раза, и в 2011 г. - в 3 раза.

Аналогичная размерная структура характерна и для популяций водохранилищ Средней Волги: в Куйбышевском, Горьковском и Чебоксарском водохранилищах в сентябре в уловах доминируют сеголетки. Однако в Средневолжских водохранилищах немалую долю среди сеголеток составляют особи менее 40 мм, тогда как в расположенном севернее Рыбинском водохранилище среди сеголетков четко выражена модальная группа размером 40-50 мм (Кияшко и др., 2006). По всей вероятности это связано с более длительным вегетационным периодом в водоемах Средней Волги и, соответственно, более продолжительным нерестовым периодом.

## 4.2. Возрастная структура, темп роста и морфология тюльки

По данным, любезно предоставленным В.И. Кияшко, осенью уловы тюльки в Рыбинском водохранилище представлены тремя возрастными группами: сеголетками, двух- и трёхлетками. Если в июне-июле основой популяции служат двухлетки, то осенью в уловах сеголетки составляли более 50%. Во всех сравниваемых вариантах сеголетки генераций 2000/08 гг., также как и двухлетки, достоверно различаются по длине и массе тела. Различия между трёхлетками были не всегда достоверны. К концу вегетационного периода длина сеголеток тюльки даже в одном поколении варьирует в значительных пределах, что, по всей вероятности, связано с порционностью икрометания и длительным нерестовым периодом. Размеры двух- и трёхлетних особей по сравнению с сеголетками изменяются в меньших пределах. Наибольшие линейные приросты тюльки наблюдаются на первом году жизни. В отличие от линейного роста, приросты массы тела максимальны в течение второго года жизни рыб. Самый высокий темп роста тюльки наблюдался в 2000 г., что, возможно, обусловлено низкой численностью популяции тюльки и отсутствием одного из основных потребителей зоопланктона в Рыбинском водохранилище - корюшки (Кияшко, Слынько, 2003). В течение последующих лет на фоне увеличения численности тюльки происходило некоторое снижение ее линейных размеров (Кияшко и др., 2006).

Рост средней массы тела тюльки в Рыбинском водохранилище в последние годы также имеет тенденцию к снижению, но в отличие от линейного роста, который снижается прямолинейно во всех возрастных группах, рост массы особенно значительно снизился у двухлеток при почти не изменившейся массе у сеголеток. Вероятно, это связано с изменениями в обеспеченности пищей, на что рыба быстро и отчетливо реагирует изменением массы, причем её изменчивость гораздо больше изменчивости длины, особенно у старших особей после достижения половой зрелости (Чугунова, 1961).

Темп линейного роста тюльки характеризуется большой вариабельностью во времени и пространстве как в пределах естественного ареала, так и во вновь освоенных водоемах. В настоящее время среди водохранилищ Средней и Верхней Волги средние размеры рыб, составляющих основу популяций - сеголеток и двухлеток, были наибольшими в Рыбинском водохранилище. На этапе увеличения численности популяции по темпу линейного роста тюлька Рыбинского водохранилища превосходила популяции других водохранилищ Волги, а также популяции из Среднего Каспия и Кременчугского водохранилища (Кияшко и др., 2006). По всей видимости, повышенный темп роста тюльки Рыбинского водохра-нилища был связан с начальным этапом становления её популяции. Наблюдаемое в настоящее время снижение темпа роста тюльки в Рыбинском водохранинилище происходит на фоне увеличения её численности. Особо надо подчеркнуть, что процесс приспособления рыбинской тюльки по рассматриваемым параметрам происходит с большей скоростью и большей отчетливостью изменений, чем в более южных водохранилищах как волжского, так и днепровского каскадов, что вполне может быть прямым следствием высокоширотной зональности Рыбинского водохранилища и более резких колебаний климатических условий.

По данным А.Н. Касьянова (2009) популяции тюльки Верхней Волги по остеологическим показателям достоверно отличаются от популяций из исторической части ареала. Также установлено, что при продвижении на север в новообразованных популяциях тюльки увеличиваются средние значения числа позвонков в туловищном отделе и общего числа позвонков. Предполагается, что выявленная горизонтально-ступенчатая клина по общему числу позвонков, по-видимому, обусловлена различиями в температурном режиме водоёмов в период морфогенеза осевого скелета. Таким образом, в процессе адаптаций популяций-вселнецев к новым условиям обитания произошли изменения и меристических признаков, что может свидетельствовать о глубоких качественных изменениях в Верхневолжских популяциях.

### 4.3. Трофоэкологическая характеристика тюльки Верхней Волги

Детально питание тюльки в водохранилищах Верхней Волги изучено В.И. Кияшко (Кияшко, Слынько, 2003; Кияшко, 2004; Кияшко и др., 2007). До середины 1990-х годов в пелагиали Верхневолжских водохранилищ доминировали 2 вида - синец и корюшка (рис. 4.1). Синец, как правило, обитал в нижних горизонтах пелагиали, корюшка - в средних (Поддубный, 1971; Пермитин, Половков, 1978). Так же как и корюшка, тюлька расселилась по водоёму неравномерно в горизонтальном и вертикальном направлениях, занимая наиболее продуктивные биотопы пелагиали (зоны аккумуляции биомасс), сконцентрировавшись в горизонте 2-7 м. Появление тюльки в водохранилище в 1994 г. совпало с развитием процесса глобального потепления, который, вероятно, стал одной из основных причин резкого снижения численности популяции северного вселенца - корюшки. Вселившись в водоём в период депрессионного состояния популяции корюшки, тюлька попала в благоприятную для нее трофическую ситуацию, заняла бывшие «корюшковые» биотопы, включившись как значимый элемент в пищевую цепь рыб-планктофагов.

В условиях Рыбинского водохранилища, подобно популяциям водохранилищ Нижней и Средней Волги, тюлька предпочитает питаться крупными пелагическими рачками (Кияшко, 2004). Наиболее часто в уловах тюлька сопряжена с молодью карповых, для которых пелагиаль представляет временное местообитание. Это факультативные планктофаги - молодь леща, плотвы, разновозрастная уклея (частота встречаемости тюльки с каждым из этих видов соответственно: 74, 62,8 и 55%). Сопряженность тюльки с другой группой типичных планктофагов - синцом, корюшкой и ряпушкой, а также факультативных планктофагов, таких как молодь окуня и судака, несколько ниже и колеблется в пределах 46-55%. Несмотря на высокую взаимосвязь, относительные величины уловов тюльки и других видов в скоплениях имеют отрицательную корреляцию, однако достоверной эта связь была у тюльки только с молодью леща и уклейкой (соответственно $r_s$= -0,55 и $r_s$= -0,33). Из этого следует, что либо тюлька при высокой её численности вытесняет с нагульных участков эти виды, либо они в силу особенностей биологии предпочитают откармливаться на разных с ней биотопах.

Данные по питанию рыб показали значительно большее сходство состава пищи тюльки с молодью карповых в тех скоплениях, где тюлька является «супердоминантом», в отличие от скоплений, где тюлька выступает в статусе «сопутствующего» вида. Кроме того, осенью, когда численность тюльки увеличивается в десятки раз, в скоплениях первого типа в кишечниках молоди карповых пища отсутствует, а в скоплениях второго типа эти виды продолжают питаться. Таким образом, тюлька как новый элемент экосистемы оказывает существенное влияние на условия нагула и спектры питания видов-аборигенов. Это действие локально и зависит от плотности тюльки в скоплении.

Иным образом складываются отношения тюльки с малочисленной пелагической группой видов - корюшкой, ряпушкой, молодью синца, судака и окуня. Независимо от статуса тюльки в скоплениях, сходство пищи у неё с этими видами в большинстве случаев высокое. Поэтому в годы при совпадении урожайных поколений происходит обострение пищевых отношение этих видов-аборигенов с тюлькой.

Замена в рыбном сообществе корюшки на тюльку резко изменила ситуацию в пелагиали Рыбинского водохранилища. Если на первом месте в рационе корюшки была *Bosmina longispina,* а на втором - *Bythotrephes, Leptodora* (Иванова, 1982), то тюлька предпочитает *Daphnia, Bythotrephes, Leptodora, Heterocope* (Кияшко, 2004). В Рыбинском водохранилище сеголетки этого вида-вселенца питаются преимущественно мелкими ветвистоусыми рачками *Bosmina, Daphnia, Chydorus* и молодью Cyclopoida, которые суммарно составляют свыше 90% массы пищевого комка. Взрослые особи избирают в пищу главным образом крупных ветвистоусых рачков - *Heterocope, L. kindtii* и *B. longimanus,* отдавая предпочтение последнему виду в осенний период. Мелкие представители планктона (*Bosmina, Chydorus,* Cyclopoida) для них служат второстепенным кормом. Возрастные особенности питания рыбинской тюльки сохраняются в различные по гидрологическим условиям годы (Кияшко и др., 2012).

Также в настоящее время можно утверждать, что тюлька, расселившись по всему Рыбинскому водохранилищу, стала важным звеном в пищевой цепи планктон - рыбы-планктофаги - хищные рыбы (Кияшко, Степанов, 2003; Степанов, Кияшко, 2008). Аналогичная роль тюльки в ихтиоценозах отмечена и для камских водохранилищ (Коняев, Костицын, 2001).

Так как тюлька является короткоцикловым, рано созревающим видом, достигшим высокой численности в Верхневолжских водохранилищах, в отдельные благоприятные годы тюлька создаёт существенную конкуренцию для рыб пелагического комплекса в целом, а особенно для молоди окунёвых и карповых. Снижение показателей линейного и весового роста тюльки можно расценивать как результат внутривидовой конкуренции за кормовые ресурсы.

В целом, популяция тюльки оказывает заметное влияние на трофоэкологическую ситуацию в пелагиали водоёма. В то же время в период расцвета популяции корюшки (1960-70-е годы) рыбы-планктофаги не сильно влияли на обилие их кормовых организмов (Половкова, Пермитин, 1981). Отрицательная (в экономическом плане) роль тюльки в водных экосистемах связана с занятием трофических ниш ценных в промысловом отношении рыб-планктофагов и молоди всех видов. С другой стороны, положительный экономический эффект от вселения тюльки связан с улучшением обеспеченностью кормом промысловых хищных рыб, что позволяет говорить о благоприятных (с рыбопромысловой точки зрения) последствиях от данной биологической инвазии.

## 4.4. Особенности воспроизводства черноморско-каспийской тюльки

Особенности воспроизводства тюльки Верхневолжских водохранилищ и в новой, и в исторической части ареала рассмотрены в работе В.В. Осипова и В.И. Кияшко (2006). Этими авторами указывается, что половозрелой тюлька Рыбинского водохранилища становится в возрасте 1+ при длине около 50 мм. На первом году жизни соотношение самцов и самок примерно равное, на втором году жизни преобладают самки. По-видимому, как и в исторической части ареала, самцы имеют более короткий жизненный цикл, чем самки. Предельный возраст тюльки Рыбинского водохранилища составляет 3,5 года.

Тюлька Рыбинского водохранилища нерестится порционно, порций икры как минимум две. В Каспийском и Азовском морях тюлька созревает на втором году жизни, выметывает 3-4 порции икры (Световидов, 1952; Михман, 1972). Диаметр икринок тюльки в Рыбинском водохранилище в среднем составляет 0,35 мм, возраст производителей в данном случае не имеет сколько-нибудь значительного влияния на размеры икры. Для сравнения, в Воткинском водохранилище диаметр икринок у самок тюльки в годовалом возрасте составляет в среднем 0,37 мм (Антонова, Пушкин, 1985), что несколько больше, чем у тюльки Рыбинского водохранилища.

Абсолютная плодовитость тюльки Рыбинского водохранилища в среднем составляет 25,4 тыс. икринок; у годовиков в среднем 8,5 тыс. икринок, у двухгодовиков в среднем 30,7 тыс. икринок.

Сравнительный анализ воспроизводительной способности тюльки разных водоёмов показал некоторые изменения абсолютной плодовитости по мере продвижения в пресные воды: наблюдается повышение относительной плодовитости тюльки при её продвижении с юга на север, вследствие чего увеличивается воспроизводительная способность самок. Для абсолютной плодовитости данная тенденция не наблюдается (Осипов, Кияшко, 2006; Dgebuadze et al., 2008).

Имеются данные по местам и срокам нереста тюльки в условиях северных водохранилищ (Степанов, 2011). На примере Шекснинского плёса Рыбинского водохранилища установлено, что нерест происходит в первой декаде июня, в вечернее время (от 21 до 23 часов). Вторая, меньшая, порция икры вымётывается, как правило, в июле. К моменту начала нереста наблюдаются наибольшие концентрации тюльки в поверхностном слое воды. Массовый нерест тюльки происходит как в открытых, так и в прирусловых участках на глубинах не менее 4 м при прогреве поверхностного слоя воды до +22°C.

## 4.5. Физиолого-биохимические особенности популяций тюльки Верхней Волги

Данных по физиологии и биохимии тюльки в новоприобретённых частях ареала очень мало. В работе, выполненной под руководством И.Л. Головановой (Голованова и др., 2007) проведено исследование активности ферментов, расщепляющих углеводные компоненты корма. Установлено, что значения общей амилолитической активности у тюльки верхневолжских водохранилищ близки к таковым у других видов рыб-планктофагов Рыбинского водохранилища. Значения общей амилолитической активности в кишечнике молоди тюльки в 1,4-2 раза выше, чем у половозрелых особей; активность сахаразы изменяется менее значительно.

Максимальный уровень общей амилолитической активности и активности сахаразы отмечен у представителей новых популяций тюльки из Рыбинского и Иваньковского водохранилищ. Не исключено, что низкий уровень ферментативной активности у тюльки Чебоксарского водохранилища и Северного Каспия обусловлен более низкой интенсивностью питания и температурой воды в момент отлова. Разная величина температурного оптимума гидролиза полисахаридов у тюльки - вселенца Верхневолжских водохранилищ (50°C) и тюльки Каспийского моря (60°C), может отражать различия температурных условий обитания в материнских и приобретенных частях ареала данного вида. Эти данные могут свидетельствовать о возникновении сложных физиологических адаптаций у рыб в новообразованных северных популяциях тюльки.

Имеются данные (Мартемьянов, Борисовская, 2010), что концентрация различных ионов в плазме и мышечной ткани тюльки существенно отличается от параметров схожей по экологии *Alosa pseudoharengus*, вселившейся в Великие Озёра из Атлантического океана (Stanley, Colby, 1971). Морские костистые рыбы отличаются от стеногалинных пресноводных видов главным образом более высоким содержанием ионов натрия во внутренней среде. Этот показатель у тюльки составил 131,8±3,3 ммоль/л. У 16 аборигенных видов стеногалинных пресноводных рыб Рыбинского водохранилища содержание ионов натрия в плазме крови поддерживается в диапазоне от 113,2±0,9 ммоль/л (уклейка) до 149,5±2,6 ммоль/л (окунь). Следовательно, концентрация натрия в плазме крови тюльки характерна для представителей стеногалинной пресноводной ихтиофауны Рыбинского водохранилища. Уровень натрия во внутренней среде тюльки существенно ниже (на 40 ммоль/л), чем у американского вселенца морского происхождения.

По сравнению с аборигенными видами рыб Рыбинского водохранилища, у тюльки в эритроцитах самое высокое содержание натрия и низкое магния. В опытах *in vitro* показано, что увеличение концентрации натрия в эритроцитах рыб ведет к увеличению кислороднесущей ёмкости крови, а снижение концентрации магния - к уменьшению сродства гемоглобина к кислороду, способствуя лучшей отдачи этого элемента тканям. Исходя из этого, можно полагать, что высокий уровень ионов натрия в эритроцитах тюльки обуславливает повышенную кислороднесущую ёмкость крови, свидетельствуя об увеличенных потребностях этого вида в кислороде. Низкий уровень ионов магния в эритроцитах тюльки характеризует высокую способность гемоглобина отдавать кислород тканям, что явно имеет адаптивное значение.

Также В.И. Мартемьяновым (устн. сообщ.) установлено, что быстрое и чрезмерное обессоливание крови в начальный период стресса на фоне замедленного действия компенсаторных процессов делает тюльку очень уязвимой к резким изменениям факторов среды. Вероятно, эта причина способствует периодической массовой гибели тюльки.

Большая работа по изучению липидного обмена тюльки проделана В.В. Халько (2007). Им установлено, что общее содержание и фракционный состав липидов в мышцах и в целом организме тюльки в Рыбинском водохранилище подвержены размерно-возрастным изменениям, характерным для этого вида и в водоёмах материнского ареала, а также и для других видов рыб в нагульный период. С увеличением размера (возраста) тюльки в её мышцах и в целом организме возрастает общее содержание липидов и триацилглицеринов, снижается содержание структурных липидов (фосфолипидов и холестерина). Такая направленность изменений показателей липидного обмена у рыб с возрастом обусловлена снижающейся по мере старения организма интенсивностью энергетического и белкового обмена, что способствует увеличению интенсивности депонирования запасных липидов и процесса жиронакопления в целом. Наряду с размерно-возрастными изменениями показателей липидного обмена у тюльки в Рыбинском водохранилище выявлены особенности его межгодовых изменений, характер которых у взрослых и молодых особей различен. У взрослых рыб происходит устойчивое снижение жирности мышечных тканей и содержания в них триацилглицеринов, что наиболее заметно в группе двухлетних особей. Это может свидетельствовать о нарастающем ухудшении физиолого-биохимического состояния половозрелых рыб в популяции тюльки и, в первую очередь, особей самой многочисленной возрастной группы, у которых осенью 2005 г. величина жировых запасов в мышцах (3,7±1,5%) приблизилась к ее критическому для сельдевых рыб значению (2-3%) (Kondo, 1974). Причина негативных изменений физиолого-биохимического состояния взрослой тюльки, возможно, заключается в несбалансированности её численности и запасов предпочитаемой ею кормовой фракции зоопланктона на современной стадии развития экосистемы Рыбинского водохранилища, характеризующейся измельчанием представителей зоопланктонного сообщества и снижением их общей биомассы (Лазарева, 2005).

В отличие от взрослых особей, уровень и структура жировых запасов, накапливаемых в теле молоди тюльки к осени, зависят от режима сработки объёма воды в Рыбинском водохранилище, оказывающего прямое влияние на внутри- и межгодовые изменения общих запасов зоопланктона в пелагиали водоема. Происходящее на современной стадии развития экосистемы Рыбинского водохранилища измельчание представителей зоопланктона и постепенное снижение биомассы (Лазарева, 2005, 2010), по-видимому, не является определяющим фактором для условий нагула молоди тюльки, основная пища которой состоит именно из мелких планктонных рачков (Кияшко, 2004). Тем не менее, можно предположить, что по мере стабилизации биомассы всех структурных компонентов в зоопланктоценозе Рыбинского водохранилища жирность потомства тюльки в последующие годы будет постепенно снижаться до некоторого предела (Халько, 2007).

На основании анализа приведённых фактов В.В. Халько (2007) предполагает, что отмеченные негативные явления в липидном обмене могут привести к увеличению естественной смертности сеголеток в течение зимовки и сокращению в результате этого объёма пополнения. Однако наблюдаемые данные по численности уловов тюльки не позволяют с твёрдой уверенностью высказаться в поддержку столь пессимистического прогноза. Численность тюльки в Рыбинском водохранилище испытывает значительные межгодовые флуктуации, одной из причин которых могут служить и особенности липидного обмена. Вероятно, экологическая пластичность вида в совокупности с высокой плодовитостью позволяет популяции быстро восстановиться после неблагоприятного периода. Возможно, наблюдаемая динамика липидного обмена тюльки в новообразованной популяции связана с адаптивными перестройками к пресноводному образу жизни (Шатуновский, 1980; El Cafsi et al., 2003).

## 4.6. Особенности распространения тюльки в северных водоёмах

Изложенные выше данные свидетельствуют о больших адаптационных перестройках у тюльки недавно возникших популяций. Эти изменения закрепились как на внутриклеточном уровне (изменение мембранной проницаемости и особенности липидного обмена), так и на уровне физиологических реакций (изменение активности пищеварительных ферментов), и даже на консервативном морфологическом уровне (остеологические параметры).

Вероятно, ключевую роль в ограничении распространения тюльки далее на север играют изменение абиотических (температура, солёность воды, гидрологический режим) и биотических факторов (количество и качество потребляемой пищи, наличие пищевых конкурентов и хищников).

С точки зрения В.И. Мартемьянова (Мартемьянов, Борисовская, 2010), при уменьшении минерализации в северных водоёмах с уменьшением концентрации натрия в воде у тюльки возрастает напряженность внутриклеточных систем, обеспечивающих натриевый гомеостаз и степень его уязвимости при резких изменениях минерализации среды. Вероятно, это обстоятельство препятствует распространению тюльки на север в пресноводные водоемы с более низким содержанием натрия в воде.

Другим возможным препятствие к распространению тюльки далее на север может служить особенность гидрологии и морфологии Белого озера. По предположению Ю.С. Решетникова (устн. сообщ.) пелагическая икра тюльки в маломинерализованном Белом озере, вероятно, постепенно погружается в придонный слой. Из-за гидрологических особенностей в этой зоне происходит постоянное взбалтывание и перемешивание грунта, что приводит к механическим повреждениям и гибели икры.

Относительно биотических факторов можно отметить снижение численности и размеров кормовых объектов в Шекснинском водохранилище по сравнению с Рыбинским и Горьковским (Волга…, 1979; Экологические…, 2001). Также в более северных водоёмах возрастает численность пищевых конкурентов - корюшки, ряпушки и молоди окуня и значительно увеличивается гнёт хищников - судака и крупного окуня.

На основании многолетних работ сотрудников ИБВВ РАН можно сделать вывод о снижении к настоящему моменту, по сравнению с начальным этапом вселения, показателей массы тела, жирности мышечной ткани и содержания в ней запасных липидов у взрослых рыб всех размерно-возрастных групп. Причиной этого могло служить ухудшение обеспеченности их предпочитаемым кормом (крупными ветвистоусыми рачками), что происходит в соответствии с выявленной В.И. Лазаревой (2005) тенденцией снижения общей биомассы и измельчания представителей зоопланктонного сообщества в пелагиали Рыбинского водохранилища.

Другим фактором, способным приводить к флуктуациям численности тюльки в Рыбинском водохранилище, могут выступать большие колебания температуры воды. Для Каспийского моря характерны значительные изменения температуры только поверхностного слоя, тогда как основная водная масса, обладая огромной теплоёмкостью, находится почти в статичном температурном режиме (Панин и др., 2005). Рыбинское водохранилище, несмотря на свои размеры, является мелководным водоёмом со значительными межсезонными и межгодовыми температурными колебаниями. Вероятно, более низкие температуры в летний сезон 2004 г. и крайне высокие - летом 2010 г., отрицательно сказались на нересте тюльки в водохранилище. Например, по сравнению с другими годами доля тюльки в осенних уловах (состоящих на 80% из сеголетков) в 2004 г. снизилась более чем в 2 раза (рис. 4.3). Это, вероятно, связано с более низкой (по сравнению со средними многолетними значениями - на 2,3°С) температурой воды.

В связи со значительным сокращением срока жизни тюльки в Верхневолжских водохранилищах по сравнению с исторической частью ареала в динамике численности популяции ключевую роль играет доля пополнения, которое составляют особи генерации текущего года. Вторым негативным фактором, обусловившим значительное депрессирование популяции тюльки, была суровая зима 2003/04 гг., сопровождавшаяся заморами рыбы. Вероятно, гибель части половозрелых особей в зимний период привела к снижению репродуктивного потенциала популяции, а холодное лето ещё более усугубило этот негативный процесс.

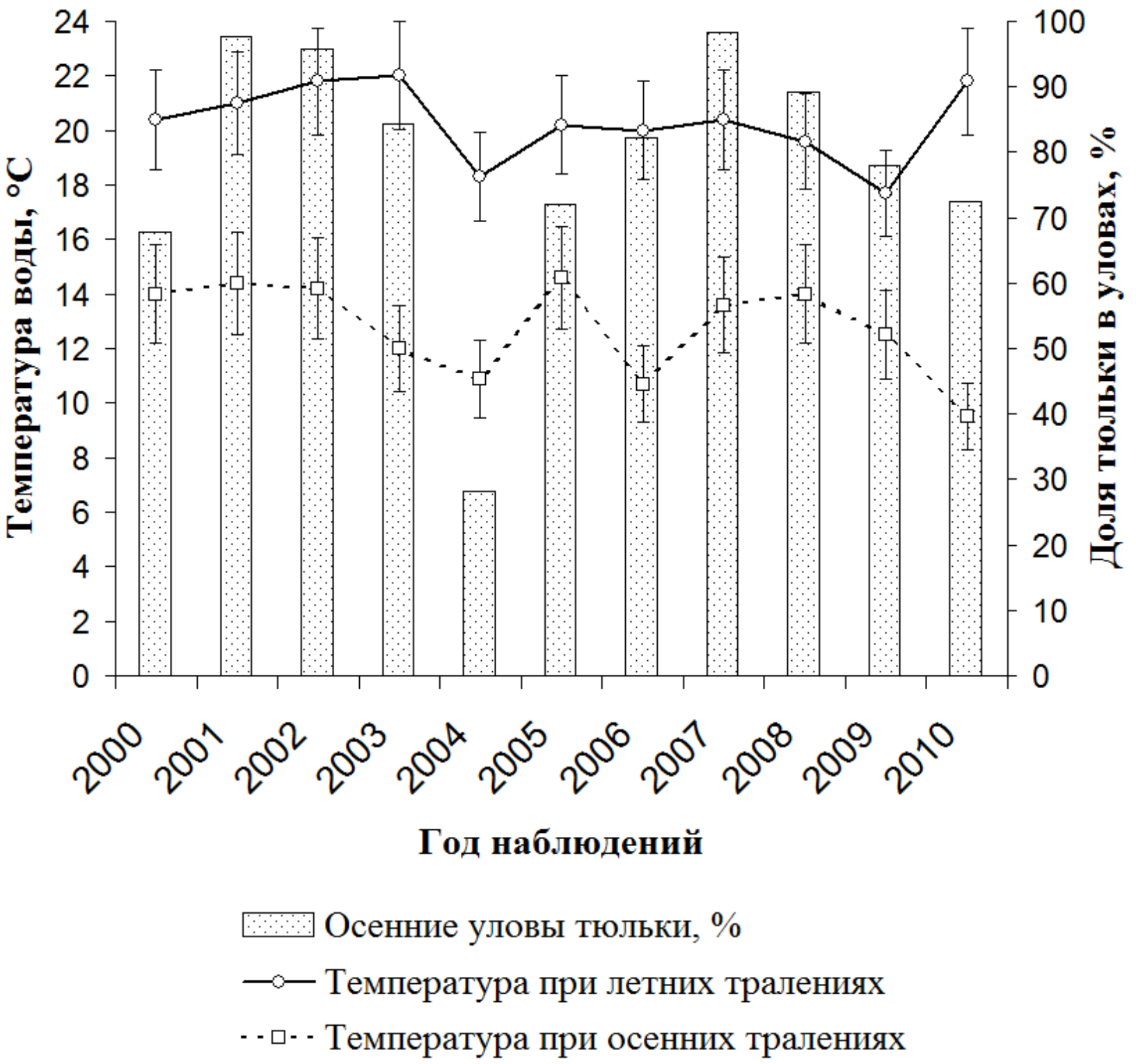

**Рисунок 4.3.** Зависимость осенних уловов тюльки от температуры поверхностного слоя воды при тралении

По данным В.И. Кияшко в 2000/02 и 2006/08 гг. наблюдалось нарастание численности тюльки до максимума, затем ее снижение до минимума в 2004 и 2010 гг. (рис. 4.4). Флуктуации численности отмечены как в летний, так и в осенний периоды, но наибольшие амплитуды колебаний выявлены осенью, когда в популяции доминировали сеголетки. Элиминация тюльки, если сравнивать осенние и летние уловы, происходила как в летний, так и в зимний периоды. Часть рыб погибала летом после первого нереста (возраст 1+). Оставшиеся особи этой возрастной группы перезимовывали и нерестились во второй раз, погибая вскоре после нереста в возрасте 2+. Однако, сравнение летних и осенних уловов даёт возможность предположить, что гибель производителей в летний период - явление не ежегодное. К сентябрю-октябрю в половине исследованных лет количество отнерестившихся производителей (возраст 1+, 2+) уменьшалось на 7–88%. В остальные годы их уловы осенью были даже несколько выше, чем летом.

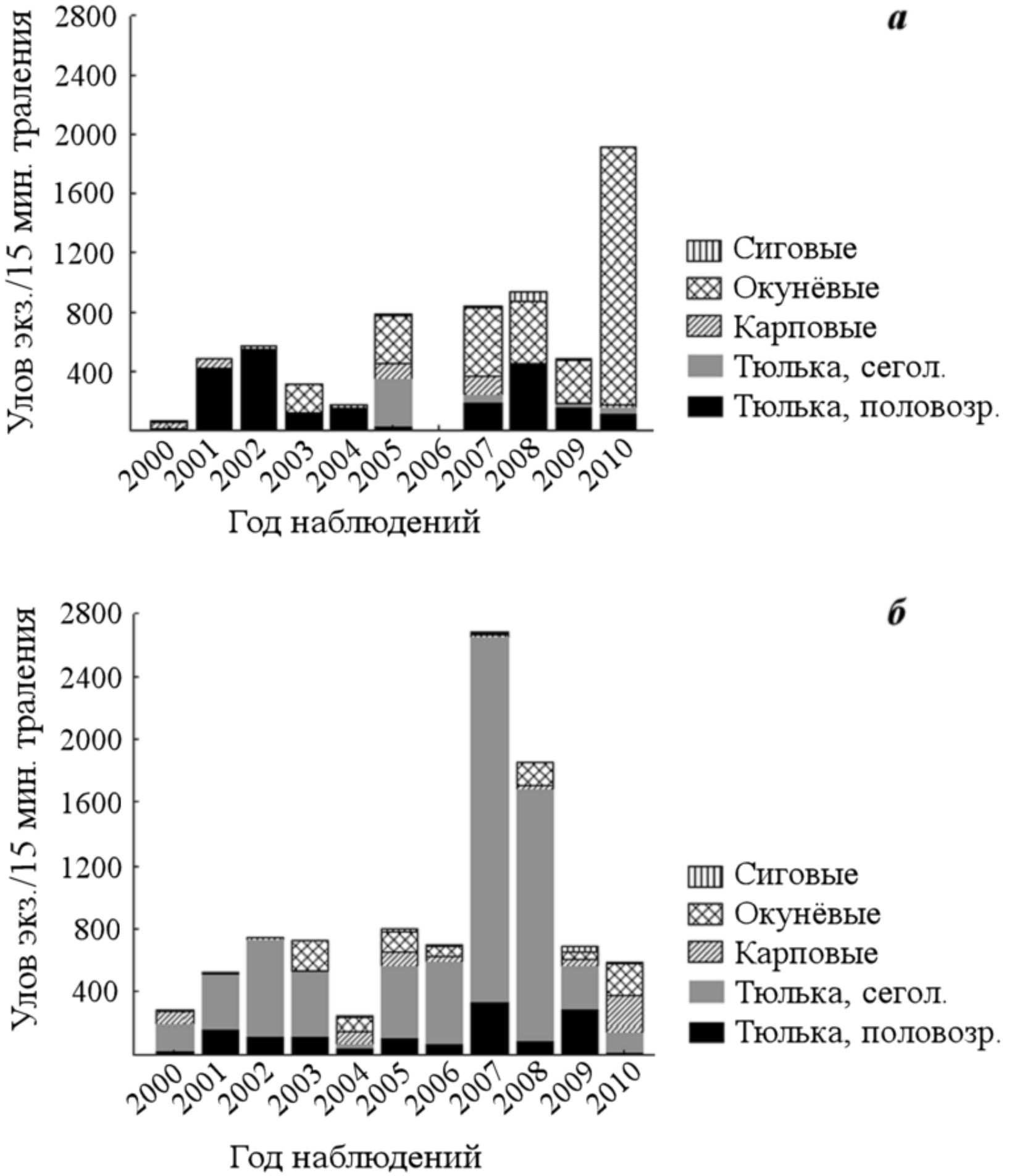

**Рисунок 4.4.** Величина и состав уловов пелагического трала в Рыбинском водохранилище в летний (***а***) и осенний (***б***) периоды (по: Кияшко и др., 2012)

В течение зимы уловы тюльки уменьшились в 3-6 раз в период 2002/05 гг. и в 3-11 раз в период 2006/09. Коэффициенты общей смертности с октября по июнь составили 66–73 % в первый период и 81–91 % во второй. В результате высокой смертности урожайных генераций в подледный сезон (например, 2007 и 2008 гг.) к лету следующего года их численность оказалась сопоставимой с таковой в

мало- и среднеурожайные годы. В мае в уловах преобладали особи с длиной тела более 50 мм: по-видимому, рыбы меньших размеров, как правило не переживают подледный сезон.

В заключение можно отметить, что занятие тюлькой важнейшей трофической ниши в сети питания пелагических рыб и её высокая численность делают крайне актуальным изучение адаптаций и осуществление прогноза динамики популяции данного вида в водохранилищах Верхней Волги.

## Глава 5.
## Генетико-биохимическая характеристика исследованных ферментных систем

Биохимическая генетика изоферментов является важным экспериментальным методом в современной экологической генетике. Это связано с тем, что наблюдаемая изменчивость биохимических маркеров отражает конкретные адаптационные особенности метаболических путей, что может служить, во многих случаях, индикатором физиологического состояния организма. Важным свойством многих ферментов служит наличие множественных молекулярных форм, катализирующих одну и ту же реакцию и, как правило, обладающих видоспецифичностью. В современной биохимической литературе под понятием «множественные молекулярные формы ферментов» (ММФФ) подразумевают все ферменты, встречающиеся у одного вида и катализирующие одну и ту же биохимическую реакцию. Термин «изофермент» применяется для тех ММФФ, которые возникают вследствие генетически обусловленных различий в первичной структуре белка, но не для форм, образованных в результате модификаций одной первичной последовательности. Таким образом, понятие «изофермент» подчеркивает генетическую основу различий и поэтому носит ограниченный характер, оно применяется только к тем множественным формам ферментов, которые кодируются определенным геном (или его аллелями). Генетические же причины, приводящие к возникновению изоферментов - множественность локусов и множественность аллелей генов. Кодирование изоферментов различными генами является весьма распространенным явлением.

В рекомендациях Международной комиссии по биохимической номенклатуре предложено употреблять термин «изофермент» в отношении ММФФ, обусловленных только генетическими причинами. Для всех остальных белков, обладающих одной и той же ферментативной активностью, остается термин ММФФ. В зарубежной литературе термин «изофермент» обычно относя к вариантам ферментов, состоящих из субъединиц и кодируемых больше, чем одним структурным геном, а для аллельных вариантов ферментов принят термин «аллозим». Для неферментативных белков, или ферментов с неустановленной генетической детерминаций, применяется термин «изоформа» (Prakash et al., 1969; Левонтин, 1978; Кирпичников, 1987; Глазко, 1988; Copeland, 2000; Schmidt, 2000). Номенклатура генетических локусов, аллелей и кодируемых ими ферментов приведена с учётом рекомендаций Дж. Шекли с соавт. (Shaklee et al., 1990).

Далее будут представлены некоторые генетико-биохимические характеристики изученных ферментных систем черноморско-каспийской тюльки *Clupeonella cultriventris* (Nordmann, 1840) и родственного морского вида - анчоусовидной тюльки *Clupeonella engrauliform*es (Borodin, 1904) из Северного Каспия.

### 5.1. Лактатдегидрогеназа, ЛДГ (LDH, E.C. 1.1.1.27)

Лактатдегидрогеназа - один из наиболее изученных ферментов. Он контролирует метаболизм лактата и пирувата при гликолизе. У высших позвоночных этот фермент имеет четвертичную структуру и является тетрамером, состоящим из двух типов субъединиц A и B (ранее обозначавшиеся как M и H). У большинства костистых рыб имеются три локуса - *A* и *B* (гомологичные соответствующим локусам млекопитающих), а также тканеспецифичный ген *C*, характерный только для рыб. В период сперматогенеза в семенниках определяется продукт независимого генного локуса *X*. В целом все изоферменты LDH различных позвоночных в значительной степени гомологичны (Holmes, Scopes, 1974; Markert et al., 1975; Глазко, 1988; Copeland, 2000). Все представители лососёвых, а также тетраплоидные карповые и некоторые осетровые имеют за счёт дупликации по 5 генов *LDH** (Слынько, 1976).

В распределении различных изоферментов LDH в тканях рыб имеется устойчивая дифференциация. Изозимы серии *A* сосредоточены преимущественно в скелетных мышцах, продукты локуса *B* больше присутствуют во внутренних органах (сердце, печени, почках, мозге), изозимы группы *C* локализуются в основном в сетчатке глаза. Такое распределение обусловлено различными функциональными свойствами этих ферментов. Мышечные изозимы лактатдегидрогеназы-*A* чувствительны к нагреву и требуют повышенных концентраций субстрата, то есть хорошо приспособлены к работе в анаэробных условиях. Изозимы серии *B* и *C* более устойчивы к нагреву и легко переносят низкие концентрации лактата, то есть адаптированы к работе в аэробных условиях (Хочачка, Сомеро, 1988; Klyachko, Ozernyuk, 2001).

У большинства рыб имеется полиморфизм по одному или нескольким локусам *LDH** (Кирпичников, 1987). Наличие у рыб нескольких генотипических вариантов *LDH** позволяет различным популяциям адаптироваться к конкретным температурным условиям среды (Colosimo et al., 2003), а при определённых условиях ещё более расширяет адаптивные возможности особей (Zietara, Skorkowski, 1995). Изменение активности LDH может служить маркером токсического воздействия на организм (Kuzmin, Kuzmina, 2001; Мещерякова, 2004).

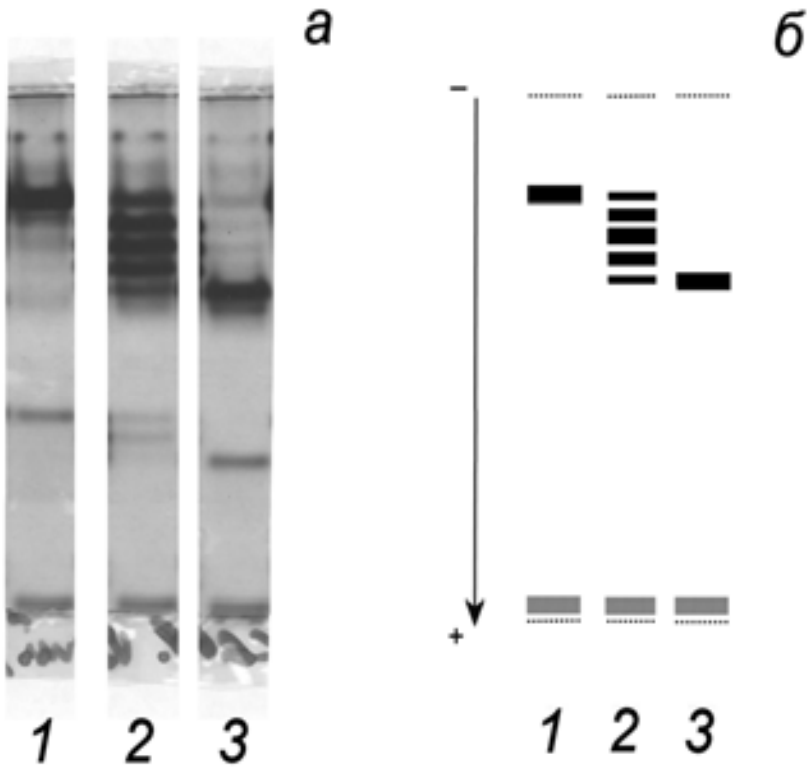

**Рисунок 5.1.** Изоферменты лактатдегидрогеназы *C. cultriventris*.
***а*** – спектр на электрофореграмме; ***б*** – схема энзимограммы.
1 – гомозиготы *LDH-A*100/100*, 2 – гетерозиготы *LDH-A*100/120*, 3 – гомозиготы *LDH-A*120/120*. Наиболее быстрая по отношению к аноду зона представлена гомотетрамерными пордуктами локуса *LDH-B** (аллельных вариантов по локусу *B** не выявлено). В центральной части электрофореграммы гетерозиготы видны гибридные молекулы между продуктами двух локусов лактатдегидрогеназы

У черноморско-каспийской тюльки во фракции водорастворимых белков белых скелетных мышц имеются продукты двух локусов - *LDH-A** и *LDH-B**, причём локус *LDH-A** представлен двумя аллельными вариантами *A*100* и *A*120* (рис. 5.1). Гибридные молекулы между продуктами локусов *LDH* A* и *B* образуются в гораздо меньшем количестве, чем между продуктами одного и того же локуса, что связано с большими различиями по первичной структуре субъединиц фермента, как это указывается и для других видов рыб (Shaklee et al., 1973; Lim et al., 1975).

Разделение изоферментов LDH в последовательной серии концентраций PAG (4-10%) позволило разделить скрытые электроморфы в гибридных зонах гетерозигот **100/120* и особенно в сложно идентифицируемых зонах *A*-B**. Криптических аллельных вариантов не выявлено.

У анчоусовидной тюльки электрофоретическая подвижность изоферментов LDH идентична таковой черноморско-каспийской тюльки. У данного вида произошла фиксация аллеля *LDH-A*100*.

У тюльки ферменты лактатдегидрогеназы устойчивы к длительному хранению и демонстрирует высокую термостабильность. Активация лактатдегидрогеназной активности при гистохимическом проявлении в пластинах PAG происходит при внесении хлорида натрия. Для пресноводных рыб, напротив, активирующее влияние оказывает сульфат магния. Возможно, это связано с изначальным эстуарным статусом волжских популяций черноморско-каспийской тюльки.

### 5.2. Малатдегидрогеназа NAD-зависимая, МДГ (MDH, E.C. 1.1.1.37)

Фермент NAD-зависимой малатдегидрогеназы в тканях животных кодируется двумя аутосомными локусами, кодирующими растворимую (*s*) и митохондриальную (*m*) формы фермента. Каждая из этих форм кодируется самостоятельным геном, у некоторых рыб двумя или тремя генами (Fisher et al., 1980). В простейшем случае MDH - димер, образованный двумя полипептидными цепями. У осетровых гены *MDH** дуплицированы, полиморфизм по ним очень сложен и спектр фермента состоит из 8-10 фракций (Слынько, 1976). У костистых рыб имеются 2 гена *MDH-A** (представлен во всех тканях) и *MDH-B** (экспрессирован преимущественно в мышцах и тканях глаза) (Wheat et al., 1973). У многих видов встречаются дупликации как по гену *MDH-A**, так и по *MDH-B**, причём любой из локусов может быть полиморфным, хотя часто этот полиморфизм является следствием конформационных изменений полипептидов, а не различий в первичной структуре молекул (Глазко, 1988; Merrit, Quattro, 2003). В общем MDH у рыб может быть отнесена к числу ферментов, образующих большое число изозимов и аллельных вариантов.

Несмотря на сложности при идентификации электрофоретических продуктов, изменение ферментативной активности MDH оказалось удачным маркером биохимических изменений у рыб в результате токсического воздействия тяжёлых металлов (Sastry et al., 1997; Levesque et al., 2002), а также при патологических изменениях (Kuzmin, Kuzmina, 2001).

Электрофоретических вариантов MDH в водорастворимой фракции белых скелетных мышц у черноморско-каспийской тюльки не обнаружено (рис. 5.2). У анчоусовидной тюльки данный фермент не изучался. Изоферменты MDH крайне чувствительны к условиям хранения проб, имеют малую теплоустойчивость и легко ингибируются даже в слабокислой среде. При гистохимическом проявлении MDH тюльки крайне чувствительна к химической чистоте субстрата (особенно к стереоизомерам малата).

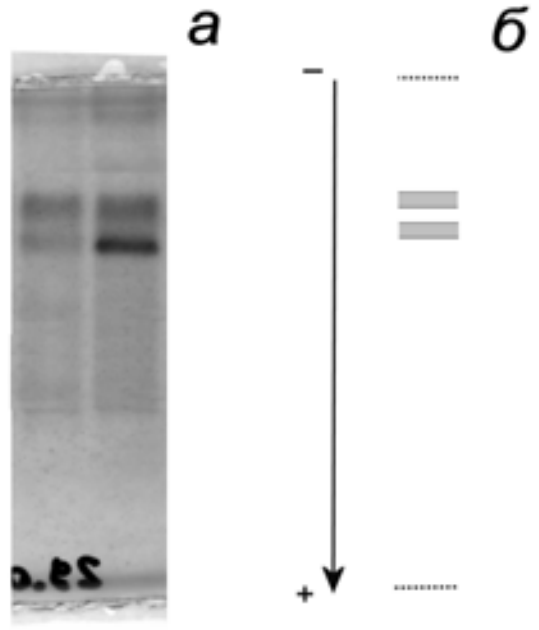

**Рисунок 5.2.** Изоферменты NAD-зависимой малатдегидрогеназы *C. cultriventris*.
***а*** – спектр на электрофореграмме; ***б*** – схема энзимограммы

### 5.3. Малатдегидрогеназа NADP-зависимая, Me (ME, E.C. 1.1.1.40)

У животных NADP-зависимая малатдегидрогеназа исторически называется малик-энзим. У позвоночных ME кодируется двумя аутосомными локусами: *s* и *m*, кодирующих растворимую и митохондриальную формы фермента соответственно. Молекулы фермента имеют тетрамерное строение, слагаются двумя субъединицами (Shows, Ruddle, 1968). Фермент представлен во всех тканях и органах рыб, играет важную роль в углеводном обмене и внутриклеточной системе производства NADP·H потенциала (Barroso et al., 2001). У некоторых высших позвоночных активность ME снижена или вовсе отсутствует в ряде форменных элементах крови (Глазко, 1988).

Имеются данные по полиморфизму *ME**. Так, у амурского чебачка *Pseudorasbora parva* имеются 2 локуса *ME**, один из которых мономорфен, а второй имеет 3 аллеля (Konishi et al., 2003). В настоящий момент накоплен большой массив данных, позволяющих использовать ME в качестве маркера интенсивности углеводного обмена и прежде всего гепато-панкреатической утилизации липидов (Shimeno et al., 1997) а также в качестве показателя обеспеченности и качества питания рыбы (Konradt, Braunbeck, 2001) и показателя токсического воздействия среды (Levesque et al., 2002).

У черноморско-каспийской тюльки ME кодируется двумя генетическими локусами: *ME-1** и *ME-2** (рис. 5.3). Продукты локуса *ME-1** отличаются малой анодной подвижностью, представлены двумя аллельными вариантами **100* и **112*, а также редким вариантом **80*, частота которого в изученных популяциях менее 1%. Продукты локуса *ME-1** и, в особенности, локуса *ME-2** крайне чувствительны к условиям хранения проб (инактивируются даже при недолгом хранении вне надлежащих условий), чистоте субстрата и требуют высокого качества кофермента. Малик-энзимная активность у тюльки активируется добавлением ионов $Mg^{2+}$. В связи со сложностью и нестабильностью результатов электрофоретического выявления продуктов локуса *ME-2**, он был исключён из популяционно-генетического анализа.

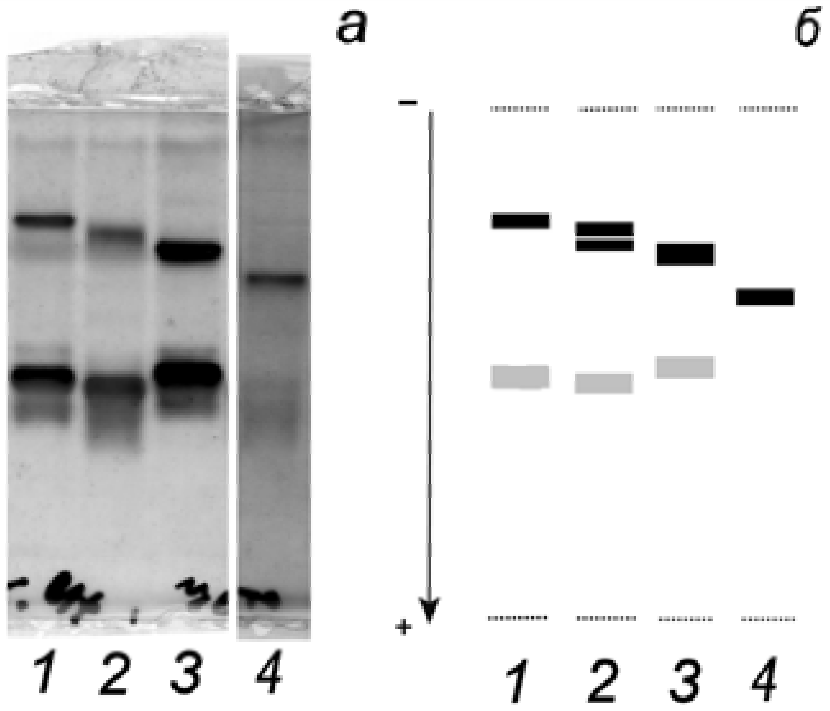

**Рисунок 5.3.** Изоферменты NADP-зависимой малатдегидрогеназы *C. cultriventris*.

***а*** – спектр на электрофореграмме; ***б*** – схема энзимограммы.

1 – гомозиготы *ME-1*100/100*, 2 – гетерозиготы *ME-1*100/112*, 3 – гомозиготы *ME-1*112/112*, 4 – мономорфная зона *ME-1*120* анчоусовидной тюльки *C. engrauliformes*. Более быстрая по отношению к аноду зона представлена продуктами локуса *ME-2**

## 5.4. α-глицерофосфатдегидрогеназа, αГФД (aGPDH, Е.С. 1.1.1.8)

α-глицерофосфатдегидрогеназа у рыб является сильно изменчивым гликолитическим ферментом. Данных по генетической и биохимической структуре этого фермента у рыб немного. Простая система двух кодоминантных аллелей одного генетического локуса *aGPDH** характерна для кеты и горбуши (Алтухов и др., 1972; Aspinwall, 1973). В мышцах окуня и плотвы обнаруживается только один изофермент aGPDH, а в печени плотвы - 5 изоформ (Мещерякова, 2004). У радужной форели имеются два полиморфных локуса, но для одного из них известен только редкий аллель; два аллеля другого локуса наследуются кодоминантно (Utter et al., 1973, цит. по: Кирпичников, 1987). У сигов рода *Coregonus* имеется два локуса *aGPDH**, причём один из них имеет 2, а другой 3 аллельных варианта (Clayton et al., 1973). Аналогично кодируется *aGPDH** у амурско-го чебачка *Pseudorasbora parva* (данные автора). Молекула фермента - димер, продукты обоих локусов независимо комбинируются друг с другом, в связи с чем идентификация изозимов крайне затруднена. В целом число аллелей, характерных для данного локуса нередко равняется 3 и даже 4 (Кирпичников, 1987). Изменение активности aGPDH (как и LDH и MDH) при токсическом загрязнении водоёмов указывает на значительную перестройку углеводного обмена рыб, причём степень изменения в активности ферментов углеводного обмена коррелирует с уровнем токсического воздействия (Мещерякова, 2004).

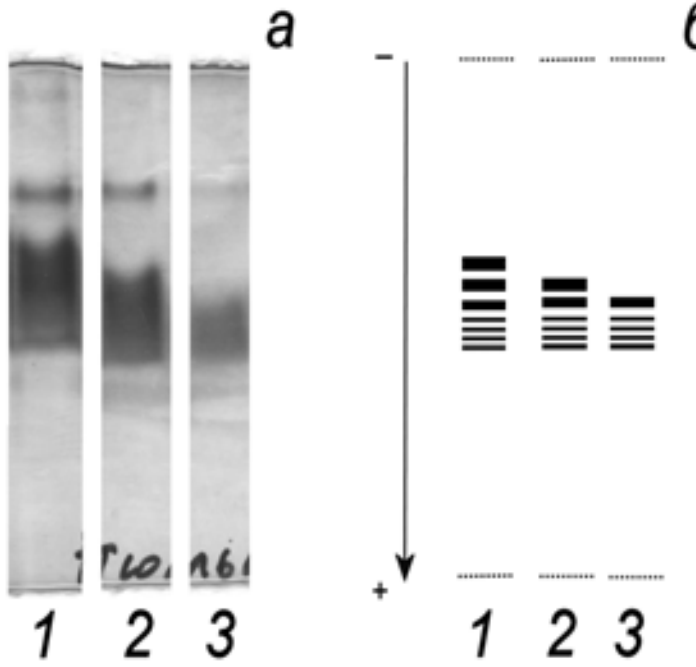

**Рисунок 5.4.** Изоферменты α-глицерофосфатдегидрогеназы *C. cultriventris*.
***а*** – спектр на электрофореграмме; ***б*** – схема энзимограммы.
1 – вариант $A^1A^2B^1B^1$ (условный вариант “гомозигота **a/a*”); 2 – вариант $A^1A^2B^2B^2$ (условный вариант “гетерозигота **a/b*”); 3 – вариант $A^1A^2B^3B^3$ (условный вариант “гомозигота **b/b*”). Наиболее близкая к катоду полоса – неспецифическое окрашивание алкогольдегидрогеназы

Разделение изоферментов aGPDH в последовательной серии концентраций PAG (4-10%) позволило разделить скрытые электроморфы в гибридных зонах между локусами *aGPDH-A** и *aGPDH-B**. Установлено, что aGPDH у черноморско-каспийской тюльки представлена продуктами двух генетических локусов, один из которых, вероятно, дуплицирован (рис. 5.4.). Для популяционно-генетического анализа использовался метод «условных вариантов» (Кирпичников, 1987; Gillespie, 1998). Изоферменты aGPDH тюльки достаточно устойчивы при длительном хранении, но крайне чувствительны к pH при гистохимическом окрашивании (оптимально значение pH 8,6), причём как повышение, так и понижение кислот-ности среды существенно ингибирует ферментативную реакцию. Зоны ферментативной активности изоформ у анчоусовидной тюльки по электрофоретической подвижности совпадают с таковыми черноморско-каспийской тюльки.

## 5.5. Глюкозо-6-фосфатдегидрогеназа, Г6ФД (G6PDH, Е.С. 1.1.1.49)

Глюкозо-6-фосфатдегидрогеназа - ключевой фермент пентозофосфатного шунта, который является определяющим при регуляции окислительно-восстановительных процессов, особенно в эритроцитах, энергетические затраты в которых невелики (Copeland, 2000; Schmidt, 2000). Вместе с тем апотомический путь распада глюкозо-6-фосфата является важным процессом продуцирования NADP·H, а также служит источником пентоз, а в случае необходимости возможен переход цикла на гликолиз, что особенно часто происходит при голодании рыб (Shimeno et al., 1997; Barroso et al., 2001; Meton et al., 2003).

Фермент G6PDH у высших позвоночных имеет тетрамерное или димерное строение. G6PDH у рыб имеет димерную структуру, у многих рыб имеентся полиморфизм по кодирующему локусу (Кирпичников, 1987). У амурского чебачка ***Pseudorasbora parva*** G6PDH мономорна (данные автора). По строению и свойствам фермент близок к 6PGDH, оба фермента, в отличие от ряда ферментов гликолиза, определяются одним геном во всех тканях. У млекопитающих имеются как полиморфные, так и мономорфные по G6PDH виды (Глазко, 1988). Имеются большой материал, иллюстрирующий важное значение пентозофосфатного пути метаболизма у рыб при голодании (Hung et al., 1997; Konradt, Braunbeck, 2001).

Важная роль G6PDH отводится в метаболизме холодноводных рыб, при температурных адаптациях у антарктических рыб (Ciardiello et al., 1995) и при сезонных изменениях метаболизма (Levesque et al., 2002). В последнее время большое внимание уделяется изучению изменения активности G6PDH при стрессовом воздействии на организм, в частности при отравлении тяжёлыми металлами и в качестве маркера злокачественных изменений в тканях печени (Kohler, Van Noorden, 1998; Winzer et al., 2002; Pandey et al., 2003).

В пресноводных популяциях черноморско-каспийской тюльки электрофоретических вариантов G6PDH в водорастворимой фракции белых скелетных мышц тюльки не обнаружено (рис. 5.5). В популяции тюльки Северного Каспия обнаружено две гетерозиготные особи с генотипом **85/100*. В популяции анчоусовидной тюльки Северного Каспия имеются лишь гомозиготы *G6PDH*110/110*. Фер-мент G6PDH тюльки довольно устойчив даже при длительном хранении. Активация фермента происходит в присутствии катионов магния.

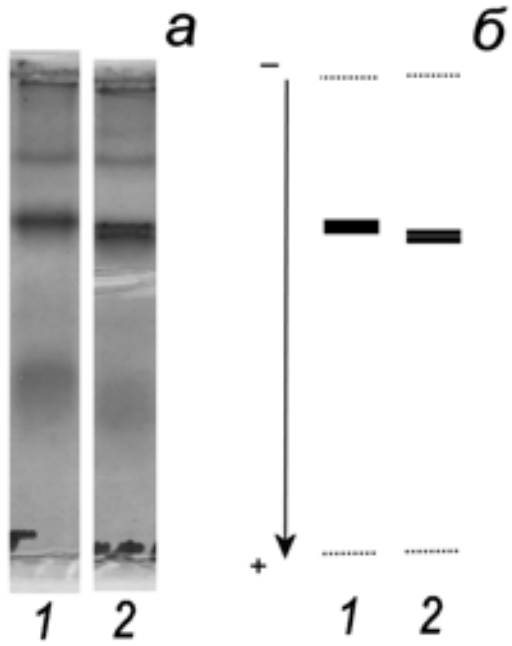

**Рисунок 5.5.** Изоферменты глюкозо-6-фосфат дегидрогеназы.
***а*** – спектр на электрофореграмме; ***б*** – схема энзимограммы.
1 – константные гомозиготы *G6PDH*100/100* черноморско-каспийской тюльки; 2 – константные гомозиготы *G6PDH*110/110* анчоусовидной тюльки *C. Engrauliformes*

**5.6. 6-Фосфоглюконатдегидрогеназа, 6ФГД (6PGDH, E.C. 1.1.1.49)**

6-Фосфоглюконатдегидрогеназа - ключевой фермент пентозофосфатного шунта. У животных определяется одним геном во всех тканях. Фермент - димер, состоит из двух субъединиц с одинаковой молекулярной массой. Как и G6PDH, 6-фосфоглюконатдегидрогеназа также участвует в восстановлении NADP·H, особенно в эритроцитах (Barroso et al., 2001).

По общему строению, субъединичной структуре, аминокислотному составу и молекулярной массе 6PGDH также крайне схожа с G6PDH. Большинство насекомых полиморфно по 6PGDH (Филиппович, Коничев, 1987). В некоторых природных популяциях высших позвоночных наблюдается полиморфизм по *6PGDH** (Глазко, 1988). У рыб имеются данные по полиморфизму у 4 видов щиповок (Ferris, Whitt, 1977) и двух видов пецилий (Leslie, Vrijenhoek, 1977), обыкновенного гольяна (Muller, Ward, 1998) и амурского чебачка (данные автора). Важное значение имеет изменение активности 6PGDH при голодании у рыб (Meton et al., 2003). Также как и для G6PDH, оценка уровня активности 6PGDH используется в качестве маркера злокачественных изменений, особенно на ранних стадиях (Maronpot et al., 1989; Kohler, Van Noorden, 1998).

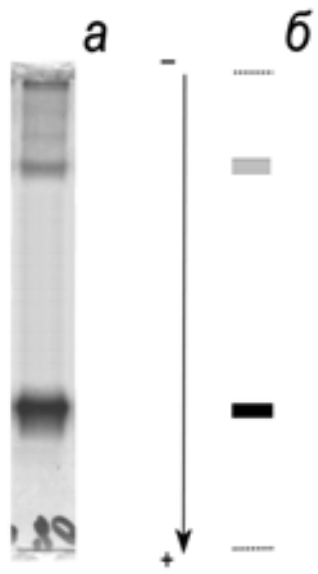

**Рисунок 5.6.** Изоферменты 6-фосфоглюконатдегидрогеназы *C. cultriventris*.
***а*** – спектр на электрофореграмме; ***б*** – схема энзимограммы

У черноморско-каспийской тюльки электрофоретических вариантов 6PGDH в водорастворимой фракции белых скелетных мышц не обнаружено (рис. 5.6). У анчоусовидной тюльки *C. engrauliformes* электрофоретическая подвижность мономорфной зоны 6PGDH соответсвует таковой *C. cultriventris*.

## 5.7. Эстеразы эфиров карбоновых кислот, Эст (EST, E.C. 3.1.1.x)

Эстеразы относятся к классу гидролаз и представлены несколькими ферментами: карбоксилэстеразой, арилэстеразой, холинэстеразой и др. Классификация этих ферментов в группе эстераз основана на использовании искусственных субстратов и ингибиторов (Shaw, Prasad, 1970). Как сывороточные, так и тканевые эстеразы являются у рыб продуктами нескольких генетических локусов, многие из которых демонстрируют индивидуальную изменчивость. Общий уровень биохимического полиморфизма эстераз очень велик как у высших позвоночных, так и у рыб (Кирпичников, 1987; Глазко, 1988). Из рыб описан полиморфизм эстераз у ближайшего родственника тюльки - океанической сельди *Clupea harengus*. У неё имеются, по крайней мере, 5 локусов эстераз, причём в каждом локусе имеется по 4-5 аллелей (Салменкова, Волохонская, 1973; Кирпичников, 1987). У обыкновенного гольяна имеются 3 генетических локуса, кодирующих карбоксилэстеразы, два из них имеют по 3 аллеля, и один двухаллельный локус (Muller, Ward, 1998). У амурского чебачка имеются два локуса эстеразы, один из которых представлен двумя аллелями.

Эстеразы обладают широкой субстратспецифичностью. Некоторые формы эстераз способны расщеплять большое количество субстратов и, возможно, кодируются одним генетическим локусом (Глазко, 1988).

Чаще всего в генетической системе эстераз преобладает один «главный» аллель, один «вспомогательный» и несколько минорных аллелей, суммарная частота которых не превышает 1%. Как правило, в генетической характеристике популяций минорные аллели из подсчётов исключаются (Ridgway et al., 1970).

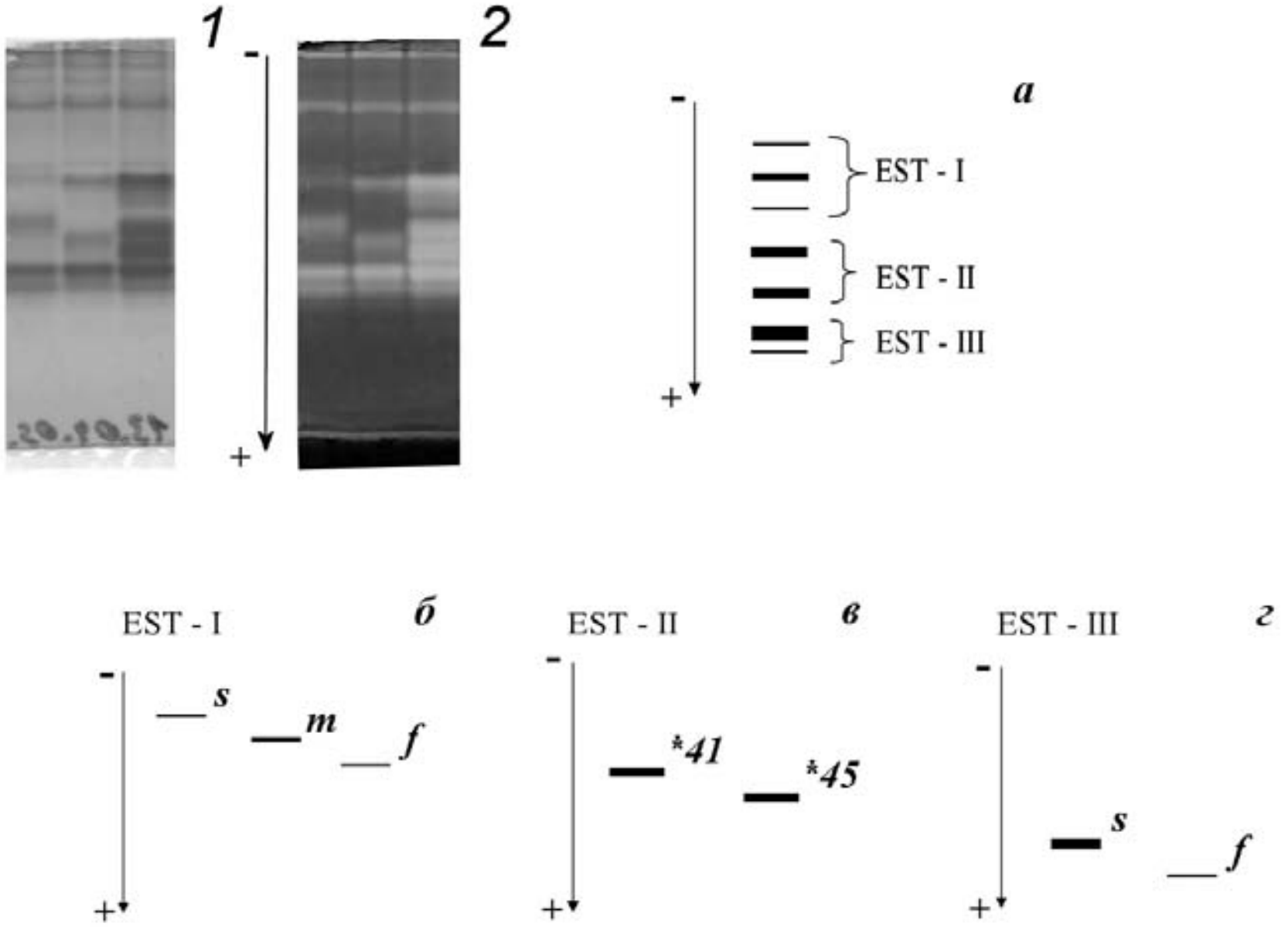

**Рисунок 5.7.** Электрофореграммы bEST (*1*) и D-EST (*2*) *C. cultriventris*. Зоны ферментативной активности этих двух локусов совпадают. EST I, II, III - условно выделенные группы зон эстеразной активности; **41*, **45* - вероятные аллельные варианты; *s* (slow), *m* (medium), *f* (fast) - условные обозначения изоформ по увеличению электрофоретической подвижности к аноду.
***а***– общая схема и место определённых зон в спектре bEST; ***б***– распределение (по увеличению анодной подвижности) и относительная активность изоформ наименее подвижной группы EST-I; ***в***– распределение и относительная активность аллелей группы EST-II; ***г***– распределение и относительная активность изоформ наиболее подвижной группы EST-III.

В данной работе изучены два фермента: 2-нафтилацетат (bEST) и 4-метилумбеллиферилацетат (D-EST) зависимые эстеразы. Возможно, в обоих случаях расщепление субстрата осуществляется одним ферментом (рис. 5.7). В области активности фермента bEST у черноморско-каспийской тюльки выделяются три условных зоны: EST-I, II, III (по увеличению подвижности к аноду, рис. 5.7). Характеристики фракций bEST представлены в табл. 5.1. Каждая фракция имеет несколько зон активности. Достоверно ген-аллельную детерминацию можно предположить лишь для зоны EST-II, в которой имеются аллели *bEST*41* и *bEST*45*. В прочих зонах эстеразной активности точно определить генетическую обусловленность конкретного трека затруднительно (Карабанов, Слынько, 2005а).

**Таблица 5.1.** Характеристика изоферментов β-эстреаз тюльки *C. cultriventris* Рыбинского водохранилища

| **Зона** | **Относительная электрофоретическая подвижность (Rf , к аноду)** | **Проявление** | **Доля в эстеразной активности, %** | |
|---|---|---|---|---|
| | | | **группы** | **общей** |
| ***EST – I*** (группа “slow”) | | | | 30 |
| Est-1***s*** | *25 | у всех особей, мономорфна | 16 | 5 |
| Est-1***m*** | *31 | у всех особей; аллельные варианты | 74 | 22 |
| Est-1***f*** | *35 | не у всех особей | 10 | 3 |
| ***EST – II*** (группа “medium”) | | | | 29 |
| Est-2***a*** | *41 | аллельные варианты | - | 29 |
| Est-2***b*** | *45 | | | |
| ***EST – III*** (группа “fast”) | | | | 40 |
| Est-3***s*** | *52 | у всех особей; мономорфна | 82 | 32 |
| Est-3***f*** | *57 | не у всех особей | 17 | 8 |

Разделение изоферментов bEST в последовательной серии концентраций PAG (4-10%) позволило разделить скрытые электроморфы в зоне EST-I*m*, где, возможно, имеются аллельные варианты. Однако достоверно установить генетическую детерминацию этого локуса не удалось. Также выявлены криптических аллельные варианты (при 4% PAG) в зоне EST-II. Здесь возможно наличие третьего аллеля, более медленного по подвижности к аноду, чем аллель **41*. Аллельные варианты, видимо, очень близки по соотношению заряд:масса. Однако для массового анализа такая концентрация геля неприемлема, да и частота этого криптического аллеля относительно невелика.

У тюльки из всех исследованных ферментов водорастворимой фракции белых скелетных мышц эстеразы наиболее устойчивы к длительному хранению и демонстрирует очень высокую термостабильность. Однако, несмотря на высокий полиморфизм и разнообразие изоформ, к использованию эстераз в популяционно-генетических исследованиях надо относиться с осторожностью. Большой набор субстратов, сложности выявления генетической обусловленности конкретных изоформ и не совсем ясное положение эстераз в метаболических цепях организма, связанное с искусственностью классификации фермента, сильно затрудняют их изучение. Электрофоретическая подвижность основных фракций эстераз у анчоусовидной тюльки соответствует таковой черноморско-каспийской тюльки.

## 5.8. Щелочная фосфатаза, ЩФ (AP, E.C. 3.1.3.1)

Характерная черта ферментов щелочных фосфатаз - слабая специ¬фичность. Они способны гидролизовать большое количество первичных фосфорных эфиров. У млекопитающих имеются два локуса AP* - один контролирует плаз¬матический фермент, другой - фосфатазу внутрен¬них органов. Эти варианты сильно отличают¬ся по термостабильности и сродству к субстрату. В целом изоферменты щелочной фосфатазы у животных изучены относительно слабо.

У насекомых имеется до 9 форм AP (Филиппович, Коничев, 1987). Полиморфизм по AP описан для некоторых животных (Makaveev, 1975; Глазко, 1988; Калинин и др., 1988). У черноморско-каспийской тюльки электрофоретических вариантов не обнаружено (рис. 5.8). Часто зона AP проявляется в зоне неспецифической активности аспартатаминотрансферазы. В популяционно-генетических исследованиях тюльки AP не использовалась в связи с неустановленной генетической детерминацией и слабой субстратспецифичностью. Ферменты AP у анчоусовидной тюльки не изучались.

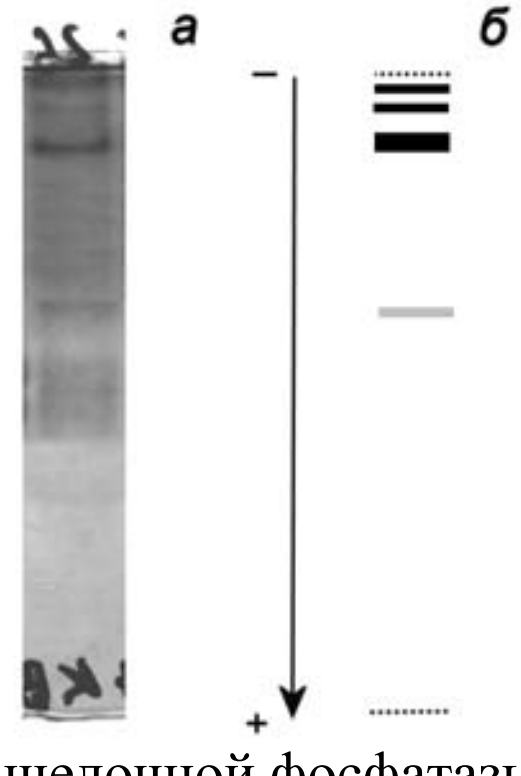

**Рисунок 5.8.** Изоформы щелочной фосфатазы *C. cultriventris*.
***а*** – спектр на электрофореграмме; ***б*** – схема энзимограммы

### 5.9. Супероксиддисмутаза, СОД (SOD, E.C. 1.15.1.1)

Супероксиддисмутаза является основным ферментом защиты тканей организма от кислорода, катализирует удаление суперксидных радикалов. У животных имеются две формы SOD: одна - димер с молекулярной массой около 31 kDa, другая - тетрамер с молекулярной массой около 90 kDa. Тетрамерный фермент характерен только для митохондриальной фракции (Asayama, Burr, 1985; Moo-Lee et al., 1985). Фермент кодируется двумя аутосомными локусами (Steiman et al., 1974). Особенно велика активность SOD в коже у рыб, что связано с барьерной функцией кожи (Nakano, 1995). У большинства рыб описан полиморфизм по одному или нескольким локусам *SOD* (Кирпичников, 1987; Ken et al., 2003).

В последнее время появилось большое количество работ, посвящённых SOD. Изменение общего уровня и активности отдельных фракций этого фермента возникает вследствие увеличения окислительной нагрузки на рыб, что связанно со всё возрастающей техногенной нагрузкой на водные экосистемы (Lopes et al., 2001; Peters et al., 2001). Имеются данные, что уровень активности SOD зависит от рациона рыбы и меняется при воздействии стрессовых факторов, в частности температуры (Parihar et al., 1996; Ruiz-Gutierrez et al., 2001).

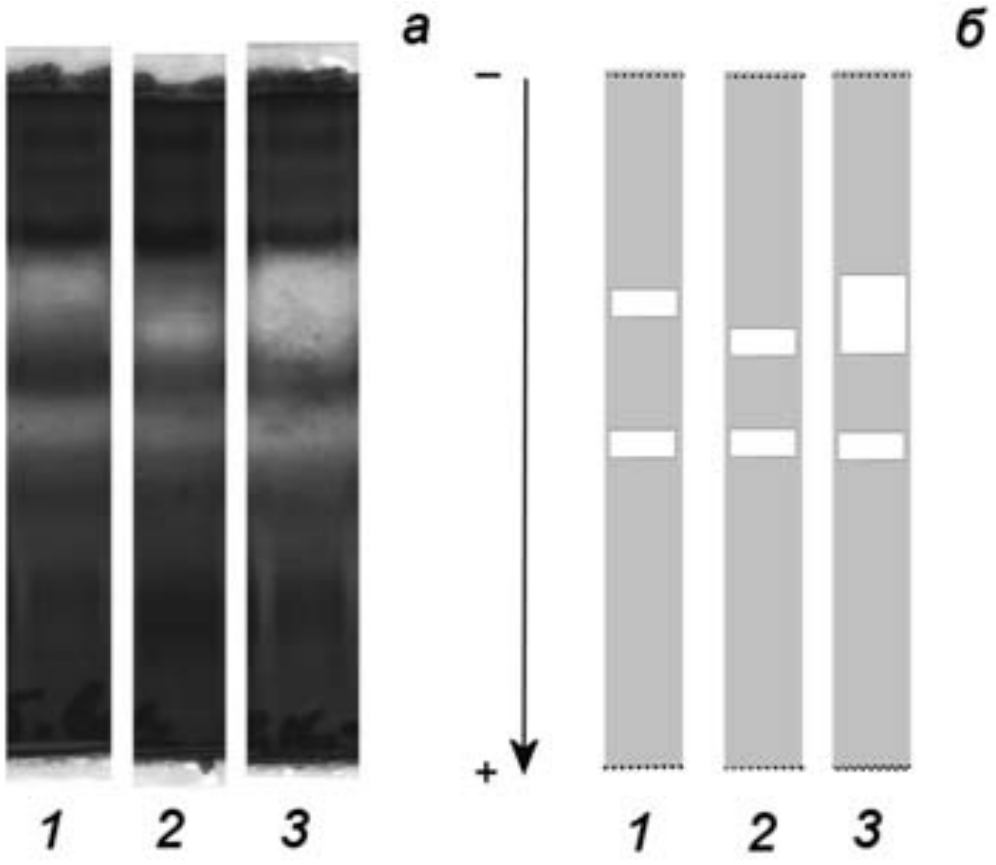

**Рисунок 5.9.** Изоферменты супероксиддисмутазы *C. cultriventris.*
***а*** – спектр на электрофореграмме; ***б*** – схема энзимограммы.
1 – рыбы с генотипом *SOD-1*100/100*; 2 – рыбы с генотипом *SOD-1*110/110*; 3 – рыбы с генотипом *SOD-1*100/110*. Более быстрая по отношению к аноду зона представлена продуктом мономорфного локуса *SOD-2**

Имеются данные, что при некоторых заболеваниях уровень SOD в организме изменяется, а также, возможно, уровень антиоксидантной активности связан с продолжительностью жизни животных (Tolmasoff et al., 1980). У черноморско-каспийской тюльки в водорастворимой фракции белых скелетных мышц SOD кодируется двумя генетическими локусами *SOD-1** и *SOD-2**. Локус *SOD-1** полиморфен, представлен двумя аллельными вариантами **100* и **110* (рис. 5.9). Изоферменты SOD тюльки чувствительны к нагреванию, активируются катионами магния $Mg^{2+}$. Активность фракций супероксиддисмутазы анчоусовидной тюльки совпадает с аналогичными фракциями черноморско-каспийской тюльки, однако удалось обнаружить только константно фиксированные варианты *SOD-1*100* и *SOD-2*100*.

## 5.10. Аспартатаминотрансфераза, AAT (AAT, E.C. 2.6.1.1)

Аспартатаминотрансфераза - один из ключевых фермент аспартат-малатного пути аминокислотного метаболизма. Фермент у животных представлен в растворимой (*s*) и митохондриальной (*m*) формах (John, Johnes, 1974). У насекомых локус *AAT** очень изменчив, обнаруживается до 10 форм (Филиппович, Коничев, 1987). Изменчивость по sAAT обнаружена у многих млекопитающих и у рыб. У сельдёвых имеется 1 полиморфный локус с 3 аллельными вариантами (Салменкова, Волохонская, 1973). У большинства рыб имеется один (реже два) локуса с одним-двумя аллельными вариантами, активность которых отличается в разных органах и на разных этапах онтогенеза (Petrovic et al., 1996; Srivastava et al., 1999). У полиплоидных видов соответствующие локусы дуплицированы (Кирпичников, 1987). В целом у рыб уровень генетической изменчивости *AAT** относительно невелик.

При исследовании электрофоретических вариантов AAT у черноморско-каспийской тюльки в водорастворимой фракции белых скелетных мышц установлено наличие одного генетического локуса с 2 аллельными вариантами *AAT *100* и **110* (рис. 5.10). Разделение изоферментов AAT в последовательной серии концентраций PAG (4-10%) позволило разделить скрытые электроморфы в зоне гомозигот, которые разделяются при крайне низких концентрациях PAG (4%), непригодных для массовых популяционно-генетических исследований. Гомозиготы **100/100* скрыты в однородной слабооформленной зоне, которую можно принять за гетерозиготную, что, при относительно редкой частоте гомозигот, очень затрудняет ана-лиз. Применение пониженной концентрации PAG (6%) вместе с использованием Трис-HCl буфера с pH 8,6 позволяет определить криптические варианты в этой зоне.

AAT тюльки чрезвычайно чувствительна к условиям хранения проб. Гистохимическое проявление фермента улучшается при добавлении в инкубационную среду малого количества хлорида натрия.

У анчоусовидной тюльки на электрофореграммах выделяются зоны активности двух аллелей *AAT *110* и **120*, причём последний характерен только для данного вида.

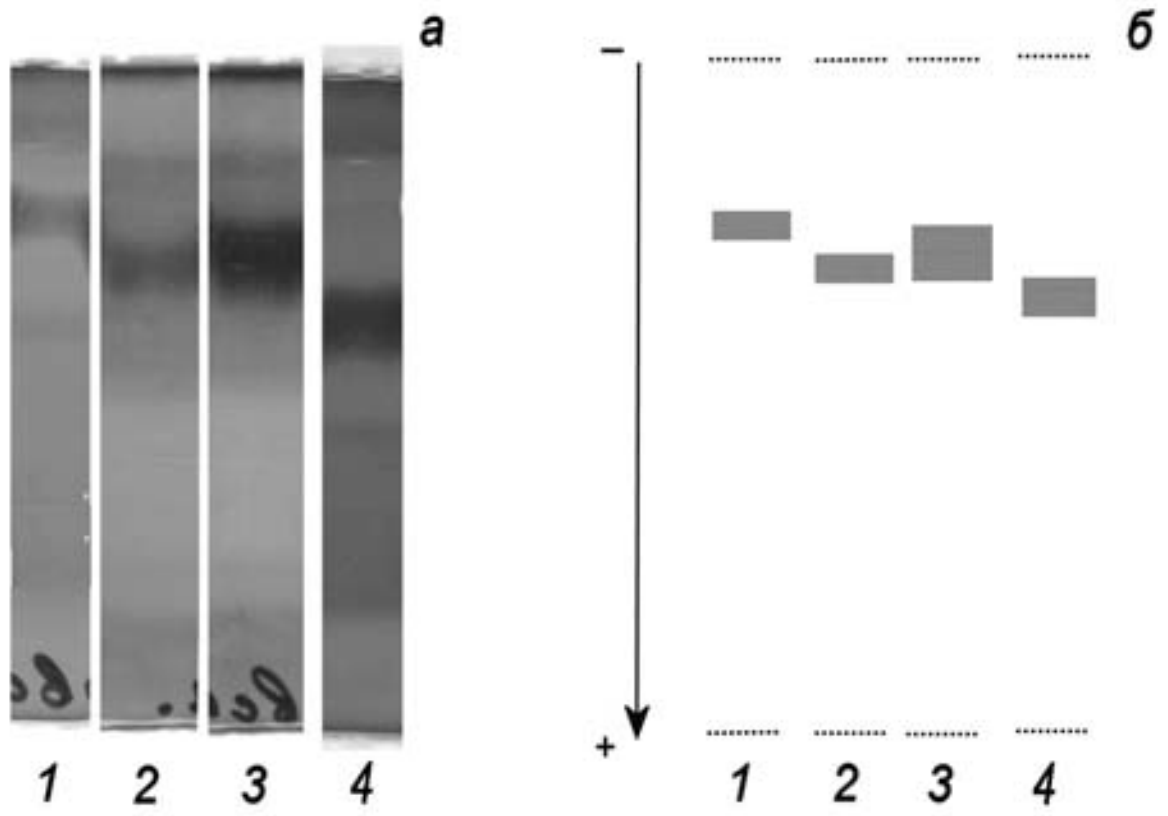

Рисунок 5.10. Изоферменты аспартатаминотрансферазы.
***а*** – спектр на электрофореграмме; ***б*** – схема энзимограммы.
1 – гомозиготы *AAT*100/100*, 2 – гомозиготы *AAT*110/110*, 3 – гетерозиготы *AAT*100/110*, 4 – зона ферментативной активности аллеля **120* (гомозигота *AAT*120/120*) анчоусовидной тюльки *C. engrauliformes*

### 5.11. Спектр общего белка, миогены, ОБ (GP)

Спектр общего белка включает в себя группу тяжёлых фракций (альбумины и преальбумины), а также миогены. У рыб мышечные белки при электрофорезе разделяются на 15-20 фракций. Большинство из них оказываются в пределах вида мономорфными, хотя у некоторых видов имеется один или несколько полиморфных локусов. Количество аллелей на локус редко больше двух (Кирпичников, 1987). Мышечные белки значительно более постоянны, чем белки сыворотки крови, что позволяет их использовать при систематических исследованиях.

Альбумины у многих видов рыб относятся к высокоизменчивым генетическим маркерам, их идентификация затруднена, а генетическая детерминация требует отдельных исследований для каждого конкретного вида животных. У черноморско-каспийской тюльки наблюдается высокая вариабельность белков в области самых «медленных» гамма-глобулиновых фракций. Подробное генетико-популяционное исследование этих изменчивых зон методом PAGE затруднено вследствие их недостаточно чёткого разделения.

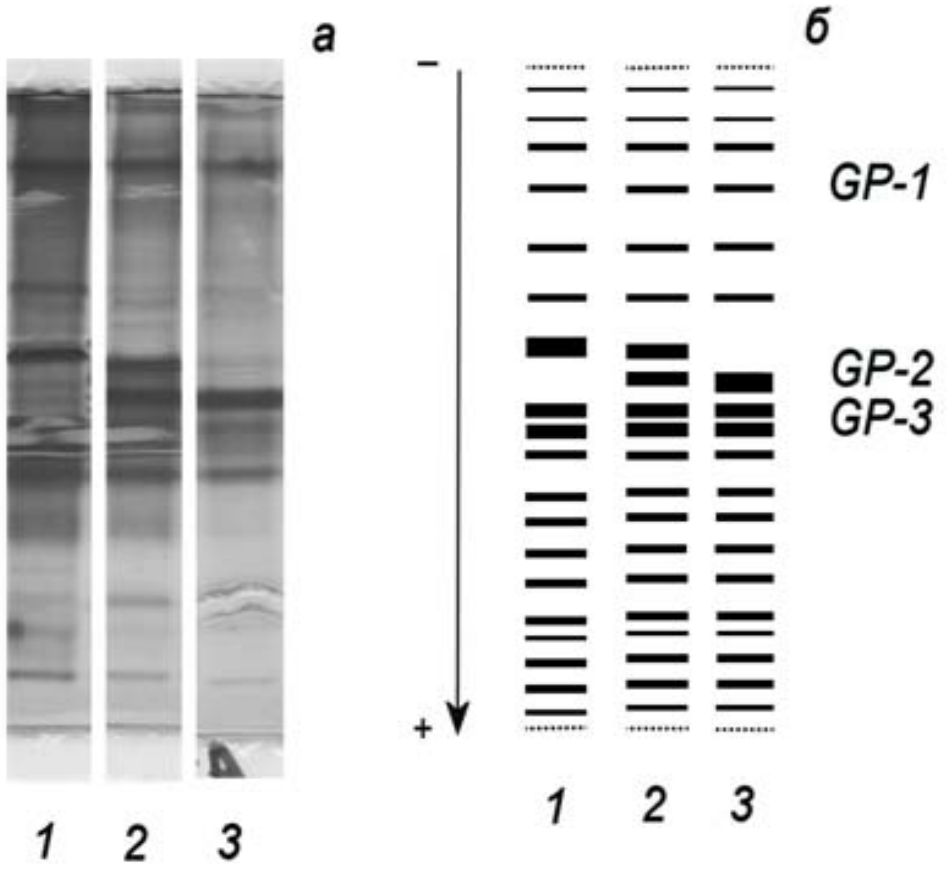

**Рисунок 5.11.** Спектр общего белка *C. cultriventris*.
***а*** – электрофореграмма, ***б*** – схема расположения фракций.
1 – гомозиготные особи *GP-2*100/100*; 2 – гетерозиготные особи *GP-2*100/115*; 3 – гомозиготные особи *GP-2 *115/115*

Применяемое нами окрашивание миогенов с использованием кумасси бриллиантового синего R-250 позволяет выявить лишь около 30% всех фракций (Smith, 2002), однако в отличие от методов радиоавтографии или окрашивания белков серебром, данный метод безопасен, быстр и не требует дорогостоящих реактивов. Фракции общего белка, обнаруживаемые с помощью кумасси, относятся к базовым группам миогенов, чётко идентифицируемым у всех видов, что позволяет их использовать в массовых популяционно-генетических исследованиях.

При исследовании электрофоретических вариантов миогенов у черноморско-каспийской тюльки в водорастворимой фракции белых скелетных мышц установлено наличие 19 базовых фракций миогенов (рис. 5.11). Обнаружена одна полиморфная фракция миогенов тюльки, обозначенная как GP-2, с двумя вероятными аллельными вариан-тами **100* и **115*. Спектр общего белка оказался крайне удачным для установления генетического родства между различными видами и расами одного вида. Между представителями рода *Clupeonella* (*C. cultriventris* и *C. engrauliformes*) нами выявлены чёткие видоспецифичные зоны (GP-1, GP-2, GP-3). Для *C. cultriventris* видоспецифичны аллели *GP-1 *100*, *GP-2 *100* и **115*, *GP-3 *100*. Для *C. Engrauliformes* отмечены варианты *GP-1 *100*, *GP-2 *110*, *GP-3 *120*.

# Глава 6.
# Популяционно-генетическая характеристика тюльки исследованных водоёмов

На основании изучения 27 популяций на протяжении с 2002 по 2011 годы по 12-17 генетическим локусам черноморско-каспийская тюлька характеризуется следующими параметрами. Доля полиморфных локусов P=0,23-0,41 (в среднем P=0,31). Локус считается полиморфным, если частота встречаемости наиболее редкого аллеля не ниже 5%. Средняя гетерозиготность, вычисленная прямым подсчётом $H_{ob}$=0,101-0,211 (по исследованным локусам). Количество аллелей на локус 1,35 (s.e. 0,05). Таким образом, черноморско-каспийская тюлька по уровню полиморфизма не сильно отличается, а по уровню гетерозиготности несколько превышает значения таковых у других сельдёвых рыб. Так, для *Clupea harengus* P=0,36, $H_{ob}$=0,07 (Andersson et al., 1981), для *Cupea pallasi* P=0,31, $H_{ob}$=0,06 (Fujio, Kato, 1979, цит. по: Nevo et al., 1984), а для *Clupeonella engrauliformes* P=0,35, $H_{ob}$=0,05 (данные автора).

## 6.1. Популяция тюльки Азовского моря

Популяционно-генетическая характеристика тюльки Таганрогского залива Азовского моря даётся на основании исследований 2003-05 гг. Пробы 2003 г. отобраны в районе Порта Кантон, в 2004-05 гг. в прибрежной части в районе Чумбур-Косы. Популяционно-генетические исследования проведены по 12 локусам (в 2005 г. - по 17 локусам). Общий уровень полиморфизма черноморско-каспийской тюльки изученной популяции составляет P=0,35 (при критерии k=0,95). Средняя гетерозиготность, вычисленная прямым подсчётом $H_{ob}$=0,145 (s.e. 0,06), ожидаемая гетерозиготность $H_{ex}$=0,131 (s.e. 0,047). Количество аллелей на локус a=1,41 (s.e. 0,12).

Основные популяционно-генетические характеристики тюльки Азовского моря даны по результатам работ 2005 г. и представлены в табл. 6.1. За трёхлетний период наблюдений установлено, что тюлька Таганрогского залива Азовского моря представляет собой генетически устойчивую популяцию, 89% локусов генетически сбалансированы, межгодовое сходство популяций ***r***=0,984±0,02 (***D***=0,016).

На основании представленных данных можно констатировать, что эта популяция генетически сбалансирована, распределение частот генотипов, не соответствующее равновесию Харди-Вайнберга, отмечено только для двух локусов (*aGPDH** и *AAT**). Однако в случае *aGPDH**, когда выделяются не истинные, а условные аллельные варианты, сложно говорить о нарушении равновесия основываясь на критерии $\chi^2$. Применение показателя λ, используемого при неодинаковом числе классов и частот распределения признака, позволяет сделать вывод о соответствии равновесному распределения частот генотипов этого локуса. Для большинства локусов (кроме *aGPDH** и *AAT**) характерен относительно небольшой дефицит гетерозигот ($D_H$<0,2). Значительный избыток гетерозигот по локусу *AAT** (*F*= -0,805) связан с выявленными криптическими аллельными вариантами (преобладание гетерозигот, выявляемое на низких концентрациях PAG).

**Таблица 6.1.** Популяционно-генетическая характеристика черноморско-каспийской тюльки популяции Таганрогского залива Азовского моря по 7 полиморфным локусам (2005 г.)

| Локус | Частоты аллелей | $\chi^2$ | λ | F | $D_H$ |
|---|---|---|---|---|---|
| *LDH-A** | *p* (**100*) = 0,75<br>*q* (**120*) = 0,25 | 1,60 | 0,47 | -0,200 | 0,185 |
| *ME-1** | *p* (**100*) = 0,45<br>*q* (**112*) = 0,55 | 0,01 | 0,05 | 0,015 | -0,028 |
| *aGPDH** | *p* (**a*) = 0,16<br>*q* (**b*) = 0,84 | 5,10* | 0,61 | 0,357 | -0,365 |
| *bEST-2** | *p* (**41*) = 0,35<br>*q* (**45*) = 0,65 | 0,31 | 0,25 | 0,107 | -0,123 |
| *SOD-1** | *p* (**100*) = 0,88<br>*q* (**110*) = 0,12 | 0,44 | 0,15 | 0,111 | -0,123 |
| *AAT** | *p* (**100*) = 0,48<br>*q* (**110*) = 0,52 | 25,89* | 2,54* | -0,805 | 0,782 |
| *GP-2** | *p* (**100*) = 0,03<br>*q* (**115*) = 0,97 | 0,03 | 0,01 | -0,026 | 0,013 |

Генетическая структура популяции на протяжении срока наблюдений оставалась стабильной, существенных изменений частот аллелей и генотипов между 2004 и 2005 годом не происходило.

## 6.2. Популяция тюльки дельты р. Дон

Популяционно-генетическая характеристика тюльки дельты Дона даётся на основании работ 2003 г. Популяционно-генетические исследования проведены по 12 локусам. Общий уровень полиморфизма черноморско-каспийской тюльки изученной популяции составляет P=0,33 (при критерии k=0,95). Средняя гетерозиготность, вычисленная прямым подсчётом $H_{ob}$=0,162 (s.e. 0,070), ожидаемая гетерозиготность $H_{ex}$=0,149 (s.e. 0,064). Количество аллелей на локус a=1,3 (s.e. 0,1).

Тюлька дельты р. Дон представляет собой генетически устойчивую популяцию, 83% локусов генетически сбалансированы. Установлено, что генетически популяция тюльки дельты Дона сходна с популяцией Таганрогского залива, сходство популяций ***r***=0,984±0,021 (***D***=0,016). Генетическое сходство по Нею (Nei, 1978) 1,0 (дистанция D менее 0,002). Основные популяционно-генетические характеристики популяции тюльки дельты Дона представлены в табл. 6.2.

**Таблица 6.2.** Популяционно-генетическая характеристика черноморско-каспийской тюльки популяции дельты р. Дон по 4 полиморфным локусам (2003 г.)

| **Локус** | **Частоты аллелей** | $\chi^2$ | **λ** | **F** | $D_H$ |
|---|---|---|---|---|---|
| *LDH-A** | *p* (**100*) = 0,75<br>*q* (**120*) = 0,25 | 1,06 | 0,39 | -0,176 | 0,159 |
| *ME-1** | *p* (**100*) = 0,33<br>*q* (**112*) = 0,67 | 0,07 | 0,12 | 0,045 | -0,060 |
| *bEST-2** | *p* (**41*) = 0,31<br>*q* (**45*) = 0,69 | 3,67 | 0,82 | 0,356 | -0,367 |
| *AAT** | *p* (**100*) = 0,47<br>*q* (**110*) = 0,53 | 5,53* | 1,17 | -0,607 | 0,554 |

На основании представленных данных можно сделать вывод о генетической сбалансированности в популяции тюльки дельты Дона (отклонение от равновесных частот генотипов имеется только по локусу *AAT**). Генетические локусы *LDH-A** и *ME-1** практически не имеют дефицита гетерозигот, распределение генотипов соответствует теоретически ожидаемому. Локус *bEST-2** испытывает дефицит гетерозигот ($D_H$>-0,3), что, вероятно, связано с воздействием отбора против гетерозигот.

### 6.3. Популяции тюльки водохранилищ Манычского каскада

Популяционно-генетическая характеристика тюльки Манычского каскада водохранилищ даётся на основании исследований 2004 г. Популяционно-генетические исследования проведены по 17 локусам. Общий уровень полиморфизма черноморско-каспийской тюльки изученных популяций составляет P=0,41 (при критерии k=0,95). Средняя гетерозиготность, вычисленная прямым подсчётом $H_{ob}$=0,193 (s.e. 0,071), ожидаемая гетерозиготность $H_{ex}$=0,163 (s.e. 0,054). Количество аллелей на локус a=1,5 (s.e. 0,1).

Все популяции водохранилищ Маныча генетически очень сильно схожи между собой; сходство популяций ***r***=0,986±0,019

(***D***=0,014). Генетическое сходство по Нею (Nei, 1978) 0,99 (D 0,001). Основные популяционно-генетические характеристики тюльки представлены в табл. 6.3. Из исследованных 7 полиморфных локусов распределение генотипов только по двум локусам (*aGPDH** и *SOD-1**) не соответствует равновесию Харди-Вайнберга (вероятно, для *aGPDH** это связано с условностью выделения изоформ).

**Таблица 6.3.** Популяционно-генетическая характеристика черноморско-каспийской тюльки популяций Манычского каскада водохранилищ по 7 полиморфным локусам (2004 г.)

| **Локус** | **Частоты аллелей** | $\chi^2$ | **λ** | **F** | $D_H$ |
|---|---|---|---|---|---|
| Пролетарское водохранилище. | | | | | |
| *LDH-A** | *p* (**100*) = 0,47<br>*q* (**120*) = 0,53 | 3,18 | 0,89 | -0,285 | 0,269 |
| *ME-1** | *p* (**100*) = 0,41<br>*q* (**112*) = 0,59 | 0,79 | 0,43 | -0,144 | -0,129 |
| *aGPDH** | *p* (**a*) = 0,45<br>*q* (**b*) = 0,55 | 18,78* | 2,14* | -0,703 | 0,681 |
| *BEST-2** | *p* (**41*) = 0,32<br>*q* (**45*) = 0,68 | 0,12 | 0,15 | 0,059 | -0,073 |
| *SOD-1** | *p* (**100*) = 0,94<br>*q* (**110*) = 0,06 | 24,75* | 0,58 | 0,998 | -0,002 |
| *AAT** | *p* (**100*) = 0,57<br>*q* (**110*) = 0,43 | 1,29 | 0,56 | -0,190 | 0,173 |
| *GP-2** | *p* (**100*) = 0,11<br>*q* (**115*) = 0,89 | 0,64 | 0,16 | -0,127 | 0,113 |
| Пролетарская ветвь Азовской ВРС. | | | | | |
| *LDH-A** | *p* (**100*) = 0,47<br>*q* (**120*) = 0,53 | 3,18 | 0,89 | 0,331 | -0,340 |
| Весёловское водохранилище. | | | | | |
| *LDH-A** | *p* (**100*) = 0,47<br>*q* (**120*) = 0,53 | 1,3 | 0,57 | -0,183 | 0,167 |
| Озеро Маныч-Гудило. | | | | | |
| *LDH-A** | *p* (**100*) = 0,75<br>*q* (**120*) = 0,25 | 0,67 | 0,31 | -0,333 | 0,222 |

По уровню гетерозиготности популяции Манычского каскада находятся в благоприятном положении: дефицит характерен только для локуса *SOD-1**, где почти полностью элиминирован второй аллель. Следует отметить, что по генетическому локусу, кодирующему лактатдегидрогеназу-А тюлька водохранилищ Маныча высокогетерозиготна: при почти равных частотах аллелей больше половины аллелофонда сосредоточено в гетерозиготах.

## 6.4. Популяция тюльки Днестровского лимана

Популяционно-генетическая характеристика тюльки Днестровского лимана даётся на основании работ 2004 г. Исследование проведено по 12 генетическим локусам.

**Таблица 6.4.** Популяционно-генетическая характеристика черноморско-каспийской тюльки популяции Днестровского лимана по 4 полиморфным локусам (2004 г.)

| Локус | Частоты аллелей | $\chi^2$ | λ | F | $D_H$ |
|---|---|---|---|---|---|
| *LDH-A** | *p* (**100*) = 0,73<br>*q* (**120*) = 0,27 | 2,21 | 0,59 | -0,238 | 0,222 |
| *ME-1** | *p* (**100*) = 0,25<br>*q* (**112*) = 0,75 | 2,22 | 0,56 | -0,333 | 0,300 |
| *bEST-2** | *p* (**41*) = 0,32<br>*q* (**45*) = 0,68 | 0,06 | 0,11 | 0,042 | -0,055 |
| *AAT** | *p* (**100*) = 0,52<br>*q* (**110*) = 0,48 | 4,22* | 1,02 | -0,419 | 0,390 |

Общий уровень полиморфизма черноморско-каспийской тюльки изученной популяции составляет P=0,40 (при критерии k=0,95). Средняя гетерозиготность, вычисленная прямым подсчётом $H_{ob}$=0,211 (s.e. 0,089), ожидаемая гетерозиготность $H_{ex}$=0,173 (s.e. 0,072). Количество аллелей на локус a=1,4 (s.e. 0,2). Основные популяционно-генетические характеристики тюльки Днестровского лимана представлены в табл. 6.4.

Из исследованных 4 полиморфных локусов распределение генотипов только по локусу *AAT** не соответствует равновесному состоянию Харди-Вайнберга (по критерию Пирсона), что связано с его высокой криптической изменчивостью. Локус *bEST-2** испытывает небольшой дефицит гетерозигот, для всех прочих локусов, в той или иной степени, характерен избыток гетерозигот.

## 6.5. Популяции тюльки р. Днепр

Популяции тюльки Днепра изучены по результатам облова в русловой части Среднего Днепра и Карачуновского водохранилища в осенне-зимний период 2005-06 г. по совокупности 16 генетических локусов. Общий уровень полиморфизма черноморско-каспийской тюльки изученных популяций составляет P=0,31 (при критерии k=0,95). Средняя гетерозиготность, вычисленная прямым подсчётом $H_{ob}$=0,120 (s.e. 0,057), ожидаемая гетерозиготность $H_{ex}$=0,113 (s.e. 0,046). Количество аллелей на локус a=1,3 (s.e. 0,1).

Обе популяции Днепра генетически подобны: сходство между выборками ***r***=0,988±0,017 (***D***=0,012), генетическое сходство по Нею (Nei, 1978) 0,99 (D 0,006). Основные популяционно-генетические характеристики представлены в табл. 6.5.

**Таблица 6.5.** Популяционно-генетическая характеристика черноморско-каспийской тюльки популяций р. Днепр по 6 полиморфным локусам (2005 г.)

| Локус | Частоты аллелей | $\chi^2$ | λ | F | $D_H$ |
|---|---|---|---|---|---|
| р. Днепр, Карачуновское водохранилище | | | | | |
| *LDH-A** | *p* (**100*) = 0,54<br>*q* (**120*) = 0,46 | 0,71 | 0,42 | -0,135 | 0,120 |
| *ME-1** | *p* (**100*) = 0,13<br>*q* (**112*) = 0,87 | 11,9* | 0,75 | 0,771 | -0,777 |
| *AAT** | *p* (**100*) = 0,50<br>*q* (**110*) = 0,50 | 26,6* | 2,53* | -0,800 | 0,778 |
| *bEST-2** | *p* (**41*) = 0,10<br>*q* (**45*) = 0,90 | 0,25 | 0,09 | -0,111 | 0,083 |
| *SOD-1** | *p* (**100*) = 0,87<br>*q* (**110*) = 0,13 | 0,44 | 0,15 | -0,152 | 0,121 |
| *GP-2** | *p* (**100*) = 0,03<br>*q* (**115*) = 0,97 | 0,03 | 0,01 | -0,026 | 0,013 |
| р. Днепр, русловая часть | | | | | |
| *LDH-A** | *p* (**100*) = 0,69<br>*q* (**120*) = 0,31 | 0,44 | 0,29 | -0,105 | 0,092 |
| *ME-1** | *p* (**100*) = 0,18<br>*q* (**112*) = 0,83 | 0,72 | 0,25 | 0,134 | -0,145 |
| *AAT** | *p* (**100*) = 0,54<br>*q* (**110*) = 0,46 | 11,70* | 1,70* | -0,548 | 0,528 |
| *bEST-2** | *p* (**41*) = 0,39<br>*q* (**45*) = 0,61 | 12,65* | 1,71* | 0,619 | -0,625 |
| *SOD-1** | *p* (**100*) = 0,89<br>*q* (**110*) = 0,11 | 5,60* | 0,47 | 0,374 | -0,382 |
| *GP-2** | *p* (**100*) = 0,00<br>*q* (**115*) = 1,00 | 0,00 | 0,00 | -0,00 | 0,00 |

При сравнении этих двух групп можно отметить б*о*льшую несбалансированность генетических характеристик популяции русловой части, чем популяции водохранилища. Так, в популяции русла Среднего Днепра при распределении частот генотипов по трём из 5 локусов (60%) по критерию $\chi^2$ наблюдается отклонение от равновесного распределения Харди-Вайнберга, а для популяции Карачуновского водохранилища распределение генотипов только двух из 6 локусов (33%) не соответствует равновесному. Частоты аллелей по всем изученным локусам (кроме *bEST-2**) в обеих популяциях при-мерно одинаковы, тогда как частоты генотипов сильно различаются, что может свидетельствовать о давлении отбора по этим локусам.

Анализ показателей гетерозиготности также демонстрирует разнонаправленные тенденции по различным локусам и в разных водоёмах. В Карачуновском водохранилище наблюдается дефицит гетерозигот лишь для локуса *ME-1**, локус *AAT** напротив, характеризуется избытком гетерозигот. Остальные локусы не испытывают значительных отклонений от равновесной частоты гетерозиготных генотипов. Аналогичная ситуация складывается и в русловой популяции, но здесь дефицит гетерозигот выражен для локуса *bEST-2** и, в меньшей степени, для локуса *SOD-1**.

Рассмотренные феномены можно объяснить значительной амплитудой колебаний абиотических факторов, характерной для данных водоёмов: русловая часть подвержена сильному влиянию паводка и колебаний уровня вышележащих водохранилищ, а Карачуновское водохранилище наоборот, сильноминерализованный и малопроточный, почти полностью изолированный от русла водоём, характеризующийся относительной стабильностью условий обитания рыб.

## 6.6. Популяция тюльки северной части Каспийского моря

Популяция тюльки Северного Каспия характеризуется на основании изучения 160 особей, выловленных во время массового нереста рыбы в мае 2005 г. в районе устья р. Сулак. Исследование проведено по 17 генетическим локусам. Общий уровень полиморфизма черноморско-каспийской тюльки изученной популяции составляет P=0,35 (при критерии k=0,95). Средняя гетерозиготность, вычисленная прямым подсчётом $H_{ob}$=0,179 (s.e. 0,074), ожидаемая гетерозиготность $H_{ex}$=0,140 (s.e. 0,050). Количество аллелей на локус a=1,5 (s.e. 0,1). Основные популяционно-генетические характеристики тюльки Северного Каспия представлены в табл. 6.6.

Распределение частот генотипов половины из исследованных локусов по критерию Пирсона (четверть локусов - по показателю Колмогорова-Смирнова) отличаются от равновесного состояния Харди-Вайнберга. Для локусов *aGPDH** и *AAT** характерны объективные причины неравновесного состояния: условность выделения изоформ (*aGPDH**) и криптическая изменчивость (*AAT**).

По показателям гетерозиготности также лишь для двух неравновесных локусов (*aGPDH** и *AAT**) характерно преобладание гетерозигот, два локуса (*bEST-2** и *SOD-1**) испытывают незначитель-ный дефицит гетерозигот, остальные локусы имеют равновесное соотношение гетеро- и гомозигот.

**Таблица 6.6.** Популяционно-генетическая характеристика черноморско-каспийской тюльки популяции Северного Каспия по 8 полиморфным локусам (2005 г.)

| Локус | Частоты аллелей | $\chi^2$ | λ | F | $D_H$ |
|---|---|---|---|---|---|
| *LDH-A** | p (**100*) = 0,81<br>q (**120*) = 0,19 | 0,01 | 0,03 | 0,056 | -0,063 |
| *ME-1** | p (**100*) = 0,60<br>q (**112*) = 0,40 | 0,78 | 0,42 | -0,071 | 0,064 |
| *aGPDH** | p (**a*) = 0,49<br>q (**b*) = 0,51 | 76,82* | 4,38* | -0,697 | 0,687 |
| *bEST-2** | p (**41*) = 0,25<br>q (**45*) = 0,75 | 6,65* | 0,96 | 0,219 | -0,225 |
| *SOD-1** | p (**100*) = 0,91<br>q (**110*) = 0,09 | 4,49* | 0,43 | 0,243 | -0,248 |
| *AAT** | p (**100*) = 0,51<br>q (**110*) = 0,49 | 60,31* | 3,88* | -0,874 | 0,862 |
| *G6PDH** | p (**85*) = 0,01<br>q (**100*) = 0,99 | 0,01 | 0,0 | -0,013 | 0,006 |
| *GP-2** | p (**100*) = 0,01<br>q (**115*) = 0,99 | 0,01 | 0,0 | -0,006 | 0,0 |

## 6.7. Популяции тюльки каскада водохранилищ р. Волги

Заселение тюлькой Волжских водохранилищ происходило на протяжении последних 60-ти лет последовательно вверх, против течения. Таким образом, должно существовать значительное генетическое единство между всеми популяциями. Это теоретическое предположение подкрепляется популяционно-генетическими исследованиями. По результатам работ 2002/11 гг. на водохранилищах Волги было исследовано более 30 выборок тюльки по 12-17 генетическим локусам. Установлено, что доля полиморфных локусов тюльки Волжских популяций составляет P=0,23-0,38 (в среднем P=0,31 при критерии k=0,95). Средняя гетерозиготность, вычисленная прямым подсчётом $H_{ob}$=0,102-0,147 (s.e. 0,066), ожидаемая гетерозиготность $H_{ex}$=0,102-0,139 (s.e. 0,053). Количество аллелей на локус a=1,4 (s.e. 0,1). Все популяции Волжского каскада водохранилищ генетически схожи между собой; сходство популяций ***r***=0,98±0,028 (***D***=0,02). Генетическое сходство по Нею (Nei, 1978) между популяциями Волги колеблется от 1,0 до 0,998 (D 0,001-0,005). Вместе с тем каждая популяция крупного водохранилища характеризуется своими популяционно-генетическими особенно-стями, обусловленными конкретными условиями среды (Карабанов, 2008а).

Анализ пространственной подразделённости тюльки 9 крупных водохранилищ Волги показал, что основную долю при формировании общей генетической изменчивости составляют межпопуляционные различия на уровне популяций разных водоемов ($F_{ST}$ 0,057 ≈ $F_{IT}$ 0,06) на фоне большой индивидуальной изменчивости ($F_{IS}$ 0,04) (табл. 6.7).

**Таблица 6.7.** Результаты расчёта F-статистики Райта для 9 популяций тюльки крупных водохранилищ Волги по 4 полиморфным генетическим локусам

| **Локус** | ***F*-коэффициент** | | |
|---|---|---|---|
| | $F_{IS}$ | $F_{IT}$ | $F_{ST}$ |
| *LDH-A** | 0,003 | 0,013 | 0,011 |
| *ME-1** | 0,036 | 0,063 | 0,028 |
| *bEST-2** | 0,217 | 0,232 | 0,2 |
| *GP-2** | -0,053 | -0,006 | 0,044 |
| Общее усреднённое значение на локус | 0,04 | 0,06 | 0,057 |

Локусы *LDH-A** и *ME-1** демонстрируют высокие значения индивидуальной изменчивости особей субпопуляции относительно всей совокупности ($F_{IT}$ 0,013 и 0,063 соответственно). Вероятно, это свидетельствует о локальной приспособленности конкретных аллельных вариантов в определённых условиях. Аллели локуса *bEST-2** демонстрируют высокие значения индивидуальной генетической изменчивости как относительно субпопуляции, так и всей популяционной структуры в целом. Возможно, такое разнообразие определяется высокой индивидуальной генетической вариабельностью по этому локусу, присущее данному виду, а также значительной изменчивости, присущей данному локусу в общем у многих животных (Глазко, 1988). Небольшие значения всех F-коэффициентов Райта вкупе с низким значением коэффициента М. Нея (Nei, 1987) $G_{ST}$ 0,003 позволяют сделать вывод о высоком генетическом родстве всех волжских популяций тюльки (Карабанов, 2008б).

### 6.7.1. Волгоградское водохранилище

Популяционно-генетическая характеристика тюльки Волгоградского водохранилища даётся на основании сбора материала в летний период 2004 г. Исследование проведено по 14 локусам. Общий уровень полиморфизма черноморско-каспийской тюльки изученной популяции составляет P=0,38 (при критерии k=0,95). Средняя гетерозиготность, вычисленная прямым подсчётом $H_{ob}$=0,133 (s.e. 0,063), ожи-даемая гетерозиготность $H_{ex}$=0,139 (s.e. 0,069). Количество аллелей на локус a=1,4 (s.e. 0,1). Генетические характеристики тюльки этой популяции представлены в табл. 6.8. В данном водоёме для всех исследованных локусов в целом соблюдается соответствие наблюдаемых и ожидаемых частот генотипов.

**Таблица 6.8.** Популяционно-генетическая характеристика черноморско-каспийской тюльки популяции Волгоградского водохранилища по 5 полиморфным локусам (2004 г.)

| Локус | Частоты аллелей | $\chi^2$ | λ | F | $D_H$ |
|---|---|---|---|---|---|
| *LDH-A** | *p* (*100) = 0,35<br>*q* (*120) = 0,65 | 0,2 | 0,2 | -0,099 | 0,071 |
| *ME-1** | *p* (*100) = 0,38<br>*q* (*112) = 0,63 | 3,48 | 0,88 | 0,467 | -0,483 |
| *bEST-2** | *p* (*41) = 0,14<br>*q* (*45) = 0,86 | 1,66 | 0,31 | 0,303 | -0,323 |
| *AAT** | *p* (*100) = 0,53<br>*q* (*110) = 0,47 | 2,95 | 0,86 | -0,417 | 0,375 |
| *GP-2** | *p* (*100) = 0,05<br>*q* (*115) = 0,95 | 0,06 | 0,02 | -0,053 | 0,026 |

Основная часть аллелофонда по локусу *AAT** и сосредоточена в гетерозиготном состоянии. Для *bEST-2** характерна практически полная элиминация аллельного варианта **41*. По показателям гетерозиготности лишь для локуса *AAT** характерно преобладание гетерозигот; локус *ME-1** испытывают небольшой дефицит гетерозигот, остальные локусы имеют примерно равновесное соотношение гетеро- и гомозигот.

### 6.7.2. р. Волга в районе г. Саратов

Популяционно-генетическая характеристика тюльки самого верхнего участка Волгоградского водохранилища (р. Волга в районе г. Саратов) даётся на основании сбора материала в летний период 2004 г. Исследование проведено по 14 генетическим локусам. Общий уровень полиморфизма черноморско-каспийской тюльки изученной популяции составляет P=0,23 (при критерии k=0,95). Средняя гетерозиготность, вычисленная прямым подсчётом $H_{ob}$=0,128 (s.e. 0,070), ожидаемая гетерозиготность $H_{ex}$=0,102 (s.e. 0,052). Количество аллелей на локус a=1,3 (s.e. 0,1). Популяционно-генетические характеристики тюльки данной популяции представлены в табл. 6.9. Для всех исследованных локусов (кроме *AAT**) в целом соблюдается соответствие наблюдаемых и ожидаемых частот генотипов.

По показателям гетерозиготности лишь для локуса *AAT** характерно преобладание гетерозигот, локус *LDH-A** испытывает небольшой дефицит гетерозигот, остальные локусы имеют практически равновесное соотношение гетеро- и гомозигот.

**Таблица 6.9.** Популяционно-генетическая характеристика черноморско-каспийской тюльки популяции р. Волги в районе г. Саратов по 4 полиморфным локусам (2004 г.)

| Локус | Частоты аллелей | $\chi^2$ | λ | F | $D_H$ |
|---|---|---|---|---|---|
| *LDH-A** | *p* (**100*) = 0,18<br>*q* (**120*) = 0,82 | 1,59 | 0,37 | 0,226 | -0,239 |
| *ME-1** | *p* (**100*) = 0,34<br>*q* (**112*) = 0,66 | 0,04 | 0,08 | -0,036 | 0,017 |
| *bEST-2** | *p* (**41*) = 0,03<br>*q* (**45*) = 0,97 | 0,04 | 0,01 | -0,034 | 0,017 |
| *AAT** | *p* (**100*) = 0,45<br>*q* (**110*) = 0,55 | 13,14* | 1,79* | -0,673 | 0,644 |

### 6.7.3. Куйбышевское водохранилище

Популяционно-генетическая характеристика тюльки Куйбышевского водохранилища даётся на основании сбора материала в летний период 2004 г. Исследование проведено по 12 генетическим локусам. Общий уровень полиморфизма черноморско-каспийской тюльки изученной популяции составляет P=0,31 (при критерии k=0,95). Средняя гетерозиготность, вычисленная прямым подсчётом $H_{ob}$=0,108 (s.e. 0,054), ожидаемая гетерозиготность $H_{ex}$=0,109 (s.e. 0,049). Количество аллелей на локус a=1,3 (s.e. 0,1).

Популяционно-генетические характеристики тюльки данного водоёма представлены в табл. 6.10. Для всех исследованных локусов в целом соблюдается соответствие наблюдаемых и ожидаемых частот генотипов.

Для локуса *LDH-A** характерно трёхкратное преобладание аллеля **120* над его альтернативным вариантом, который, однако, сохраняется и присутствует в популяции даже в виде редких гомозигот. Для локуса *bEST-2** также характерна потеря аллеля **41*, который, всё же сохраняется в популяции от полной утраты. Аналогичная ситуация характерна и для локуса *ME-1**. Локус *AAT** находится в константном состоянии, сохранение редких генотипов происходит за счёт скрещивания гетерозигот.

По показателям гетерозиготности лишь для локуса *AAT** характерно преобладание гетерозигот, остальные локусы испытывают их незначительный дефицит.

**Таблица 6.10.** Популяционно-генетическая характеристика черноморско-каспийской тюльки популяции Куйбышевского водохранилища по 4 полиморфным локусам (2004 г.)

| Локус | Частоты аллелей | $\chi^2$ | λ | F | $D_H$ |
|---|---|---|---|---|---|
| *LDH-A** | *p* (**100*) = 0,25<br>*q* (**120*) = 0,75 | 0,29 | 0,20 | 0,088 | -0,101 |
| *ME-1** | *p* (**100*) = 0,16<br>*q* (**112*) = 0,84 | 1,65 | 0,34 | 0,208 | -0,219 |
| *bEST-2** | *p* (**41*) = 0,17<br>*q* (**45*) = 0,83 | 1,44 | 0,33 | 0,201 | -0,211 |
| *AAT** | *p* (**100*) = 0,39<br>*q* (**110*) = 0,61 | 3,83 | 0,95 | -0,322 | 0,304 |

### 6.7.3. Чебоксарское водохранилище

Популяционно-генетическая характеристика тюльки Чебоксарского водохранилища даётся на основании сбора материала в летний период 2003 г. Исследование проведено по 12 локусам. Общий уровень полиморфизма черноморско-каспийской тюльки изученной популяции составляет P=0,31 (при критерии k=0,95). Средняя гетерозиготность, вычисленная прямым подсчётом $H_{ob}$=0,128 (s.e. 0,066), ожидаемая гетерозиготность $H_{ex}$=0,113 (s.e. 0,052). Количество аллелей на локус a=1,3 (s.e. 0,1).

Популяционно-генетические характеристики тюльки данной популяции представлены в табл. 6.11. Для всех исследованных локусов (кроме *AAT** и *bEST-2**, по критерию Пирсона) в целом соблюдается соответствие наблюдаемых и ожидаемых частот генотипов. Отклонение от равновесных соотношений генотипов, возможно, объясняется небольшим объёмом выборки либо действием отбора.

**Таблица 6.11.** Популяционно-генетическая характеристика черноморско-каспийской тюльки популяции Чебоксарского водохранилища по 4 полиморфным локусам (2003 г.)

| Локус | Частоты аллелей | $\chi^2$ | λ | F | $D_H$ |
|---|---|---|---|---|---|
| *LDH-A** | *p* (**100*) = 0,26<br>*q* (**120*) = 0,74 | 0,10 | 0,12 | 0,050 | -0,063 |
| *ME-1** | *p* (**100*) = 0,24<br>*q* (**112*) = 0,76 | 1,30 | 0,42 | -0,183 | 0,168 |
| *bEST-2** | *p* (**41*) = 0,11<br>*q* (**45*) = 0,89 | 7,14* | 0,52 | 0,439 | -0,447 |
| *AAT** | *p* (**100*) = 0,56<br>*q* (**110*) = 0,44 | 4,39* | 1,03 | -0,524 | 0,476 |

По показателям гетерозиготности лишь для локуса *AAT** характерно преобладание гетерозигот, локус *bEST-2** испытывает дефи-цит гетерозигот, другие локусы имеют почти равновесное соотношение гетеро- и гомозигот.

### 6.7.4. Горьковское водохранилище

Популяционно-генетическая характеристика тюльки Горьковского водохранилища даётся на основании сбора материала в летний период 2005 г. и осенью 2011 г. Сбор материала проводился в верхней - русловой (г. Волгореченск) и нижней - озёрной (г. Чкаловск) частях водоёма. Исследование проведено по 14 локусам. Общий уровень полиморфизма черноморско-каспийской тюльки изученной популяции составляет P=0,31 (при критерии k=0,95). Средняя гетерозиготность, вычисленная прямым подсчётом $H_{ob}$=0,101 (s.e. 0,055), ожидаемая гетерозиготность $H_{ex}$=0,107 (s.e. 0,051). Количество аллелей на локус a=1,3 (s.e. 0,1). Обе популяции Горьковского водохранилища генетически идентичны: ***r***=0,995±0,001 (***D***=0,003), генетическое сходство по Нею (Nei, 1978) 1,0.

Популяционно-генетические характеристики тюльки данной популяции представлены в табл. 6.12. Для более корректного сравнения с другими водохранилищами Волги здесь представлены данные за 2005 г. Популяцинно-генеческие параметры популяции тюльки данного водоёма в 2011 г. практически не изменились.

**Таблица 6.12.** Популяционно-генетическая характеристика двух популяций черноморско-каспийской тюльки популяции Горьковского водохранилища по 4 полиморфным локусам (2005 г.)

| Локус | Частоты аллелей | $\chi^2$ | λ | F | $D_H$ |
|---|---|---|---|---|---|
| г. Чкаловск | | | | | |
| *LDH-A** | *p* (**100*) = 0,21<br>*q* (**120*) = 0,79 | 1,78 | 0,44 | 0,214 | -0,224 |
| *ME-1** | *p* (**100*) = 0,27<br>*q* (**112*) = 0,73 | 3,13 | 0,70 | 0,283 | -0,292 |
| *bEST-2** | *p* (**41*) = 0,09<br>*q* (**45*) = 0,91 | 11,25* | 0,54 | 0,530 | -0,536 |
| *AAT** | *p* (**100*) = 0,51<br>*q* (**110*) = 0,49 | 5,47* | 1,17 | -0,390 | 0,371 |
| г. Волгореченск | | | | | |
| *LDH-A** | *p* (**100*) = 0,29<br>*q* (**120*) = 0,71 | 0,29 | 0,22 | 0,085 | -0,091 |
| *ME-1** | *p* (**100*) = 0,21<br>*q* (**112*) = 0,79 | 5,37* | 0,76 | 0,371 | -0,379 |
| *bEST-2** | *p* (**41*) = 0,08<br>*q* (**45*) = 0,92 | 3,01 | 0,25 | 0,278 | -0,276 |
| *AAT** | *p* (**100*) = 0,58<br>*q* (**110*) = 0,42 | 1,66 | 0,63 | -0,295 | 0,261 |

Несколько различное распределение генотипов при почти одинаковой частоте аллелей определяется индивидуальными особенностями особей, видимо, связанных с локальными условиями ($F_{IS}$ 0,044 и $F_{IT}$ 0,05 для всех локусов), при этом нельзя говорить об особой субпопуляционной структуре тюльки Горьковского водохранилища ($F_{ST}$ 0,006). Для всех исследованных локусов по критерию Колмогорова-Смирнова соблюдается соответствие наблюдаемых и ожидаемых частот генотипов. Критерий Пирсона показывает отклонение наблюдаемых частот генотипов от теоретических по двум локусам (*bEST-2** и *AAT**) для популяции приплотинной части водохранилища и по одному локусу (*ME-1**) - для речной части.

По показателям гетерозиготности лишь для локуса *AAT** характерен некоторый избыток гетерозигот, остальные локусы испытывают их незначительный дефицит.

### 6.7.5. Рыбинское водохранилище

Рыбинское водохранилище - крупнейший водоём Верхней Волги. Акватория водохранилища делится на 4 плёса (Волжский, Моложский, Шекснинский, Центральный). Мониторинг популяционно-генетической структуры черноморско-каспийской тюльки ведётся с момента её заселения в 2001 г. В настоящей работе отражены данные по генетическому разнообразию тюльки Рыбинского водохранилища в период 2003/11 гг. Можно отметить, что за период наблюдений с 2003 по 2011 гг. популяция тюльки Рыбинского водохранилища представляет собой устойчивую панмиктичную популяцию. На первоначальном этапе вселения тюлька Волго-Шекснинских водохранилищ характеризовалась отсутствием значимой внутрипопуляционной дифференциации (Слынько, Лапушкина, 2003). К настоящему моменту можно условно выделить несколько крупных стад тюльки, примерно соответствующих четырём плёсам Рыбинского водохранилища, однако низкое значение инбридингового коэффициента $F_{ST}$=0,002 свидетельствует о высокой степени обмена генами между особями из всех частей популяции. Межгодовое сходство в популяции тюльки Рыбинского водохранилища ***r***=0,997±0,006 (***D***=0,003). Генетическое сходство (по данным 2006, 2007 и 2011 гг.) четырёх крупных стад тюльки (выделенных по соответствующим плёсам) по Нею (Nei, 1978) 1,0.

Изучение популяционно-генетической структуры тюльки данного водоёма проведено в общей сложности по 17 генетическим локусам. Общий уровень полиморфизма черноморско-каспийской тюльки изученной популяции составляет P=0,31 (при критерии k=0,95). Средняя гетерозиготность, вычисленная прямым подсчётом $H_{ob}$=0,147 (s.e. 0,075), ожидаемая гетерозиготность $H_{ex}$=0,121 (s.e. 0,055). Количество аллелей на локус a=1,4 (s.e. 0,1).

Основные популяционно-генетические характеристики тюльки Рыбинского водохранилища представлены в табл. 6.13. Следует отметить, что приведённые результаты даны для всей популяции в целом и в разных плёсах могут различаться. В 2006 г. практически по всем полиморфным локусам распределение генотипов отличается от теоретического (по критерию Пирсона). Через пять лет, в 2011 г., наоборот, по большинству локусов наблюдалось соответствие равновесным частотам генотипов.

**Таблица 6.13.** Популяционно-генетическая характеристика черноморско-каспийской тюльки популяции Рыбинского водохранилища по 5 полиморфным локусам на примере 2006 и 2011 гг.

| Локус | Частоты аллелей | $\chi^2$ | λ | F | $D_H$ |
|---|---|---|---|---|---|
| 2006 г. | | | | | |
| *LDH-A** | *p* (*100) = 0,25<br>*q* (*120) = 0,75 | 4,91* | 0,83 | 0,158 | -0,160 |
| *ME-1** | *p* (*100) = 0,36<br>*q* (*112) = 0,64 | 5,18* | 1,04 | -0,165 | 0,162 |
| *bEST-2** | *p* (*41) = 0,14<br>*q* (*45) = 0,86 | 7,27* | 0,64 | 0,197 | -0,199 |
| *AAT** | *p* (*100) = 0,53<br>*q* (*110) = 0,47 | 20,09* | 2,24* | -0,528 | 0,524 |
| *GP-2** | *p* (*100) = 0,001<br>*q* (*115) = 0,999 | 0,001 | 0,001 | -0,003 | 0,0 |
| 2011 г. | | | | | |
| *LDH-A** | *p* (*100) = 0,20<br>*q* (*120) = 0,80 | 0,06 | 0,07 | -0,017 | 0,014 |
| *ME-1** | *p* (*100) = 0,28<br>*q* (*112) = 0,73 | 1,38 | 0,47 | 0,097 | -0,100 |
| *bEST-2** | *p* (*41) = 0,17<br>*q* (*45) = 0,83 | 31,89* | 1,59* | 0,417 | -0,419 |
| *AAT** | *p* (*100) = 0,52<br>*q* (*110) = 0,48 | 20,1* | 2,24* | -0,728 | 0,724 |
| *GP-2** | *p* (*100) = 0,001<br>*q* (*115) = 0,999 | 0,001 | 0,001 | -0,003 | 0,0 |

Показатели гетерозиготности в популяции также сильно варьируют по годам, что, вероятно, связано с воздействием отбора на гетерозиготы у разных локусов. Редкий аллель *GP-2*115* не оказывает влияния на генетические характеристики популяции из-за своей пренебрежительно низкой концентрации.

### 6.7.6. Угличское и Иваньковское водохранилища

Два верхних Волжских водохранилища - Угличское и Иваньковское - небольшие водоёмы речного типа. Тюлька по ним расселилась относительно недавно, популяции генетически практически идентичны: ***r***=0,998±0,002 (***D***=0,001), генетическое сходство по Нею (Nei, 1978) 1,0.

**Таблица 6.14.** Популяционно-генетическая характеристика двух популяций черноморско-каспийской тюльки популяций Угличского и Иваньковского водохранилищ по 4 полиморфным локусам (2003 г.)

| Локус | Частоты аллелей | $\chi^2$ | $\lambda$ | F | $D_H$ |
|---|---|---|---|---|---|
| Угличское водохранилище | | | | | |
| *LDH-A** | *p* (**100*) = 0,26<br>*q* (**120*) = 0,74 | 0,22 | 0,18 | -0,076 | 0,062 |
| *ME-1** | *p* (**100*) = 0,32<br>*q* (**112*) = 0,68 | 0,55 | 0,32 | -0,118 | 0,104 |
| *bEST-2** | *p* (**41*) = 0,18<br>*q* (**45*) = 0,82 | 4,45* | 0,61 | 0,347 | -0,356 |
| *AAT** | *p* (**100*) = 0,59<br>*q* (**110*) = 0,41 | 11,99* | 1,67* | -0,636 | 0,614 |
| Иваньковское водохранилище | | | | | |
| *LDH-A** | *p* (**100*) = 0,24<br>*q* (**120*) = 0,76 | 0,79 | 0,32 | 0,110 | -0,117 |
| *ME-1** | *p* (**100*) = 0,25<br>*q* (**112*) = 0,75 | 0,28 | 0,20 | 0,063 | -0,071 |
| *bEST-2** | *p* (**41*) = 0,15<br>*q* (**45*) = 0,85 | 6,74* | 0,65 | 0,322 | -0,327 |
| *AAT** | *p* (**100*) = 0,44<br>*q* (**110*) = 0,56 | 5,62* | 1,17 | -0,292 | 0,282 |

Исследование проведено в 2003 г. в общей сложности по 12 генетическим локусам. Общий уровень полиморфизма черноморско-каспийской тюльки изученных популяций составляет P=0,31 (при критерии k=0,95). Средняя гетерозиготность, вычисленная прямым подсчётом для Угличского водохранилища $H_{ob}$=0,143 (s.e. 0,071), ожидаемая гетерозиготность $H_{ex}$=0,123 (s.e. 0,055). Для Иваньковского водохранилища эти показатели несколько ниже: средняя гетерозиготность, вычисленная прямым подсчётом, соответствует ожидаемым значениям $H_{ob}$=$H_{ex}$=0,114 (s.e. 0,055). Количество аллелей на локус a=1,3 (s.e. 0,1).

Основные популяционно-генетические характеристики тюльки Угличского и Иваньковского водохранилищ представлены в табл. 6.14. В обеих популяциях по двум локусам (*AAT** и *bEST-2**) имеет-ся достоверное отклонение от равновесных частот генотипов, что, вероятно, связано с их крайним географическим положением относительно естественноисторического ареала тюльки и малым сроком от момента вселения (возможно, последствия «эффекта основателя»). По показателям гетерозиготности популяция тюльки Угличского водохранилища характеризуется небольшим избытком гетерозигот по всем локусам, за исключением *bEST-2**. Популяция Иваньковского водохранилища, наоборот, испытывает незначительный дефицит гетерозигот (кроме локуса *AAT**).

## 6.8. Популяция тюльки Шекснинского водохранилища

Шекснинское водохранилище - самое северное из всех водоёмов, освоенных черноморско-каспийской тюлькой. Популяция тюльки Шекснинского водохранилища генетически схожа с популяцией Рыбинского водохранилища: ***r***=0,99±0,01 (***D***=0,001), генетическое сходство по Нею (Nei, 1978) 0,998 (D 0,002). Исследование проведено в 2005 г. в общей сложности по 17 генетическим локусам.

**Таблица 6.15.** Популяционно-генетическая характеристика черноморско-каспийской тюльки популяции Шекснинского водохранилища по 4 полиморфным локусам (2005 г.)

| Локус | Частоты аллелей | $\chi^2$ | λ | F | $D_H$ |
|---|---|---|---|---|---|
| *LDH-A** | *p* (**100*) = 0,28<br>*q* (**120*) = 0,72 | 2,45 | 0,63 | -0,391 | 0,348 |
| *ME-1** | *p* (**100*) = 0,14<br>*q* (**112*) = 0,86 | 0,27 | 0,12 | -0,158 | 0,105 |
| *bEST-2** | *p* (**41*) = 0,19<br>*q* (**45*) = 0,81 | 0,85 | 0,28 | -0,231 | 0,192 |
| *AAT** | *p* (**100*) = 0,60<br>*q* (**110*) = 0,40 | 4,44* | 1,01 | -0,667 | 0,583 |

Общий уровень полиморфизма черноморско-каспийской тюльки изученной популяции составляет P=0,31 (при критерии k=0,95). Средняя гетерозиготность, вычисленная прямым подсчётом $H_{ob}$=0,155 (s.e. 0,074), ожидаемая гетерозиготность $H_{ex}$=0,114 (s.e. 0,052). Количество аллелей на локус a=1,3 (s.e. 0,1). Основные популяционно-генетические характеристики представлены в табл. 6.15. Популяция тюльки Шекснинского водохранилища очень малочисленная, объём выборки составил лишь 16 особей. По критерию Колмогорова-Смирнова все локусы находятся в равновесном состоянии.

Для всех локусов характерна невысокая частота одного из аллелей (*LDH-A*100*, *ME-1*100*, *bEST-2*41*, *AAT*110*), которые сохраняются лишь в гетерозиготном состоянии. По критерию гетерозиготности для данной популяции характерно преобладание гетерозигот по всем исследованным локусам. Вероятно, это связано с краевым северным положением популяции, а давление отбора благоприятствует сохранению аллелей в составе гетерозигот.

## 6.9. Популяция тюльки Камского водохранилища

Популяционно-генетическая характеристика тюльки Камского водохранилища даётся на основании сбора материала в осенний период 2011 г. Исследование проведено по 8 локусам. Общий уровень полиморфизма черноморско-каспийской тюльки изученной популяции составляет P=0,41 (при критерии k=0,95). Средняя гетерозиготность, вычисленная прямым подсчётом $H_{ob}$=0,236 (s.e. 0,066), ожидаемая гетерозиготность $H_{ex}$=0,194 (s.e. 0,052). Количество аллелей на локус a=1,5 (s.e. 0,1). Основные популяционно-генетические характеристики представлены в табл. 6.16.

**Таблица 6.16.** Популяционно-генетическая характеристика черноморско-каспийской тюльки популяции Камского водохранилища по 4 полиморфным локусам (2011 г.)

| Локус | Частоты аллелей | $\chi^2$ | λ | F | $D_H$ |
|---|---|---|---|---|---|
| *LDH-A** | *p* (**100*) = 0,15<br>*q* (**120*) = 0,85 | 1,28 | 0,28 | -0,169 | 0,156 |
| *ME-1** | *p* (**100*) = 0,54<br>*q* (**112*) = 0,46 | 1,06 | 0,51 | -0,144 | 0,133 |
| *bEST-2** | *p* (**41*) = 0,18<br>*q* (**45*) = 0,82 | 2,59 | 0,47 | 0,240 | -0,248 |
| *AAT** | *p* (**100*) = 0,50<br>*q* (**110*) = 0,50 | 14,63* | 1,95* | -0,610 | 0,590 |

Следует отметить, что данной в выборке тюльки из Камского водохранилища отсутствуют гомозиготы *LDH-A*100/100*, а сам аллель **100* сохраняется лишь у редких гетерозигот. Из изученных локусов дефицит гетерозигот характерен лишь для *bEST-2**.

## 6.10. Популяционная генетика родственного морского вида - анчоусовидной тюльки *Clupeonella engrauliform*es (Borodin, 1904) из северной части Каспийского моря

Анчоусовидная тюлька (килька) - облигатный морской вид, обитающий в северной части Каспийского моря не далее границы опреснения. Из всех представителей рода *Clupeonella* анчоусовидная тюлька по экологии наиболее близка к черноморско-каспийской тюльке.

Исследование северокаспийской популяции анчоусовидной тюльки проведено в 2004-05 годах в общей сложности по 12 генетическим локусам. Общий уровень полиморфизма анчоусовидной тюльки составляет P=0,35 (при критерии k=0,95). Средняя гетерозиготность, вычисленная прямым подсчётом $H_{ob}$=0,052 (s.e. 0,02), ожидаемая гетерозиготность $H_{ex}$=0,073 (s.e. 0,03). Количество аллелей на локус a=1,4 (s.e. 0,1). Основные популяционно-генетические характеристики анчоусовидной тюльки Северного Каспия представлены в табл. 6.17.

**Таблица 6.17.** Популяционно-генетическая характеристика анчоусовидной тюльки *C. engrauliformes* Северного Каспия по 7 полиморфным локусам (2005 г.)

| Локус | Частоты аллелей | $\chi^2$ | λ | F | $D_H$ |
|---|---|---|---|---|---|
| *LDH-A** | *p* (*100) = 0,99<br>*q* (*120) = 0,01 | 0,01 | 0,01 | -0,013 | 0,007 |
| *ME-1** | *p* (*112) = 0,05<br>*q* (*120) = 0,95 | 3,24 | 0,18 | 0,208 | -0,213 |
| *bEST-2** | *p* (*41) = 0,72<br>*q* (*45) = 0,28 | 5,34* | 0,94 | 0,269 | -0,274 |
| *AAT** | *p* (*110) = 0,05<br>*q* (*120) = 0,95 | 3,36 | 0,18 | 0,209 | -0,214 |
| *aGPDH** | *p* (*a) = 0,10<br>*q* (*b) = 0,90 | 2,82 | 0,31 | 0,204 | -0,210 |
| *SOD-1** | *p* (*100) = 0,82<br>*q* (*110) = 0,18 | 5,90* | 0,73 | 0,443 | -0,445 |
| *GP-2** | *p* (*100) = 0,06<br>*q* (*115) = 0,94 | 3,26 | 0,21 | 0,293 | -0,246 |

Следует отметить, что популяционно-генетические характеристики *C. engraulifor*mes по показателям полиморфизма и гетерозиготности ближе к морским сельдям (Nevo et al., 1984), чем к *C. cultirventris* и пресноводных рыб. Анчоусовидная тюлька чётко дифференцируется от черноморско-каспийской по локусам *GP-3**, *G6PDH**, *ME-1** и *AAT**. У анчоусовидной тюльки практически полностью отсутствует аллель *LDH-A*120*, который встречается только в составе крайне редких гетерозигот. Также анчоусовидная тюлька характеризуется высокой частотой аллеля *GP-2 *100*, аналог которого в значимом количестве встречается лишь в популяциях черноморско-каспийской тюльки Манычских водохранилищ. Для данной популяции характерно преобладание гетерозигот по всем исследованным локусам, кроме *LDH-A**.

# Глава 7.
# Генетическая дифференциация *Clupeonella cultriventris* на всём протяжении ареала

Одним из основных вопросов, активно обсуждавшимся на протяжении многих лет, служит таксономический статус различных крупных географических популяций тюльки. У черноморско-каспийской тюльки в связи с большим географическим ареалом раньше выделяли 3 подвида: черноморская тюлька *C. delicatula delicatula*, каспийская тюлька *C. delicatula caspia*, азовская тюлька *C. delicatula azovi* и отдельная пресноводная форма - чархальская тюлька *C. delicatula* m. *tscharchalensis*, обитавшая в оз. Чархал (Владимиров, 1950; Световидов, 1952). В дальнейшем, от разделения вида на подвиды было решено отказаться (Аннотированный …, 1999). Вместе с тем, как упоминалось выше, в ряде работ (Kottelat, 1997; Богуцкая, Насека, 2004; Kottelat, Freyhof, 2007) обосновывается выделение азово-черноморской, каспийской, волжской и манычской популяций тюльки как отдельных подвидов или видов. Таким образом, для популяционно-генетических исследований крайне важным является вопрос достоверности отнесения различных популяций тюльки к одному и тому же виду. В противном случае любые сравнения, например морских и пресноводных тюлек, становятся некорректными.

## 7.1. Таксономический статус крупных групп популяций тюльки на основании генетико-биохимических данных

Для определения иерархической структуры была проведена UPGMA-кластеризация крупных групп популяций тюльки из разных бассейнов на основании дистанций Нея (Nei, 1972, 1975), с использованием в качестве репера данных по каспийской популяции анчоусовидной тюльки. Представленные результаты свидетельствуют, что на всем протяжении своего современного ареала черноморско-каспийская тюлька представляет собой совокупность популяций *Clupeonella cultriventris* sensu lato, надежно дифференцированную от родственного вида - анчоусовидной тюльки (рис. 7.1). Значимых генетических различий межвидового уровня (Ayala, 1975) между азово-черноморской, каспийской и манычской тюлькой не наблюдается (дистанция 0,046), тогда как между черноморско-каспийской и анчоусовидной тюлькой различия достигают величины D 0,69, что может характеризовать их как надежные таксономические виды.

В пределах кластера черноморско-каспийской тюльки можно выделить 2 субкластера: Понто-Каспийский и Волжско-Камско-Манычский в статусе популяционных групп (генетическая дистанция около 0,045). Не исключено, что формирование этих субкластеров вызвано длительным обитанием тюльки в морских и пресноводных водоёмах. Подобная генетическая дифференциация между экологическими формами широко распространена у многих рыб (Ryman et al., 1979; Ryman, Stahl, 1981; Kornfield et al., 1982).

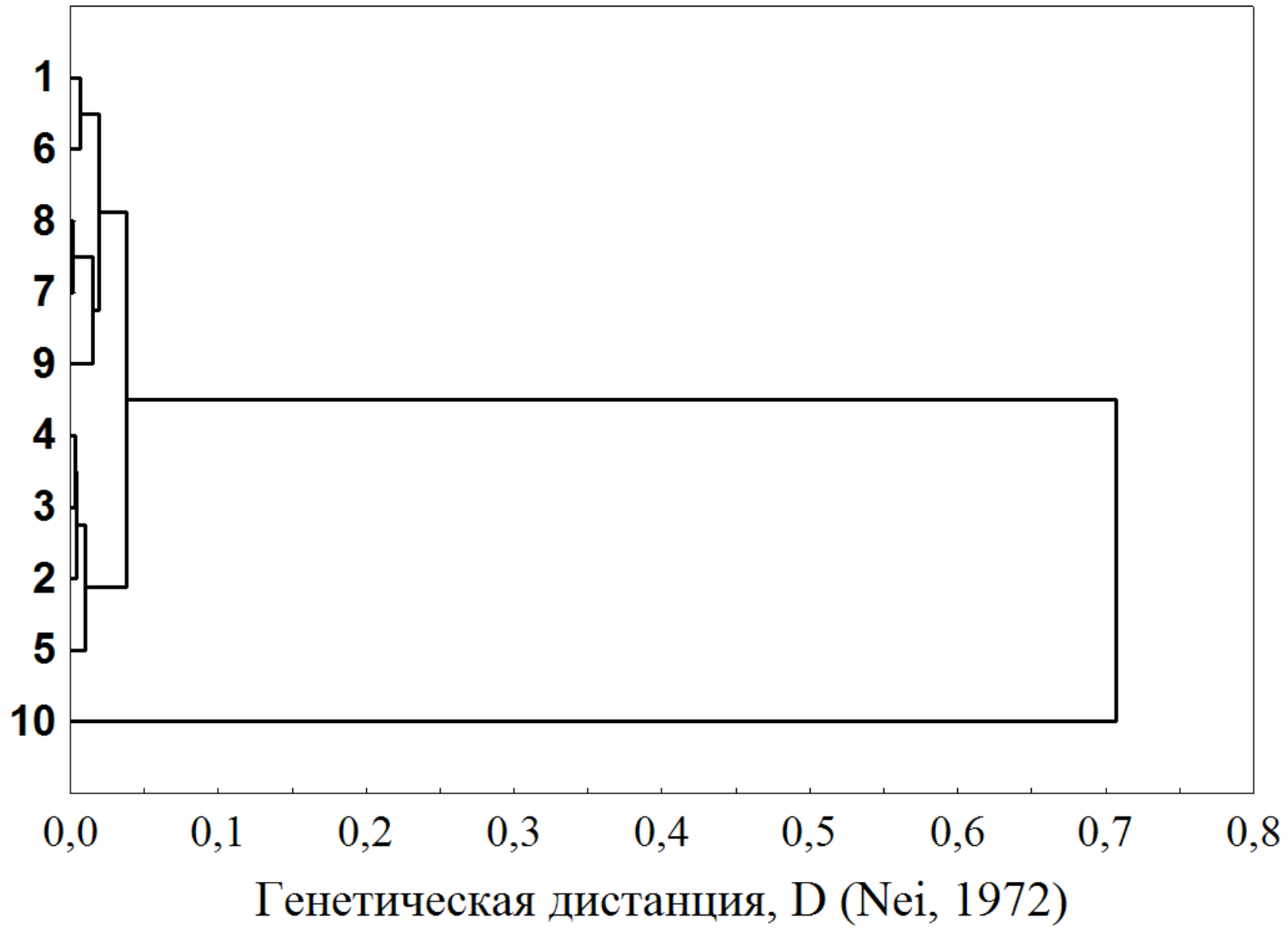

**Рисунок 7.1.** Иерархическая структура основных групп популяций *C. cultriventris* с реперным видом *C. engrauliformes* на основании популяционно-генетического анализа 12 локусов. Результат иерархической кластеризации (UPGMA) на основании генетических дистанций Нея (Nei, 1972).

Популяции: **1**–Северный Каспий, **2**–Волгоградское водохранилище, **3**–Горьковское водохранилище, **4**–Рыбинское водохранилище, **5**–Камское водохранилище, **6**–Азовское море, **7**–Днестровский лиман, **8**–р. Днепр, **9**–Маныч, **10**–*C. engrauliformes* (Каспийское море)

Отсутствие значимых различий пресноводных популяций Днепра в сравнении с лиманными черноморскими популяциями, вероятно, может быть объяснено исторической молодостью этих пресноводных групп, в которых изменения частот аллелей ещё не достигли существенных значений.

Таким образом, популяционно-генетическая структура вида *C. cultriventris* определяется не столько удалённостью новообразованных популяций от исторического ареала, сколько, вероятно, временем существования в пресных водах. Все представленные данные позволяют говорить о возможном наличии физиологических рас тюльки, обитающей в высоко- и низко-минерализованных водах, определяемых экологической селективностью ряда аллелей маркерных генетических локусов, но не о существовании независимых таксонов (Слынько и др., 2010).

Немаловажным вопросом, ответ на который даёт изучение популяционной структуры тюльки, является происхождение популяций Маныча. Существует мнение (Доманевский и др., 1964; Биологические…, 2004) что заселение тюлькой Манычских водохранилищ произошло из Волги после создания Волго-Донского канала через Цимлянское водохранилище. Представленные данные предлагают выдвинуть другую гипотезу. При анализе распределения популяционно-генетических показателей при многомерном шкалировании (Zar, 1999) популяция Маныча находится несколько отдельно как от волжских, так и от морских групп (рис. 7.2). В области Пространств 1-2 Манычская популяция единственная занимает III четверть. В Пространстве 1 популяция Маныча располагается в значениях <0, что сближает её с пресноводной тюлькой Волжской группы. При рассмотрении расположения Манычской популяции относительно других групп в совокупности Пространств 1-2-3 можно отметить некоторую близость её к тюльке Азовского моря и, в меньшей степени - Каспия (рис. 7.2).

Если предположить, что тюлька проникла в Маныч из Волги после строительства Волго-Донского канала, то её популяционно-генетические характеристики должны быть схожими с таковыми Волги с поправкой на изменение частот аллелей в локальных условиях (более высокая минерализация воды в Маныче). Но, в таком случае, скорость микроэволюционных преобразований именно в этом водоёме должна быть очень высокой, чтобы примерно за полвека настолько сильно изменить генетические характеристики популяции. Против столь большой скорости генетических преобразований у тюльки свидетельствует значительное сходство между популяциями Днепра и Черноморских лиманов. Приток мигрантов в Днепр из лиманов сильно затруднён, так что наблюдаемые различия между популяциями будут следствием различной стратегии адаптации к местным условиям, т.е. результатом микроэволю-ционных преобразований в популяциях. Вместе с тем, проявление различий между популяциями Днепра и лиманов не столь велико (рис. 7.1. и 7.2). Возможно, что тюлька проникла в Маныч по системе каналов из Таганрогского залива через Дон. Но и в этом случае за срок около 70 лет должно произойти столь же существенное перераспределение частот аллелей. Вполне вероятно, что в исторический период существовал обмен рекрутами между популяциями Маныча и Азовского моря, который в настоящее время маловероятен в связи с высокой изоляцией водохранилищ Манычского каскада (Карабанов, 2009).

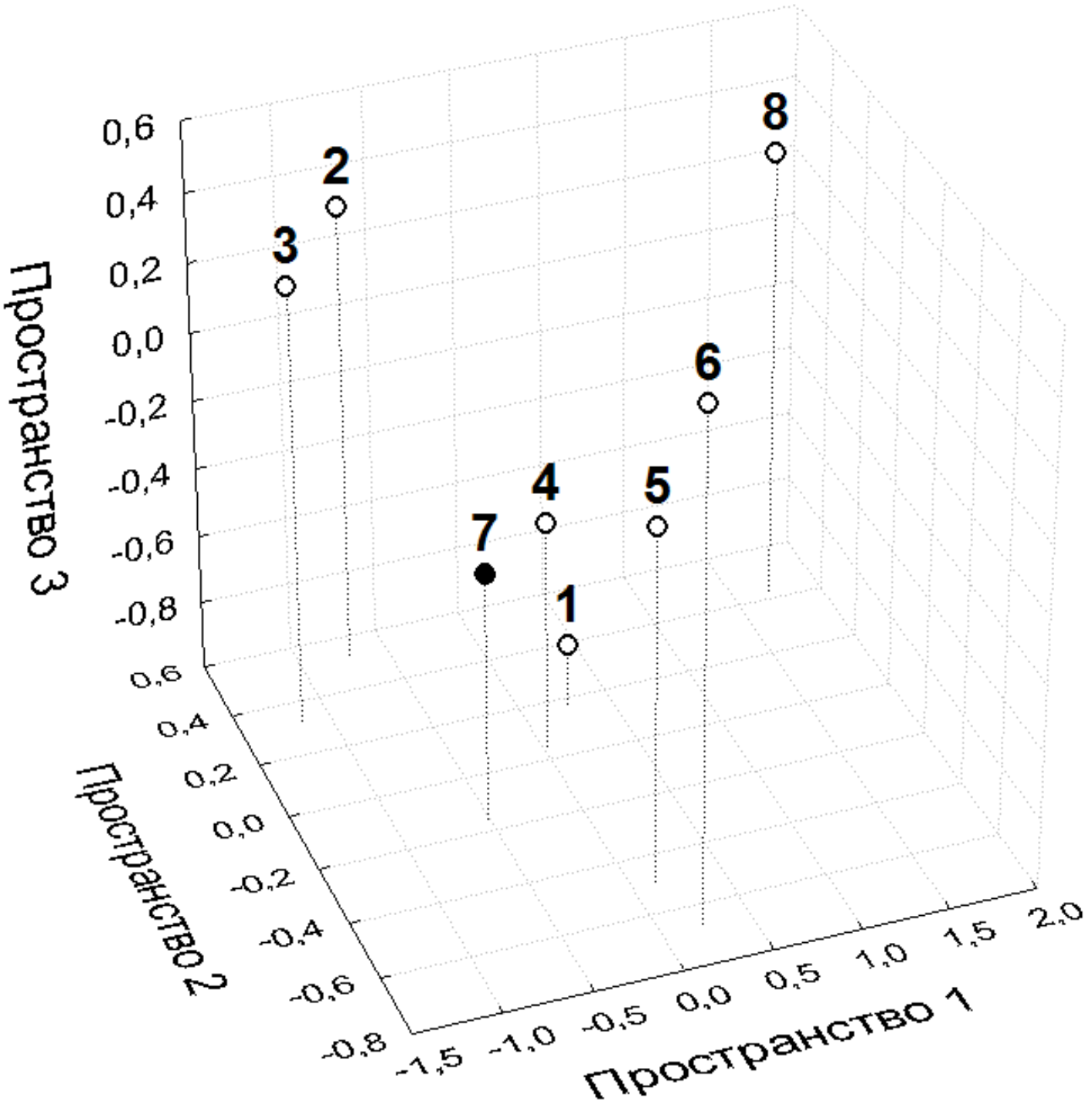

**Рисунок 7.2.** Расположение Манычской популяции тюльки при многомерном шкалировании относительно основных групп популяций *C. cultriventris* с маркерным видом *C. engrauliformes* в результате популяционно-генетического анализа 12 локусов. Построение на основании генетических дистанций Нея (Nei, 1978).

Популяции: **1**–Северный Каспий, **2**–Волгоградское водохранилище, **3**–Рыбинское водохранилище, **4**–Азовское море, **5**–Днестровский лиман, **6**–р. Днепр, **7**–Маныч, **8**–*C. engrauliformes* (Каспийское море)

Рассмотрение популяционно-гененетических данных с применением факторного анализа (Zar, 1999) позволяет выяснить особенности распределения генетических характеристик от факторов среды и предположить вероятные адаптивные преимущества для конкретных популяций вселенца.

**Таблица 7.1.** Результаты факторного анализа по основным группам популяций тюльки

| Переменная | Фактор 1 | Фактор 2 | Фактор 3 |
|---|---|---|---|
| Сев. Каспий | -0,99409 | 0,10793 | 0,01141 |
| Азовское море | -0,99810 | 0,06081 | 0,00902 |
| Рыбинское водохранилище | -0,99965 | -0,02361 | -0,01132 |
| Волгоградское водохранилище | -0,99993 | -0,00171 | -0,01044 |
| р. Маныч | -0,99972 | 0,02077 | -0,00968 |
| *C. engrauliformes* | 0,98610 | 0,16574 | -0,01125 |

При анализе факторных нагрузок можно предположить, что Фактор 1 отражает межвидовую дифференциацию тюлек. Популяции *C. cultriventris* располагаются в положительной зоне, тогда как *C. engrauliformes* - в отрицательной. Вероятно, Фактор 2 показывает различия между популяциями разных частей ареала тюльки. В данном случае Манычская популяция демонстрирует промежуточное положение как относительно морских, так и Волжских популяций тюльки, что подтверждается и результатами Многомерного шкалирования. Фактор 3, предположительно, отражает частные адаптации к среде обитания - для морских популяций нагрузка больше 0, а для пресноводных и солоноватоводных - меньше. Данный факт можно объяснить вероятной селективностью ряда локусов по отношению к общей минерализации воды.

Таким образом, представленные данные более свидетельствуют о реликтовом статусе популяций Маныча, вероятно оставшихся после высыхания пролива между Черным и Каспийским морями, чем о вселении тюльки из Волги в Маныч.

## 7.2. Происхождение и современная структура Волжских популяций тюльки

Другим немаловажным вопросом, разрешить который можно с применением методов генетико-биохимических исследований, является происхождение и быстрая экспансия Волжских популяций тюльки. Для этого проведён популяционный анализ частоты встречаемости генотипов и аллелей 12 локусов с применением метода многомерного шкалирования (рис. 7.3,*а*) и метода Главных компонент (для локуса *LDH-A**, рис. 7.3,*б*).

Полученные данные демонстрируют генетическое единство всех популяций Волги. В общем массиве можно условно выделить группу нижневолжских (рис. 7.3,*а*, п. 3,4,5) и верхневолжских (рис. 7.3,*а*, п. 6,7,9) популяций. Из общей структуры выделяется только тюлька Иваньковского водохранилище (рис. 7.3,*а*, п. 10), а популяция Рыбинского водохранилище (рис. 7.3,*а*, п. 8) занимает промежуточное положение.

Вероятно, в Иваньковском водохранилище, исследованном в 2003 г., тюлька находилась на самом раннем этапе вселения, поэтому генетические характеристики популяции находились в неустойчивом состоянии (для неё характерен высокий дефицит гетерозигот $D_H$= -0,44), что и привлекло к значительным различиям по сравнению с популяциями других водохранилищ Верхней Волги. «Промежуточное» положение тюльки Рыбинского водохранилища можно объяснить его большой площадью с высоким разнообразием биотопов и объёмом водных масс, что приводит к образованию локальных стад рыб и большой вариации генетических характеристик при рассмотрении водоёма в целом (возможно, следствие «эффекта Валунда»).

Для выяснения филогенеза тюльки Волги полезно рассмотреть распределение частот генотипов и аллелей локуса *LDH-A** (рис. 7.3,*б*). Как было показано ранее (Глава 6), аллель *LDH-A*100* превалирует в морских популяциях и, вероятно, сцеплен с генетическим комплексом, ответственным за адаптации к морским условиям. В свою очередь, аллель *LDH-A*120* доминирует в популяциях пресноводных водохранилищ Волги. Изучение распределения аллелей *LDH-A** позволяет по вкладу в ГК 1 чётко отделить *C. cultriventris* (<0) от *C. engrauliformes* (>0). Все пресноводные популяции тюльки в пространстве Главных компонент сосредоточены в IV четверти (рис. 7.3,*б*), тогда как морские популяции находятся в III четверти. «Пограничное» положение тюльки Волгоградского водохранилища, вероятно, объясняется притоком мигрантов из дельты Волги (Карабанов, 2012б).

Максимальная концентрация аллеля *LDH-A*120* наблюдается в Волге у г. Саратова, что разительно отличает эту популяцию от всех остальных. Вероятно, это связано с остатками реликтовой популяции пресноводной жилой тюльки, формировавшей пресноводные популяции пойменных озёр и ильменей Нижней Волги (Берг, 1948; Казанчеев, 1963). Возможно, что обитавшая до создания водохрани-лищ тюлька Саратовских затонов представляла собой жилую пресноводную физиологическую расу черноморско-каспийской тюльки. Это предположение позволяет снять ряд вопросов, связанных с экспансией вселенца вверх по Волге. Так, длительное существование пресноводной популяции позволило выработать комплекс адаптаций к обитанию и размножению в низкоминерализованных волжских водах. Вероятно, закрепление данных адаптаций было сопряжено с увеличением концентрации аллеля *LDH-A*120*, которая и достигла наблюдаемого высокого уровня. Таким образом, после создания каскада Волжских водохранилищ наличие адаптаций для обитания в пресных водах позволило тюльке быстро освоить образовавшиеся новые биотопы и осуществить экспансию по всей Волге.

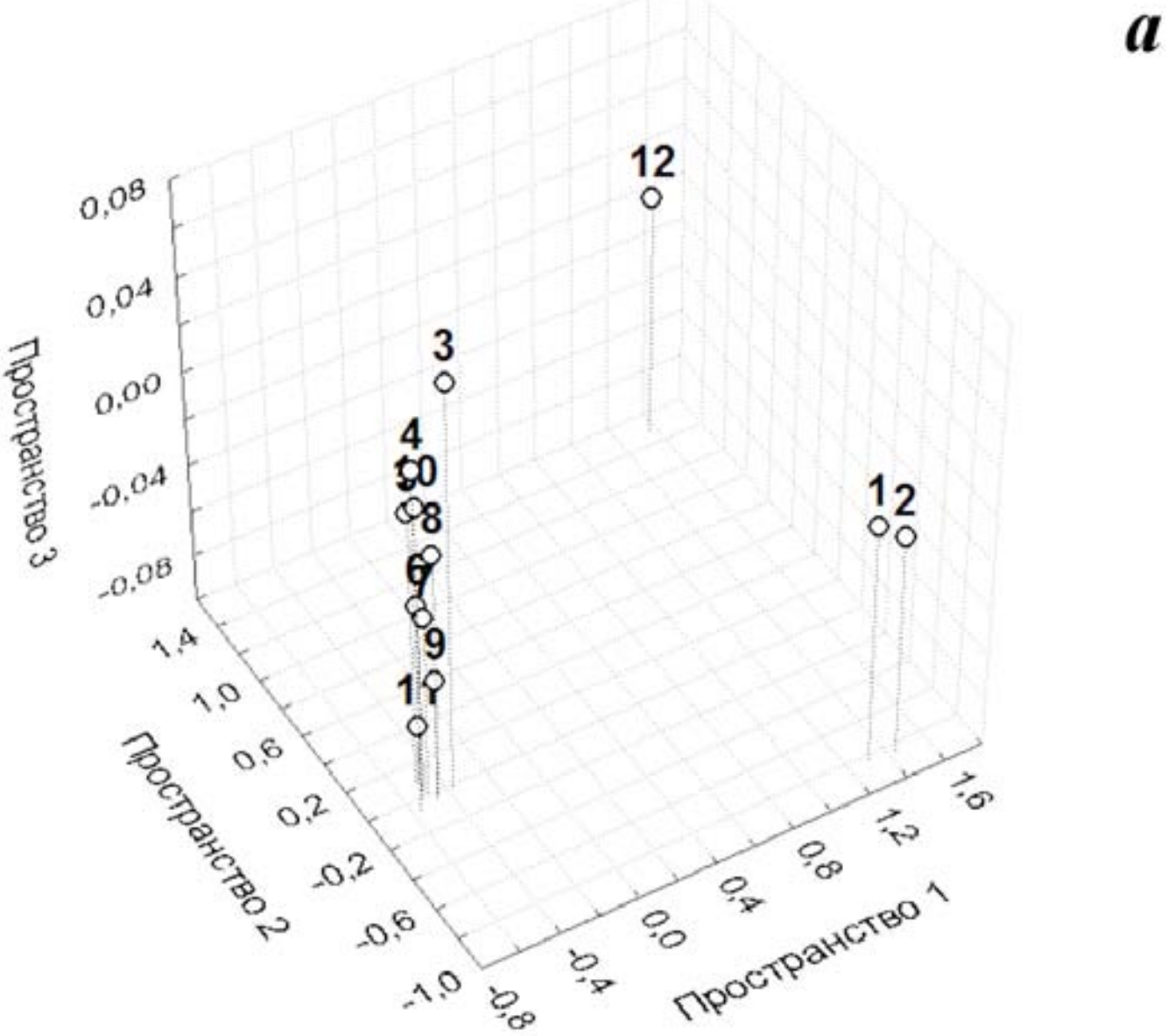

**Рисунок 7.3.** Генетическая структура Волжских популяций *C. cultriventris*. Для оценки генетических расстояний приведены две морские популяции и родственный вид *C. engrauliformes*. ***а*** – Многомерное шкалирование по 12 генетическим локусам.

Популяции: **1**–Азовское море, **2**–Северный Каспий, **3**–Волгоградское водохранилище, **4**–р. Волга в районе г. Саратов, **5**–Куйбышевское водохранилище, **6**–Горьковское водохранилище, **7**–Чебоксарское водохранилище, **8**–Рыбинское водохранилище, **9**–Угличское водохранилище, **10**–Иваньковское водохранилище, **11**–Шекснинское водохранилище, **12**–*C. engrauliformes* (Каспийское море)

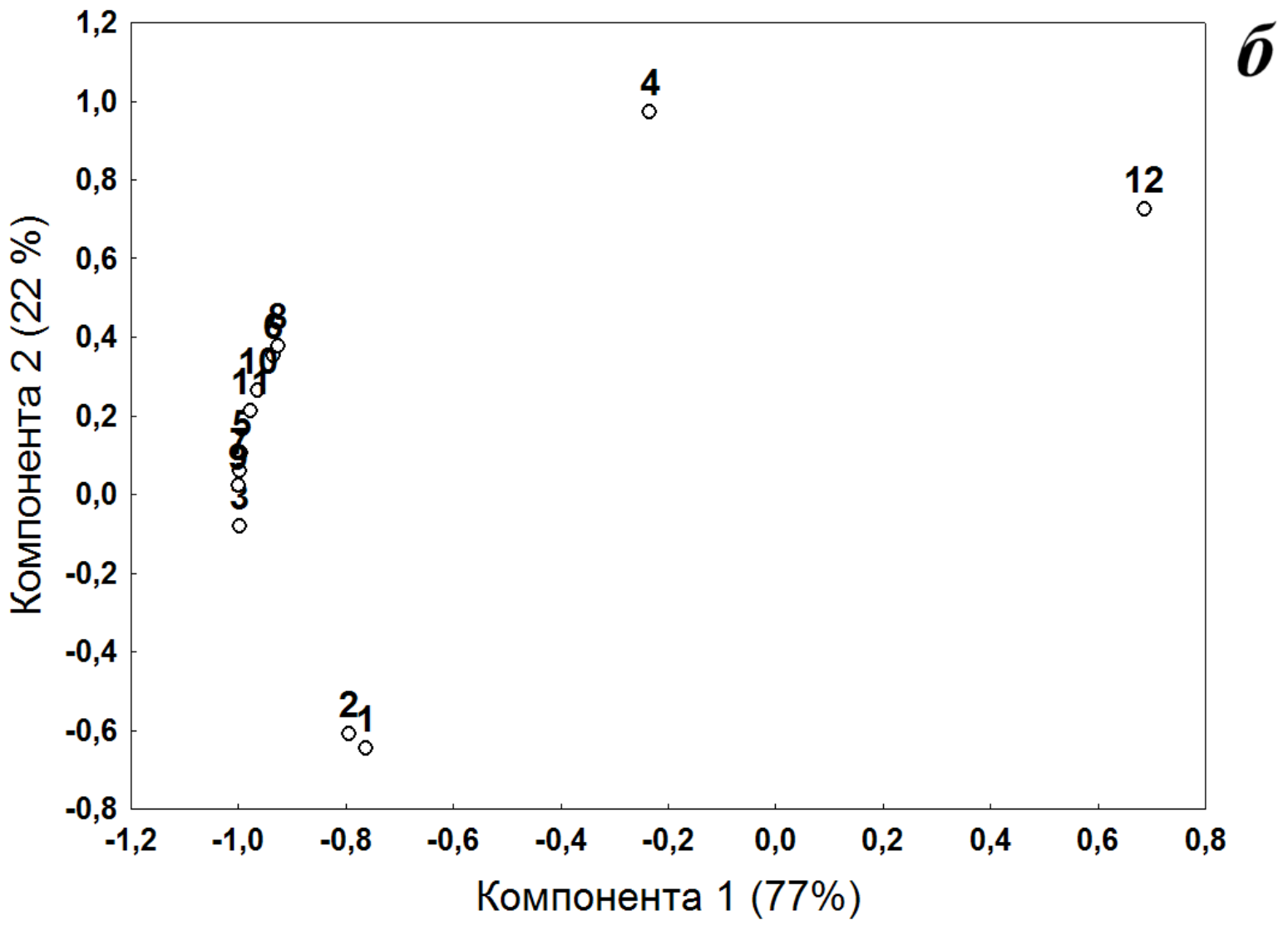

**Рисунок 7.3.** Продолжение. ***б*** – распределение частот встречаемости генотипов и аллелей локуса *LDH-A** в пространстве Главных компонент.

Популяции: **1**–Азовское море, **2**–Северный Каспий, **3**–Волгоградское водохранилище, **4**–р. Волга в районе г. Саратов, **5**–Куйбышевское водохранилище, **6**–Горьковское водохранилище, **7**–Чебоксарское водохранилище, **8**–Рыбинское водохранилище, **9**–Угличское водохранилище, **10**–Иваньковское водохранилище, **11**–Шекснинское водохранилище, **12**–*C. engrauliformes* (Каспийское море)

## 7.3. Особенности географического распределения некоторых аллелей в ареале *C.cultriventris*

Изучение клинальной изменчивости частот аллелей является важнейшим компонентом популяционной биологии рыб. В большинстве случаев пространственная генетическая неоднородность обусловлена пониженной вероятностью генетического обмена из-за удалённости между разными популяциями животных (Beacham et al., 1989; Shaklee, Varnavskaya, 1994; Simoes et al., 2008). Зависимость частоты встречаемости определённых аллелей от географического положения популяции описано для американского угря (Williams et al., 1973), лососевых (Алтухов и др., 1997) и ряда других рыб (Кирпичников, 1987; Алтухов, 2003). Очень часто клинальность связана с экологическими особенностями существования вида. Так, для чукучана (по локусу *EST**) и нерки (по локусу *LDH-$B_1$**) различия в температурной устойчивости аллозимов соответствуют направлению клин аллелей соответствующих локусов (Koehn, 1969; Кирпичников, 1987). Наличие географических особенностей распределения аллелей может быть связано не только с экологическими особенностями, но и возникнуть в результате вторичного контакта между генетическими расами одного вида. Примером этого служит клинальность распределения аллеля гемоглобина *HB-1** у балтийской трески, который является маркером разных физиологических рас с узкой зоной смешивания (Sick, 1965; Mork et al., 1982). Некоторые авторы (Powell, Taylor, 1979; Голубцов, 1988) полагают, что выбор оптимальной среды существования генетически различными особями является основным фактором поддержания белкового полиморфизма в популяциях животных самой разной систематической принадлежности.

Для черноморско-каспийской тюльки закономерности в распределении аллелей трёх генетических локусов не обнаружено (рис. 7.4). Однако при анализе этих данных была установлена особенность распределения аллей локуса лактатдегидрогеназы-А. Аллель *LDH-A*100* значительно преобладает в популяциях тюльки Каспийского ($p$=0,81) и Азовского ($p$=0,75) морей. В Днестровском лимане Чёрного моря, подверженного большому опреснению из Днестра, частота этого аллеля несколько ниже ($p$=0,73). Для речных популяций тюльки (кроме Волжских и Камских) концетрация варианта *LDH-A*100* ниже, чем для морских популяций, но всё равно составляет не менее 50%.

Принципиально иная картина наблюдается в водохранилищах Волги (рис. 7.4). Здесь значительно преобладает аллельный вариант *LDH-A*120* (Karabanov, Slynko, 2010). Лишь для самого нижнего, Волгоградского водохранилище, его доля чуть менее двух третей ($p$=0,65). В Волге у г. Саратов частота встречаемости аллеля *LDH-A*120* максимальна ($p$=0,82) и несколько снижается при продвижении вверх по каскаду водохранилищ. Столь существенные различия в аллельных частотах могут свидетельствовать о наличии частичной репродуктивной изоляции между популяциями. Вместе с тем, как показано на работах с Cichlidae (McKaye et al., 1984), отсутствие фиксации альтернативных аллелей указывает на относительно недавнее прекращение генетических обменов, либо на отсутствие абсолютной изоляции, что в полной мере справедливо по отношению к тюльке (Карабанов, 2006).

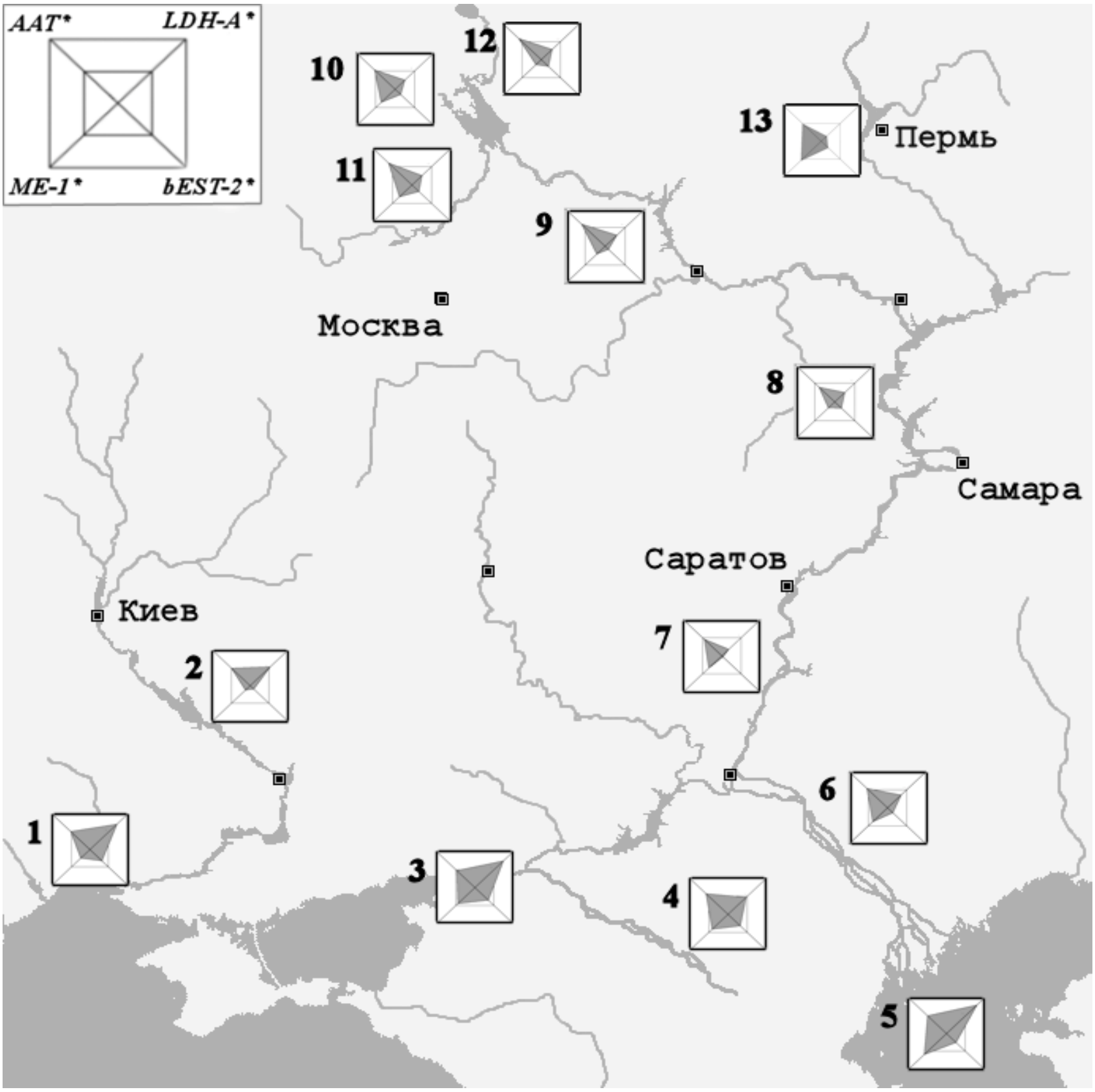

**Рисунок 7.4.** Географическое распределения аллелей некоторых генетических локусов черноморско-каспийской тюльки по основным частям ареала. Указана доля наиболее «медленного» аллеля.

**1**–акватория Днестровского лимана, район г. Белгород-Днестровский; **2**–акватория Среднего Днепра: район устья р. Самара, Карачуновкое водохранилище, р. Днепр; **3**–акватория Азовского моря: район пос. Порт Кантон, Чумбур-коса, дельта р. Дон, тоня «Сельдёвая»; **4**–акватория Манычского каскада водохранилищ: канал ветви Азовской водораспределительной системы, Весёловское водохранилище, Пролетарское водохранилище, оз. Маныч-Гудило; **5**–северная часть Каспийского моря, район устья р. Сулак; **6**–средняя зона Волгоградского водохранилища; **7**–верхняя часть Волгоградского водохранилища у г. Саратов; **8**–Камский плёс Куйбышевского водохранилище; **9**–Чебоксарское водохранилище возле г. Космодемьянск; **10**–Рыбинское водохранилище (14 станций отбора проб); **11**–Угличское водохранилище: г. Углич и г. Калязин, Иваньковское водохранилище: г. Дубна и г. Конаково; **12**–Шекснинское водохранилище, Сизменское расширение; **13**–Камское водохранилище, г. Пермь

Особенности распределения аллелей в популяциях тюльки Волги, вероятно, кроется в микрофилогенезе волжских популяций тюльки. На основании ДНК-фингерпринтинга с применением RAPD-маркеров установлено, что филогенетически волжская тюлька происходит от тюльки Каспийского моря (Столбунова, Слынько, 2005; Слынько и др., 2010). Характерное лишь для Волги преобладание аллеля *LDH-A*120* можно объяснить либо особенным физиологическим статусом пресноводных тюлек, либо особым происхождением этих популяций.

В случае существования пресноводной физиологической расы, маркером которой служит аллель *LDH-A*120*, следовало бы ожидать и преобладания этого аллеля в другом пресноводном водоёме - р. Днепр. Однако, для популяций тюльки среднего течения Днепра частота встречаемости аллеля *LDH-A*120*, $p$=0,31. Это существенно ниже, чем в водохранилищах Волги. Другое объяснение, больше соответствующее наблюдаемым фактам, заключается в происхождении волжской тюльки от жилой формы Саратовских затонов. До зарегулирования Волги данная малоизученная пресноводная форма обитала в затонах и ильменях у г. Саратов (Световидов, 1952; Казанчеев, 1963). Согласно предположению И.В. Шаронова (1971) после создания каскада водохранилищ именно эта пресноводная тюлька могла получить возможность расселиться по акватории Саратовского водохранилища, а в дальнейшем и по всей Волге.

Если принять предположение о происхождении волжских популяций тюльки от формы, адаптированной к условиям пресных вод, то становится понятным и высокая скорость её расселения. И генетически, и морфологически черноморско-каспийская тюлька на всём своём ареале представляет один вид. Таким образом, вероятно, жилая тюлька Саратовских затонов представляла собой пресноводную физиологическую расу *C. cultriventris*. Маркером этой расы может служить аллель *LDH-A*120*. Косвенно такое предположение подтверждается максимальной концентрацией данного аллеля именно в популяции тюльки, выловленной у г. Саратова ($p$=0,82). Таким образом, географические особенности генетической структуры тюльки являются следствием их происхождения, что было показано и в работах по дрозофилам (Simoes et al., 2008). Аналогичная ситуация складывается и при рассмотрении особенностей распределения аллелей локуса *bEST-2**. В водохранилищах Волги частота встречаемости аллеля *bEST-2*41* не превышает 17% (минимальна в саратовской популяции - 3%), тогда как в остальных частях ареала она гораздо больше. Для остальных исследованных генетических локусов такой закономерности не наблюдается.

Чтобы подтвердить гипотезу об особом происхождении Волжских популяций тюльки, следует проверить наличие генетических маркеров, характерных именно для этих популяций. В качестве «волжских» маркеров предполагается принять аллельные варианты *LDH-A*120* и *bEST-2*45*. Таким образом, следует определить их селективное или нейтральное значение, а также установить факторы, обусловливающие особенности их распределения.

# Глава 8.
# Эколого-генетический анализ популяций тюльки бассейна Волги

По теории нейтральности различия между популяциями связаны со случайными процессами и служат отражением генетических расхождений между относительно изолированными популяциями (Оно, 1973; Кимура, 1985; Nei, 1987). Данный подход удобен при изучении филогенетического родства между крупными таксонами и при популяционно-генетических работах с применением ДНК. При изучении изоферментов, как правило, наблюдается большое количество примеров селективности тех или иных аллельных вариантов, которые могут служить маркерами физиологических или географических рас (Кирпичников, 1972, 1987; Ayala, 1974; Watt, 1985; Алтухов, 2003).

## 8.1. Проверка исследованных генетических систем на селективность

Чёткими аллельными вариантами, представленными в достаточной мере во всех исследованных популяциях тюльки, обладают локусы *LDH-A**, *bEST-2** и *ME-1**. Остальные исследованные локусы либо мономорфны (*MDH**, *6PGDH**), либо имеют низкую частоту альтернативных вариантов (ряд фракций спектра общего белка *GP**), либо не являются генетически детерминированными аллельными вариантами (*AP**, *aGPDH**) или имеется недостаточный для достоверной интерпретации материал (*SOD**).

Мерой для оценки возможной селективности аллеля определённого локуса служило изучение относительной приспособленности генотипов в исследованных популяциях. Подробная генетическая характеристика исследованных популяций приведена в Главе 6. Опираясь на эти данные и используя критерий приспособленности как меры давления отбора, можно построить диаграмму множественных приспособленностей С. Райта (Алтухов, 2003) с «пиками» в области низкого давления отбора и «впадинами» в районах высокого давления отбора (рис. 8.1).

Для популяций тюльки Волжского каскада водохранилищ характерна довольно высокая степень приспособленности. Давление отбора минимально для популяций Таганрогского залива, Маныча и Северного Каспия (рис. 8.1, п. 4,6,7). В прочих группах имеется давление отбора различной интенсивности по всем трём локусам. Эти данные могут свидетельствовать об умеренном числе селективно нейтральных полиморфных локусов, выделяемых методом изоферментного анализа.

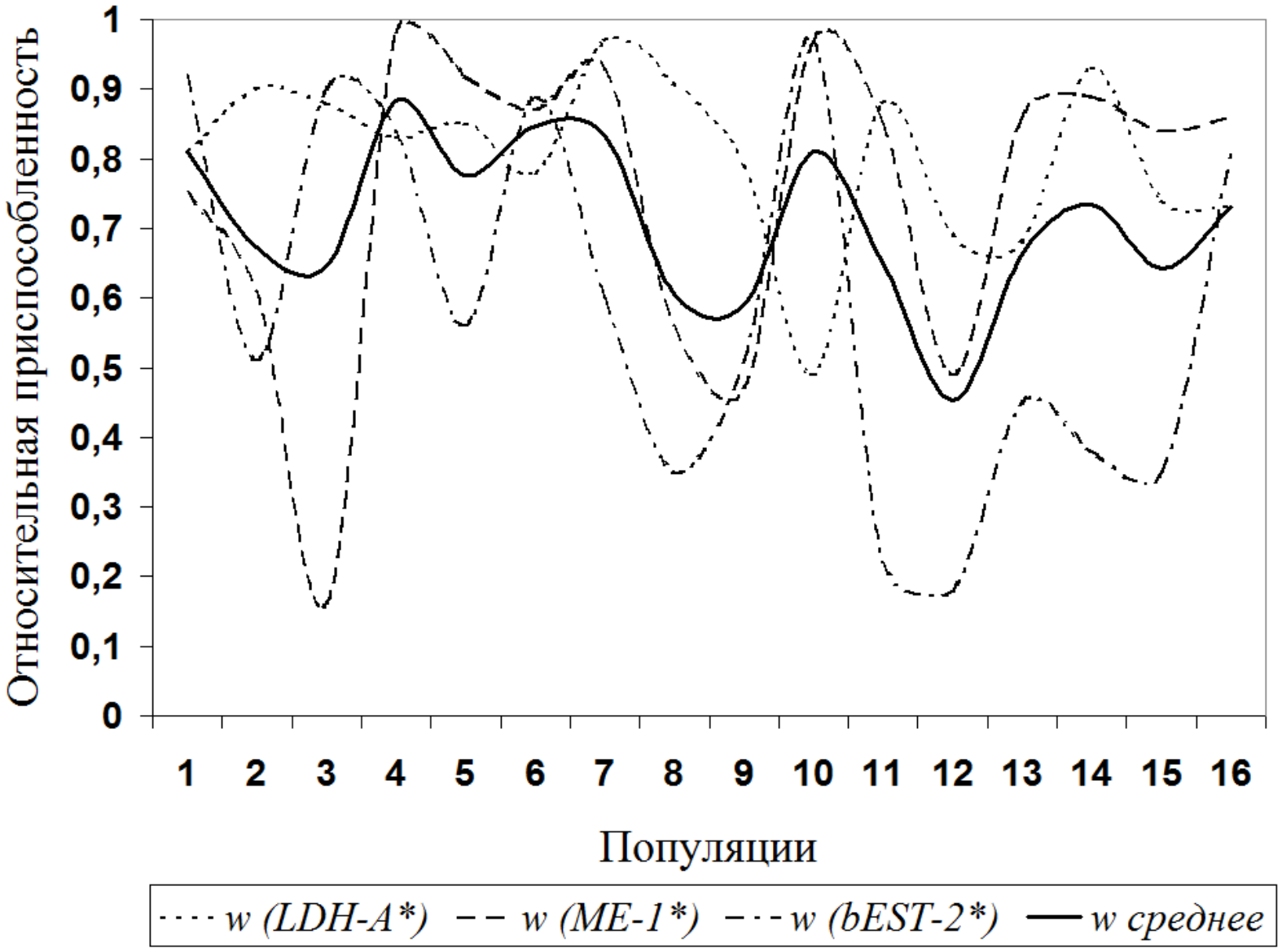

**Рисунок 8.1.** Диаграмма относительной приспособленности (*w*) по генетическим локусам *LDH-A**, *ME-1** и *bEST-2**.

По оси абсцисс - исследованные популяции: **1**–Днестровский лиман, **2**–русловая часть р. Днепр, **3**–Карачуновское водохранилище, **4**–Таганрогский залив Азовского моря, **5**–дельта р. Дон, **6**–Манычские водохранилища, **7**–Северный Каспий, **8**–Волгоградское водохранилище, **9**–р. Волга в районе г. Саратов, **10**–Куйбышевское водохранилище, **11**–Чебоксарское водохранилище, **12**–Горьковское водохранилище, **13**–Рыбинское водохранилище, **14**–Угличское водохранилище, **15**–Иваньковское водохранилище, **16**–Шекснинское водохранилище. По оси ординат - относительная приспособленность

Для черноморско-каспийской тюльки, несмотря на относительно небольшой материал, примерная доля локусов, подверженных балансирующему отбору составляет 36%, дизруптивному - 46% и 18% селективно нейтральных из исследованных локусов. Для лососей Дальнего Востока России доля локусов, подверженных балансирующему отбору составляет 38-50%, дизруптивному - 38-48%, селективно нейтральны - 14-23% локусов (Алтухов и др., 1997; Алтухов, 2003). Для лососей Аляски и Британской Колумбии для всех трёх групп локусов примерно равна, что, видимо, связано с более мягкими условиями существования в этих районах (Wilmot et al., 1994). В целом для лососевых уровень селективно нейтральных аллелей составляет около 20% (Динамика…, 2004). По данным М. Кимуры (1985), основанных на изучении белкового полиморфизма, доля нейтральных локусов у разных видов составляет от 13 до 21%, а по данным об SNP-полиморфизме у человека - около 20% (Fay et al., 2001).

Известно, что для географически и исторически связанных популяций должны обнаруживаться эффекты изоляции расстоянием (Carvalho, 1993; Slatkin, 1993). Такой анализ был проведён Ю.П. Алтуховым с сотр., где было показано как наличие связи между генетическим и географическим расстоянием, так и её отсутствие (Алтухов и др., 1997). Аналогичная работа была проведена для популяций тюльки Волжских водохранилищ. Генетические дистанции были получены из попарного сравнения межпопуляционных величин $F_{ST}$ коэффициентов Райта (Ней, Кумар, 2004). Данное сравнение для популяций тюльки более корректно, чем использование параметров межпопуляционных эффективных численностей популяций, выраженных с учётом процента миграции ($N_em$) в связи со сложностью прямого определения этого показателя. Результаты, рассчитанные для 8 популяций черноморско-каспийской тюльки Волжского каскада водохранилищ на основании 12 полиморфных локусов, показывают отсутствие достоверной связи между географическим и генетическим расстояниями: $r^2$=0,023 при $p>0,13$. Эти данные могут свидетельствовать о селективной составляющей в пространственной изменчивости ряда аллозимных локусов в популяциях исследованного вида (Динамика…, 2004).

Среди всех изученных локусов тюльки особое место занимает лактатдегидрогеназа-А. Гетерогенность частот аллелей по локусу *LDH-A** среди изученных популяций достаточно высока и не может быть объяснена воздействием случайных факторов. Регрессионный анализ показал сильную зависимость частоты аллеля *LDH-A*100* от общей минерализации водоема (рис. 8.2). Коэффициент регрессии по Спирмену $r_S$=0,87 в доверительном интервале при $p<0,05$. Относительно других факторов окружающей среды (растворенный кислород, pH, средняя температура водоема) достоверной зависимости их влияния на распределение частот аллелей *LDH-A** не установлено (Карабанов, 2011).

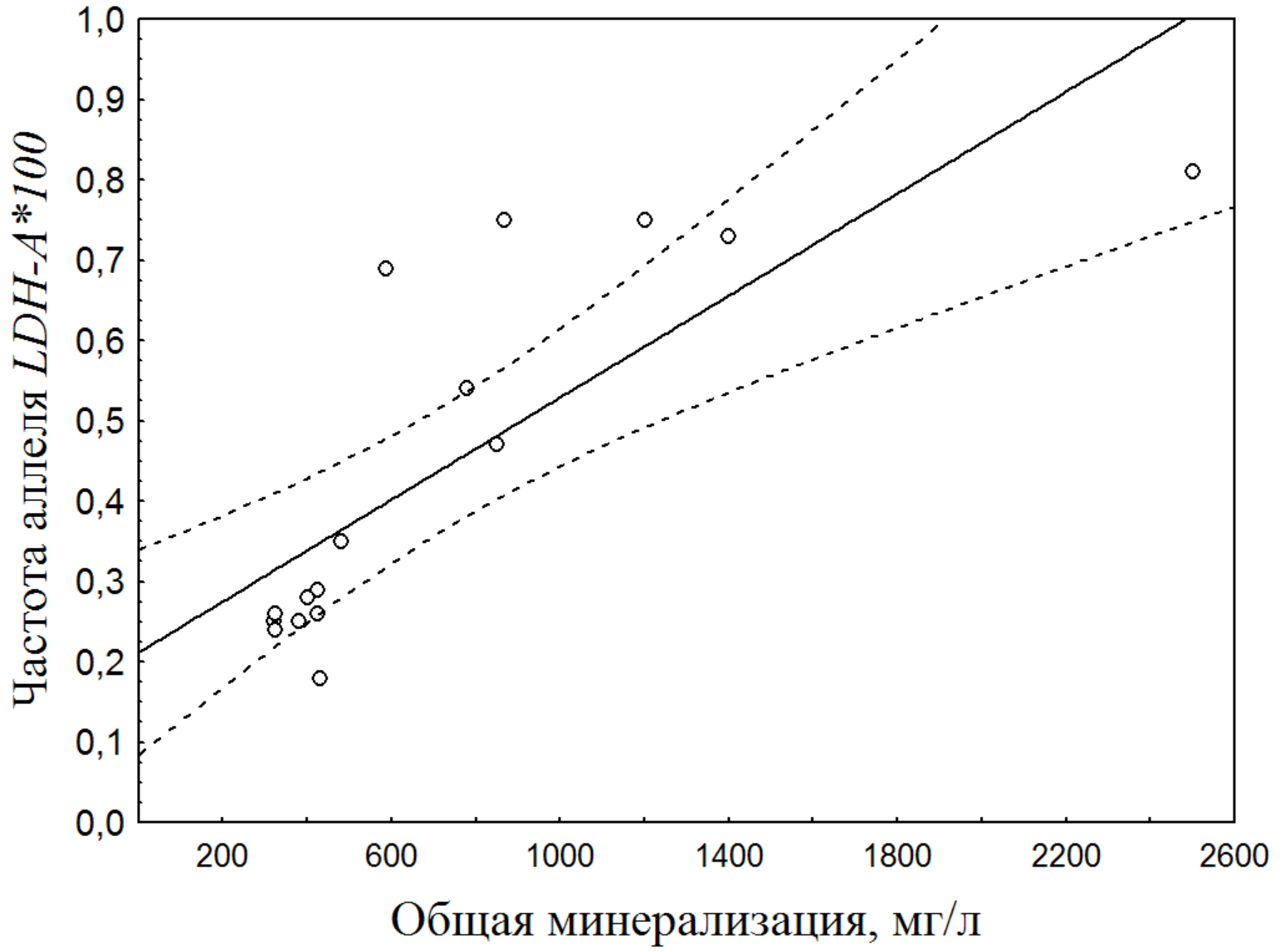

**Рисунок 8.2.** Зависимость частоты аллеля *LDH-A*100* от общей минерализации исследованных водоёмов. Пунктиром отмечена зона доверительного интервала при уровне значимости $p<0,05$

Вероятно, наблюдаемая особенность распределения частоты *LDH-A*100* связана с историческим «морским» происхождением этого аллеля. Косвенно данное предположение можно доказать крайне высокой концентрацией аналогичного варианта в популяции облигатно морского родственного вида *C. engrauliformes*, а также атлантической и тихоокеанской сельдей *Clupea harengus* и *C. pallasi* (Jorstad, 2004). Приняв это предположение, можно объяснить довольно высокие частоты *LDH-A*100* в днепровских популяциях как следствие их филогенетической молодости, когда частоты «морского» аллеля ещё не приведены к «оптимальному» значению. Также не исключено, что на генный баланс популяций водохранилищ Нижнего Днепра значительное влияние оказывает обмен рекрутами с черноморскими лиманами.

Наблюдаемую низкую частоту аллеля *LDH-A*100* в Волжских популяциях можно объяснить как следствие их происхождения от жилой пресноводной формы. Реликтовые популяции тюльки в затонах под Саратовом, вероятно, сохранились со времён Хвалынской трансгрессии и последующего отступления Каспия около 20-40 тыс. л. назад (Панин и др., 2005; Бардюкова, 2007). Вероятно, за этот период произошло перераспределение аллельных частот до современного уровня, в результате чего аллель *LDH-A*120* можно назвать «волжским» маркером.

Примером аналогичной пространственной генетической дифференциации, связанной с расселением из рефугиумов, служит послеледниковое распространение лососей. Следы этого процесса маркируются генетическими особенностями популяций как на уровне аллозимов (Withler, 1985; Hamilton et al., 1989; Makhrov et al., 2002), так и на уровне ДНК-маркеров (Makhrov et al., 2005).

Таким образом, наблюдаемое распределение аллелей ряда генетических локусов тюльки, не соответствующее теории нейтральности, можно объяснить непосредственным влиянием окружающей среды. Для определения конкретного влияния факторов среды на изоферменты тюльки, и связанную с этим адаптивную ценность различных аллелей требуется экспериментальное изучение реакции изоферментных систем на воздействия тех или иных факторов (Кирпичников, 1987).

## 8.2. Влияние абиотических факторов на изоферменты тюльки

Селективное значение аллелей любого генетического локуса связано с конкретными физиологическими преимуществами продуцируемых аллозимов в данных условиях. В научной литературе (Semeonoff, Robertson, 1968) имеются данные о связи параметров белкового полиморфизма с биологическими особенностями животных. Регуляция процессов метаболизма возможна как путём изменения соотношения аллельных вариантов, так и путём непосредственной регуляция активности аллозимов. При вселении тюльки в пресноводные водоёмы произошло изменение качественного состава среды обитания гидробионта и связанное с этим изменение физиологических характеристик ферментов. Для изучения селективности продуктов аллелей - маркеров пресноводных тюлек, были проведены эксперименты *in vitro* по влиянию солей, карбамида и температуры на изоферменты черноморско-каспийской тюльки (Карабанов, 2012).

### 8.2.1. Влияние различных солей на аллозимы тюльки

Большинство аллозимов отличаются друг от друга по своим функциональным особенностям. Возможным показателем дивергенции изоферментов могут служить различия в их устойчивости к воздействию солей при различном качественном составе действующего агента и разной экспозиции, а также, как показателя функциональных свойств фермента - скорости специфического окрашивания геля при добавлении различных солей в инкубационную смесь.

В результате экспериментов *in vitro* установлены существенные различия от воздействия активных агентов - сульфатов $SO_4^{2-}$ ($MgSO_4 \cdot 7H_2O$) и хлоридов $Cl^-$ (NaCl) на экспрессию изоферментов LDH-A и bEST-2. Во всех случаях отмечается достоверное снижение активности и увеличение времени окрашивания изученных аллозимов. При повышении концентрации соответствующего агента негативные эффекты от «просаливания» увеличиваются. Для LDH-A и bEST-2 можно отметить общую закономерность: происходит репрессия всех аллоформ, при этом число изоферментов не изменяется. Для лактатдегидрогеназы-А характерно значительное репрессирование для гетерополимеров, в отличие от гомополимеров. Возможно, данный эффект объясняется разрушением водородных связей, относительно более слабых в гетеротерамере, и сопряженным с этим нарушением четвертичной структуры фермента (Gilbert, 2000).

Для всех изученных изоферментов добавление как индивидуальных солей, так и их смесей в инкубационную среду (в нелетальных концентрациях) увеличивает скорость окрашивания. Это явление, вероятно, связано с повышением концентрации катионов (прежде всего в случае $Mg^{2+}$), необходимых для работы ферментов. Ион металла связывается с ферментом в аллостерическом центре, в результате чего происходит изменение электрического заряда полипептидной глобулы или изменение взаимной ориентации разных участков молекулы белка (или окружающих молекул) (Gilbert, 2000). При воздействии солей на изоферменты bEST-2 установлено отсутствие каких-либо различий между аллельными вариантами (Карабанов, Слынько, 2005а).

Также было изучено *in vitro* влияние искусственной морской воды «океанического» и «каспийского» типа (Алекин, 1952) на некоторые аллозимы тюльки. В обоих случаях происходит репрессирование ферментативной активности; хотя в воде «каспийского» состава этот процесс и менее выражен, но различия недостоверны ($p>0,05$). Исходя из этих данных можно сделать предположение, что для изоферментов тюльки при сочетанном влиянии неорганических солей отсутствует компенсаторный эффект и их воздействие также деструктивно, как и у отдельных ионов.

Для аллелей *LDH-A** установлена следующая закономерность: продукты аллеля *LDH-A*100* гораздо более устойчивы к воздейст-вию хлоридов, чем продукты варианта *LDH-A*120* (рис. 8.3). Не исключено, что данная физиологическая адаптация связана с изменением ионного баланса рыбы: продукты аллеля *LDH-A*100* более приспособлены к работе в условиях повышенной минерализации в морях с высокой хлоридностью воды, а продукты аллеля *LDH-A*120* могут быть адаптированы для работы в условиях пресных вод.

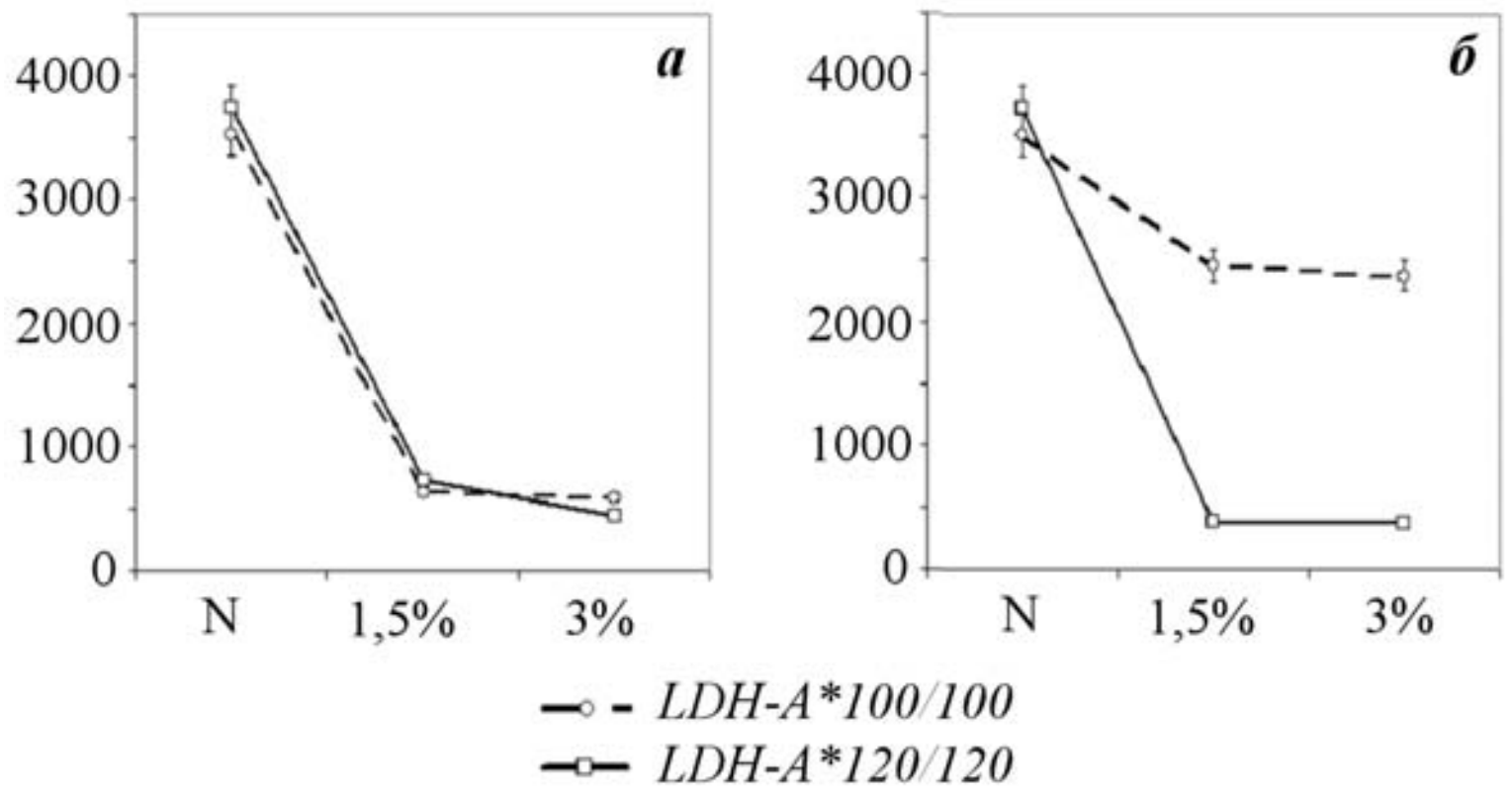

**Рисунок 8.3.** Влияние $MgSO_4$ (***а***) и NaCl (***б***) разной концентрации (1,5 % и 3 %) на аллозимы лактатдегидрогеназы-А черноморско-каспийской тюльки.

По оси абсцисс - концентрация соли, по оси ординат - ферментативная активность, выраженная через относительную оптическую плотность (D). На рисунках показано стандартное отклонение при $p=0{,}05$

### 8.2.2. Влияние карбамида на аллозимы тюльки

Черноморско-каспийская тюлька относится к категории эврибионтных видов, в том числе по параметрам солености и температуры (Приходько, 1979). Эврибионтность любого организма во многом определяется широкими оптимумами тканевого и внутриклеточного метаболизма, а также, нередко, высокими уровнями генетического разнообразия. Переход морского вида к обитанию в пресных водах должен вести к значительным изменениям в системах осморегуляции и азотистого обмена. Морские костистые рыбы - типичные аммонотелические организмы, но способные выдерживать значительные концентрации карбамида в тканях, хотя и в несравненно меньших концентрациях, чем хрящевые рыбы (Шмидт-Ниельсон, 1982). Сообразно этому изучена устойчивость некоторых изоферментов черноморско-каспийской тюльки к высоким концентрациям карбамида *in vitro* и сравнение данной харак-теристики с таковой у типичного пресноводного обитателя - синца *Ballerus ballerus* Рыбинского водохранилища.

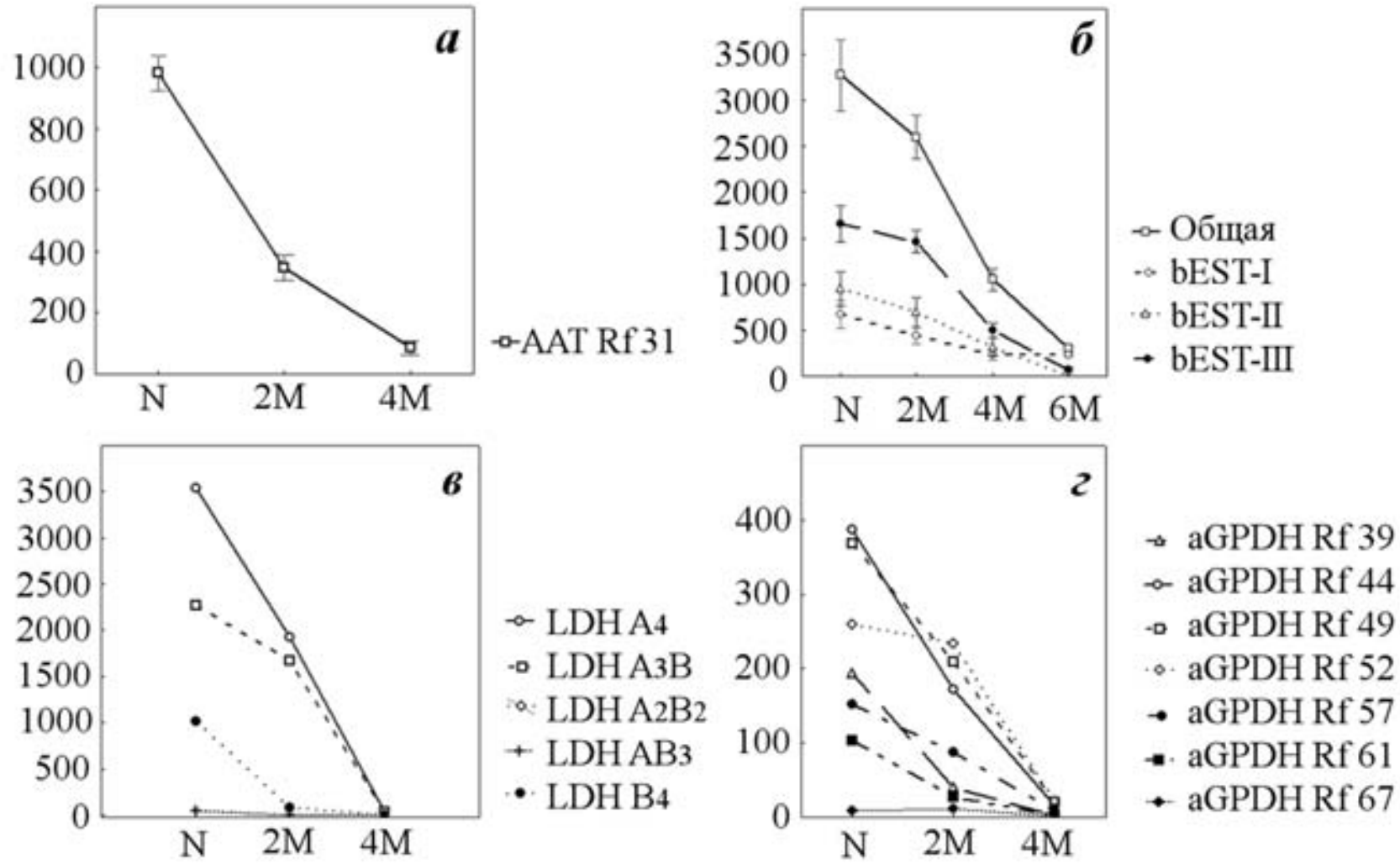

**Рисунок 8.4.** Изменение активности изоформ изученных ферментов при различных концентрациях карбамида. ***а*** – AAT, ***б*** – bEST, ***в*** – LDH, ***г*** – aGPDH.
По оси абсцисс - активность изоферментов в норме (N) и при различных концентрациях карбамида (2 М и 4 М); по оси ординат - ферментативная активность, выраженная через относительную оптическую плотность (D). На рисунках показано стандартное отклонение при p=0,05

При воздействии карбамида на изоферменты синца уже при 2 М концентрации полностью репрессировались все изученные ферментные системы. Различные изоферменты тюльки демонстрируют разную устойчивость к карбамиду (рис. 8.4). Для AAT (рис. 8.4,*а*) характерна линейная зависимость снижения ферментативной активности при повышении концентрации агента. Самыми резистентными оказались изоферменты bEST - даже при концентрации карбамида в 2 М суммарная активность снижается не критически. Наибольшая роль в этом процессе принадлежит фракции EST-III. При дальнейшем повышении концентрации карбамида все эстеразы снижают активность в разы, хотя для фракции EST-I характерно сохранение остаточной активности даже при концентрации 6 М (рис. 8.4,*б*). Лактатдегидрогеназа является тетрамером. В данной работе исследованы продукты гомозигот *LDH-A*100/100* и константных гомозигот *LDH-B*100/100*. В эксперименте бóльшую толерантность к агенту демонстрируют изозимы, содержащие субъединицу лактатдегидрогеназы-А (рис. 8.4,*в*), причём при снижении ее доли устойчивость всего фермента снижается. Среди множественных форм aGPDH выделить общие черты реакции на карбамид не удалось, можно лишь отметить, что в данном случае имеются примеры форм и с высокой, и с низкой толерантностью (рис. 8.4,*г*) (Карабанов, 2007).

### 8.2.3. Функциональные различия в теплоустойчивости и активности некоторых изоферментов черноморско-каспийской тюльки

Успешное заселение тюлькой речных водохранилищ, в том числе северных, должно быть обусловлено эвригалинностью и эвритермностью данного вида. Температурным различиям аллозимов посвящено множество работ, основной вывод из которых заключается в значительной адаптационной роли большинства таких изменений (Кирпичников, 1977; Redding, Schreck, 1979; Zietara, Skorkowski, 1995; Colosimo et al., 2003). Так как расселение тюльки шло в северном направлении, то можно ожидать наличие и температурной селективности аллозимов у тюльки.

Для всех исследованных ферментов при повышении температуры отмечено снижение общей ферментативной активности, однако у различных ферментов имеются свои особенности этого процесса. В порядке возрастания теплоустойчивости исследованные ферменты распределились в следующей последовательности - AAT, aGPDH, bEST (рис. 8.5,*а*).

Для AAT характерна линейная зависимость снижения ферментативной активности при повышении температуры (рис. 8.5,*б*). Все основные изоферменты bEST при температуре 5°С имеют практически одинаковую активность. Однако при повышении температуры до 35°С в значительной степени репрессируются ферменты фракции EST-II, а ферменты фракции EST-I снижают свою активность на 30-60 % (у различных особей) (рис. 8.5,*в*). Однако этот процесс компенсируется возрастанием активности ферментов EST-III, за счёт чего суммарная эстеразная активность снижается не столь занчительно (рис. 8.5,*е*). Этот компенсационный механизм работает вплоть до повышения температуры до 50°С. Начиная с 65°С, как правило, все ферменты bEST инактивируются. В целом, подобная картина зависимости изменения эстеразной активности от температуры была описана ранее (Koehn et al., 1971) для палевого нотрописа *Notropis stramineus*.

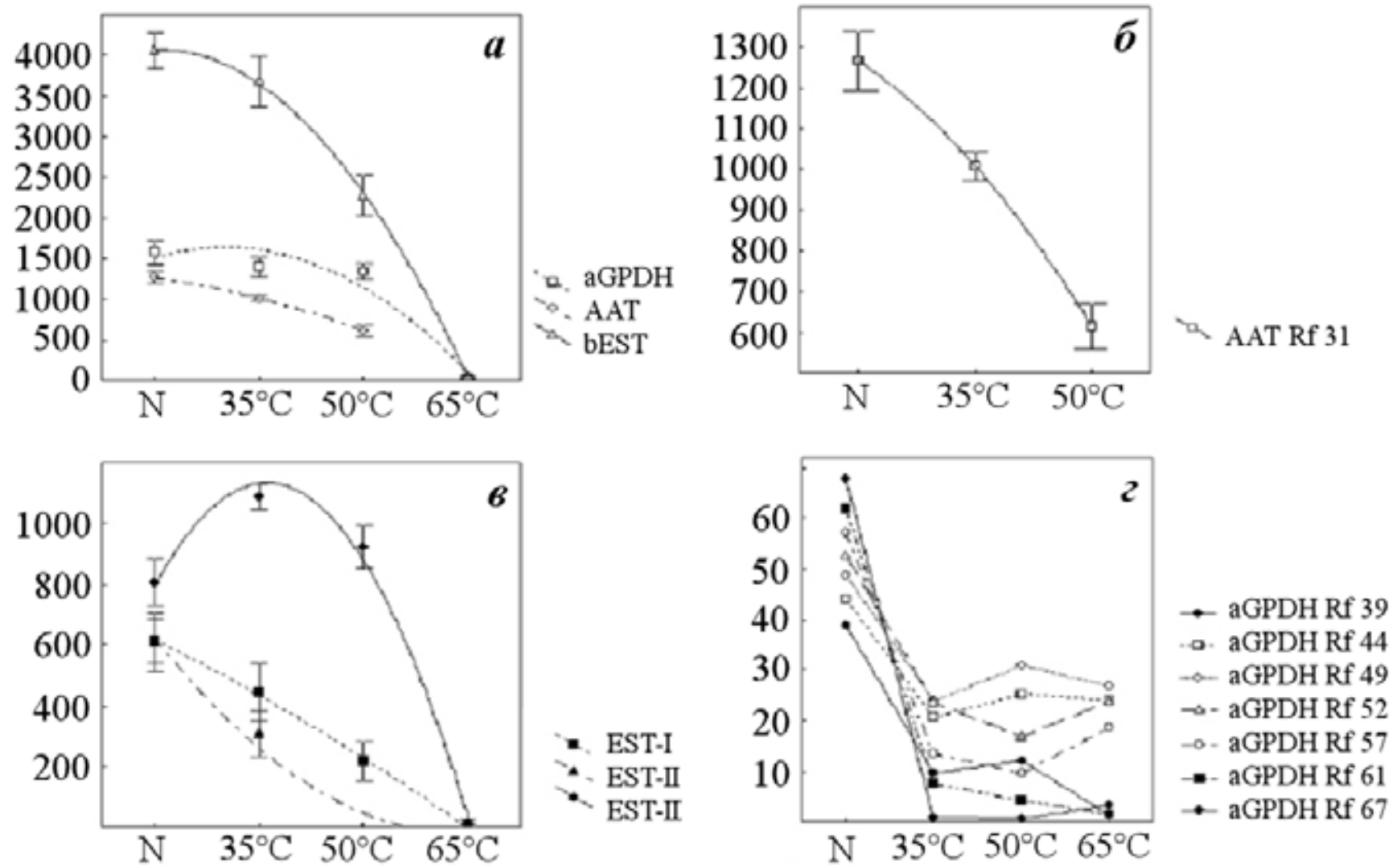

**Рисунок 8.5.** Изменение активности изоформ изученных ферментов при повышении температуры (Карабанов, 2012а).
***а*** – изменение суммарной ферментативной активности; ***б*** – линейная зависимость снижения ферментативной активности ААТ; ***в*** – компенсационный механизм поддержания нормального уровня эстеразной активности за счёт увеличения активности терморезистентной фракции EST-III; *г* – распределение активности изоформ aGPDH при повышении температуры на «северную» (темные маркеры) и «южную» (светлые маркеры) группы. По оси абсцисс - активность изоферментов в норме (N) и при различной интенсивности фактора (температура). По оси ординат - ферментативная активность, выраженная через относительную оптическую плотность, D. На рисунках показано стандартное отклонение при p=0,05

Более сложная зависимость наблюдается для изоферментов aGPDH (рис. 8.5,*г*). Эти изоферменты мы разделяем на две условные группы, обозначенные «южная» (четыре изофермента с Rf 44, 49, 52, 57) и «северная» (три изофермента, представленыt наиболее подвижными (Rf 61, 67) и наименее подвижными (Rf 39) фракциями). Данная сегрегация особенно заметна при рассмотрении активности ферментов при сублетальной температуре (65°С). Изоферменты «южной» группы более активны при высокой температуре (45-60°С), а изоферменты «северной» группы - при пониженных температурах (5°С). Изоферменты «южной» группы, вероятно, являются гетерополимерами, чем, по-видимому, и объясняется их повышенная толерантность к высоким температурам.

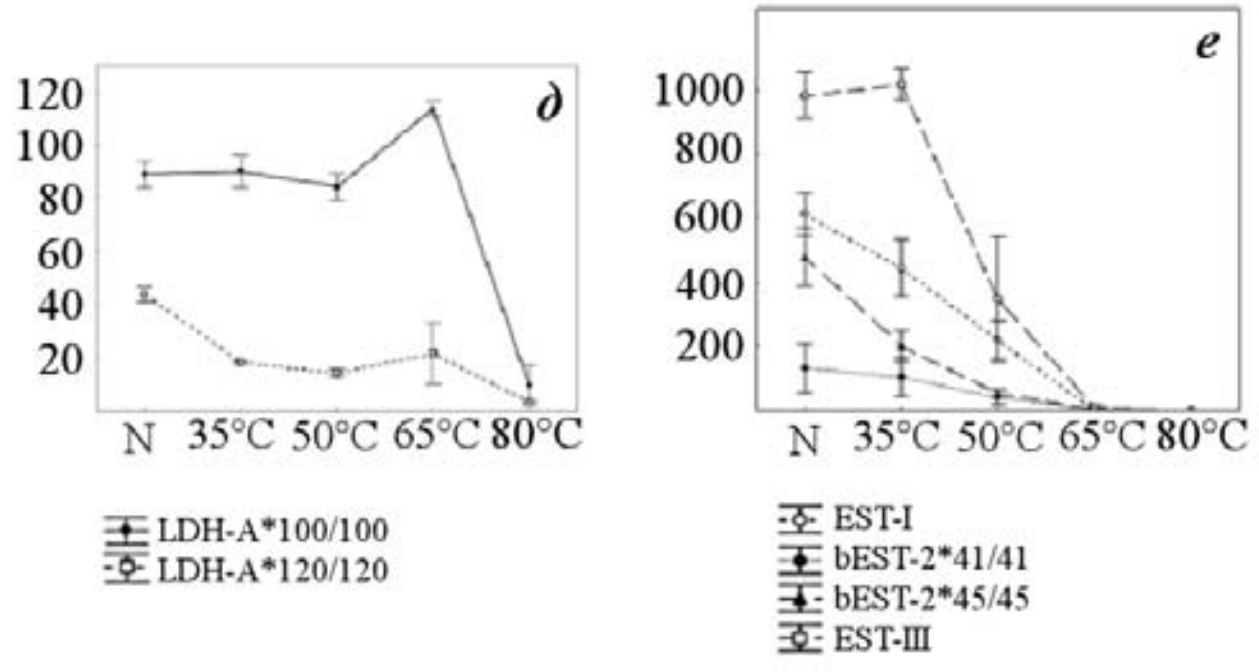

**Рисунок 8.5.** Продолжение. Различия в теплоустойчивости индивидуальных аллозимов LDH-A (***д***) и bEST (***е***) черноморско-каспийской тюльки. По оси абсцисс - активность изоферментов в норме (N) и при различной интенсивности фактора (температура). По оси ординат - ферментативная активность, выраженная через относительную оптическую плотность, D. На рисунках показано стандартное отклонение при p=0,05

Продукты аллеля лактатдегидрогеназы-А **100* демонстрируют большую теплоустойчивость, чем аллозимы - продукты аллеля *LDH-A*120*, однако и продукт аллеля **120* сохраняет, хоть и в меньшей степени, свою активность во всём температурном диапазоне (рис. 8.5,*д*). В динамике изменения активности при повышении температуры этих изоферментов также наблюдаются существенные различия. Так, аллозимы - гомозиготы лактатдегидрогеназы-А **100/100* незначительно изменяют свою активность при повышенных температурах, тогда как для их альтернативных гомозигот характерно сильное равномерное снижение активности при повышении температуры. Вероятно, такие отличия связаны с приспособленностью продуктов аллеля *LDH-A*100* к функционированию в условиях повышенных температур. Это предположение косвенно подтверждается более высокой частотой встречаемости *LDH-A*100* в южных популяциях тюльки. Аналогичная связь между генотипом особей и устойчивостью к повышенным температурам описана для чёрного окуня *Micropterus salmonoides* из Северной Америки (Smith et al., 1983).

# Глава 9.
# Формирование и развитие эколого-генетических адаптаций *Clupeonella cultriventris* в условиях Рыбинского водохранилища

Для выяснения направленности адаптационных преобразований тюльки при экспансии Волжских водохранилищ следует рассмотреть наличие половой, онтогенетической и сезонной зависимости в распределении аллелей.

Модельным водоёмом для этих наблюдений было выбрано Рыбинское водохранилище, которое обладает рядом преимуществ: вселение тюльки произошло относительно недавно и адаптационные процессы можно наблюдать в «реальном времени». Водохранилище имеет большие размеры, обладает разнообразными биотопами и богатым животным миром, что позволяет проследить влияние как биотических, так и абиотических факторов и возможность пространственной дифференцировки популяции. Также немаловажна близость водоёма к ИБВВ РАН, чем обеспечивается возможность многолетнего и межсезонного мониторинга популяции тюльки. Генетическими локусами для исследования выбраны *LDH-A**, *ME-1** и *bEST-2**. Они обладают хорошо различающимися аллелями, выполняют известную биохимическую функцию, по ним накоплен обширный материал сравнения, а также, как было показано выше, их аллели обладают селективным значением и могут служить маркерами адаптационных преобразований при расширении ареала тюльки.

## 9.1. Пространственная подразделённость в популяции Рыбинского водохранилища

На любую природную популяцию постоянно действует ряд элементарных эволюционных факторов, таких как мутационный процесс, миграции, дрейф генов и естественный отбор. Эти эволюционные факторы, влияя на генетическую структуру популяции, по сути, являются внешними по отношению к ней самой. Однако подобное рассмотрение не может быть полным без учёта внутренней организации популяции. Природные популяции, как правило, не являются панмиктическими группами, а представляют собой исторически сложившуюся совокупность частично изолированных субпопуляций. Изоляция может быть вызвана как физическими (географическая изоляция препятствиями или расстоянием), так и физиологическими причинами. Субпопуляции постоянно обменива-ются друг с другом генетическим материалом и испытывают влияние всех тех же факторов, что и популяция в целом. Изучение пространственной структуры позволяет не только установить факторы, обеспечивающие внутрипопуляционную дифференциацию, но и прогнозировать дальнейшие генетические процессы в целой популяции.

Для изучения внутрипопуляционной структуры используется сравнение доли внутригрупповых скрещиваний по отношению к общему числу скрещиваний в популяции в целом. На генетическом уровне это выражается в повышении доли гомозигот за счёт уменьшения доли гетерозигот. В таком случае подразделённость популяции на отдельные скрещивающиеся группы формально эквивалентна отсутствию инбридинга во всей популяции, а степень такой дифференциации прямопропорциональна размаху различий генных частот (Ли, 1978; Динамика…, 2004).

Большой вклад в описание локальной дифференциации частот генов в подразделённой популяции принадлежит С. Райту (Wright, 1943), предложившему использовать определение трёх коэффициентов корреляции, определяющими частоту объединений двух гамет относительно субпопуляции ($F_{IS}$) и относительно общей популяции ($F_{IT}$), а $F_{ST}$ - как корреляция между двумя гаметами, случайно взятыми из каждой субпопуляции. Таким образом, коэффициент $F_{ST}$ показывает степень генетической дифференциации между субпопуляциями. В общем плане все эти индексы отражают отклонение в популяции от панмиксии и определяются соотношением гомозиготных и гетерозиготных генотипов.

Вместе с тем, концепция корреляций объединяющихся гамет плохо приложима к природным условиям. Естественные популяции (особенно пелагических, сильно подвижных рыб), как правило, не являются случайными выборками из множества равноудалённых друг от друга популяций. Для преодоления этих проблем М. Ней (Nei, 1987; Ней, Кумар, 2004) предложил переопределить F-коэффициенты с использованием ожидаемых и наблюдаемых гетерозиготностей в исследуемых популяциях. В результате этого стало возможно использование F-статистики применительно к любым ситуациям, вне зависимости от отбора, так как коэффициенты определяются в терминах имеющихся аллельных и генотипных частот. Однако при анализе природных популяций не стоит безоговорочно доверять выводам, полученным с применением теоретических методов математической генетики. Использование F-статистики может служить важным подспорьем в популяционно-генетических исследованиях, но может и привносить дополнительные ошибки и сложности при интерпретации результатов иследования (Beaumont, Nichols, 1996; Beaumont, 2005; Holsinger, Weir, 2009). Так, индексы фиксации, определённые через гетерозиготности часто представляют случайные величины, так как популяционная структура может сильно меняться между поколениями, что особенно характерно для короткоцикловых видов. Ещё следует отметить, что влияние отбора в разных частях местообитания популяции также может существенно изменять частоты генотипов, тем самым маскируя действительную внутрипопуляционную подразделённость. Увеличение же выборки и числа анализируемых локусов в этом случае не приведёт к увеличению достоверности выводов, так как полученные данные в конечном счёте будут отражать не внутрипопуляционную структуру, а индивидуальные особенности обитающих в данном месте особей (Горбачев, 2011).

**Таблица 9.1.** Уровень межпопуляционной дифференциации по ряду полиморфных локусов у тюльки Рыбинского водохранилища

| **Локус** | **Год наблюдений** | | | | | | | | |
|---|---|---|---|---|---|---|---|---|---|
| | **2002** | | | **2006** | | | **2011** | | |
| | $F_{IS}$ | $F_{IT}$ | $F_{ST}$ | $F_{IS}$ | $F_{IT}$ | $F_{ST}$ | $F_{IS}$ | $F_{IT}$ | $F_{ST}$ |
| *LDH-A** | 3,9 | 4,6 | 0,8 | -3,1 | -1,3 | 1,8 | -2,1 | -1,7 | 0,4 |
| *ME-1** | -1,9 | 0,5 | 2,3 | 11,2 | 13,2 | 2,2 | 8,6 | 9,3 | 0,8 |
| *bEST-2** | 1,2 | 1,7 | 0,6 | 4,4 | 5,5 | 1,1 | 4,1 | 4,4 | 0,7 |
| Среднее (на локус) | 1,07 | 2,27 | *1,23* | 4,17 | 5,80 | *1,70* | 3,53 | 4,01 | *0,63* |

Примечание: F-коэффициенты представлены в процентных долях.

Относительно черноморско-каспийской тюльки, обитающей в Рыбинском водохранилище, все эти ограничения вполне справедливы. Так, тюлька - короткоцикловый пелагический вид, совершающий нерестовые и кормовые миграции, пути которых не определены. Популяции тюльки на периферии ареала постоянно испытывают как межгодовые, так и посезонные сильные колебания численности. Поддержание довольно высокого белкового полиморфизма, несомненно, связано с действием отбора (по крайней мере, на значительную часть изученных локусов). Таким образом следует крайне осторожно относиться к интерпретации результатов F-статистики при изучении внутрипопуляцонной подразделённости тюльки Рыбинского водохранилища.

На протяжении 2002-11 годов для тюльки Рыбинского водохранилища проводился мониторинг генетического состояния популя-ции по всем четырём плёсам (Центральному, Волжскому, Шекснинскому и Моложскому). Стандартная сеть станций и объём выборки даны в Главе 2.

На основании полученных данных определён уровень пространственной подразделённости для популяции тюльки Рыбинского водохранилища (табл. 9.1) на различных этапах её существования. В период, последовавший непосредственно за вселением (2002 г.) внутрипопуляционная дифференциация отсутствовала ($F_{IT} \approx F_{IS} » F_{ST}$). В последующий период (до 2006 г.) тоже отмечается значительный индивидуальный вклад отдельных особей в генетическое разнообразие всей системы, при этом, вероятно, уже начинается образование внутрипопуляционных группировок, что отражает возросший коэффициент $F_{ST}$. Это позволило сделать предположение о начале образования локальных стад тюльки (Карабанов, 2010). Однако в дальнейшем произошло снижение значений $F_{ST}$, и на современном этапе вновь наблюдается отсутствие внутрипопуляционных различий в водохранилище ($F_{IT}>F_{IS}>F_{ST}$). По-видимому, значительные межгодовые колебания F-коэффициентов в данном случае подтверждают точку зрения Л. Джоста (Jost, 2008) что $F_{ST}$, $G_{ST}$ и аналогичные популяционно-генетические параметры оценивают вовсе не генетические расстояния между группировками особей, а их генетическую изменчивость.

Применение даже комплексной статистической обработки популяционно-генетических данных не позволяет установить сколько-нибудь чёткую внутрипопуляционную подразделённость у тюльки Рыбинского водохранилища. Так, применение метода многомерного шкалирования хотя и позволяет разделить наиболее удалённые локальные стада тюльки, но со временем взаимное расположение стад меняется (рис. 9.1), что не позволяет с уверенностью говорить о продолжительном существовании локальных группировок тюльки.

Представленные данные объясняются биологическими особенностями вида. Тюлька - стайный пелагический вид, совершающий большие пищевые и нерестовые миграции. Расстояния акватории Рыбинского водохранилища не являются достаточным препятствием для стабильного межгруппового скрещивания. Некоторая пространственная дифференцировка между наиболее удалёнными группировками рыб, вероятно, связана с наличием локальных нерестилищ и тяготеющих к ним местных стад, что характерно и для многих других рыб (Wilmot et al., 1994; Алтухов, 2003).

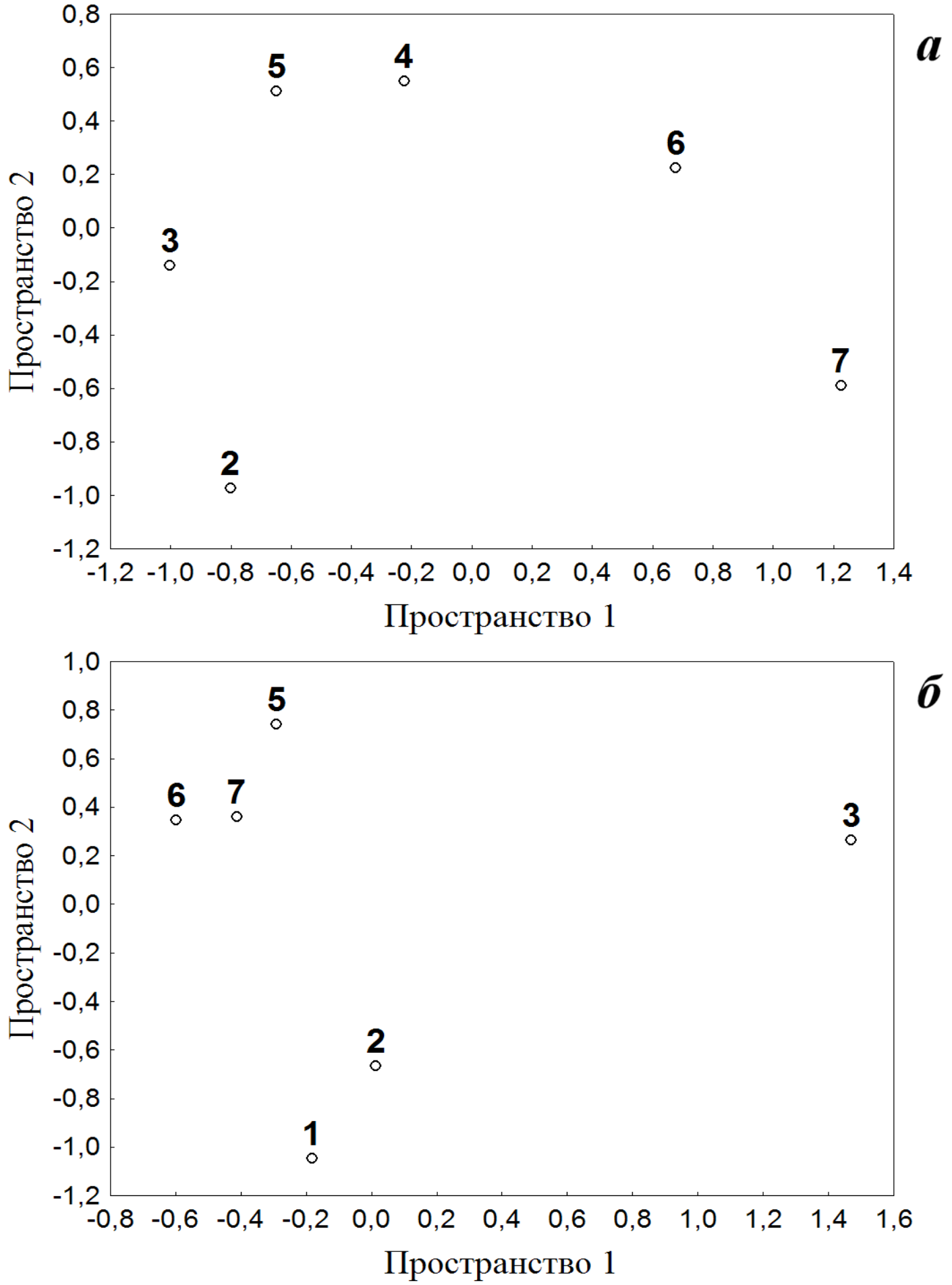

**Рисунок 9.1.** Пространственная структура некоторых локальных группировок тюльки Рыбинского водохранилища в 2002 (***а***), и 2006 (***б***) годах по результатам многомерного шкалирования (12 генетических локусов).
Места отбора проб: **1**–Глебово (Волжский плёс); **2**–Переборы (Волжский плёс); **3**–Центральный мыс (Центральный плёс); **4**–Брейтово (Центральный/Моложский плёс); **5**–Первомайка (Моложский плёс); **6**–Ягорба (Центральный/Шекснинский плёс); **7**–Любец (Шекснинский плёс)

Также возможно, что внутрипопуляционная подразделённость обусловлена выбором оптимальной среды существования генетиче-ски различными особями, мигрирующими в локации с теми или иными условиями (Powell, Taylor, 1979). Хотя, относительно тюльки такая закономерность и не установлена, вовсе исключать её нельзя.

## 9.2. Межполовые различия

У многих организмов особи различного пола по-разному реагируют на факторы окружающей среды. Считается, что у разнополых организмов, например у птиц и млекопитающих, различия в давлении отбора приводят к морфологическим, физиологическим поведенческим и фенотипическим отличиям (Frelinger, 1972; Nadeau, Baccus, 1981). Хотя половой диморфизм часто связывают с половым отбором, есть виды, у которых жизнеспособность особей различного пола различна (Хедрик, 2003).

Для тюльки на первом году жизни в целом соблюдается равенство полов, тогда как у старших возрастных групп преобладают самки (Кияшко и др., 2006). Возможная причина этого кроется в различной интенсивности отбора по отношению к разным полам. Результаты работ по изучению особенностей генетической организации у различных полов тюльки Рыбинского водохранилища в 2004 и 2007 гг. представлены в табл. 9.2.

Возможно, отмечаемые достоверные изменения частот аллелей в популяции тюльки Рыбинского водохранилища за наблюдаемый период связаны с внутрипопуляционной подразделённостью, а не с непосредственным действием отбора. Однако в данной работе сбор материала проводился на одной станции (Ягорба), самцы и самки отбирались из одной выборки. Маловероятно, что данная выборка по происхождению является гибридной из нескольких локальных популяций.

Если графически отобразить действие отбора для самцов и самок по двум локусам для периода 2004/07 годов (рис. 9.2), то можно отметить значительные отличия между ними. Так, для локуса лактатдегидрогеназы-А характерен отбор по аллелю *LDH-A*100*, тогда как ожидаемая жизнеспособность особей обоих полов по локусу *bEST-2** находится в зоне стабильного равновесия (Карабанов, 2012в).

На основании представленных данных можно обозначить, что со временем давление отбора на самцов значительно возросло, тогда как воздействие отбора на самок осталось на том же уровне. В результате этих процессов в популяции возросла концентрация аллеля *LDH-A*100* и снизилась частота аллеля *bEST-2*41*.

**Таблица 9.2.** Генетические различия между полами в популяции тюльки Рыбинского водохранилища в 2004 и 2007 годах

| | | | | | | |
|---|---|---|---|---|---|---|
| **2004 г.** | | | | | | |
| Локус ***LDH-A**** | | | | Локус ***bEST-2**** | | |
| m | *p*(*100)<br>=0,22<br>*p*(*120)<br>=0,78 | *p*(*100/100)=0,02<br>*p*(*100/120)=0,39<br>*p*(*120/120)=0,59 | $S_m$=<br>0,24 | *p*(*41)<br>=0,19<br>*p*(*45)<br>=0,81 | *p*(*41/41)=0,06<br>*p*(*41/45)=0,26<br>*p*(*45/45)=0,68 | $S_m$=<br>0,29 |
| f | *p*(*100)<br>=0,24<br>*p*(*120)<br>=0,76 | *p*(*100/100)=0,06<br>*p*(*100/120)=0,37<br>*p*(*120/120)=0,57 | $S_f$=<br>0,02 | *p*(*41)<br>=0,16<br>*p*(*45)<br>=0,84 | *p*(*41/41)=0,05<br>*p*(*41/45)=0,22<br>*p*(*45/45)=0,73 | *$S_f$=<br>0,35 |
| Σ | *p*(*100)<br>=0,23<br>*p*(*120)<br>=0,77 | *p*(*100/100)=0,04<br>*p*(*100/120)=0,38<br>*p*(*120/120)=0,58 | S=<br>0,09 | *p*(*41)<br>=0,17<br>*p*(*45)<br>=0,83 | *p*(*41/41)=0,15<br>*p*(*41/45)=0,66<br>*p*(*45/45)=0,19 | S=<br>0,36 |
| **2007 г.** | | | | | | |
| Локус ***LDH-A**** | | | | Локус ***bEST-2**** | | |
| m | *p*(*100)<br>=0,27<br>*p*(*120)<br>=0,73 | *p*(*100/100)=0,13*<br>*p*(*100/120)=0,28*<br>*p*(*120/120)=0,58* | *$S_m$=<br>0,31 | *p*(*41)<br>=0,10<br>*p*(*45)<br>=0,90 | *p*(*41/41)*=0,03*<br>*p*(*41/45)*=0,14*<br>*p*(*45/45)*=0,83* | *$S_m$=<br>0,45 |
| f | *p*(*100)<br>=0,28<br>*p*(*120)<br>=0,72 | *p*(*100/100)=0,19<br>*p*(*100/120)=0,19*<br>*p*(*120/120)=0,53* | $S_f$=<br>0,02 | *p*(*41)<br>=0,16<br>*p*(*45)<br>=0,84 | *p*(*41/41)*=0,03*<br>*p*(*41/45)*=0,03*<br>*p*(*45/45)*=0,93* | *$S_f$=<br>0,35 |
| Σ | *p*(*100)<br>=0,28<br>*p*(*120)<br>=0,72 | *p*(*100/100)=0,16*<br>*p*(*100/120)=0,24*<br>*p*(*120/120)=0,60* | *S=<br>0,43 | *p*(*41)<br>=0,13<br>*p*(*45)<br>=0,87 | *p*(*41/41)=0,03<br>*p*(*41/45)=0,09<br>*p*(*45/45)=0,88 | *S=<br>0,63 |

* - значение $\chi^2$ выше табличного, наблюдается отклонение от равновесия Харди-Вайнберга. m - самцы, f - самки. *p* - частоты аллелей или генотипов; $S_m$ , $S_f$ , S - коэффициенты отбора соответственно против самцов, самок и всей группы в целом

В научной литературе имеется ряд работ, свидетельствующих о большей стабильности белкового полиморфизма самок по сравнению с самцами (Christiansen et al., 1977; Nygren, 1981; Алтухов, Варнавская, 1983; Ильин, Голубцов, 1985). Значительное давление отбора именно на самцов, вероятно, связано с их ролью в качестве «носителей изменчивости». В случае черноморско-каспийской тюльки самцы обладают большим генотипическим разнообразием, однако существенная их часть погибает уже на первом году жизни, что и сдвигает соотношение полов у 2-3-летних рыб к преобладанию самок. Аналогичные различия в изменчивости аллельных частот у разных полов описаны и для других животных (Christiansen et al., 1977; Berry et al., 1981; Голубцов, Ильин, 1985). В случае достоверных генетических различий между полами, как правило, существует и наличие специфических для животных разного пола особенностей временн*о*й динамики параметров белкового полиморфизма (Голубцов, 1988).

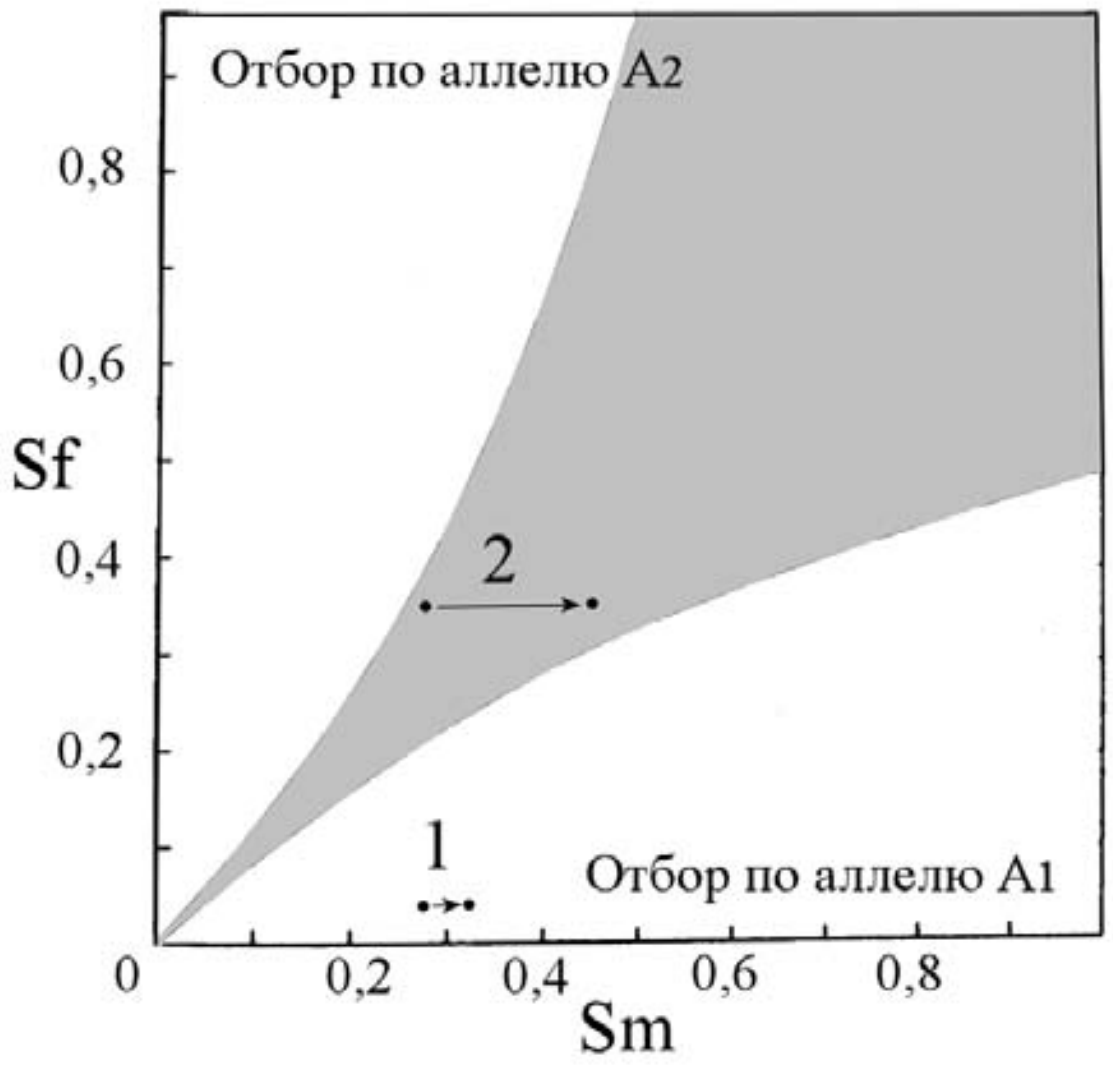

**Рисунок 9.2.** Изменение интенсивности отбора по локусам *LDH-A** (**1**) и *bEST-2** (**2**) в период с 2004 по 2007 гг. (направление сдвига указано стрелкой). Область стабильного равновесия затемнена. $S_m$ и $S_f$ - коэффициенты отбора против самцов и самок соответственно (по: Kidwell et al., 1977 из: Хедрик, 2003, с изм.)

Для объяснения механизма межполовых различий следует обратиться к данным по географическому распределению аллелей в ареале тюльки. Как следует из данных, представленных в Главах 6 и 7, аллель *LDH-A*100* характерен для морских популяций тюльки исторической части ареала. Можно предположить, что максимальная приспособленность и величина «удачных» комбинаций аллелей существует в историческом центре ареала (Майр, 1974). Косвенным подтверждением этой гипотезы служит б*о*льшая продолжительность жизни морских тюлек (до 6-7 лет в море), а также б*о*льший размер и высокий темп роста в море по сравнению с пресноводными популяциями. В таком случае возможным маркером «успешности» особей может служить аллель *LDH-A*100*. Регулярные флуктуации частот аллелей и генотипов лактатдегидрогеназы-А в довольно молодой популяции тюльки Рыбинского водохранилища могут быть следствием различного действия отбора. Для локуса *bEST-2**, напротив, характерно снижение частоты преобладающего в морях аллельного варианта *bEST-2*41*. Вероятно, такое снижение связано с особенностями липидного обмена тюльки северных популяций, когда «пресноводный» аллель оказывается селективно значимым для данных условий.

## 9.3. Генетическая характеристика разных возрастных групп тюльки

По данным многолетних наблюдений, значительную часть популяции тюльки Рыбинского водохранилище составляют особи возрастом до 2+. Особенно большое количество сеголетков встречается в осенних уловах, когда их численность достигает 60-80%. При такой высокой доле молоди важным вопросом служит изучение селективности тех или иных генотипов в различных возрастных группах.

Для определения степени отбора в разновозрастных группах тюльки применён метод оценки интенсивности отбора через обратный расчёт относительной приспособленности генотипов (Ли, 1977; Хедрик, 2003). Полученные результаты сведены в табл. 9.3 и графически показаны на рис. 9.3. Анализ представленных данных позволяет выявить все возможные варианты влияния отбора в процессе онтогенеза тюльки. Так, для локусов *LDH-A** и *bEST-2** интенсивность отбора относительно медленно мигрирующих аллельных вариантов в процессе онтогенеза снижается. Для локуса *ME-1** характерен обратный процесс, а для локуса *AAT** относительные приспособленности генотипов в онтогенезе тюльки практически не изменяются.

К сожалению, прямая оценка селективной ценности особей с определённым генотипом для тюльки практически неосуществима в связи с невозможностью искусственного размножения, содержания и подроста молоди этого вида. В результате различной селективной ценности изменяются не только частоты аллелей в разных возрастных группах тюльки, но и частоты конкретных генотипов (табл. 9.3).

В группе неполовозрелых рыб по локусу *LDH-A** относительная приспособленность гетерозигот максимальна. Это, вероятно, связано с меньшими затратами гетерозиготных животных, по сравнению с гомозиготными, на основной обмен веществ, вследствие чего особи с более высоким уровнем мультилокусной гетерозигот-ности быстрее растут, и более активно потребляют пищу (Sigh, Zouros, 1978; Garton et al., 1984).

**Таблица 9.3.** Генетические характеристики у разных возрастных групп в популяции тюльки Рыбинского водохранилища

| | ***LDH-A**** | | ***ME-1**** | |
|---|---|---|---|---|
| Juv. | *p*(**100*)=0,2<br>*p*(**120*)=0,8 | *p*(**100/100*)= 0,01<br>*p*(**100/120*)= 0,37<br>*p*(**120/120*)=0,61 | *p*(**100*)=0,25<br>*p*(**112*)=0,75 | *p*(**100/100*)=0,04<br>*p*(**100/112*)=0,32<br>*p*(**112/112*)=0,63 |
| Ad. | *p*(**100*)=0,25<br>*p*(**120*)=0,75 | *p*(**100/100*)=0,07<br>*p*(**100/120*)=0,35<br>*p*(**120/120*)=0,58 | *p*(**100*)=0,25<br>*p*(**112*)=0,75 | *p*(**100/100*)=0,02<br>*p*(**100/1112*)=0,36<br>*p*(**112/112*)=0,61 |
| | ***bEST-2**** | | ***AAT**** | |
| Juv. | *p*(**41*)=0,13<br>*p*(**45*)=0,87 | *p*(**41/41*)=0,03<br>*p*(**41/45*)=0,19<br>*p*(**45/45*)=0,78 | *p*(**100*)=0,46*<br>*p*(**110*)=0,54* | *p*(**100/100*)=0,02<br>*p*(**100/110*)=0,89<br>*p*(**110/110*)=0,05 |
| Ad. | *p*(**41*)=0,14<br>*p*(**45*)=0,86 | *p*(**41/41*)=0,03<br>*p*(**41/45*)=0,23<br>*p*(**45/45*)=0,74 | *p*(**100*)=0,50*<br>*p*(**110*)=0,50* | *p*(**100/100*)=0,5<br>*p*(**100/110*)=0,86<br>*p*(**110/110*)=0,8 |

* - значение $\chi^2$ выше табличного; наблюдается нарушение равновесия Харди-Вайнберга. Juv. - неполовозрелые особи, Ad. - половозрелые особ. *p* - частоты аллелей или генотипов

У взрослых особей тюльки более приспособленными оказываются гомозиготы по аллелю *LDH-A*100*. Вероятно, данные особи способны более успешно осваивать ресурсы среды или обладают большим сроком жизни, что и объясняет довольно высокую наблюдаемую частоту аллеля *LDH-A*100* на протяжении всего периода существования тюльки в водохранилище. Однако для ювенильных особей, вероятно, большее значение имеет выживание «в данный момент». Возможно, именно высокий уровень гетерозиготности у ювенильных особей обуславливает их более широкую норму реакции по отношению ко многим абиотическим факторам. В частности это выражается в более широком диапазоне летальных температур молоди по сравнению с половозрелыми рыбами (Смирнов, 2005), что также, возможно, связано с большим уровнем гетерозиготности у молоди (Redding, Schreck, 1979).

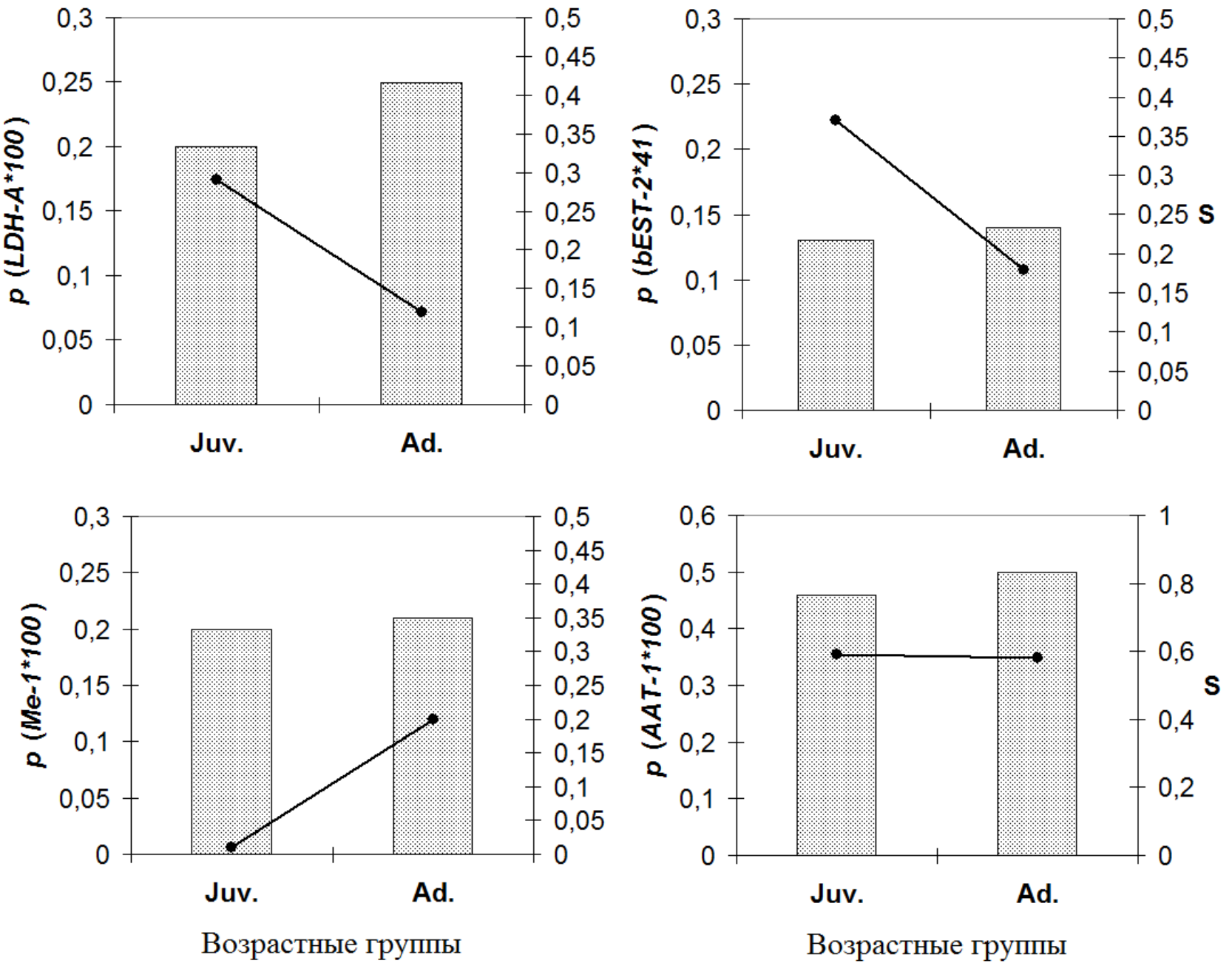

**Рисунок 9.3.** Интенсивность отбора по отношению к одному из аллелей у тюльки разных возрастов: ***а*** – *LDH-A*100*; ***б*** – *bEST-2*41*; ***в*** – *ME-1*100*; *г* – *AAT-1*100*.

*p* - частота аллеля (столбчатые диаграммы, основная ось ординат), *S* - интенсивность отбора (линейный график, вспомогательная ось ординат), Juv. - неполовозрелые особи возрастом 0+, Ad. - взрослые особи возрастом 2+ – 3+

## 9.4. Особенности питания тюльки разных генетических групп

Успешность выживания и размножения тюльки в условиях северных водохранилищ не в последнюю очередь связана с преодолением самого критического этапа - первой зимовкой. Успешность зимовки напрямую обусловлена накоплением необходимого минимального количества резервных питательных веществ (Шатуновский, 1980). Для черноморско-каспийской тюльки в исторической части ареала имеется чёткая связь между интенсивностью питания, динамикой жирности, ростом и воспроизводством рыб (Световидов, 1957; Луц, Рогов, 1978). Особенности динамики липидного обмена в зависимости от питания и связанная с этим вероятность успешной зимовки также характерна и для Рыбинского водохранилища (Халь-ко, 2007). Для изучения особенностей питания тюльки разных генетических групп проведен биологический анализ, изучение спектров питания и генетическое маркирование индивидуальных особей тюльки по результатам двух суточных тралений (ст. Ягорба) 07-08.08 в 2003 г. (по локусу *LDH-A**) и 30.06-01.07 в 2004 г. (по локусам *LDH-A** и *bEST-2**). Трофологический анализ выполнен к.б.н. В.И. Кияшко (ИБВВ РАН). Всего проанализировано 336 особей тюльки (примерно равное количество самцов и самок). В дальнейшем изложении расчёты проведены с использованием выраженных в процентном отношении восстановленных весов пищевых объектов. В качестве меры интенсивности питания применён индекс наполнения пищеварительного тракта рыбы (Кияшко, 2004).

В результате проведённых исследований установлены различия в суточной интенсивности питания для половозрелых самцов и самок тюльки (рис. 9.4).

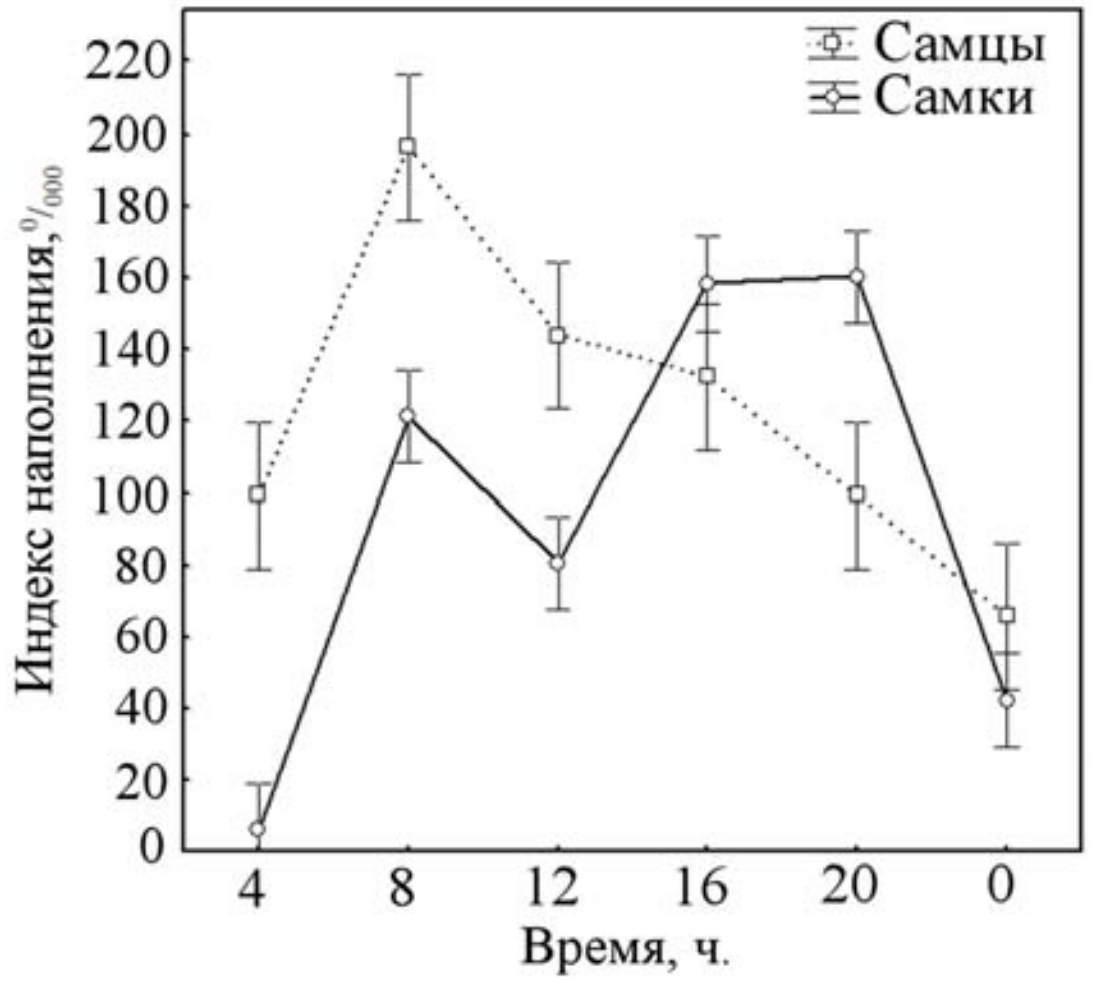

**Рисунок 9.4.** Суточная интенсивность питания тюльки Рыбинского водохранилища. На графике приведены средние значения и ошибка среднего по группам самцов и самок в течение суток

В основном рыбы питаются в светлое время суток (с 8 до 20 часов). Для самок характерно два пика питания - утренний (около 8 часов) и вечерний (с 16 до 20 часов). Самцы активно питаются лишь в утренние часы, постепенно снижая интенсивность питания к тёмному времени суток. Особенности такой пищевой активности у самок можно объяснить суточной активностью пищевых организмов и физиологическим состоянием рыб. В период наблюдений самки на-ходились в преднерестовом состоянии, когда пластический и энергетический обмен у рыб очень высокий (Шатуновский, 1980). Для обеспечения достаточного уровня энергетической обеспеченности организма, самки тюльки вынуждены питаться чаще. Вместе с тем, половые продукты занимают значительный объём в полости тела самок, что механически затрудняет наполнение кишечника рыбы. Для самцов эти факторы, вероятно, оказывают меньшее влияние, поэтому суточная цикличность питания для них менее выражена.

Частоты аллелей в группах самцов и самок были примерно равны: у самок по локусу *LDH-A** наблюдается небольшое преобладание аллеля **100* (*p*=0,25 против *p*=0,22 у самцов), а у самцов - небольшое преобладание аллеля *bEST-2*41* (*p*=0,19 против *p*=0,15 у самок).

Чтобы исключить вероятность изучения двух сцепленных локусов *LDH-A** и *bEST-2** был проведён анализ независимого распределения генотипов этих аллелей по методу ранговых корреляций Спирмена и Кендалля (Ивантер, Коросов, 2003). На основании представленного материала можно сделать вывод о независимом распределении генотипов рассмотренных локусов. Оба показателя (Ts=0,036 и Tk=0,035) гораздо ниже табличных значений, что позволяет сделать вывод об отсутствии достоверного отличия коэффициента от ноля. На основании этого можно утверждать об отсутствии достоверной зависимости в распределении генотипов локусов *LDH-A** и *bEST-2**.

Для уменьшения количества исследуемых факторов все организмы, обнаруженные в пищевом комке тюльки, были разделены на две условные группы: «крупные» - размером более 1 мм, и «мелкие» - размером менее 1 мм. В первую группу вошли представители родов *Daphnia*, *Leptodora*, *Bythotrephes* и *Heterocope*. Во вторую - *Bosmina*, *Diaptomus*, а также ряд Cyclopoida и Chydorida. Избирательность питания оценивалась по индексу Ивлева (Методическое…, 1974) на основе данных В.И. Кияшко по питанию тюльки и В.И. Лазаревой по зоопланктону на данной станции. Трофологическая характеристика тюльки разных полов и различных генотипов по группам кормовых организмов представлена табл. 9.4 и 9.5.

Для определения особенности питания различных групп тюльки (половая и дифференциация по разным генотипам) применялся метод главных эффектов множественного дисперсионного анализа (MANOVA).

**Таблица 9.4.** Трофологическая характеристика тюльки различных генотипов. Приведены средневзвешенные значения и ошибка показателей для «крупных» пищевых организмов (MANOVA)

| **Фактор** | **Эффект** | **N** | **Индекс наполнения** | | ***Daphnia*** | | ***Leptodora*** | | ***Bythotrephes*** | | ***Heterocope*** | |
|---|---|---|---|---|---|---|---|---|---|---|---|---|
| | | | **M** | **± m** | **M** | **± m** | **M** | **± m** | **M** | **± m** | **M** | **± m** |
| пол | самка | 133 | *127,805* | *7,450* | *8,799* | *2,815* | *3,843* | *1,132* | *1,254* | *1,254* | *49,674* | *3,119* |
| | самец | 130 | *145,785* | *8,966* | *16,788* | *3,589* | *2,105* | *0,635* | *7,175* | *2,816* | *44,416* | *1,743* |
| _LDH-A*_ | _*100/100_ | 13 | 134,581 | 11,908 | 0,010 | 0,010 | 0,266 | 0,168 | 0,000 | 0,000 | *98,247* | *0,276* |
| | _*100/120_ | 102 | 146,242 | 47,846 | 16,654 | 4,296 | 2,406 | 0,957 | 2,084 | 1,663 | *48,709* | *5,618* |
| | _*120/120_ | 148 | 131,543 | 12,219 | 11,176 | 2,776 | 3,421 | 0,901 | 5,535 | 2,294 | *43,872* | *4,437* |
| _bEST-2*_ | _*41/41_ | 8 | *202,173* | *9,649* | 4,950 | 2,485 | 0,728 | 0,482 | *0,010* | *0,010* | *64,850* | *14,588* |
| | _*41/45_ | 22 | *138,441* | *33,665* | 12,086 | 5,377 | 3,207 | 1,302 | *0,802* | *0,802* | *47,672* | *7,389* |
| | _*45/45_ | 77 | *129,415* | *15,850* | 13,759 | 2,792 | 3,152 | 0,828 | *5,589* | *2,126* | *45,050* | *4,087* |

N - количество исследованных рыб с наполненным кишечником; M - среднее значение, m - ошибка среднего. Эффекты, уровень значимости для которых $p<0,05$ выделены *курсивом*

**Таблица 9.4.** Продолжение. Трофологическая характеристика тюльки различных генотипов. Приведены средневзвешенные значения и ошибка показателей для «мелких» пищевых организмов (MANOVA)

| **Фактор** | **Эффект** | **N** | ***Bosmina*** | | ***Diaptomus*** | | **Cyclopoida** | | **Chydorida** | |
|---|---|---|---|---|---|---|---|---|---|---|
| | | | **M** | **± m** | **M** | **± m** | **M** | **± m** | **M** | **± m** |
| пол | самка | 133 | 6,040 | 1,838 | 1,519 | 0,885 | *28,525* | *4,049* | 0,002 | 0,002 |
| | самец | 130 | 8,255 | 2,014 | 0,003 | 0,003 | *21,238* | *2,515* | 0,022 | 0,016 |
| *LDH-A** | **100/100* | 13 | 0,886 | 0,583 | 0,000 | 0,000 | *0,600* | *0,600* | 0,000 | 0,000 |
| | **100/120* | 102 | 6,751 | 1,820 | 1,338 | 0,962 | *22,055* | *3,028* | 0,000 | 0,000 |
| | **120/120* | 148 | 7,631 | 1,927 | 0,488 | 0,488 | *27,584* | *3,396* | 0,019 | 0,013 |
| *bEST-2** | **41/41* | 8 | 3,924 | 3,376 | 0,000 | 0,000 | 25,538 | 10,704 | 0,001 | 0,001 |
| | **41/45* | 22 | 5,135 | 1,705 | 0,000 | 0,000 | 31,089 | 5,809 | 0,008 | 0,006 |
| | **45/45* | 77 | 8,043 | 1,790 | 1,068 | 0,624 | 23,087 | 2,700 | 0,014 | 0,011 |

N - количество исследованных рыб с наполненным кишечником; M - среднее значение, m - ошибка среднего. Эффекты, уровень значимости для которых $p<0,05$ выделены *курсивом*

**Таблица 9.5.** Избирательность питания тюльки различных генотипов. Приведены значения средних и ошибки средних индекса избирательности питания Ивлева для «крупных» пищевых организмов (MANOVA)

| **Фактор** | **Эффект** | **N** | **Индекс Ивлева** | | | | | | | |
|---|---|---|---|---|---|---|---|---|---|---|
| | | | ***Daphnia*** | | ***Leptodora*** | | ***Bythotrephes*** | | ***Heterocope*** | |
| | | | **M** | **± m** | **M** | **± m** | **M** | **± m** | **M** | **± m** |
| *LDH-A** | **100/100* | 13 | -1,000 | 0,000 | -0,689 | 0,183 | -1,000 | 0,000 | 0,979 | 0,010 |
| | **100/120* | 102 | -0,490 | 0,102 | -0,486 | 0,116 | -0,895 | 0,073 | 0,628 | 0,120 |
| | **120/120* | 148 | -0,629 | 0,065 | -0,415 | 0,091 | -0,824 | 0,069 | 0,526 | 0,094 |
| *bEST-2** | **41/41* | 8 | -0,693 | 0,133 | -0,644 | 0,225 | -1,000 | 0,001 | 0,704 | 0,245 |
| | **41/45* | 22 | -0,622 | 0,118 | -0,279 | 0,151 | -0,913 | 0,087 | 0,506 | 0,179 |
| | **45/45* | 77 | -0,572 | 0,067 | -0,475 | 0,083 | -0,821 | 0,065 | 0,58 | 0,083 |

N - количество исследованных рыб с наполненным кишечником; M - среднее значение, m - ошибка среднего

**Таблица 9.5.** Продолжение. Избирательность питания тюльки различных генотипов. Приведены значения средних и ошибки средних индекса избирательности питания Ивлева для «мелких» пищевых организмов (MANOVA)

| **Фактор** | **Эффект** | **N** | **Индекс Ивлева** | | | | | |
|---|---|---|---|---|---|---|---|---|
| | | | ***Bosmina*** | | ***Diaptomus*** | | **Cyclopoida** | |
| | | | **M** | **± m** | **M** | **± m** | **M** | **± m** |
| *LDH-A** | **100/100* | 13 | -0,699 | 0,183 | -1,000 | 0,000 | -0,952 | 0,048 |
| | **100/120* | 102 | -0,230 | 0,098 | -0,927 | 0,051 | -0,099 | 0,075 |
| | **120/120* | 148 | -0,307 | 0,073 | -0,978 | 0,022 | 0,166 | 0,062 |
| *bEST-2** | **41/41* | 8 | -0,622 | 0,213 | -1,000 | 0,001 | -0,270 | 0,212 |
| | **41/45* | 22 | -0,327 | 0,13 | -1,000 | 0,001 | -0,061 | 0,100 |
| | **45/45* | 77 | -0,247 | 0,067 | -0,945 | 0,031 | -0,232 | 0,057 |

N - количество исследованных рыб с наполненным кишечником; M - среднее значение, m - ошибка среднего

Для категориальных факторов устанавливались противоположные значения контрастов исходя из альтернативности проявления биологических явлений. Так, для к.ф. «пол» устанавливались контрасты +1/-1 между эффектами самцы-самки, для генотипов *LDH-A** и *bEST-2** контрасты определялись как +2/+1/0 для гомозигот, гетерозигот и альтернативных гомозигот соответственно. В дальнейшем анализе рассмотрены только эффекты категориальных факторов, уровень значимости для которых $p<0,05$ (табл. 9.4, 9.5).

Изучение спектра питания тюльки разных полов (рис. 9.5) позволяет отметить более высокий коэффициент наполнения кишечника у самцов, что связано с их более высокой интенсивностью питания в светлое время суток (рис. 9.4). Однако самки проявляют большую «разборчивость» в питании, предпочитая потреблять более крупные пищевые организмы (*Heterocope* и *Leptodora*), а также большое количество Cyclopoida. Вероятно, самки предпочитают активно выбирать более крупные пищевые объекты, тогда как самцы, имея более широкий спектр питания, используют всю доступную пищу.

При рассмотрении трофологических особенностей тюльки различных генотипов локуса лактатдегидрогеназы-А можно отметить значительные различия в качественном составе пищи для разных особей с разным генотипом, при относительно равных индексах наполнения (рис. 9.6). Особи с генотипом *LDH-A*100/100* проявляют высокую селективность в питании, выбирая крупных *Heterocope*. Избирательность потребления *Heterocope* гомозиготами **100/100* подтверждается высоким значением индекса Ивлева (0,979), который значимо превышает таковой для гетерозигот и альтернативных гомозигот (0,628 и 0,526 соответственно) (табл. 9.5). Вероятно, у этих рыб некоторая потеря в количестве потреблённых пищевых организмов компенсируется их качественным составом, достаточным для удовлетворения потребностей метаболизма рыбы. Альтернативные гомозиготы *LDH-A*120/120* напротив, в меньших количествах потребляют «крупный» планктон, обычно поглощая большое количество мелких ракообразных, в основном Cyclopoida. Индекс избирательности для таких гомозигот **120/120* по этой группе пищевых организмов (0,166) выше, чем для других генотипов (табл. 9.5).

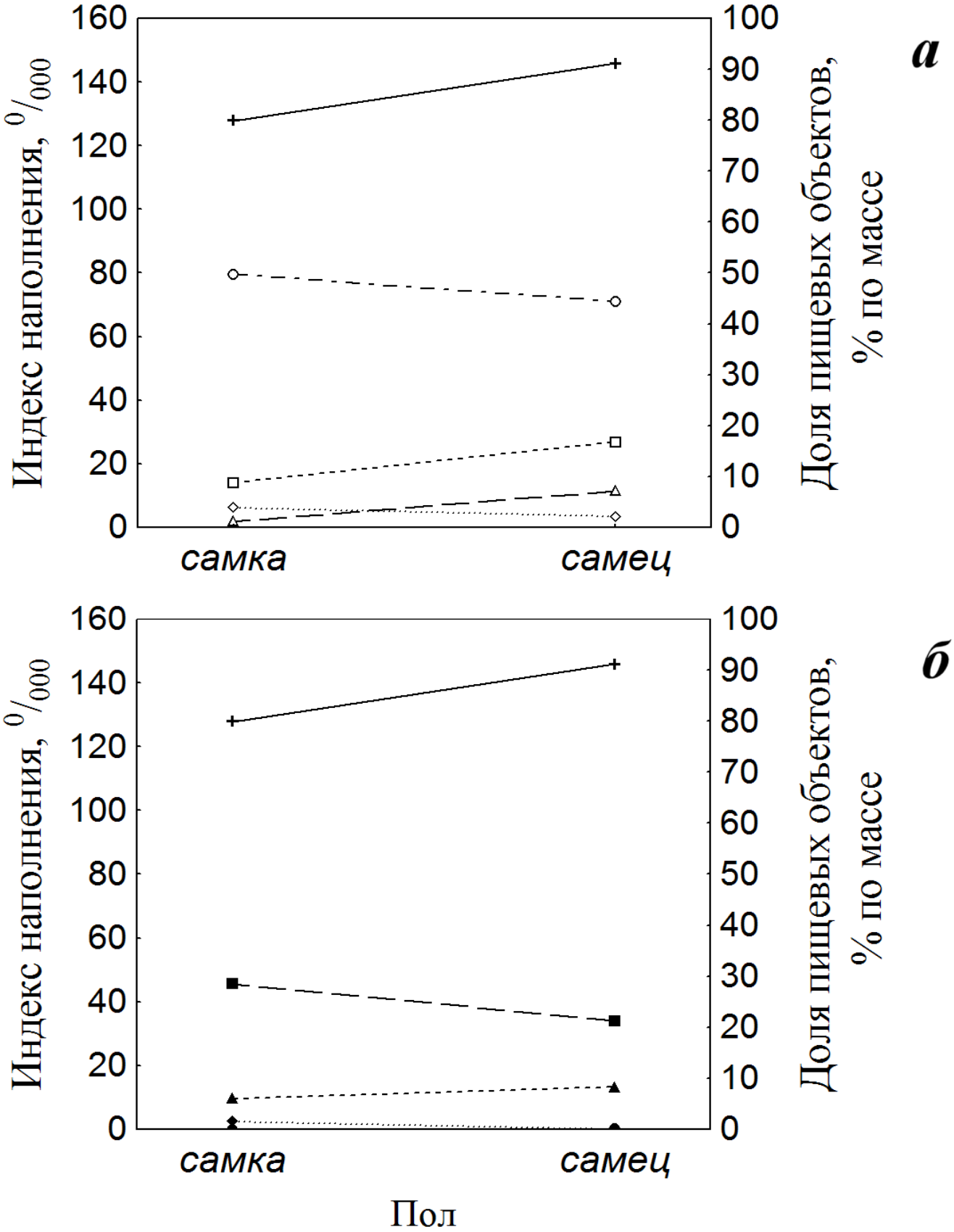

**Рисунок 9.5.** Особенности питания тюльки в зависимости от размера пищевых объектов: различия между полами. Группа «крупных» (***а***) и «мелких» (***б***) пищевых организмов. MANOVA, приведены значения средневзвешенных средних.

Условные обозначения: —+— - индекс наполнения; «крупные» пищевые организмы (***а***): —□— *Daphnia*; ◇ *Leptodora*; —△— *Bythotrephes*; —○— *Heterocope*; «мелкие» пищевые организмы (***б***) : —▲— *Bosmina*; ◆ *Diaptomus*; —■— Cyclopoida; —●— Chydorida

Гетерозиготы обладают широким спектром питания, за счёт чего их индекс наполнения выше, чем у обеих гомозигот. Следует отметить, что для черноморско-каспийской тюльки, обитающей в Каспийском море, основу питания также составляют крупные *Heterocope* (Барышева, 1952; Kozlovsky, 1991), при этом в популяции Каспийского моря доминирует аллель *LDH-A*100*. Возможно, что комплекс адаптаций, сцепленный с аллельным вариантом *LDH-A*100* и направленный на потребление крупных кормовых организмов имеет «морское» происхождение и связан с адаптациями тюльки, возникшими и закреплёнными в исторической части ареала. Нельзя исключить, что наблюдаемое в середине 2000-х годов некоторое увеличение среднего размера тюльки (Глава 4) также связано с возрастанием доли аллеля *LDH-A*100* в популяции.

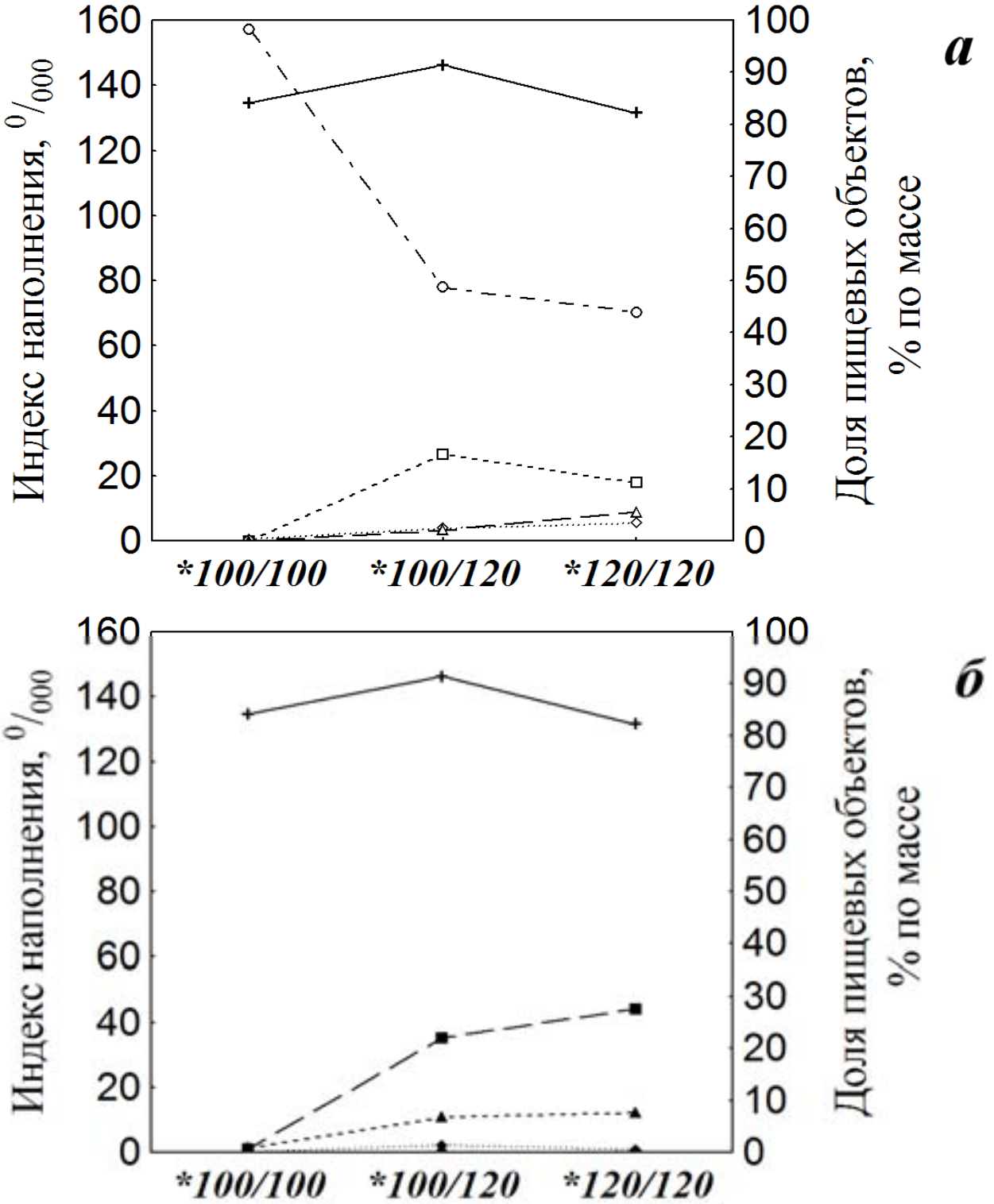

**Рисунок 9.6.** Особенности питания тюльки в зависимости от размера пищевых объектов: различия между генотипами *LDH-A**. Группа «крупных» (***а***) и «мелких» (***б***) пищевых организмов. MANOVA, приведены значения средневзвешенных средних.

Условные обозначения: —+— - индекс наполнения; «крупные» пищевые организмы (***а***): —□— *Daphnia*; ◇ *Leptodora*; —△— *Bythotrephes*; —○— *Heterocope*; «мелкие» пищевые организмы (***б***) : —▲— *Bosmina*; ◆ *Diaptomus*; —■— Cyclopoida; —●— Chydorida

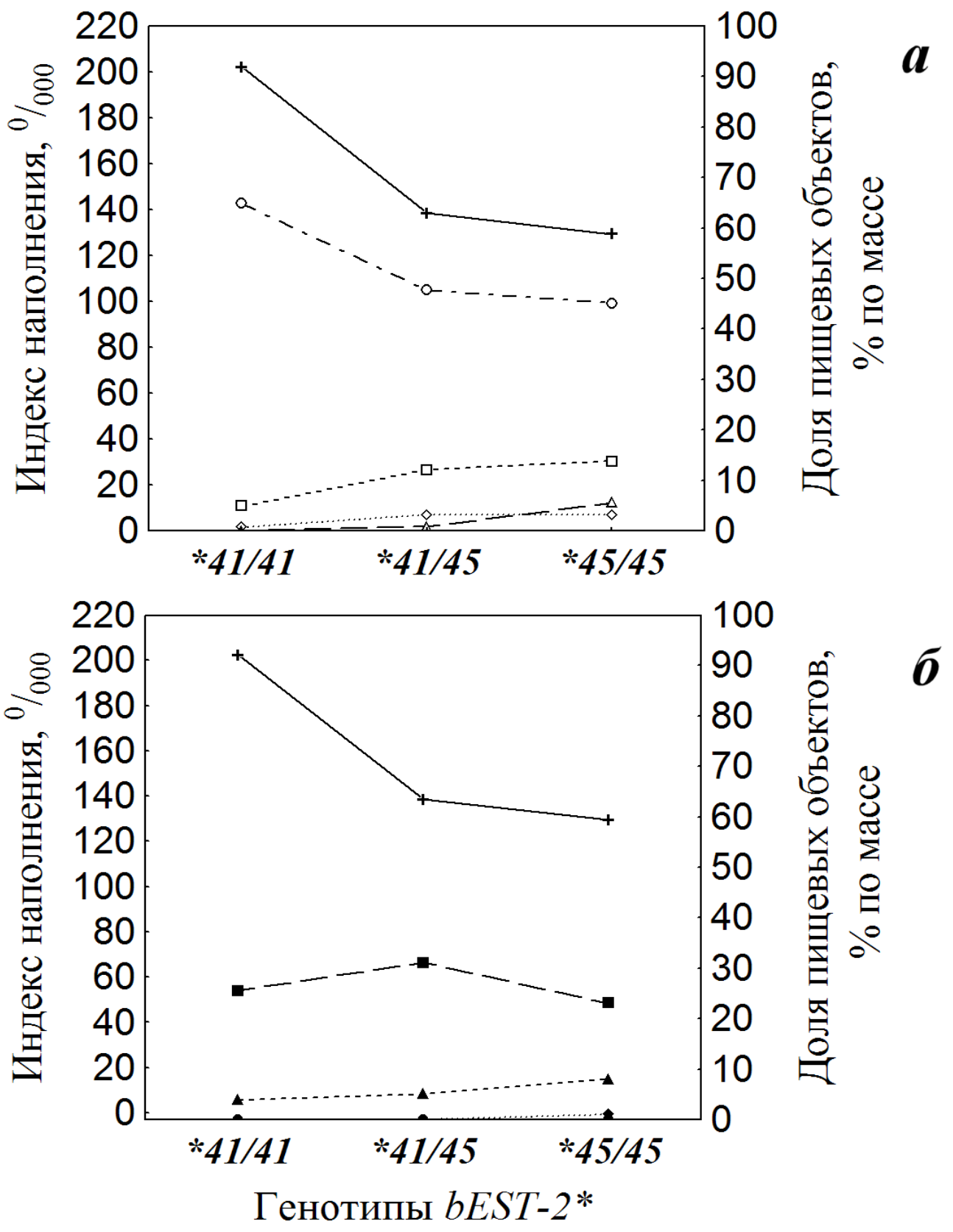

**Рисунок 9.7.** Особенности питания тюльки в зависимости от размера пищевых объектов: различия между генотипами *bEST-2** (***в***). Группа «крупных» (***а***) и «мелких» (***б***) пищевых организмов. MANOVA, приведены значения средневзвешенных средних.
Условные обозначения: - индекс наполнения; «крупные» пищевые организмы (***а***): *Daphnia*; *Leptodora*; *Bythotrephes*; *Heterocope*; «мелкие» пищевые организмы (***б***) : *Bosmina*; *Diaptomus*; Cyclopoida; Chydorida

На примере различного трофического статуса особей, маркированных по аллелям локуса *LDH-A** прослеживаются разные пути адаптаций в добывании пищи у этой стайной рыбы. Тогда как одни особи сосредотачиваются в основном на потреблении крупных пищевых объектов, другие особи активно поедают мелкий планктон. Маловероятно, что гомозиготы *LDH-A*120/120* «отказываются» от поедания крупных организмов. Скорее всего, особи с генотипом **100/100,* а также (в немного меньшей степени) и гетерозиготы активно выедая крупных рачков, обедняют скопления планктонных ракообразных, к которым приурочена данная стая рыб. В результате гомозиготам **120/120* остаётся потреблять большее количество мелких организмов. Возможно, особенности добывания пищи, маркируемые аллелем *LDH-A*100* связаны с более активным плаванием, свойствами зрениея либо поведенческими реакциями, направленными на добывание пищи (или какими-то другими адаптациями, сцепленными с данным аллелем).

Вместе с тем, специализация гомозигот *LDH-A*100/100* на питание одним видом крупных ракообразных таит в себе значительную опасность: зимой, когда количество пищевых организмов значительно снижается, а планктон представлен в основном мелкими формами, такие особи в первую очередь подвергаются элиминации. Именно этот процесс и отражается в сезонном колебании частот аллелей *LDH-A**, что будет рассмотрено далее. В осенне-зимний период, когда количество и биомасса планктона значительно уменьшается, гомозиготы **100/100* начинают испытывать дефицит пищи. Гетерозиготы, а особенно - альтернативные гомозиготы, за счёт широкого спектра питания и потребления мелких кормовых организмов способны обеспечить себя достаточным количеством пищи и успешно пережить зиму. Однако в летний период за счёт скрещивания между гетерозиготами происходит сезонное повышение доли гомозигот **100/100* и увеличение их численности за счёт преимущества в питании крупными ракообразными.

Для рыб с разными генотипами по локусу *bEST-2** также прослеживаются значительные различия в их трофологическом статусе (рис. 9.7). Для особей, гомозиготных по аллелю *bEST-2*41* характерно самое высокое значение индекса наполнения. Это, прежде всего, связано с активным потреблением крупных *Heterocope*. Индекс избирательности для гомозигот **41/41* по этому пищевому объекту выше, чем для других генотипов (табл. 9.5). Другим важным компонентом в питании данных гомозигот составляют различные ветвистоусые ракообразные. Альтернативные гомозиготы по аллелю *bEST-2*45* также потребляют довольно большое количество *Heterocope*, но в их питании уже немалую долю занимают крупные *Daphnia* и *Bythotrephes*, а из группы «мелких» кормовых организмов велика роль Cyclopoida и *Bosmina*. Однако индекс наполнения у этих особей ниже, чем для гомозигот **41/41*. Геторизиготы **41/45* имеют значения индекса наполнения примерно схожие с гомозиготами *45/45*. Основу их пищевого комка составляют крупные *Heterocope* и различные Cyclopoida.

На примере локуса *bEST-2** можно выдвинуть предположение о разной селективной ценности аллелей этого локуса. Особи, гомозиготные по аллелю *bEST-2*45* обладают хоть и меньшим индексом наполнения, но самым широким спектром питания. Так как частота этого аллеля в процессе существования Рыбинской популяции тюльки сохраняется на очень высоком уровне (табл. 6.13), то можно выдвинуть предположение, что участие соответствующего аллозима в липидном обмене тюльки даёт некоторые преимущества в условиях этого северного водохранилища. Значительное повышение частоты аллеля *bEST-2*41* происходит лишь в летне-осенний период, когда в популяции очень велика доля сеголетков, для которых относительная приспособленность генотипа **41/41* максимальна (табл. 9.3). Однако уже к зиме частота этого аллеля падает (см. далее), а в генеративной части популяции к периоду размножения достигает наблюдаемых значений.

## 9.5. Сезонная динамика генетических показателей

Изменение генетических характеристик в течение года в популяциях позвоночных животных - распространённое явление (Динамика…, 2004). Если для проходных и полупроходных рыб такие данные имеются, то для жилых форм рыб этот вопрос не очень изучен (Алтухов, 2003). В подавляющем большинстве случаев существует наличие прямой связи между динамикой параметров белкового полиморфизма и сезонными изменениями среды обитания, либо с циклическими изменениями численности животных (Голубцов, 1988). Сезонная динамика в случае тюльки интересна ещё и тем, что это короткоцикловый вид, срок жизни основной массы особей в Рыбинском водохранилище, как правило, не превышает 3-х лет (Кияшко и др., 2006). Таким образом, даже небольшие преимущества при перезимовке рыбы, особенно сеголеток, сильно увеличивают шансы на успешное размножение в следующем сезоне.

Для наблюдения за сезонной динамикой частот аллелей в сезон 2005/06 гг. на станции Коприно (Рыбинское водохранилище, Волжский плёс) проводился лов тюльки с последующим генетико-биохимическим анализом проб. Полученные результаты представлены на рис. 9.8 и 9.9.

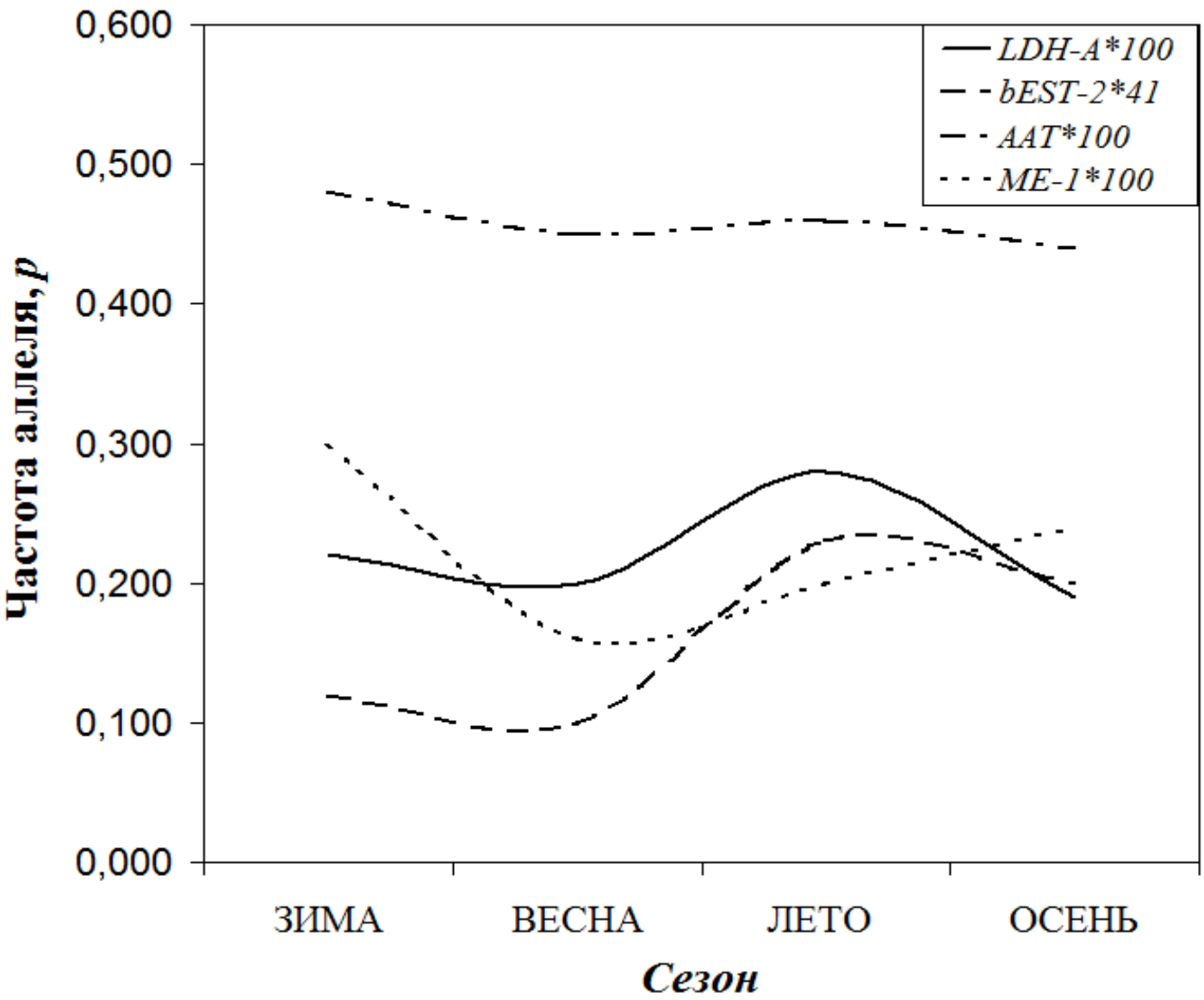

**Рисунок 9.8.** Сезонные изменения частот медленномигрирующих (по отношению к аноду) аллельных вариантов некоторых генетических локусов тюльки

Представленные данные позволяют установить несколько возможных вариантов динамики аллельных частот. Для локуса аспартатаминотрансферазы характерна значительная константность частот аллелей. Достоверных изменений этого параметра между сезонами не наблюдается ($p>0,05$). Вероятно, значимого влияния на зимовку и размножение данный локус не оказывает, хотя частоты аллелей несколько изменяются в осеннее-зимний период. Иная картина наблюдается для аллелей локуса лактатдегидрогеназы-А. В зимний период его частота невелика, а к началу весеннего сезона ещё более снижается ($\chi^2=11,19$; $p<0,05$). В весенне-летний период частота аллеля *LDH-A*100* возрастает ($\chi^2=10,13$; $p<0,05$) и к концу лета достигает своего годового максимума. К началу осени частота аллеля **100* вновь начинает достоверно снижаться ($\chi^2=4,81$; $p<0,05$).

Дифференциальное выживание особей, различных по генотипам *LDH** описано для многих видов рыб (Кирпичников, Иванова, 1977; Johnson, 1971; Redding, Schreck, 1979; Mork, Sundnes, 1985). Вероятно, для тюльки сезонные колебания частоты аллеля *LDH-A*100* в предзимний период связаны с элиминацией мальков первой и второй порций нереста. Наибольшее давление отбора по локусу

*LDH-A** наблюдается при смене летнего сезона на осенний, когда происходит наибольшая смертность сеголетков тюльки перед зимовкой.

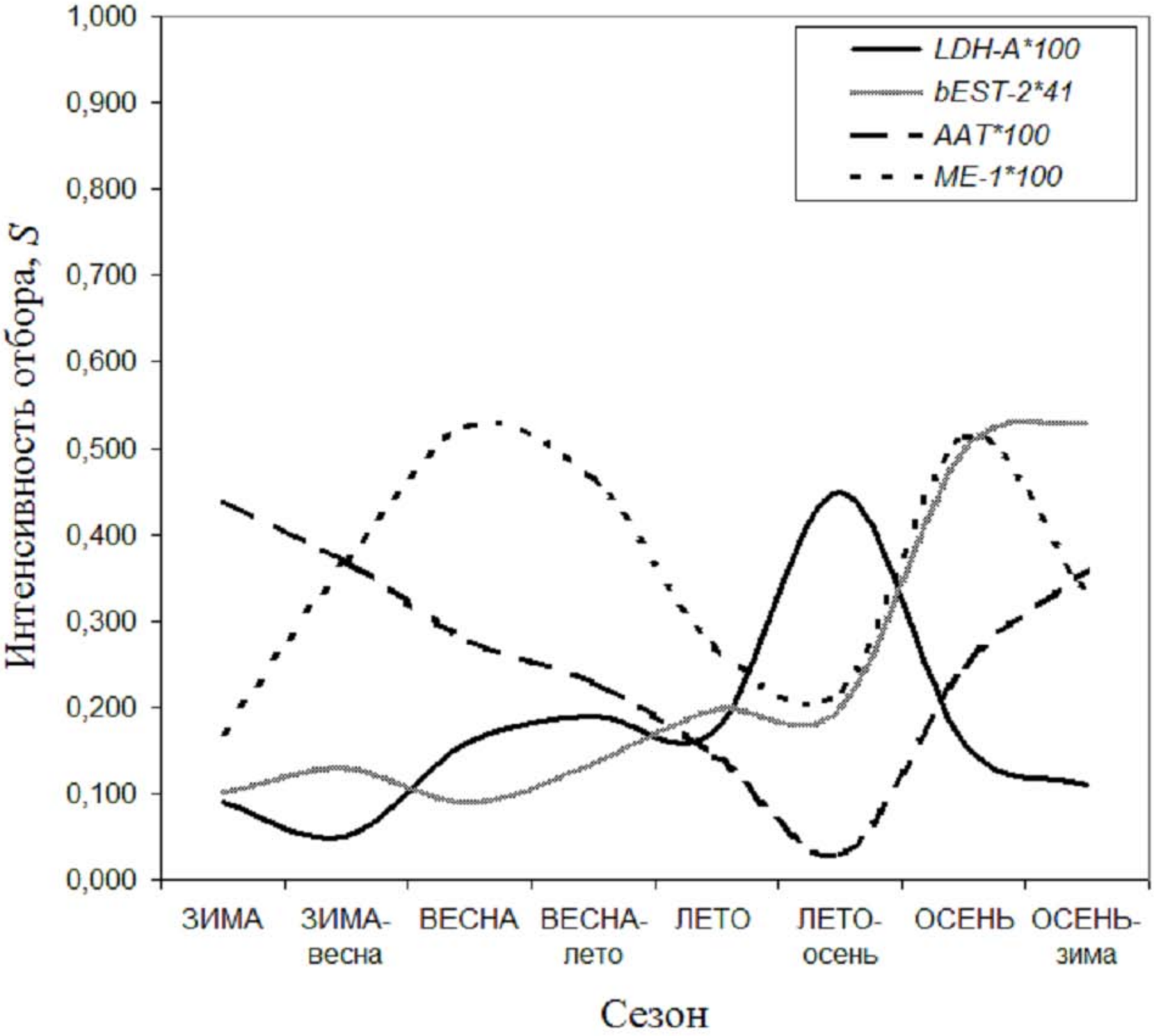

**Рисунок 9.9.** Сезонные изменения интенсивности отбора по некоторым генетическим локусам черноморско-каспийской тюльки (Рыбинское водохранилище, сезон 2005/06 гг.)

Для локуса эстеразы также характерно летнее увеличение частоты аллеля *bEST-2*41*. К осени его частота незначительно снижается, а существенное достоверное уменьшение наступает лишь в зимний период ($\chi^2$=13,47; $p<0,05$). Вероятно, наибольшее влияние отбора на локус *bEST-2** сказывается поздней осенью, когда происходит похолодание и смена кормовой базы.

Для локуса *ME-1** начиная с летнего периода характерно постепенное увеличение аллеля *ME-1*100*, достоверное уменьшение которого происходит лишь поздней зимой и в начале весны ($\chi^2$=17,07; $p<0,05$). Для этого генетического локуса характерны два всплеска давления отбора - весной и осенью (рис. 9.9), что может быть связано со сменой приоритета от выживания к размножению, как это происходит у многих животных (Айала, 1981).

При сравнении типа и интенсивности отбора в течение года для разных генетических локусов тюльки наблюдается различная динамика относительных приспособленностей генотипов (рис. 9.10).

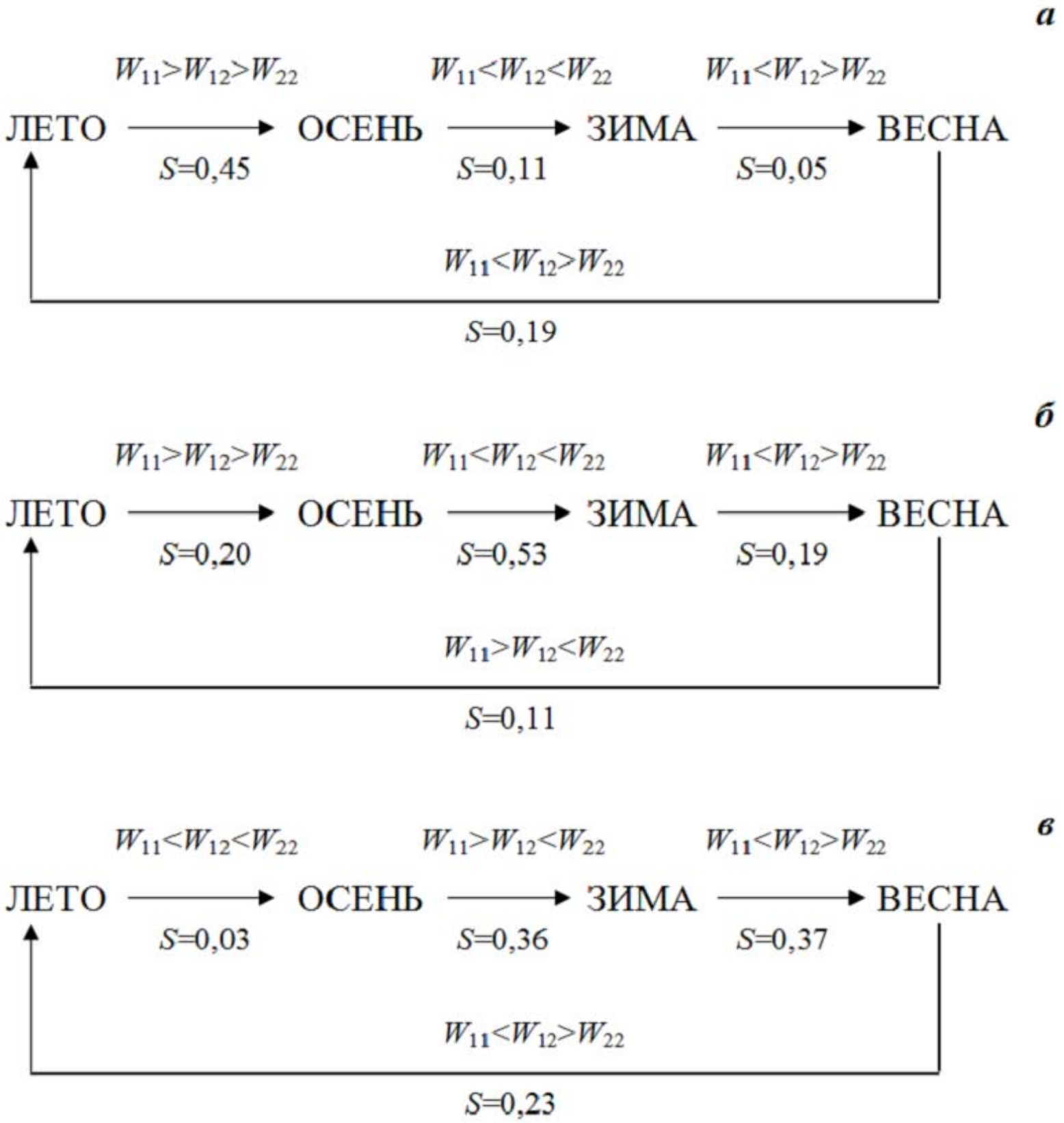

**Рисунок 9.10.** Схема изменения интенсивности и типа отбора в популяции тюльки в течение сезона 2005/06 гг. (рыбинское водохранилище, Волжский плёс) по трём генетическим локусам: ***а*** – *LDH-A**; ***б*** – *bEST-2**; ***в*** – *ME-1**. Условные обозначения: $S$ - интенсивность отбора; $W_{11}$, $W_{12}$, $W_{22}$ - относительные приспособленности генотипов

Для локуса лактатдегидрогеназы-А характерно сильное изменение относительных приспособленностей генотипов и давления отбора в течение сезона (рис. 9.10,*а*). Наиболее сильно отбор действует при переходе к зиме, когда радикально изменяется относи-тельная приспособленность гомозигот. В данном случае отбор оказывает направленное воздействие: максимальная приспособленность гомозигот по аллелю *LDH-A*100* в летний период, зимой меняется на максимальную приспособленность альтернативных гомозигот **120/120*. В весенний период отбор носит характер балансирующего, а наибольшая приспособленность перемещается на гетерозиготы.

Для локуса *bEST-2** характерны изменения селективной ценности различных генотипов, аналогичные локусу *LDH-A**, с той лишь разницей, что при переходе с весны к лету отбор меняется с балансирующего на дизруптивный (рис. 9.10,*б*). В данном случае в летний период под воздействием дизруптивного отбора в популяции тюльки возникает большое генотипическое разнообразие по этому локусу. При переходе к зиме происходит элиминация большого числа особей, прежде всего сеголетков и старых особей, однако накопленное разнообразие, вероятно, способствует сохранению части удачных для данного сезона комбинаций.

Для генетического локуса *ME-1** также характерна смена всех типов отбора, причём значительное изменение частот аллелей происходит лишь в зимний период, когда меняется тип отбора с дизруптивного на балансирующий (рис. 9.10,*в*).

Наблюдаемые колебания генетических показателей свидетельствую о сложных сезонных перестройках в генетической структуре популяции, связанными с изменениями возрастного состава и условиями зимовки основной части популяции - сеголетков и особей возраста 1+, что происходит как непосредственно под влиянием сезонных меняющихся абиотических факторов, так и, вероятно, со сменой качественного и количественного состава пищевых организмов.

## Заключение

Представленные данные позволяют определить стратегические пути генетико-бихимических адаптаций тюльки при расселении по речным водохранилищам. Рассмотрение адаптационных преобразований в случае биологических инвазий позволяет изучить этот процесс «здесь и сейчас» в природных условиях, благодаря чему достигается лучшее понимание микроэволюционных преобразований в краевых популяциях, чем при теоретических построениях, основанных на лабораторных исследованиях.

Для тюльки, расселившейся по всей Волге, наблюдаемое генотипическое разнообразие организмов определяется в биохимических адаптациях на уровне изоферментов. Это прослеживается для аллозимов мышечной лактатдегидрогеназы-А. Аллельный вариант *LDH-A*100*, вероятнее всего, имеет исторически «морское» происхождение. На популяционном уровне такая гипотеза подтверждается особенностями географического распределения аллелей локуса *LDH-A**. В морях частота аллозима *LDH-A*100* максимальна, а у родственного, облигатно морского вида - *C. engrauliformes* изофермент представлен только продуктом аналогичного аллеля. В популяциях атлантической и тихоокеанской сельдей *Clupea harengus* и *C. pallasi* также преобладает аналогичный аллозим лактатдегидрогеназы-А (Jorstad, 2004).

Частота варианта *LDH-A*120* возрастает в пресноводных экосистемах, что прослеживается на популяциях тюльки Днепра, Волги и Камы. При рассмотрении генетических характеристик эти групп популяций становится возможным проследить скорость микроэволюционных преобразований в условиях частичной изоляции. Необычность Волжско-Камских популяций тюльки (существенное преобладание аллеля *LDH-A*120*) можно объяснить происхождением их от пресноводной формы, обитавшей в затонах у г. Саратов. Вероятно, эта жилая форма возникла в результате Хвалынской трансгрессии и последующего отступления Каспия около 20-40 тыс. л. назад. За этот период произошли значительные адаптации к обитанию в условиях пресных вод речных экосистем, что отразилось в существенном перераспределении аллельных частот *LDH-A**. В Днепре таких процессов не было, а продвижение тюльки по реке затрудняли Днепровские пороги. После начала гидростроительства на Днепре в начале XX века гидрологический режим реки сильно изменился: пороги исчезли и возникли водохранилища. В результате этого произошло расселение тюльки из солоноватоводных лиманов в Днепр. Так как эволюционный возраст этих популяций мал и такого срока, вероятно, недостаточно для существенного изменения генетических показателей, то и наблюдаемая дивергенция между популяциями тюльки лиманов и Днепра невелика и не достигает таких значений, как для Волги с Камой и Каспийского моря.

При изучении стратегии биохимических адаптаций интересно исследование эколого-генетических особенностей тюльки изолированных популяций в высокоминерализованных водоёмах (Маныч, Карачуновское водохранилище). Изоляция Карачуновского водохранилища, образованного в устьевой части р. Славянка произошла более века назад. Вероятно, с той же поры существует локальная популяция тюльки этого водоёма. Группа популяций тюльки Кумо-Манычской котловины, вероятно, возникла в результате разделения Понто-Каспия. По-видимому, эти популяции сохранились до настоящего времени как реликты. Длительное существование таких локальных стад рыб в условиях высокой минерализации вод привело к наблюдаемому паритетному соотношению частот аллелей *LDH-A**. Таким образом, и в случае изоляции происходят адаптационные изменения к условиям обитания, что выражается в сдвиге генетических характеристик и изменении физиологического оптимума организмов. При длительной изоляции эколого-генетические характеристики популяции сдвигаются к оптимальным значениям для данного водоёма, а со временем обеспечивается максимально возможное соответствие генотипа к условиям среды и очень долгое устойчивое состояние системы, как это произошло в случае популяций тюльки из водохранилищ Маныча.

Таким образом, современная филогеографическая ситуация, сложившаяся в результате саморасселения тюльки по рекам Европы определяется как широкой экологической пластичностью, так и значительными миграционными способностями вида. Это подтверждается популяционно-генетическими данными по изолятам, а также структурой «переходных» популяций низовий Волгоградского водохранилища и Нижнего Днепра. Судя по балансам аллелей изученных генетических локусов, эти популяции испытывают приток мигрантов из эстуариев. Между всеми популяциями Волжского каскада водохранилищ также, вероятно, существует постоянный обмен мигрантами, что обеспечивает генетическое единство вида во всём ареале. На основании генетико-биохимических и данных RAPD-PCR пересмотр таксономического статуса различных крупных географических групп популяций тюльки безоснователен.

Важную роль в распределении генотипов тюльки в новообразованной популяции играют сезонные изменения абиотических факторов и смена качественного и количественного состава пищевых организмов. Сезонное изменение интенсивности и типа отбора приводит к возникновению высокого генотипического разнообразия внутри популяции. Создание такой генетической гетерогенности обеспечивает поддержание широкой нормы реакции у тюльки в условиях северных водохранилищ, что позволяет этому виду успешно пережить наиболее критический период существования - зимовку.

В целом в случае биологических инвазий при адаптогенезе возможно как быстрое изменение генотипа вселенцев, так и возникновение адаптаций без структурных перестроек генома (Prentis et al., 2008). Приведённые в настоящей работе данные о генетических процессах в новообразованных популяциях черноморско-каспийской тюльки хорошо согласуются с представлением о системах стабилизации генофондов (Артамонова, Махров, 2008). При заселении тюлькой новых водоёмов на начальных этапах наблюдаются процессы дестабилизации генетической структуры (значительное влияние отбора, выраженное в достоверных отклонениях от равновесных частот аллелей большинства генетических локусов). Данный процесс прослеживался в начале 2000-х годов при заселении тюлькой Рыбинского водохранилища. В дальнейшем, происходит стабилизация генетических показателей (частоты аллелей приближаются к равновесным), начинают формироваться локальные стада (внутрипопуляционные группировки), тюлька становится органичным компонентом экосистемы водохранилища. Поддержание устойчивого состояния популяции обеспечивается балансирующим отбором, сохраняющим довольно высокий уровень полиморфизма в северных популяциях тюльки.

Полученные данные хорошо согласуются с адаптационной «гипотезой Красной королевы» Ван-Велена – Левонтина (Левонтин, 1981). В северных водоёмах, значительно отличающихся по всем параметрам от водоёмов исторической части ареала тюльки, происходят регулярные изменения условий существования рыб, в первую очередь связанные с зимовкой. Соответственно, постоянно изменяется адаптивная ценность различных генотипов. Довольно высокая гетерозиготность поддерживается наличием в некоторые сезоны балансирующего отбора, сменяющегося на дизруптивный, что обеспечивает большое генотипическое разнообразие в популяции, которое потом может изменяться направленным отбором по определённому генотипу. Таким образом, в популяции поддерживается состояние высокой приспособленности к условиям существования, а широкая норма реакции обеспечивается за счёт менее приспособленных к данным условиям, но потенциально адаптивных генотипов - носителей возможных модификаций, позволяющих организму выжить при изменении воздействия факторов среды вне зоны физиологического оптимума (Шишкин, 1984).

## Список литературы

Адели Ю. Изучение биологии инвазийного гребневика (*Mnemiopsis leidyi*) на Иранском побережье Каспийского моря // Чужеродные виды в Голарктике (Борок-2). Тез. докл. Рыбинск-Борок, 2005. С. 111–112.

Айала Ф. Механизмы эволюции / Эволюция. М.: Мир, 1981. С. 33–65.

Алекин О.А. Гидрохимия. Л.: ГИМИЗ, 1952. – 162 с.

Алтухов Ю.П. Генетические процессы в популяциях. М.: Академкнига, 2003. – 431 с.

Алтухов Ю.П., Салменкова Е.А., Омельченко В.Т., Сачко Г.Д., Слынько В.И. О числе мономорфных и полиморфных локусов в популяции кеты *Oncorhynchus keta* Walb. - одного из тетраплоидных лососевых // Генетика. 1972. Т.8. №2. С. 67–75.

Алтухов Ю.П., Варнавская Н.В. Адаптивная генетическая структура и её связь с внутрипопуляционной дифференциацией по полу, возрасту и скорости роста у тихоокеанского лосося - нерки *Oncorhynchus nerka* (Walb.) // Генетика. 1983. N.19. №5. С. 696–807.

Алтухов Ю.П., Салменкова Е.А., Омельченко В.Т. Популяционная генетика лососевых рыб. М.: Наука, 1997. – 288 с.

Амброз А.И. Рыбы Днепра, Южного Буга и Днепровско-Бугского лимана. Киев: АН УССР, 1956. – 405 с.

Аннотированный каталог круглоротых и рыб континентальных вод России. Ред. Решетников Ю.С. М.: Наука, 1998. – 220 с.

Антонов П.И., Козловский С.В. О самопроизвольном расширении ареалов некоторых понто-каспийских видов по каскадам водохранилищ // Инвазии чужеродных видов в Голарктике. Борок: ИБВВ РАН-ИПЭЭ РАН, 2003. С. 18–25.

Антонова Е.Л., Пушкин Ю.А. Изменчивость морфобиологических показателей тюльки средней Камы // Биология водоемов западного Урала. Проблемы воспроизводства и использования ресурсов. Пермь: ПермГУ, 1985. С. 52–57.

Артамонова В.С., Махров А.А. Генетические системы как регуляторы процессов адаптации и видообразования (к системной теории микроэволюции) // Современной проблемы биологической эволюции: труды конференции. К 100-летию Гос. Дарвинского муз. М., 2008. С. 381–403.

Атлас пресноводных рыб России. Ред. Решетников Ю.С. Т.1. М.: Наука, 2003. – 379 с.

Бардюкова Е.Н. Возраст Хвалынских трансгрессий Каспийского моря // Океанология. 2007. №3. С. 432–438.

Барышева К.П. Питание обыкновенной кильки в Среднем Каспии // Тр. Мосрыбвтуза. 1952. Вып.4. С. 108–130.

Берг Л.С. Рыбы пресных вод СССР и сопредельных стран. Ч.1. М.-Л.: Изд. АН СССР, 1948. – 468 с.

Бердичевский Л.С., Яблонская Е.А., Астахова Т.В., Беляева В.Н., Маилян Р.А. Биологическая продуктивность Каспия (современное состояние, мероприятия по ее повышению и задачи научных исследований) // Биологические ресурсы Каспийского моря. Астрахань, 1972. С. 4–23.

Бернстон М. Гистохимия ферментов. М.: Мир, 1965. – 464 с.
Бикбулатова Е.С., Бикбулатов Э.С. Органическое вещество в водохранилищах Средней и Нижней Волги / Гидрохимические исследования Волжских водохранилищ. Рыбинск, 1982. С. 101–113.
Биологические инвазии в водных и наземных экосистемах. Ред. Алимов А.Ф., Богуцкая Н.Г. М: КМК, 2004. – 436 с.
Богуцкая Н.Г., Насека А.М. Каталог бесчелюстных и рыб пресных и солоноватых вод России с номенклатурными и таксономическими комментариями. М.: КМК, 2004. – 389 с.
Булахов В.Л. Обогащение ихтиофауны Ленинского водохранилища путём акклиматизации полупроходных видов рыб. Автореф. дисс. … канд. биол. наук. Днепропетровск, 1966. – 24 с.
Буторин Н. В. Гидрологические процессы и динамика водных масс в водохранилищах волжского каскада. – Л.: Наука, 1969. – 319 с.
Былинкина А.А., Трифонова Н.А. Особенности гидрохимического режима Иваньковского водохранилища в связи с объёмом и распределением водного стока / Гидрохимические исследования Волжских водохранилищ. Рыбинск, 1982. С. 3–20.
Витковский А.З. Современное состояние ихтиофауны водохранилищ Манычского каскада. Автореф. дисс. … канд. биол. наук. Ставрополь, 2000. – 24 с.
Витковский А.З., Богачев А.Н. Распространение инвазионных видов рыб в Азово-Донском бассейне // Чужеродные виды в Голарктике (Борок-2). Тез. докл. Рыбинск-Борок, 2005. С. 139–140.
Владимиров В.И. О систематическом положении азовской и черноморской тюльки *Clupeonella delicatula* (Nordmann) // ДАН СССР. 1950. Т.70. №1. С. 125–128.
Владимиров В.И. Дивергенция тюльки [*Clupeonella delicatula delicatula* (Nordm.)] бассейна Днестра и её причины // Зоол. ж. 1951. Т.30. №6. С. 578–585.
Владимиров В.И., Сухойван П.Г., Бугай К.С. Размножение рыб в реке Днепре в условиях зарегулированного стока и охрана рыбных запасов // Изв. ГосНИОРХ. 1964. Вып.57. С. 15–24.
Волга и её жизнь. Ред. Буторин Н.В., Мордухай-Болтовской Ф.Д. Л.: Наука, 1979. – 348 с.
Воловик С.П. Продуктивность и проблемы управления экосистемой Азовского моря. Автореф. дисс. … докт. биол. наук. М., 1985. – 50 с.
Воронков П.П. Основные черты формирования химического состава вод искусственных водоёмов и расчёт их минерализации // Тр. лаб. Озероведения АН СССР. 1958. №7. С. 121–129.
Вылканов А., Петрова В., Рождественский А., Маринов Т., Найденов В. Черноморские лиманы и лагуны / Черное море. Сборник. Л.: Гидрометеоиздат, 1983. С. 282–306.
Гааль Э., Медьеши Г., Верецкеи Л. Электрофорез в разделении биологических макромолекул. М.: Мир, 1982. – 448 с.
Генетика изоферментов. М.: Наука, 1977. – 275 с.

Георгиев Ж., Александрова-Колеманова К. Рыбы / Черное море. Сборник. Л.: Гидрометеоиздат, 1983. С. 132–163

Гидробиологический режим Днестра и его водоёмов. Ред. Брагинский Л.П. Киев: Наук. думка, 1992. – 356 с.

Глазко В.И. Генетика изоферментов сельскохозяйственных животных. Итоги науки и техн. ВИНИТИ, 1988. Сер. Общ. генетика. – 212 с.

Голицын Г.С., Панин Г.Н. Ещё раз об увеличении уровня Каспийского моря // Вестн. АН СССР. 1989. №9. С. 59–63.

Голованова И.Л., Карабанов Д.П., Слынько Ю.В. Активность пищеварительных карбогидраз тюльки *Clupeonella cultriventris* из различных частей ареала // Вопр. рыболовства. 2007. Т.8. №1(29). С. 110–119.

Голубцов А.С. Внутрипопуляционная изменчивость животных и белковый полиморфизм. М.: Наука, 1988. – 168 с.

Голубцов А.С., Ильин И.И. Структура внутрипопуляционной и генетической изменчивости у ротана (Eleotridae, Pisces): различия между самцами и самками // Фенетика популяций. М.: Наука, 1985. С. 149–150.

Горбачев В.В. Новое ограничение микросателлитных маркеров для их применения в популяционных исследованиях (на примере панмиктических популяций) // Вавиловский журн. генетики и селекции. 2011. Т.15. №4. С. 746–749.

Горин А.Н. Закономерности распределения молоди тюльки (*Clupeonella cultriventris* Nordmann) Воткинского водохранилища. Автореф. дисс. … канд. биол. наук. М., 1991. – 21 с.

Грант В. Эволюция организмов. М.: Мир, 1980. – 408 с.

Двинских С.А., Китаев А.Б. Гидрология камских водохранилищ. Пермь: Изд.-во Пермск. университета, 2008. – 266 с.

Дгебуадзе Ю.Ю. Проблемы инвазий чужеродных организмов // Экологическая безопасность и инвазии чужеродных организмов. М.: Всеросс. конф. по экол. безопасности, 2002. С. 11–14.

Денисова А.И. Влияние каскадного расположения водохранилищ на их гидрохимический режим // Гидробиол. ж. 1971. Т.7. №5. С. 15–25.

Денисова А.И. Формирование гидрохимического режима водохранилищ Днепра и методы его прогнозирования. Киев: Наук. думка, 1979. – 292 с.

Дехтяр М. Н. Экологические сукцессии литоральной зоны водохранилищ Днепра // Гидробиол. ж. 1985. Т.21. №2. С. 24–29.

Динамика популяционных генофондов при антропогенных воздействиях. Ред. Алтухов Ю.П. М.: Наука, 2004. – 619 с.

Доманевский Л.Н., Дронов Л.Г. Ткачева Н.С. Пелагические рыбы Цимлянского водохранилища // Известия ГосНИИ озер. и реч. рыбн. хоз. 1964. Т. 57. С. 161–167.

Досон Р., Эллиот Д., Эллиот У., Джонс К. Справочник биохимика. М.: Мир, 1991. – 544 с.

Дубровин Л.И., Матарзин Ю.М., Печеркин И.А. Камское водохранилище. Пермь: Пермск. книжн. изд.-во, 1959. – 159 с.

Егерева С.И. Тюлька // Тр. Татарск. отд. ГосНИОРХ. Вып. 11. С. 106–108.

Ельченкова О.Н., Светлакова Э.И. Состояние рыбного промысла на водоемах Пермской области в 2000 г. // Рыбные ресурсы Камско-Уральского региона и их рациональное использование. Пермь: Пермск. гос. ун.-т, 2001. С. 36–39.

Животовский Л.А. Интеграция полигенных систем в популяциях. М.: Наука, 1984. – 183 с.

Завадский К.М. Структура вида / Современные проблемы эволюционной теории. Л.: Наука, 1967. С. 243–295.

Земля донская. Ростов-на-Дону: Ростовское нижн. Изд.-во, 1975. – 288 с.

Зенкевич Л.А. Моря СССР, их флора и фауна. М.: ГУПИ МП РСФСР, 1951. – 368 с.

Зимбалевская Л. Н. Гидробиологические исследования Днепра и его водохранилищ // Гидробиол. ж. 1990. Т.26. №3. С. 9–20.

Зиновьев Е.А. Ихтиофауна Пермкого края / Состояние и охрана окружающей среды Пермского края в 2007 году. Пермь: Упр.-е по охране окр. среды МПР Пермского края, 2007. С.57–62.

Зонн И.С. Каспий: иллюзии и реальность. М.: ТОО «Коркис», 1999. – 468 с.

Иванова М.Н. Популяционная изменчивость пресноводных корюшек. Рыбинск: ИБВВ АН СССР, 1982. – 148 с.

Ивантер Э.В., Коросов А.В. Введение в количественную биологию. Петрозаводск: ПетрГУ, 2003. – 304 с.

Ильин И.И., Голубцов А.С. Электрофоретический анализ изоферментов супероксиддисмутазы и октанолдегидрогеназы в скрещиваниях ротана *Perccottus glehni* Dyb. Сообщение 1. наследование электроморф и зависимость выживаемости от генотипов особей // Генетика. 1985. Т.21. №9. С. 1542–1547.

Казаков Б.А., Ломадзе Н.Х. Веселовское водохранилище / Водно-болотные угодья России. Т.6. Водно-болотные угодья Северного Кавказа. М.: Wetlands International, 2006. С. 40–50.

Казанчеев Е.Н. Рыбы Каспийского моря. М.: Рыбн. хоз., 1963. – 180 с.

Калинин В.В., Калинина О.В., Алтухов Ю.П. Связь приспособленности с уровнем индивидуальной гетерозиготности и полом у европейского анчоуса (*Engraulis encrasicholus* L.) // Генетика. 1988. Т.24. №3. С. 48–54.

Карабанов Д.П. Географическая изменчивость частот аллелей лактатдегидрогеназы (К.Ф. 1.1.1.27) черноморско-каспийской тюльки *Clupeonella cultriventris* (Nordm., 1840) // «Биология внутренних вод». Мат. XIII межд. шк.-конф. Рыбинск: Дом печати, 2006. С. 88–93.

Карабанов Д.П. Особенности устойчивости некоторых изоферментов черноморско-каспийской тюльки (*Clupeonella cultriventris* (Nordm., 1840)) к высоким концентрациям карбамида *in vitro* // Ихтиологические исследования на внутренних водоёмах. Саранск: МордвГУ, 2007. С. 68–70.

Карабанов Д.П. Современная структура и происхождение волжских популяций тюльки *Clupeonella cultriventris* (Nordmann, 1840) на основании эколого-генетических данных // Водные экосистемы: трофические уровни и проблемы поддержания биоразнообразия. Вологда: ВЛ ГосНИОРХ, ВГПУ, 2008*а*. С. 297–299.

Карабанов Д.П. Современная популяционная структура черноморско-каспийской тюльки *Clupeonella cultriventris* (Nordm., 1840) (CLUPEIDAE) // Совр. пробл. науки и образования. 2008*б*. №6(2). С. 28.

Карабанов Д.П. Популяционная структура черноморско-каспийской тюльки *Clupeonella cultriventris* (Nordm., 1840) черноморского бассейна на основании генетико-биохимических данных // Рибне господарство. 2009. В.67. С. 72–75.

Карабанов Д.П. Пространственная подразделённость в популяции черноморско-каспийской тюльки *Clupeonella cultriventris* (Nordm., 1840) Рыбинского водоранилища // «Актуальные проблемы экологии и эволюции в исследованиях молодых учёных». Мат. конф. мол. сотр. и асп. ИПЭЭ РАН. М: КМК, 2010. С. 141–144.

Карабанов Д.П. Влияние степени минерализации водоемов на процесс генетико-биохимической адаптации костистых рыб // Вода: химия и экология. 2011. №4. С. 50–53.

Карабанов Д.П. Функциональные различия в устойчивости к абиотическим факторам некоторых изоферментов костистых рыб // Вода: химия и экология. 2012*а*. №7. С. 50–58.

Карабанов Д.П. Происхождение и современная структура волжских популяций тюльки *Clupeonella cultriventris* (Nordmann, 1840) (Clupeiformes, Clupeidae) // «Актуальные проблемы экологии и эволюции в исследованиях молодых учёных». Тез. докл. конф. молодых сотр. и асп. ИПЭЭ РАН. М.: Тов.-во научн. изданий КМК, 2012*б*. С. 23.

Карабанов Д.П. Генетические различия особей разных полов в популяции тюльки *Clupeonella cultriventris* (Nordmann, 1840) Рыбинского водохранилища как результат микроэволюционных процессов // Современные проблемы эволюции. Ульяновск: УлГПУ, 2012*в*. С. 217–222.

Карабанов Д.П., Слынько Ю.В. Функциональные различия между изоформами и аллоформами эстераз черноморско-каспийской тюльки *Clupeonella cultriventris* Nordmann, 1840 Рыбинского водохранилища // Современные проблемы физиологии и биохимии водных организмов. Петрозаводск: ИБ КарелНЦ РАН, 2005*а*. С. 63–70.

Карабанов Д.П., Слынько Ю.В. Некоторые особенности использования метода PAGE при изучении популяций рыб Южного Региона России // Современные технологии мониторинга и освоения природных ресурсов южных морей России. Ростов-на-Дону: ЮжНЦ РАН, 2005*б*. С. 81–82.

Карабанов Д.П., Кияшко В.И., Кодухова Ю.В., Лаврова Е.И., Слынько Ю.В. Изменения в пелагическом комплексе рыб Рыбинского водохранилища после вселения черноморско-каспийской тюльки (*Clupeonella cultriventris* Nordmann, 1840) // Научные основы экологического мониторинга водохранилищ. Хабаровск: ИВЭП ДВО РАН. 2010. С. 77–79.

Карпевич А.Ф. Избранные труды. Т.1. Эколого-физиологические особенности гидробионтов. М.: ВНИРО, 1998. – 922 с.

Каспийское море: гидрология и гидрохимия. М.: Наука, 1986. – 261 с.

Касьянов А.Н. Изучение некоторых меристических признаков у черноморско-каспийской тюльки *Clupeonella cultriventris* (Clupeidae), вселившейся в Рыбинское водохранилище // Вопр. ихтиологии. Т.49. №5. С. 661–668.

Касымов А.Г. Каспийское море. Л.: Гидрометеоиздат, 1987. – 152 с.

Кимура М. Молекулярная эволюция: теория нейтральности. М.: Мир, 1985. – 398 с.

Кирпичников В.С. Биохимический полиморфизм и проблема так называемой недарвиновской эволюции // Успехи совр. биол. 1972. Т.74. №2. С. 231–246.

Кирпичников В.С. Генетика и селекция рыб. Л.: Наука, 1987. – 520 с.

Кирпичников В.С. Селективный характер биохимического полиморфизма у камчатский нерки *Oncorhynchus nerka* (Walb.) // Основы классификации и филогении лососевидных рыб. Л.: Наука, 1977. С. 53–60.

Кирпичников В.С., Иванова И.М. Изменчивость частот аллелей локусов лактатдегидрогеназы и фосфоглюкомутазы в локальных популяциях, различных возрастных группах и последовательных поколениях нерки (*Oncorhynhus nerka* Walb.) // Генетика. 1977. Т. 13. №7. С. 1183–1193.

Китаев А.Б., Рочев А.В. Гидрохимический режим приплотинной части Камского водохранилища // Географический вестник. 2008. №2(8). С. 1–18.

Кияшко В.И. Трофоэкологическая характеристика тюльки *Clupeonella cultriventris* в водохранилищах Средней и Верхней Волги // Вопр. ихтиол. 2004. Т.44. №6. С. 811–820.

Кияшко В.И., Слынько Ю.В. Структура пелагических скоплений рыб и современная трофологическая ситуация в открытых плёсах Рыбинского водохранилища после вселения черноморско-каспийской тюльки // Инвазии чужеродных видов в Голарктике. Борок:ИБВВ РАН-ИПЭЭ РАН, 2003. С. 259–271.

Кияшко В.И., Осипов В.В., Слынько Ю.В. Размерно-возрастные характеристики и структура популяции тюльки *Clupeonella cultriventris* при ее натурализации в Рыбинское водохранилище // Вопр. ихтиол. 2006. Т.46. №1. С. 68–76.

Кияшко В.И., Степанов М.В. Изменения в трофических цепях Рыбинского водохранилища, вызванные вселением черноморско-каспийской тюльки // Трофические связи в водных сообществах и экосистемах. Борок: ИБВВ РАН, 2003. С. 54–55.

Кияшко В.И., Халько Н.В., Лазарева В.И. О суточном ритме и элективности питания тюльки *Clupeonella cultriventris* (Nordmann, 1840) в Рыбинском водохранилище // Вопр. ихтиол. 2007. Т.44. №3. С. 389–398.

Кияшко В.И., Карабанов Д.П., Яковлев В.Н., Слынько Ю.В. Становление и развитие популяции черноморско-каспийской тюльки *Clupeonella cultriventris* (Nordmann, 1840) в Рыбинском водохранилище // Вопр. ихтиологии. 2012. Т. 52. № 5. С. 571–580.

Клевакин А.А. Динамика расселения чужеродных видов рыб в Чебоксарском водохранилище // Чужеродные виды в Голарктике (Борок-2). Тез. докл. Рыбинск-Борок, 2005. С. 152–154.

Козловский С.В. Особенности естественного воспроизводства тюльки в Куйбышевском водохранилище / Биологическая продуктивность и качество воды Волги и её водохранилищ. М.: Наука, 1984. С. 211–214.

Козловский С.В. Экология кильки *Clupeonella delicatula caspia* m. *tscharchalensis* (Borodin) и её роль в экосистеме Куйбышевского водохранилища. Автореф. дисс. … канд. биол. наук. Тольятти, 1987. – 23 с.

Комплексный анализ воздействия регулирования стока реки Волга на экосистемы поймы и дельты. Астрахань-Волгоград: ЮНЕСКО, 2004. – 36 с.

Коняев В.П., Костицын В.Г. К биологии хищных рыб Камского водохранилища // Рыбные ресурсы Камско-Уральского региона и их рациональное использование. Пермь: Пермск. гос. ун.-т, 2001. С. 67–70.

Криксунов Е.А. Теория пополнения и интерпретация динамики популяций рыб // Вопр. ихтиол. 1995. Т.35. №3. С. 301–329.

Крыжановский С.Г. Развитие тюльки *Clupeonella delicatula* (Nordmann) / Материалы по развитию сельдёвых рыб. Тр. ИМЖ. 1956. Вып.17. С. 245–253.

Кузнецов В.А. Особенности размножения и роста тюльки - нового компонента ихтиофауны Куйбышевского водохранилища // Научн. докл. высш. шк. Биол. науки. 1973. №6. С. 23–25.

Лазарева В.И. Сравнительный анализ состава и обилия летнего зоопланктона Рыбинского водохранилища в 1987–1988 гг. и 1997–2004 гг. // Биологические ресурсы пресных вод: беспозвоночные. Рыбинск: Рыбинский дом печати, 2005. С. 182–224.

Лазарева В.И. Структура и динамика зоопланктона Рыбинского водохранилища. М.: Тов-во научн. изд-в КМК, 2010. – 183 с.

Левонтин Р.К. Адаптация / Эволюция. Ред. Мина М.В. М.: Мир, 1981. С. 240–264.

Левонтин Р. Генетические основы эволюции. М.: Мир, 1978. – 351 с.

Ли Ч. Введение в популяционную генетику. М.: Мир, 1978. – 555 с.

Луц Г.И. Экология азовской тюльки и рациональное использование её запасов. Автореф. дисс. … канд. биол. наук. М., 1978. – 24 с.

Луц Г.И. Размерно-возрастная структура популяции тюльки *Clupeonella cultriventris* (Nordman) Азовского моря // Вопр. ихтиол. 1981. Т.21. №5. С. 946–949.

Луц Г.И., Рогов С.Ф. Динамика жирности и формирование запасов тюльки и хамсы в азовском море в зависимости от термического режима зимы // Гидробиол. журн. 1978. Т.14. №2. С. 31–35.

Майр Э. Популяции, виды и эволюция. М.: Мир, 1974. – 460 с.

Майский В.Н., Миндер Л.П., Дорменко В.В. Биология тюльки / Тюлька Азовского моря. Симферополь: Крымиздат, 1950. С. 5–19.

Мартемьянов В.И., Борисовская Е.А. Показатели водно-солевого обмена у вселившейся в Рыбинское водохранилище тюльки *Clupeonella cultriventris* (Clupeiformes, Clupeidae) в сравнении с аборигенными и морскими видами // Росс. Журн. биол. инвазий. 2010. №2. С. 37–46.

Мельников Г.Б. Гидробиологический режим Днепровского водохранилища после его восстановления // Вестн. НИИ гидробиологии Днепропетровского ун.-та. 1955. №11. С. 3–17.

Методическое пособие по изучению питания и пищевых отношений рыб в естественных условиях. М.: Наука, 1974. – 254 с.

Мещерякова О.В. Динамика активности изоферментов лактатдегидрогеназы, малатдегидрогеназы и α-глицерофосфатдегидрогеназы в процессе адаптаций рыб к различным факторам окружающей среды. Автореф. дисс. … канд. биол. наук. Петрозаводск, 2004. – 23 с.

Мина М.В., Клевезаль Г.А. Рост животных. М.: Наука, 1976. – 291 с.

Миноранский В.А. Животный мир Ростовской области (состав, значение, сохранение биоразнообразия). Ростов-на-Дону: Изд-во ООО ЦВВР, 2002. – 360 с.

Миноранский В.А. Дельта Дона / Водно-болотные угодья России. Т.6. Водно-болотные угодья Северного Кавказа. М.: Wetlands International, 2006. С. 28–40.

Миноранский В.А., Хохлов А.Н., Ильюх М.П. Озеро Маныч-Гудило / Водно-болотные угодья России. Т.6. Водно-болотные угодья Северного Кавказа. М.: Wetlands International, 2006. С. 50–63.

Михман Ф.С. Влияние естественных факторов на численность азовской тюльки // Тр. ВНИРО. 1970. Т.71. С. 167–179.

Михман А.С. Закономерности колебаний численности азовской тюльки // Тр. ВНИРО. 1972. Т.83. С. 235–247.

Мордухай-Болтовской Ф.Д. Основные трофические связи в волжских водохранилищах / Труды ин-та биологии водохранилищ АН СССР. 1963. Вып.5(8). С. 45–57.

Ней М., Кумар С. Молекулярная эволюция и филогенетика. Киев: Квiц, 2004. – 418 с.

Оно С. Генетические механизмы прогрессивной эволюции. М.: Мир, 1973. – 227 с.

Осипов В.В. Изменчивость роста и жизненных циклов тюльки *Clupeonella cultriventris* (Nordmann, 1840) в связи с её вселением в пресноводные экосистемы. Автореф. дисс. … канд. биол. наук. М., 2006. – 24 с.

Осипов В.В., Кияшко В.И. Особенности воспроизводства тюльки *Clupeonella cultriventris* (Clupeiformes, Clupeoidei) при вселении в пресноводные водоёмы // Вопр. ихтиол. 2006. Т.46. №4. С. 574–576.

Остерман Л.А. Методы исследования белков и нуклеиновых кислот. М.: Наука, 1981. – 286 с.

Павлов Д.С., Дгебуадзе Ю.Ю., Слынько Ю.В., Столбунова В.В., Карабанов Д.П., Папченков В.Г. Проникновение чужеродных генов в нативные популяции. Разработка системы мониторинга и оценка последствий // «Динамика генофондов растений, животных и человека». Матер. отч. конфер. М.: ИОГен РАН, 2005. С. 137-138.

Панин Г.Н., Мамедов Р.М., Митрофанов И.В. Современное состояние Каспийского моря. М.: Наука, 2005. – 356 с.

Пермитин И.Е., Половков В.В. Особенности образования и динамика структуры скоплений пелагических рыб // Теоретические аспекты рыбохозяйственных исследований водохранилищ. Л.: Наука, 1978. С. 78–106.

Пианка Э. Эволюционная экология. М.: Мир, 1981. – 400 с.

Поддубный А.Г. Экологическая топография популяций рыб в водохранилищах. Л.: Наука, 1971. – 306 с.

Половкова С.Н., Пермитин И.Е. Об использовании кормового зоопланктона нагульными скоплениями рыб-планктофагов / Внутрипопуляционная изменчивость питания и роста рыб. Ярославль: ЯГТУ, 1981. С. 3–35.

Полянинова А.А. Кормовая рыбопродуктивность // Научные основы устойчивого рыболовства и рационального распределения промысловых объектов Каспийского моря. М.: ВНИРО, 1998. С. 30–43.

Попов А.И. Бионвазийные виды зоопланктона в Саратовском и Куйбышевском водохранилищах // Чужеродные виды в Голарктике (Борок-2). Тез. докл. Рыбинск-Борок, 2005. С. 97–98.

Правдин И.Ф. Руководство по изучению рыб (преимущественно пресноводных). М.: Пищевая промышл., 1966. – 376 с.

Природные условия и естественные ресурсы Ростовской области. Ростов-на-Дону: Рост. гос. университет, 2002. – 432 с.

Приходько Б.И. Экологические черты каспийских килек (Род *Clupeonella*) // Вопр. ихтиологии. 1979. Т.19. №5. С. 801–812.

Пушкин Ю. А., Светлакова Э.И. Рыбное хозяйство Западно-Уральского региона (бас. р. Камы) // Биол. ресурсы камск. водохран. и их использ. Пермь: Перм. гос. ун-т., 1992. С. 5–27.

Пушкин Ю.А. Воткинское водохранилище // Изв. ГосНИИ озер. и реч. рыбн. хоз. Т.102. 1975. С.161–175.

Пушкин Ю.А., Антонова Е.Л. Тюлька *Clupeonella delicatula caspia* morpha *tscharchalensis* (Borodin) как новый компонент ихтиофауны камских водохранилищ // Тр. Пермск. Лаб. ГосНИОРХ. 1977. Т.1. С. 30–47.

Райдер К., Тэйлор К. Изоферменты. М.: Мир, 1983. – 112 с.

Ресурсы живой природы. Ч. 1. Водные животные. Ростов-на-Дону: Изд-во Ростовского ун-та, 1980. – 296 с.

Решетников Ю.С., Богуцкая Н.Г., Васильева Е.Д., Дорофеева Е.А., Насека А.М., Попова О.А., Савваитова К.А., Сиделева В.Г., Соколов Л.И. Список рыбообразных и рыб пресных вод России // Вопр. ихтиологии. 1997. Т.37. №6. С. 723–771.

Рождественский А., Цветков Л. Азовское море / Черное море. Сборник. Л.: Гидрометеоиздат, 1983. С. 306–319.

Рыбинское водохранилище и его жизнь. Л.: Наука, 1972. – 364 с.

Салменкова Е.А., Волохонская Л.Т. Биохимический полиморфизм в популяциях диплоидных и тетраплоидных видов рыб / Биохимическая генетика рыб. Л.: Наука, 1973. С. 54–61.

Световидов А.Н. О каспийских и черноморских сельдёвых из рода *Caspiolosa* и *Clupeonella* и об условиях их формирования // Зоол. ж. 1943. Т.22. №4. С. 222–233.

Световидов А.Н. О каспийской и черноморско-азовской тюльке [*Clupeonella delicatula* (Nordmann)] // ДАН СССР. 1945. Т.46. №5. С. 226–228.

Световидов А.Н. О причинах различия в росте каспийских и черноморско-азовских сельдевых // Зоол. ж. 1957. Т.36. №11. С. 1735–1745.

Световидов А.Н. Сельдёвые (Clupeidae). Фауна СССР. Рыбы. Т.2. Вып.1. М.-Л.: Изд. АН СССР, 1952. – 333 с.

Семеняк Л.В. Эколого-химические закономерности формирования биологической полноценности водной среды Каспийского моря. Автореф. дисс. … докт. биол. наук. М., 1996. – 48 с.

Симпсон Дж.Г. Темпы и формы эволюции. М.: Гос. издат. иностр. лит., 1948. – 358 с.

Слынько В.И. Множественные молекулярные формы малат- и лактатдегидрогеназы русского осетра (*Acipenser guldenstadti* Br.) и белуги (*Huso huso* L.) // ДАН СССР. 1976. Т.228. №2. С. 470–472.

Слынько Ю.В., Кияшко В.И., Яковлев В.Н. Список видов рыбообразных и рыб бассейна реки Волги / Каталог растений и животных водоёмов бассейна Волги. Ярославль: ЯГТУ, 2000. С. 252–277.

Слынько Ю.В., Лапушкина Е.Е. Генетические стратегии ареальной экспансии пелагических видов рыб в речной экосистеме // Инвазии чужеродных видов в Голарктике. Борок: ИБВВ РАН-ИПЭЭ РАН, 2003. С. 281–288.

Слынько Ю.В., Карабанов Д.П., Столбунова В.В. Генетический анализ внутривидовой структуры черноморско-каспийской тюльки *Clupeonella cultriventris* (Nordmann, 1840) (Actinopterygii: Clupeidae) // Доклады Академии наук. 2010. Т. 433. № 2. С. 283–285.

Смирнов А.И. Изменчивость тюльки - *Clupeonella delicatula* (Nordmann) в придунайских водоёмах // Зоол. ж. 1967. Т.46. №7. С. 1081–1087.

Смирнов А.К. Сезонная и возрастная динамика верхних летальных температур у карповых и окуневых видов рыб. Автореф. дисс. … канд. биол. наук. Борок: ИБВВ РАН, 2005. – 25 с.

Степанов М.В., Кияшко В.И. Роль тюльки (*Clupeonella cultriventris* Nordmann) в питании хищных рыб Рыбинского водохранилища // Биол. внутр. вод. 2008. №4. С. 86–89.

Степанов М.В. Морфо-биологическая характеристика черноморско-каспийской тюльки *Clupeonella cultriventris* (Nordmann, 1840) в Рыбинском водохранилище. Автореф. дисс. … канд. биол. наук. Борок, 2011. – 23 с.

Столбунова В.В., Слынько Ю.В Микрофилогенез черноморско-каспийской тюльки по результатам RAPD-анализа // Чужеродные виды в Голарктике (Борок-2). Тез. докл. Рыбинск-Борок, 2005. С. 173–174.

Сухойван П.Г., Вятчанина Л.И. Рыбное население и его продуктивность // Беспозвоночные и рыбы Днепра и аго водохранилищ. Киев: Наук. думка, 1989. С. 136–173.

Троицкий С.К., Цуникова Е.П. Рыбы бассейнов Нижнего Дона и Кубани. Руководство по определению видов. Ростов-на-Дону: Ростовск. книжн. изд., 1988. – 112 с.

Тучковенко Ю.С., Доценко С.А., Дятлов С.Е., Нестерова Д.А., Скрипник И.А., Кирсанова Е.В. Влияние гидрологических условий на изменчивость гидрохимических и гидробиологических характеристик вод Одесского региона северо-западной части Черного моря // Морьск. екологiч. ж. 2004. Т.3. №4. С. 75–86.

Устарбеков А.К., Аджимурадов К.А. Промыслово-биологические исследования обыкновенной кильки по Дагестанскому побережью Каспия // Биологические ресурсы Дагестанского прибрежья Каспийского моря. Махачкала: ДагФ АН СССР, 1982. С. 92–96.

Устарбекова Д.А. Морфологическая изменчивость анчоусовидной кильки *Clupeonella engrauliformes* (Borodin, 1904) в современных условиях Каспийского моря. Автореф. дисс. … канд. биол. наук. М., 2011. – 26 с.

Филиппович Ю.Б., Коничев А.С. Множественные формы ферментов насекомых и проблемы сельскохозяйственной энтомологии. М.: Наука, 1987. – 168 с.

Фортунатов М.А. О некоторых проблемах изучения Волги и водоёмов Волжского бассейна // Волга-1. Матер. 1-ой конфер. по изуч. водоёмов басс. Волги. Куйбышев, 1971. С. 11–18.

Халько В.В. К вопросу о физиолого-биохимическом состоянии тюльки *Clupeonella cultriventris* (Clupeidae, Clupeiformes) в Рыбинском водохранилище // Вопр. ихтологии. 2007. №3. С. 406–417.

Хедрик Ф. Генетика популяций. М.: Техносфера, 2003. – 592 с.

Хозяйкин А.А. Влияние слабого теплового воздействия на популяционно-динамические характеристики массовых видов планктонных ракообразных (на примере водоема-охладителя Пермской ГРЭС). Автореф. дисс. … канд. биол. наук. Борок, 2011. – 23 с.

Хочачка П., Сомеро Дж. Биохимическая адаптация. М.: Мир, 1988. – 568 с.

Циплаков Э.П. Тюлька / Тр. Татарск. отд. ГосНИОРХ. 1972. Вып. 12. С. 175–177.

Циплаков Э.П. Расширение ареалов некоторых видов рыб в связи с гидростроительством на Волге и акклиматизационными работами // Вопр. ихтиологии. 1974. Т.14. №3 (86). С. 396–405.

Чижов Н.И., Абаев Ю.И. Рыбы водоёмов Краснодарского края. Краснодар: Краснод. книжн. изд.-во, 1968. – 95 с.

Чугунова Н.И. О закономерности роста рыб и их значение в динамике популяций // Тр. совещ. Ихтиол. комм. АН СССР. 1961. Вып.13. С. 94–107.

Шаронов И.В. Расширение ареала некоторых рыб в связи с зарегулированием Волги // Волга-1. Тр. 1-ой конфер. по изуч. водоёмов басс. Волги. Куйбышев, 1971. С. 226–232.

Шаронов И.В. Проникновение северных и южных форм рыб в Куйбышевское водохранилище / Тр. Татарск. отд. ГосНИОРХ. 1972. Вып. 12. С. 178–179.

Шатуновский М.И. Экологические закономерности обмена веществ морских рыб. М.: Наука, 1980. – 283 с.

Шашуловский В.А., Ермолин В.А. Инвазийные виды в ихтиофауне Волгоградского водохранилища // Чужеродные виды в Голарктике (Борок-2). Тез. докл. Рыбинск-Борок, 2005. С. 184–185.

Шевченко П.Г. Эколого-морфологическая характеристика тюльки *Clupeonella cultriventris* (Nordmann) и её роль в экосистеме Днепровских водохранилищ. Автореф. дисс. … канд. биол. наук. Киев, 1991. – 18 с.

Шишкин М.А. Фенотипические реакции и эволюционный процесс (еще раз об эволюционной роли модификаций) / Экология и эволюционная теория. Л.: Наука, 1984. С. 196–216.

Шкорбатов Г.Л. Эколого-физиологические аспекты микроэволюции водных животных. Харьков: Издат. Харьковск. унив., 1973. – 200 с.

Шмидт-Ниельсон К. Физиология животных. Приспособление и среда. Т.1. М.: Мир, 1982. – 416 с.

Экологические проблемы Верхней Волги. Ред. Копылов А.И. Ярославль: ЯГТУ, 2001. – 427 с.

Экологический вестник Дона. Вып. «О состоянии окружающей среды и природных ресурсов Ростовской области в 2002 году». Ростов-на-Дону, 2003. – 291 с.

Элтон Ч. Экология нашествий животных и растений. М.: Иностр. литература, 1960. – 230 с.

Яблоков А.В., Юсуфов А.Г. Эволюционное учение. М.: Высш. школа, 2006. – 310 с.
Andersson L., Ryman N., Rosenberg R., Stahl G., Genetic variability in Atlantic herring (*Clupea harengus harengus*): discription of protein loci and population data // Hereditas. 1981. V.95. P. 69–78.
Asayama K., Burr I.M. Rat superoxid dismetases // J. Biol. Chem. 1985. V.260. P. 2112–2217.
Aspinwall N. Inheritance of alphglycerophosphate dehydrogenase in the pink salmon *Oncorhynchus gorbuscha* (Walb.) // Genetics. 1973. V.73. №4. P. 639–643.
Ayala F.J. Biological selection: natural selection or random walk? // Amer. Sci. 1974. V.62. P. 692–701.
Ayala F.J. Genetic differentiation during speciation process // Evol. Biol. 1975. V.8. P. 1–78.
Ayala F.J. Genetic variation in natural populations: Problem of electrophretically cryptic alleles // Poc. Nat. Acad. Sci. USA. 1982. V.79. P. 550–554.
Barroso J.B., Peragon J., Garcia-Salguero L., de la Higuera M., Lupianez J.A. Carbohydrate deprivation reduces NADPH-production in fish liver but not in adipose tissue // Int. J. Biochem. Cell Biol. 2001. V.33. P. 785–796.
Beacham T.D., Murray C.B., Withler L.E. Age, morphology and biochemical genetic variation of Yukon river chinook salmon // Trans. Amer. Fish. Soc. 1989. V.118. P. 46–63.
Beaumont M.A. Adaptacion and speciation: what can $F_{ST}$ tell us? // TRENDS Ecol. Evol. 2005. V.20. №8. P. 435–440.
Beaumont M.A., Nichols R.A. Evaluating loci for use in the genetic analysis of population structure // Proc. R. Soc. Lond. B. 1996. V.263. P. 1619–1626.
Berry R.J., Sage R.D., Lidicker W.Z., Jackson W.B. Genetical variation in three Pacific house mouse populations // Ibid. 1981. V.193. N3. P. 391–404.
Carvalho G.R. Evolutionary aspects of fish distribution: genetic variability and adaptation // J. Fish. Biol. 1993. V.43. P. 53–73.
Christiansen F.B., Frydenberg O., Simonsen V. Genetics of Zoarces populations. X. Selections component analysis of the Est III polymorphism using samples of successive cohorts // Ibid. 1977. V.87. N2. P. 129–150.
Ciardiello M.A., Camardella L., di Prisco G. Glucose-6-phosphate dehydrogenase from the blood cells of two Antarctic teleosts: correlation with cold adaptation // Biochem. Biophys. Acta. 1995. V.1250. P. 76–82.
Clayton J.W., Franzin W.G., Tretiak D.N. Genetics of glycerol-3-phosphate dehydrogenase isizymes in white muscle of lake whitefish (*Coregonus clupeaformis*) // J. Fish. Res. B. Can. 1973. V.30. №2. P. 187–193.
Colosimo A., Giuliani A., Maranghi F., Brix O., Thorkildsen S., Fischer T., Knust R., Poertner H.O. Physiological and genetical adaptation to temperature in fish populations // Cont. Shelf. Res. 2003. V.23. P. 1919–1928.
Copeland R.A. Enzymes: a practical introduction to structure, mechanism, and data analysis. Wiley-VCH, Inc., 2000. P. 1–390.
Copp G.H. Is fish condition correlated with water conductivity? // J. Fish. Biol. 2003. V.63. P. 263–266.

Daskalov G.M., Mamedov E.V. Integrated fisheries assessment and possible causes for the collapse of anchovy kilka in the Caspian Sea // ICES J. Mar. Sci. 2007. V.64. P. 503–511.
Davis B.J. Disc-electrophoresis. II. Method and application to human serum proteins // Ann. N.Y. Acad. Sci. 1964. V.124. P. 404–427.
Dgebuadze Yu.Yu., Kiyashko V.I., Osipov V.V. Life-history variation in invasive populations of Caspian Kilka, *Clupeonella cultriventris* (Clupeidae, Pisces) in the Volga River basin // NEOBIOTA. 2008. V.7. P. 153–159.
El Cafsi M., Romdhane M.S., Chaouch A., Masmoudi W., Kheriji S., Chanussot F., Cherif A. Qualitative needs of lipids by mullet, *Mugil cephalus*, fry during freshwater acclimation // Aquaculture. 2003. V.225. P. 233–241.
Fay J.G., Wyckhoff G.J., Chung-I Wu. Positive and negative selection on the human genome // Genetics. 2001. V.158. P. 1227–1234.
Ferris S.D., Whitt G.S. Duplicate gene expression in diploid and tetraploid loaches (Cypriniformes, Cobitidae) // Biochem. Genet. 1977. V.15. №11–12. P. 1097–1112.
Fisher S.E., Shaklee J.B., Ferris S.D., Whitt G.S. Evolution of five isozyme systems in the chordates // Genetica. 1980. V.52–53. P. 73–85.
Frelinger J.A. The maintenance of transferrin polymorphism in pigeons // Proc. Nat. Ac. Sci. USA. 1972. V.69. P. 326–329.
Fuller P.L., Nico L.G., Williams J.D. Nonindigenous fishes introduced into inland waters of the United States. USGS. Bethesda, Maryland, 1999. – 613 p.
Garton D.W., Koehn R.K., Scott T.M. Multiple-locus heterozygosity and the physiological energetics of growth in the coot clam, *Mulinia lateralis*, from a natural population // Genetics. 1984. V.108. P. 445–455.
Gilbert H.F. Basic concepts in biochemistry. McGraw-Hill Com., Inc., 2000. P. 1–331.
Gillespie J.H. Population Genetics. A Concise Guide. Baltimore: The Johns Hopkins Univ. Press, 1998. P. 1–169.
Gollasch S., Rosenthal H., Botnen H., Crncevic M., Gilbert M., Hamer J., Hulsmann N., Mauro C., McCann L., Minchin D., Ozturk B., Robertson M., Sutton C., Villac M.C. Species richness and invasion vectors: sampling techniques and biases // Biol. Invas. V.5. P. 365–377.
Graziani C., Moreno C., Villarroel E., Orta T., Lodeiros C., De Donato M. Hybridization between the freshwater prawns *Macrobrachium rosenbergii* (De Man) and *M. carcinus* (L.) // Aquaculture. 2003. V.217. P. 81–91.
Hamilton K.A., Ferguson A., Taggart J.B., Tomasson T., Walker A., Fahy E. Postglacial colonization of brown trout, *Salmo trutta* L.: LDH-5 as a phylogeographic marker locus // J. Fish Biol. 1989. V.35. P. 651–664.
Heger T., Trepl L. Predicting biological invasions // Biol. Invas. 2003. V.5. P. 313–321.
Holmes R.S., Scopes R.K. Immunochemical homologies among vertebrate lactate dehydrogenases // J. Biochem. 1974. V.43. P. 167–177.
Holsinger K.E., Weir B.S. Genetics in geographically structured populations: defining, estimating and interpreting $F_{ST}$ // Nat. Rev. Genet. 2009. V.10. 639–650.
Huber M., Knutti R. Antropogenic and natural warming inferred from changes in Earth's energy balance // Nature Geoscience. 2012. V.5. P. 31–36.

Hufbauer R.A., Torchin M.E. Integrating Ecological and Evolutionary Theory of Biological Invasions / Biological Invasions (Nentwig W., ed.). Springer-Verlag: Berlin, Heidelberg, 2007. P. 79–96.

Hung S.S.O., Liu W., Li H., Storebakken T., Cui Y. Effect of starvation on some morphological and biochemical parameters in white sturgeon, *Acipenser transmontanus* // Aquaculture. 1997. V.151. P. 357–363.

John R., Johnes R. The nature of multiple forms of cytoplasmic aspartate aminotransferase from pig ans sheep heart // J. Biochem. 1974. V.141. P. 401–406.

Johnson M.S. Adaptive lactate dehydrogenase variation in the crested blenny, *Anoplachus* // Heredity. 1971. V.27. P. 205–226.

Johnson T., Gerrish P.J. The fixation probability of a beneficial allele in a population dividing by binary fission // Genetica. 2002. V.115. P. 283–287.

Jorstad K.E. Evidence for two highly differentiated herring groups at Goose Bank in the Barents Sea and the genetic relationship to Pacific herring, *Clupea pallasi* // Envir. Biol. of Fishes. 2004. V.69. P. 211–221.

Jost L. $G_{ST}$ and its relatives do not measure differentiation // Mol. Ecol. 2008. V.17. P. 4015–4026.

Karabanov D.P., Slynko Yu.V. Genetic-biochemistry adaptations of *Clupeonella cultriventris* Nordmann, 1840 at expansion in Volga river basin // “Alien species in Holarctic (Borok-2)”. Book of Abstr. II Int. Symp. Russia, Borok, 2005. P. 195–196.

Karabanov D.P., Slynko Y.V. Change of alleles frequencies of lactate dehydrogenase of kilka (*Clupeonella cultriventris* Nordm., 1840) at expansion in water reservoirs of various types // “Invasion of alien species in Holartic. Borok-3”. Book of Abstr. III Int. Symp. Russia, Borok-Myshkin, 2010. P. 52.

Ken C.F., Lin C.T., Shaw J.F., Wu J.L. Characterization of fish Cu/Zn–superoxide dismutase and its protection from oxidative stress // Mar. Biotechnol. 2003. №5. P. 167–173.

Kiyashko V.I., Karabanov D.P., Slynko Yu.V. The number and species diversity dynamics of fish’s pelagic assemblages in Rybinsk reservoir after invasion of *Clupeonella cultriventris* Nordm. (Clupeiformes, Clupeidae) // “Invasion of alien species in Holartic. Borok-3”. Book of Abstr. III Int. Symp. Russia, Borok-Myshkin, 2010. P. 54–55.

Klyachko O.S., Ozernyuk N.D. Different functional and structural properties of lactate dehydrogenase isozymes at different stages of *Danio rerio* ontogenesis // Rus. J. Develop. Biol. 2001. V.32. №5. P. 310–312.

Koehn R.K. Esterase heterogeneity: dynamic of a polymorphism // Science. 1969. V.163. P. 943–944.

Koehn R.K., Peretz J.E., Merritt R.B. Esterase enzyme function and genetical structure of populations of the freshwater fish, *Notopis strameneus* // Amer. Natur. 1971. V.105. №941. P. 51–68.

Kohler A., Van Noorden C.J.F. Initial velocities *in situ* of G6PDH and PGDH and expression of proliferating cell nuclear antigen (PCNA): sensitive diagnostic markers of environmentally induced hepatocellular carcinogenesis in a marine flatfish *(Platichthys flesus* L.) // Aquat. Toxicol. 1998. V.40. P. 233–252.

Kondo H. Studies on the lipid of herring // Bull. Fac. Fish. Hokkaido Univ. V.25. №1. P. 68–77.
Konishi M., Hosoya K., Takata K. Natural hybridization between endangered and introduced species of *Pseudorasbora*, with their genetic relationships and characteristics inferred from allozyme analyses // J. Fish Biol. 2003. V.63. P. 213–231.
Konradt J., Braunbeck T. Alterations of selected metabolic enzymes in fish following long-term exposure to contaminated streams // J. Aquat. Ecosyst. Stress Recov. 2001. V.8. P. 299–318.
Kornfield I.L., Smith D.C., Gagnon P.S., Taylor J.N. The cichlid fish of Cuarto Ceinegas, Mexico: different evidence of conspecificity among distinct trophic morphs // Evolution. 1982. V.36. P. 658–664.
Kottelat M. European freshwater fishes. An heuristic checklist of the freshwater fishes of Europe (exclusive of former USSR), with an introduction for non-systematists and comments on nomenclature and conservation // Biologia. 1997. V.52 Suppl.5. P. 1–271.
Kottelat M., Freyhof J. Book of European freshwater fishes. Switzerland, Delemont: UICN, 2007. P. 1–646.
Kozlovsky S.V. *Clupeonella cultriventris caspia* (Svetovidov, 1941) / The Freshwater Fishes of Europe (H.Hoestlandt, ed.). V.2. Clupeidae, Angullidae. Verlag Wiesbaden. AULA, 1991. P. 55–86.
Kuzmin E.V., Kuzmina O.Yu. Activities of lactate and malate dehydrogenase of the sterlet and Russian sturgeon under conditions of massive spread of muscle pathology // Rus. J. Evol. Biochem. Physiol. 2001. V.37. №1. P. 35–42.
Largiader R. Hybridization and Introgression Between Native and Alien Species / Biological Invasions (Nentwig W., ed.). Springer-Verlag: Berlin, Heidelberg, 2007. P. 275–292.
Leppakoski E., Olenin S., Gollash S. The Baltic Sea - a fielkd laboratory for invasion biology / Invasive aquatic species of Europe. Distribution, Impacts and Management (Leppakoski E., Olenin S., Gollasch S., eds.). Kluwer Ac. Publ.: Dordecht-Boston-London, 2002. P. 253–259.
Leslie J.F., Vrijenhoek R.C. Genetic analysis of natural populations of *Poeciliopsis monacha*: allozyme inheritance and pattern of mating // Heredity. 1977. V.68. №5. P. 301–306.
Levesque H.M., Moon T.W., Campbell P.G.C., Hontela A. Seasonal variation in carbohydrate and lipid metabolism of yellow perch (*Perca flavescens*) chronically exposed to metals in the field // Aquat. Toxicol. 2002. V.60. P. 257–267.
Lim S.T., Kay R.M., Bailey G.S. Lactate dehydrogenase isozymes of salmonid fese: evidence for unique and rapid functional divergence of duplicated $H_4$ lactate dehydrogenases // J. Biol. Chem. 1975. V.250. №5. P. 1790–1800.
Lopes P.A., Pinheiro T., Santos M.C., Mathias M.L., Collares-Pereira M.J., Viegas-Crespo A.M. Response of antioxidant enzymes in freshwater fish populations (*Leuciscus alburnoides* complex) to inorganic pollutants exposure // Science Tech. Env. 2001. V.280. P. 153–163.
Makaveev T. Genetic polymorphism and enzyme activity of plasmic alkaline phosphotase in the water buffalo // Ibid. 1975. V.8. №4. P. 101–103.

Makhrov A.A., Skaala O., Altukhov Yu.P. Alleles of sAAT-1,2* isoloci in brown trout: potential diagnostic marker for tracking routes of post-glacial colonization in northern Europe // J. Fish Biol. 2002. V.61. P. 842–846.
Makhrov A.A., Verspoor E., Artamonova V.S., O'Sullivan M. Atlantic salmon colonization of the Russian Arctic coast: pioneers from North America // J. Fish Biol. 2005. V.67. P. 68–79.
Manchenko G.P. Handbook of detection of enzymes on electrophoretic gels. CRC Press, 2003. P. 1–553.
Marco D.E., Paez S.A., Cannas S.A. Species invasiveness in biological invasions: a modelling approach // Biol. Invas. 2002. V.4. P. 193–202.
Markert C.L., Shaklee J.B., Whitt G.S. Evolution of a gene // Science. 1975. V.189. P. 102–114.
Maronpot R.R., Pitot H.C., Peraino C. Use of rat liver altered focus models for testing chemicals that have completed two-year carcinogenicity studies // Toxicol. Pathol. 1989. V.17. №10. P. 651–662.
McKaye K.R., Kocher T., Reinthal P., Kornifield I. Genetic evidence for allopatric and sympatric differentiation among color morph of a Lake Malawi cichlid fish // Evolution. 1984. V.1. P. 215–219.
Merrit T.J.S., Quattro J.M. Evolution of the vertebrate cytosolic malate dehydrogenase gene family: duplication and divergence in Actinopterygian fish // J. Mol. Evol. 2003. V.56. P. 265–276.
Meton I., Fernandez F., Baanante I.V. Short- and long-term effects of refeeding on key enzyme activities in glycolysis–gluconeogenesis in the liver of gilthead seabream (Sparus aurata) // Aquaculture. 2003. V.225. P. 99–107.
Moo-Lee W., Friedman D.J., Ayala F.J. Superoxid dismetase: en evolutionare puzzle // Proc. Nat. Acad. Sci. USA. 1985. V.82. P. 824–828.
Mork J., Reyterwall C., Ryman N., Stahl G. Genetic variation in Atlantic cod (*Gadus morhua* L.): a quantitative estimate from a Norwegian coastal population // Hereditas. 1982. V.96. №1. P. 55–61.
Mork J., Sundnes G. O-group cod (*Gadus morhua*) in captivity: differential survival of certain genotypes // Ibid. 1985. V.39. N1. P. 63–70.
Muller G., Ward P. Genetic variability in the European minnow, *Phoxinus phoxinus* (L.) // Hydrobiologia. 1998. V.364. 183–188.
Muller J. Invasion history and genetic population structure of riverine macroinvertebrates // Zoology. 2001. V.104. №3. P. 1–10.
Nadeau J.H., Baccus R. Selection components of four allozymes in natural populations of *Peromyscus maniculatus* // Evolution. 1981. V.35. P. 11–20.
Nakano T., Sato M., Takeuchi M. Unique molecular properties of superoxide dismutase from teleost fish skin // FEBS Lett. 1995. V. 360. P. 197–201.
Nei M. Genetic distance between populations // Amer. Nat. 1972. V.106. P. 283–292.
Nei M. Estimation of average heterozygosity and genetic distance from a small number of individuals // Genetics. 1978. V.89. P. 583–590.
Nei M. Molecular evolutionary genetics. Columbia Univ. Press. N.Y., USA. 1987. P. 1– 512.
Nevo E., Beiles A., Ben-Sholomo R. The evolutionary significance of genetic diversity: ecological, demographic and life history correlates / Evolutionary dynamics of genetic diversity. 1984. V.53. Inst. of Evolution. Haifa, Israel. P. 13–213.

Nygren J. Allozyme variation in natural populations of fieldvole (*Microtus agrestis* L.). II. Survey of an isolated island population // Hereditas. 1980. V.1. P. 107–114.
Nyquist W.E. Statistical Genetics, with a focus on Animal and Plant Breeding // Agronomy. №615. Purdue University. 1990. P. 1–1206.
Ornstein L. Disc-electrophoresis. I. Background and theory // Ann. N.Y. Acad. Sci. 1964. V.121. P. 321–337.
Pandey S., Parvez S., Sayeed I., Haque R., Bin-Hafeez B., Raisuddin S. Biomarkers of oxidative stress: a comparative study of river Yamuna fish *Wallago attu* (Bl. & Schn.) // Science T. Env. 2003. V. 309. P. 105–115.
Parihar M.S., Dubey A.K., Javeri T., Prakash P. Changes in lipid peroxidation, superoxide dismutase activity, ascorbic acid and phospholipid content in liver of freshwater catfish *Heteropneustes fossilis* exposed to elevated temperature // J. Termal. Biol. 1996. V.21. №5/6. P. 323–330.
Peters L.D., Porte C., Livingstone D.R. Variation of antioxidant enzyme activities of sprat *(Sprattus sprattus)* larvae and organic contaminant levels in mixed zooplankton from the southern North Sea // Mar. Pollut. Bull. 2001. V.42. P. 1087–1095.
Petrovic S., OzreticB., Krajnovic-Ozretic M. Cytosolic aspartate aminotransferase from the grey mullet *(Mugil auratus* Risso) red muscle: isolation and properties // J. Biochem. Cell Biol. 1996. V.28. №8. P. 873–881.
Powell J.R., Taylor C.E. Genetic variation in ecologically diverse environments // Amer. Sci. 1979. V.67. P. 590–596.
Prakash S., Lewontin R.C., Hubby J.L. A molecular approach to the study of genetic heterozygosity in natural population. IV. Patterns of genetic variation in central marginal and isolated population of *Drosophila pseudoobscura* // Genetics. 1969. V. 61. №4. P. 841–858.
Prentis P.J., Wilson J.R.U., Dormontt E.E., Richardson D.M., Lowe A.J. Adaptive evolution in invesive species // TRENDS Plant Sci. 2008. Vol.13. №6. P. 288–294.
Redding J.M., Schreck C.B. Possible adaptive significance of certain enzyme polymorphisms in steel trout (*Salmo gairdneri*) // J. Fish. Res. Board. Canad. 1979. V.36. P. 544–551.
Reinitz G.L. Tests for association of transferrine and lactate dehydrogenase phenotypes with weight gain in rainbow trout (*Salmo gairdneri*) // J. Fish. Res. B. Canada. 1977. V.34. №12. P. 2333–2337.
Ridgway G.J., Sherburne S.W., Lewis R.D. Polymorphism in the esterase of Atlantic herring // Trans. Amer. Fish. Soc. 1970. V.99. №1. P. 147–151.
Rivers of Europe (Tockner K., Uehlinger U., Robinson C.T. eds.). Elsevier/Academic Press. 2009. P. 1–700.
Robert A., Couvet D., Sarrazin F. Bottlenecks in large populations: the effect of immigration on population viability // Evol. Ecol. 2003. V.17. P. 213–231.
Ruiz-Gutierrez V., Vazquez C.M., Consuelo S.M. Liver lipid composition and antioxidant enzyme activities of spontaneously hypertensive rats after ingestion of dietary fats (fish, olive and high-oleic sunflower oils) // Biosci. Rep. 2001. V.21. №3. P. 271–285.
Ryman N., Allendorf F.W., Stahl G. Reproductive isolation with little genetic divergence in sympatric population of brown trout (*Salmo trutta*) // Genetics. 1979. V.92. P. 247–262.

Ryman N., Stahl G. Genetic perspectives of the identifications and conservations of Scandinavian stocks of fishes // Canad. J. Fish. Aquat. Sci. 1981. V.38. P. 1562–1575.

Sastry K.V., Sachdeva S., Rathee P. Chronic toxic effects of cadmium and copper, and their combination on some enzymological and biochemical parameters in *Channa punctatus* // J. Environ. Biol. 1997. V.18. P. 291–303.

Schindler D.W., Parker B.R. Biological pollutants: alien fishes in mountain lakes / Water, Air, and Soil Pollution: Focus. N.2. Kluwer Ac. Publ.: Dordecht-Boston-London, 2002. P. 379–397.

Schmidt F. Biochemistry. Part I. IDG Books Worldwide, Inc. 2000. P. 1–176.

Semeonoff R., Robertson F.W. A biochemical and ecological study of plasma esterase polymorphism in natural populations of the fieldvole, *Microtus agrestis* L. // Ibid. 1968. V.1. N3. P. 205–227.

Shaklee J.B., Kepes K.L., Whitt G.S. Specialized lactate dehydrogenase isozymes: the molecular and genetic basis for the unique eye and liver LDH-s teleost fishes // J. Exp. Zool. 1973. V.185. №2. P. 217–240.

Shaklee J.B., Allendorf F.W., Morizot D.C., Witt G.S. Gene nomenclature for protein-coding loci in fish // Trans. Amer. Fish. Soc. 1990. Vol.119. №1. P. 2–15.

Shaklee J.B., Varnavskaya N.V. Electrophoretic characterization of odd-year pink salmon (*Oncorhynchus gorbuscha*) populations from the Pacific coast of Russia, and comparison with selected North American populations // Canad. J. fish. Aquat. Sci. 1994. V.51. P. 158–170.

Shaw C.R., Prasad R. Starch gel electrophoresis of enzymes - a compilation of recipes // Biochem. Genet. 1970. V.4. №2. P. 297–320.

Shiganova T.A. Impact of the invaders ctenophores *Mnemiopsis leidyi* and *Beroe ovata* on the foodweb and biodiversity of the Black sea // Alien species in Holarctic (Borok-2). Book of abstr. Borok, Russia. 2005. P. 127–128.

Shimeno S., Shikata T., Hosokawa S., MasumotoT., Kheyyali D. Metabolic response to feeding rates in common carp, *Cyprinus carpio* // Aquaculture. 1997. V.151. P. 371–377.

Shows T.B., Ruddle F.H. Malate dehydrogenase: evidence for tetrameric structure in *Mus musculus* // Science. 1968. V.160. P. 1356–1357.

Sick K. Hemoglobin polymorphism of cod in the Baltic and the Danish Belt sea // Hereditas. 1965. V.54. P. 19–48.

Sigh S.M., Zouros E. Genetic variation associated with growth rate in the American oyster (*Crassostrea virginica*) // Evolution. 1978. V.32. P. 342–353.

Simoes P., Santos J., Fragata I., Mueller L.D., Rose M.R., Matos M. How repeatable is adaptive evolution? The role of geographical origin and founder effects in laboratory adaptation // Evolution. 2008. V.62. P. 1817–1829.

Slatkin M. Isolation by distance in equilibrium and non-equilibrium populations // Evolution. 1993. V.47. P. 264–279.

Slynko U.V., Korneva L.G., Rivier I.K., Papchenkov V.G., Scherbina G.Kh., Orlova M.I., Terriaut T.V. The Caspian-Volga-Baltic Invasion corridor / Invasive aquatic species of Europe. Distribution, Impacts and Management. Kluver Ac. Publ. Dordecht-Boston-London. 2002. P. 399–411.

Smith J.B. Quantification of proteins on polyacrylamide gels / The Protein Protocols Handbook. (Walker J.M., ed.). Totowa, NJ: Humana Press Inc., 2002. P. 57–60.
Smith M.W., Smith M.H., Scott S.L., Liu E.H., Jones J.C. Rapid evolution in a post-thermal environment // Copeia. 1983. №1. P. 182–193.
Srivastava A.S., Oohara I., Suzuki T., Singh S.N. Activity and expression of aspartate aminotransferase during the reproductive cycle of a fresh water fish, *Clarias batrachus* // Fish Physiol. Biochem. 1999. V.20. P. 243–250.
Stanley J., Colby P. Effect of temperature on electrolyte balance and osmoregulation in the alervife (*Alosa pseudoharengus*) in fresh and sea water // Trans. Am. Fish. Soc. 1971. V.100. №4. P. 624–638.
Steiman A.M., Navik V.R., Abernathy J.L., Hill R.L. Bovine erythrocyte superoxid dismetase complete amino acid sequence // J. Biol. Chem. 1974. V.249. P. 7326.
Suchentrunk F., Mamuris Z., Sfougaris A.I., Stamatis C. Biochemical genetic variability in brown hares (*Lepus europaeus*) from Greece // Biochem. Genet. 2003. V.41. P. 127–140.
Svetovidov A.N. Clupeidae / Check-list of the Fishes of the North-Eastern Atlantic and of the Mediterranean (Hureau J.-C., Monod T., eds.). Vol.1. Paris: UNESCO, 1973. P. 99–109.
Swaminathan M.S. Bio-diversity: an effective safety net against environmental pollution // Envir. Pollut. 2003. V.126. P. 287–291.
Szumiec M.A., Bialowas H. Effect of genetics and temperature on carp juvenile survival in ponds of the temperate climate // Aquacult. International. 2003. V.11. P. 349–356.
Tolmasoff J.M., Ono T., Cutler R.G. Superoxid dismetase: correlation with life-span and specific metabolic rate in primate species // Proc. Nat. Acad. Sci. USA. 1980. V.77. P. 2777–2781.
Vesterberg O., Hansen L. New procedure for concentration and analytical isoelectric focusing of proteins // Biochem. Biophys. Acta. 1978. №2. P. 269–373.
Walker J. M. Nondenaturing polyacrylamide gel electrophoresis of proteins / The Protein Protocols Handbook. (Walker J.M., ed.). Totowa, NJ: Humana Press Inc., 2002. P. 57–60.
Watt W.B. Bioenergetics and evolutionary genetics: opportunities for new synthesis // Amer. Natur. 1985. V.125. P. 118–143.
Wheat T.E., Whitt G.S., Childers W.F. Linkage relationships of six enzyme loci in interspecific sunfish hybrids (genus *Lepomis*) // Genetics. 1973. V.74. №2. P. 343–350.
Williams G.S., Koehn R.K., Mitton J.B. Genetic diffenrentiation without isolation in the American eel, *Anguilla rostrata* // Evolution. 1973. V.27. P. 192–201.
Wilmot R.L., Everett R.J., Spearman W.J., Baccus R., Varnavskaya N.V., Putivkin S.V. Genetic stock structure of western Alaska chum salmon and comparison with Russian Far East stock // Can. J. Fish. Aquat. Sci. 1994. V.51. Suppl.1. P. 84–94.
Winzer K., Van Noorden C.J.F., Kohler A. Glucose-6-phosphate dehydrogenase: the key to sex-related xenobiotic toxicity in hepatocytes of European flounder (*Platichthys flesus* L.)? // Aquat. Toxicol. 2002. V.56. P. 275–288.
Withler R.E. LDH-4 allozyme variability in North American sockeye salmon (*Oncorhynchus nerka*) populations // Canad. J. Zool. 1985. V.63. P. 2924–2932.

Wright S. Isolation by distance // Genetics. 1943. V.24. P. 114–138.
Zar J.H. Biostatistical analysis, 4th ed. New Jersey: Prentice Hall. 1999. P. 1–672.
Zietara M.S., Skorkowski R.F. Thermostability of lactate dehydrogenase LDH-$A_4$ isoenzyme: effect of heat shock protein DnaK on the enzyme activity // Int. J. Biochem. Cell. Biol. 1995. V.27. P. 1169–1174.

**Карабанов Дмитрий Павлович**

# ГЕНЕТИЧЕСКИЕ АДАПТАЦИИ ЧЕРНОМОРСКО-КАСПИЙСКОЙ ТЮЛЬКИ *CLUPEONELLA CULTRIVENTRIS* (NORDMANN, 1840) (ACTINOPTERYGII: CLUPEIDAE)

Монография